\documentclass[aps,prb,reprint]{revtex4-2}

\usepackage[tbtags]{amsmath}
\usepackage{amssymb}
\usepackage[dvipdfmx]{graphicx}
\usepackage[FIGTOPCAP,nooneline]{subfigure}
\subcapraggedrighttrue
\usepackage{dcolumn}
\usepackage{bm}
\usepackage{braket}
\usepackage[version=3]{mhchem}
\usepackage{hyperref}
\usepackage{xcolor}
\hypersetup{
    colorlinks=true,
    citecolor=blue,
    linkcolor=blue,
    urlcolor=blue,
}
\usepackage{url}
\usepackage{mathrsfs}
\allowdisplaybreaks
\usepackage{multirow}   
\usepackage{float}       
\usepackage{cancel}
\usepackage{longtable}
\usepackage{mathtools}

\begin{document}

\preprint{APS/123-QED}

\title{Nonlinear Magnetoelectric Effect under Magnetic Octupole Order: Its Application to a $d$-Wave Altermagnet and a Pyrochlore Lattice with All-In/All-Out Magnetic Order}
\author{Jun {\=O}ik\'{e}} 
\email{oike.jun.32y@st.kyoto-u.ac.jp}
\author{Koki Shinada} 
\email{shinada.koki.64w@st.kyoto-u.ac.jp}
\author{Robert Peters}
\affiliation{Department of Physics, Kyoto University, Kyoto 606-8502, Japan}

\date{\today}% It is always \today, today,

\begin{abstract}

Extensive investigation has recently been conducted into a new class of antiferromagnetic order known as magnetic octupole order.
However, the high rank of octupoles makes it difficult to detect and manipulate them by using conventional methods such as the anomalous Hall effect.
In this paper, we propose the nonlinear magnetoelectric effect (NMEE), a second-order response to an electric field that induces a spontaneous magnetization, as a finite response under magnetic octupole order.
First, we classify the magnetic point groups to identify antiferromagnets with such order, and derive the NMEE tensor using quantum kinetic theory.
Then, we confirm the effectiveness of the NMEE through model calculations for two specific examples: a $d$-wave altermagnet and a pyrochlore lattice with all-in/all-out magnetic order.
In particular, the intrinsic NMEE exhibits a large response in a magnetic Weyl semimetal phase of the pyrochlore lattice.
This enhanced response is explained by the fact that the response tensor involves the quantum metric, which is enhanced near Weyl points.
Furthermore, our results show that the NMEE has a sizeable value that can be detected by the magneto-optical Kerr effect.

\end{abstract}


\maketitle

%%%%%%%%%%%%%%%%%%%%%%%%%%%%%%%%%%%%%%%%%%%%%%%%%%%%%%%%%%%%%%%%%%%%%%%%%%%%%%%%%%%%%%%%%%%%%%%%%%%%%%%%%%%%%%%%%%%%%%%%%%%%%%%%%%%%%%%%%%%%%%%%%%%%%%%%%%%%%%%%%%%%%%%%%%%%%%%%%%%%%%%%%%%%%%%%
%%%%%%%%%%%%%%%%%%%%%%%%%%%%%%%%%%%%%%%%%%%%%%%%%%%%%%%%%%%%%%%%%%%%%%%%%%%%%%%%%%%%%%%%%%%%%%%%%%%%%%%%%%%%%%%%%%%%%%%%%%%%%%%%%%%%%%%%%%%%%%%%%%%%%%%%%%%%%%%%%%%%%%%%%%%%%%%%%%%%%%%%%%%%%%%%
%%%%%%%%%%%%%%%%%%%%%%%%%%%%%%%%%%%%%%%%%%%%%%%%%%%%%%%%%%%%%%%%%%%%%%%%%%%%%%%%%%%%%%%%%%%%%%%%%%%%%%%%%%%%%%%%%%%%%%%%%%%%%%%%%%%%%%%%%%%%%%%%%%%%%%%%%%%%%%%%%%%%%%%%%%%%%%%%%%%%%%%%%%%%%%%%
%Sec I

\section{Introduction} \label{sec:Introduction}

Atomic-scale magnetic multipoles unify multiple degrees of freedom of electrons in solids and describe phenomena such as magnetic anisotropy and unconventional magnetic ordering.
Such magnetic multipoles have been mainly observed in $f$-electron systems, where the orbital is coupled to the spin through the relativistic spin-orbit coupling (SOC)~\cite{Kuramoto2009-ln,Santini2009-er,Mydosh2020-gm}.
In addition to these relativistic magnetic multipoles, a nonrelativistic magnetic multipole has recently been discovered in unconventional collinear antiferromagnets (AFMs)~\cite{Bhowal2024-vj, McClarty2024-eo}.
These AFMs exhibit a nonrelativistic spin-splitting in $\bm{k}$-space~\cite{Noda2016-wt,Naka2019-uk,Naka2020-pm,Hayami2019-zk,Ahn2019-cy,Yuan2020-al,Smejkal2020-zu,Yuan2021-xz,Mazin2021-ea,Egorov2021-uo}
and have been dubbed altermagnets to distinguish them from ferromagnets and conventional AFMs~\cite{Smejkal2022-xr,Smejkal2022-xx,Mazin2022-nf}.
In particular, $d$-wave altermagnets, such as \ce{RuO2}~\cite{Bose2022-bp,Bai2022-lx,Karube2022-mg,Feng2022-jm,Fedchenko2024-lw,Lin2024-qk}, where the spin-splitting fulfills $d$-wave symmetry, are known to exhibit a ferroic ordering of magnetic octupoles~\cite{Bhowal2024-vj, McClarty2024-eo}.

%%%%%%%%%%%%%%%%%%%%%%%%%%%%%%%%%%%%%%%%%%%%%%%%%%%%%%%%%%%%%%%%%%%%%%%%%%%%%%%%%%%%%%%%%%%%%%%%%%%%%%%%%%%%%%%%%%%%%%%%%%%%%%%%%%%%%%%%%%%%%%%%%%%%%%%%%%%%%%%%%%%%%%%%%%%%%%%%%%%%%%%%%%%%%%%%

Furthermore, by summing the atomic-scale magnetic multipoles within a cluster, one can describe a magnetic multipole that spans multiple atomic sites~\cite{Hayami2016-xs,Suzuki2017-ve,Suzuki2018-kz,Suzuki2019-ek,Huebsch2021-vx}.
Such cluster-scale magnetic multipoles can explain complex spin structures, such as noncollinear and noncoplanar configurations.
For example, the spin configuration of a chiral AFM, \ce{Mn3\textit{Z}} (\textit{Z}=Sn,Ge)~\cite{Nakatsuji2015-ll,Kiyohara2016-wl,Nayak2016-tl}, and the all-in/all-out (AIAO) magnetic configuration~\cite{Zhao2011-uc,Ishikawa2012-mf,Sagayama2013-xe,Tomiyasu2012-cy} of pyrochlore iridates, \ce{\textit{R}2Ir2Al20} (\textit{R}=rare-earth)~\cite{Matsuhira2007-rt,Matsuhira2011-ca,Witczak-Krempa2014-rt}, are interpreted as a cluster-scale magnetic octupole~\cite{Suzuki2017-ve,Arima2013-mp}.

%%%%%%%%%%%%%%%%%%%%%%%%%%%%%%%%%%%%%%%%%%%%%%%%%%%%%%%%%%%%%%%%%%%%%%%%%%%%%%%%%%%%%%%%%%%%%%%%%%%%%%%%%%%%%%%%%%%%%%%%%%%%%%%%%%%%%%%%%%%%%%%%%%%%%%%%%%%%%%%%%%%%%%%%%%%%%%%%%%%%%%%%%%%%%%%%

Here, we focus on the relationship between magnetic octupole order and response phenomena.
In this context, magnetic octupole order refers to a ferroic ordering of atomic-scale or cluster-scale magnetic octupoles.
Under magnetic octupole order, distinct responses appear depending on whether the octupole order is \textit{the lowest-rank magnetic octupole order} or not.
Note that here we define systems in which the lowest-rank nonvanishing magnetic multipole order is the octupole order as systems with lowest-rank magnetic octupole order.
For example, AFMs with magnetic dipole and octupole orders exhibit the anomalous Hall effect (AHE)~\cite{Smejkal2020-zu,Feng2022-jm,Nakatsuji2015-ll,Kiyohara2016-wl,Nayak2016-tl,Ueda2017-jr,Ueda2018-wq,Kim2020-er,Li2021-zv,Ghosh2023-hl} because the dipole order activates the AHE~\cite{Nagaosa2010-we}.
Given the difficulty of detecting and controlling the N\'{e}el vector of AFMs by external fields, these AFMs are strong candidates for antiferromagnetic spintronics~\cite{Jungwirth2016-td,Baltz2018-or}.
On the other hand, AFMs with lowest-rank magnetic octupole order do not exhibit low-rank responses such as the AHE without further symmetry reduction~\cite{Smejkal2020-zu,Feng2022-jm,Ueda2017-jr,Ueda2018-wq,Kim2020-er,Li2021-zv,Ghosh2023-hl}.
Furthermore, these AFMs do not exhibit magneto-optical effects that are the optical analogs of the AHE, making domain imaging and control challenging.
Therefore, finding a finite response under lowest-rank magnetic octupole order, i.e., \textit{a magnetic octupole response}, remains an important task.

%%%%%%%%%%%%%%%%%%%%%%%%%%%%%%%%%%%%%%%%%%%%%%%%%%%%%%%%%%%%%%%%%%%%%%%%%%%%%%%%%%%%%%%%%%%%%%%%%%%%%%%%%%%%%%%%%%%%%%%%%%%%%%%%%%%%%%%%%%%%%%%%%%%%%%%%%%%%%%%%%%%%%%%%%%%%%%%%%%%%%%%%%%%%%%%%

In this paper, we propose the nonlinear magnetoelectric effect (NMEE) as a magnetic octupole response.
The NMEE is a second-order response to an external electric field $\bm{E}$ that induces a spontaneous magnetization $\bm{M}$~\cite{Xiao2022-xr,Xiao2023-yu,Baek2023-aq,Feng2024-gj,Guo2024-vh,Oike2024-ms,Hu2024-gc}:
\begin{align}
    \label{eq:def_NMEE}
    M_i=\zeta^{(2)}_{i;jk}E_jE_k,
\end{align}
where $i,j,k$ label a Cartesian component.
The NMEE tensor $\zeta^{(2)}_{i;jk}$ is a rank-$3$, time-reversal ($\mathcal{T}$)-odd axial tensor with identical symmetry as magnetic octupoles~\cite{Urru2022-ww}, which suggests the potential effectiveness of the NMEE.

%%%%%%%%%%%%%%%%%%%%%%%%%%%%%%%%%%%%%%%%%%%%%%%%%%%%%%%%%%%%%%%%%%%%%%%%%%%%%%%%%%%%%%%%%%%%%%%%%%%%%%%%%%%%%%%%%%%%%%%%%%%%%%%%%%%%%%%%%%%%%%%%%%%%%%%%%%%%%%%%%%%%%%%%%%%%%%%%%%%%%%%%%%%%%%%%

\begin{table*}
    \caption{Classification of MPGs with magnetic octupole order based on whether they exhibit finite magnetic dipole or quadrupole orders.
             The symbol ``$\mathcal{P}\bigcirc/\mathcal{P}\times$" indicates the presence ($\mathcal{P}\bigcirc$) or absence ($\mathcal{P}\times$) of the inversion center, and ``$\checkmark/-$" denotes whether a certain type of multipole order is allowed ($\checkmark$) or forbidden ($-$).
             The classification is performed by using MTENSOR of the Bilbao Crystallographic Server~\cite{Elcoro2019-yl}.
             Note that type-I MPGs can be further classified by the presence or absence of magnetic quadrupole order, but this classification is not included in this paper.
             Furthermore, we list lowest-rank responses that characterize each category. 
             Each response is also applicable to the lower categories (e.g., the NMEE is applicable to type-I and type-II MPGs), but it does not characterize these lower categories.} 
    \vspace{1mm}
    \label{tab:Classification}
    \centering
    \tabcolsep = 0.3mm
    \renewcommand\arraystretch{1.3}
    \begin{tabular}{ccc|cccc}
      \hline \hline
       \multicolumn{3}{c|}{\multirow{2}{*}{MPGs}} & Magnetic & Magnetic & Magnetic & Lowest-rank  \\
      &&& dipole & quadrupole & octupole & responses \\
      \hline 
      \multicolumn{1}{c|}{\multirow{3}{*}{Type-I}}&($\mathcal{P}\bigcirc$) & $\bar{1}$, $2/m$, $2'/m'$, $m'm'm$, $4/m$, $4/mm'm'$, $\bar{3}$, $\bar{3}m'$, $6/m$, $6/mm'm'$ & \multirow{3}{*}{\checkmark} & \multirow{3}{*}{-/\checkmark} & \multirow{3}{*}{\checkmark} & \multirow{3}{*}{AHE} \\
      \multicolumn{1}{c|}{}&\multirow{2}{*}{($\mathcal{P}\times$)} & $1$, $2$, $2'$, $m$, $m'$, $2'2'2$, $m'm2'$, $m'm'2$, $4$, $\bar{4}$, $42'2'$, $4m'm'$, $\bar{4}2'm'$, & & & & \\
      \multicolumn{1}{c|}{}&& $3$, $32'$, $3m'$, $6$, $\bar{6}$, $62'2'$, $6m'm'$, $\bar{6}m'2'$ (31 groups)& & & &  \\ \hline
      \multicolumn{1}{c|}{\multirow{2}{*}{Type-II}}&\multirow{2}{*}{($\mathcal{P}\times$)} & $mm2$, $222$, $4'$, $\bar{4}'$, $422$, $4'22'$, $4mm$, $4'm'm$, $\bar{4}2m$, $\bar{4}'2'm$, $\bar{4}'2m'$, & \multirow{2}{*}{-} & \multirow{2}{*}{\checkmark} & \multirow{2}{*}{\checkmark} & LME \\ 
      \multicolumn{1}{c|}{}&& $32$, $3m$, $\bar{6}'$, $622$, $6mm$, $\bar{6}'m'2$, $\bar{6}'m2'$, $23$, $\bar{4}'3m'$ (20 groups) & & & & INHE   \\ \hline 
      \multicolumn{1}{c|}{\multirow{3}{*}{Type-III}}&\multirow{2}{*}{($\mathcal{P}\bigcirc$)} & $mmm$, $4'/m$, $4/mmm$, $4'/mm'm$,  & \multirow{3}{*}{-} & \multirow{3}{*}{-} & \multirow{3}{*}{\checkmark} & PME  \\
      \multicolumn{1}{c|}{}&& $\bar{3}m$, $6'/m'$, $6/mmm$, $6'/m'mm'$, $m\bar{3}$, $m\bar{3}m'$  & & & &  NMEE \\ 
      \multicolumn{1}{c|}{}& ($\mathcal{P}\times$) &  $6'$, $6'22'$, $6'mm'$, $\bar{6}m2$, $4'32'$ (15 groups) & & & & TNHE \\
      \hline \hline
    \end{tabular}
\end{table*}

%%%%%%%%%%%%%%%%%%%%%%%%%%%%%%%%%%%%%%%%%%%%%%%%%%%%%%%%%%%%%%%%%%%%%%%%%%%%%%%%%%%%%%%%%%%%%%%%%%%%%%%%%%%%%%%%%%%%%%%%%%%%%%%%%%%%%%%%%%%%%%%%%%%%%%%%%%%%%%%%%%%%%%%%%%%%%%%%%%%%%%%%%%%%%%%%

Here, we confirm the actual effectiveness of the NMEE as follows:
First, we classify the magnetic point groups (MPGs), examine which multipole orders are activated, and find potential AFMs with lowest-rank magnetic octupole order.
In particular, we focus on a $d$-wave altermagnet and a pyrochlore lattice with AIAO magnetic order.
Then, we derive the NMEE tensor using quantum kinetic theory and demonstrate through model calculations that the NMEE takes a finite value in these systems.
Notably, the intrinsic NMEE exhibits a large response in a magnetic Weyl semimetal phase of the pyrochlore lattice~\cite{Wan2011-sf,Witczak-Krempa2012-ro,Witczak-Krempa2013-bv,Yamaji2014-fh,Varnava2018-ub}.
This enhanced response is explained by the fact that the response tensor involves the quantum metric, which is enhanced near Weyl points. 
Finally, we discuss experimental realization and show that the NMEE has a sizeable value that can be detected by the magneto-optical Kerr effect.

%%%%%%%%%%%%%%%%%%%%%%%%%%%%%%%%%%%%%%%%%%%%%%%%%%%%%%%%%%%%%%%%%%%%%%%%%%%%%%%%%%%%%%%%%%%%%%%%%%%%%%%%%%%%%%%%%%%%%%%%%%%%%%%%%%%%%%%%%%%%%%%%%%%%%%%%%%%%%%%%%%%%%%%%%%%%%%%%%%%%%%%%%%%%%%%%

The rest of this paper is organized as follows:
Section~\ref{subsec:Classification} shows the classification result of the MPGs, and Sec.~\ref{subsec:PMAs} introduces the example systems for which we calculate the NMEE later. 
In Sec.~\ref{sec:Formulation}, we derive the NMEE tensor and explain its relation to quantum geometry.
Section~\ref{sec:Model_calculation_altermagnet} and a part of Sec.~\ref{subsec:Numerical_calculation} show numerical results of the NMEE for the $d$-wave altermagnet and the pyrochlore lattice with AIAO magnetic order, respectively.
In the rest of Sec.~\ref{subsec:Numerical_calculation} and Sec.~\ref{subsec:Analytical_calculation}, we discuss the origin of the enhanced response. 
Finally, we conclude this work and discuss the possible experimental realization of the NMEE in Sec.~\ref{sec:Conclusion}.

%%%%%%%%%%%%%%%%%%%%%%%%%%%%%%%%%%%%%%%%%%%%%%%%%%%%%%%%%%%%%%%%%%%%%%%%%%%%%%%%%%%%%%%%%%%%%%%%%%%%%%%%%%%%%%%%%%%%%%%%%%%%%%%%%%%%%%%%%%%%%%%%%%%%%%%%%%%%%%%%%%%%%%%%%%%%%%%%%%%%%%%%%%%%%%%%
%%%%%%%%%%%%%%%%%%%%%%%%%%%%%%%%%%%%%%%%%%%%%%%%%%%%%%%%%%%%%%%%%%%%%%%%%%%%%%%%%%%%%%%%%%%%%%%%%%%%%%%%%%%%%%%%%%%%%%%%%%%%%%%%%%%%%%%%%%%%%%%%%%%%%%%%%%%%%%%%%%%%%%%%%%%%%%%%%%%%%%%%%%%%%%%%
%%%%%%%%%%%%%%%%%%%%%%%%%%%%%%%%%%%%%%%%%%%%%%%%%%%%%%%%%%%%%%%%%%%%%%%%%%%%%%%%%%%%%%%%%%%%%%%%%%%%%%%%%%%%%%%%%%%%%%%%%%%%%%%%%%%%%%%%%%%%%%%%%%%%%%%%%%%%%%%%%%%%%%%%%%%%%%%%%%%%%%%%%%%%%%%%
%Sec II

\section{Classification of the magnetic point groups} \label{sec:Classification}

We first derive magnetic multipoles and then review their ferroic orderings, i.e., magnetic multipole orders.
Magnetic multipoles are derived from the spatial gradient expansion of the interaction energy ${E_{\mathrm{int}}}=-\int \bm{\mu}(\bm{r}) \cdot \bm{H}(\bm{r}) d\bm{r}$ between a magnetic field $\bm{H}(\bm{r})$ and a magnetization density $\bm{\mu}(\bm{r})$~\cite{Bhowal2024-vj,Spaldin2008-uv,Spaldin2013-vj}:   
\begin{align} 
    E_{\mathrm{int}}&=-\int \bm{\mu}(\bm{r})\cdot \bm{H}(\bm{0})d\bm{r}-\int r_i \mu_j(\bm{r}) \partial_i H_j(\bm{0}) d\bm{r} \notag \\
    & \quad -\frac{1}{2}\int r_i r_j \mu_k(\bm{r}) \partial_i \partial_j H_k(\bm{0}) d\bm{r}+\cdots,
\end{align}
where $\partial_i=\partial/\partial r_i$.
The first term represents a magnetic dipole, $\bm{m}=\int \bm{\mu}(\bm{r})d\bm{r}$, which acts as an order parameter for ferromagnets.
The second term denotes a magnetic quadrupole, $q_{ij}=\int r_i \mu_j(\bm{r})d\bm{r}$, which can serve as an order parameter for noncentrosymmetric magnets because of an odd number of position coordinates.
The third term describes a magnetic octupole, $\mathcal{O}_{ijk}=\int r_i r_j \mu_k(\bm{r})d\bm{r}$, which acts as the lowest-rank order parameter when both magnetic dipole and quadrupole orders are absent.
In the following, we classify the MPGs in terms of possible multipole orders and find AFMs with lowest-rank magnetic octupole order.

%%%%%%%%%%%%%%%%%%%%%%%%%%%%%%%%%%%%%%%%%%%%%%%%%%%%%%%%%%%%%%%%%%%%%%%%%%%%%%%%%%%%%%%%%%%%%%%%%%%%%%%%%%%%%%%%%%%%%%%%%%%%%%%%%%%%%%%%%%%%%%%%%%%%%%%%%%%%%%%%%%%%%%%%%%%%%%%%%%%%%%%%%%%%%%%%
%%%%%%%%%%%%%%%%%%%%%%%%%%%%%%%%%%%%%%%%%%%%%%%%%%%%%%%%%%%%%%%%%%%%%%%%%%%%%%%%%%%%%%%%%%%%%%%%%%%%%%%%%%%%%%%%%%%%%%%%%%%%%%%%%%%%%%%%%%%%%%%%%%%%%%%%%%%%%%%%%%%%%%%%%%%%%%%%%%%%%%%%%%%%%%%%

\subsection{Magnetic point groups with lowest-rank magnetic octupole order} \label{subsec:Classification}

Table~\ref{tab:Classification} summarizes our classification result. 
Magnetic point groups with magnetic octupole order are classified into three categories: type-I, type-II, and type-III.
Type-I and type-II MPGs support lowest-rank magnetic dipole and quadrupole orders, respectively; thus, their magnetic octupole order is not the lowest-rank.
On the other hand, type-III MPGs allow lowest-rank magnetic octupole order because of the absence of magnetic dipole and quadrupole orders.
Thus, type-III MPGs are the focus of this paper.
Note that Table~\ref{tab:Classification} is consistent with the comprehensive classification based on group theory~\cite{Yatsushiro2021-af}.

%%%%%%%%%%%%%%%%%%%%%%%%%%%%%%%%%%%%%%%%%%%%%%%%%%%%%%%%%%%%%%%%%%%%%%%%%%%%%%%%%%%%%%%%%%%%%%%%%%%%%%%%%%%%%%%%%%%%%%%%%%%%%%%%%%%%%%%%%%%%%%%%%%%%%%%%%%%%%%%%%%%%%%%%%%%%%%%%%%%%%%%%%%%%%%%%

Each category is characterized by its lowest-rank responses, which have response tensors with identical symmetry and rank as the lowest-rank multipole.
For example, type-I MPGs activate the AHE as one of their lowest-rank responses.
Indeed, AFMs belonging to this category exhibit the AHE~\cite{Smejkal2020-zu,Feng2022-jm,Nakatsuji2015-ll,Kiyohara2016-wl,Nayak2016-tl,Ueda2017-jr,Ueda2018-wq,Kim2020-er,Li2021-zv,Ghosh2023-hl}, which we confirm by reviewing previous AHE measurements of some centrosymmetric AFMs in Appendix~\ref{app:F}. 
Type-II MPGs include lowest-rank magnetic quadrupole order, which is an odd-parity multipole order.
Therefore, their lowest-rank responses are emergent phenomena such as the linear magnetoelectric effect (LME)~\cite{Hayami2018-ip,Watanabe2018-xc} and the intrinsic nonlinear Hall effect (INHE)~\cite{Kirikoshi2023-ii}. 
On the other hand, type-III MPGs require responses that have both $\mathcal{T}$-odd axial and at least rank-3 response tensors. 
In particular, their lowest-rank responses fall into two types: a linear response with rank-2 input and rank-1 output fields, and a nonlinear response with rank-1 input and rank-1 output fields.
A typical linear response is the piezomagnetic effect (PME)~\cite{Bhowal2024-vj,McClarty2024-eo,Aoyama2024-nz}, which induces a spontaneous magnetization by applying a mechanical strain.
On the other hand, typical nonlinear responses are the NMEE and the third-order nonlinear Hall effect (TNHE).
Indeed, Refs.~\cite{Fang2023-xl,Farajollahpour2024-yf} and~\cite{Sorn2023-be} theoretically demonstrated that the TNHE is effective for $d$-wave altermagnets and Pr-based heavy-fermion compounds, \ce{Pr\textit{T}2Al20} (\textit{T}=Ti,V), with ferro-octupole order, respectively.
Unlike the PME, the NMEE and TNHE may lead to a practical device application because they are electrically controllable without imposing large mechanical strain.
Of these, the NMEE may be the most suitable for magnetic octupole responses because it is a lower-rank response with respect to electric fields.

%%%%%%%%%%%%%%%%%%%%%%%%%%%%%%%%%%%%%%%%%%%%%%%%%%%%%%%%%%%%%%%%%%%%%%%%%%%%%%%%%%%%%%%%%%%%%%%%%%%%%%%%%%%%%%%%%%%%%%%%%%%%%%%%%%%%%%%%%%%%%%%%%%%%%%%%%%%%%%%%%%%%%%%%%%%%%%%%%%%%%%%%%%%%%%%%

Table~\ref{tab:Classification} allows us to find potential AFMs with lowest-rank magnetic octupole order.
In this paper, we focus on a $d$-wave altermagnet and a pyrochlore lattice with AIAO magnetic order.
Other interesting candidates are listed in Table~\ref{tab:candidates} at the end of this paper.

%%%%%%%%%%%%%%%%%%%%%%%%%%%%%%%%%%%%%%%%%%%%%%%%%%%%%%%%%%%%%%%%%%%%%%%%%%%%%%%%%%%%%%%%%%%%%%%%%%%%%%%%%%%%%%%%%%%%%%%%%%%%%%%%%%%%%%%%%%%%%%%%%%%%%%%%%%%%%%%%%%%%%%%%%%%%%%%%%%%%%%%%%%%%%%%%
%%%%%%%%%%%%%%%%%%%%%%%%%%%%%%%%%%%%%%%%%%%%%%%%%%%%%%%%%%%%%%%%%%%%%%%%%%%%%%%%%%%%%%%%%%%%%%%%%%%%%%%%%%%%%%%%%%%%%%%%%%%%%%%%%%%%%%%%%%%%%%%%%%%%%%%%%%%%%%%%%%%%%%%%%%%%%%%%%%%%%%%%%%%%%%%%

\subsection{Example systems: $d$-wave altermagnet and pyrochlore lattice with AIAO magnetic order} \label{subsec:PMAs}

We first focus on $d$-wave altermagnetism, which is characterized by $C_{4}\mathcal{T}$ symmetry with a $90^\circ$ rotation ($C_4$) and a spin flip ($\mathcal{T}$) [Fig.~\ref{fig:Altermagnet}].
In particular, the crystal rotation distinguishes the altermagnetism from conventional antiferromagnetism~\cite{Smejkal2022-xr,Smejkal2022-xx}, which is characterized by $t_{1/2}\mathcal{T}$ symmetry with a half-unit cell translation ($t_{1/2}$) or $\mathcal{P}\mathcal{T}$ symmetry with spatial inversion ($\mathcal{P}$)~\cite{Smejkal2022-el}.
This feature leads to novel phenomena such as efficient spin-current generation~\cite{Bose2022-bp}, spin-splitting torque~\cite{Bai2022-lx,Karube2022-mg}, and lifted Kramers degeneracy~\cite{Fedchenko2024-lw,Lin2024-qk}.
However, $C_{4}\mathcal{T}$ symmetry prohibits the AHE~\cite{Fang2023-xl}, and realizing a finite AHE requires further symmetry reduction~\cite{Feng2022-jm}, which is achieved by applying a magnetic field [see Appendix~\ref{app:F}].
In Table~\ref{tab:Classification}, $d$-wave altermagnets, \ce{RuO2} and \ce{MnF2}, indeed belong to a type-III MPG $4'/mm'm$ with $C_{4}\mathcal{T}$ symmetry~\cite{Smejkal2020-zu,Yuan2020-al}.

%%%%%%%%%%%%%%%%%%%%%%%%%%%%%%%%%%%%%%%%%%%%%%%%%%%%%%%%%%%%%%%%%%%%%%%%%%%%%%%%%%%%%%%%%%%%%%%%%%%%%%%%%%%%%%%%%%%%%%%%%%%%%%%%%%%%%%%%%%%%%%%%%%%%%%%%%%%%%%%%%%%%%%%%%%%%%%%%%%%%%%%%%%%%%%%%

The second system is a pyrochlore lattice [Fig.~\ref{fig:Pyrochlore}] with AIAO magnetic order [Fig.~\ref{fig:AIAO}].
Pyrochlore lattices host cubic crystalline symmetry~\cite{Matsuhira2007-rt,Matsuhira2011-ca,Witczak-Krempa2014-rt}, which forbids the AHE~\cite{Yang2014-iz}.
Furthermore, AIAO magnetic order also preserves cubic symmetry \cite{Yamaura2012-zd,Sagayama2013-xe}, and thus realizing a finite AHE requires further symmetry reduction~\cite{Ueda2017-jr,Ueda2018-wq,Kim2020-er,Li2021-zv,Ghosh2023-hl}, which is achieved by applying a magnetic field or strain [see Appendix~\ref{app:F}].
In Table~\ref{tab:Classification}, pyrochlore iridates, \ce{\textit{R}2Ir2Al20}, forming AIAO magnetic order below the N\'{e}el temperature indeed belongs to a type-III MPG $m\bar{3}m'$. 
Meanwhile, pyrochlore lattices form a Luttinger semimetal state with fourfold degenerate quadratic band crossings at the $\varGamma$ point due to $\mathcal{P}$ and $\mathcal{T}$ symmetries.
This state can change into a topologically nontrivial quantum phase through a symmetry-breaking perturbation~\cite{Moon2013-mr,Kondo2015-dr}.
For example, some theories predict the emergence of a magnetic Weyl semimetal phase driven by AIAO magnetic order~\cite{Wan2011-sf,Witczak-Krempa2012-ro,Witczak-Krempa2013-bv,Yamaji2014-fh,Varnava2018-ub}. 
Therefore, pyrochlore lattices with AIAO magnetic order are good candidates for examining the relationship between the NMEE and quantum geometry.

%%%%%%%%%%%%%%%%%%%%%%%%%%%%%%%%%%%%%%%%%%%%%%%%%%%%%%%%%%%%%%%%%%%%%%%%%%%%%%%%%%%%%%%%%%%%%%%%%%%%%%%%%%%%%%%%%%%%%%%%%%%%%%%%%%%%%%%%%%%%%%%%%%%%%%%%%%%%%%%%%%%%%%%%%%%%%%%%%%%%%%%%%%%%%%%%

\begin{figure}[t]
    \setcounter{figure}{1} 
    \begin{minipage}[t]{{0.38\columnwidth}}
    \centering
    \subfigure[]{
        \includegraphics[height=2.5cm,width=1.0\hsize]{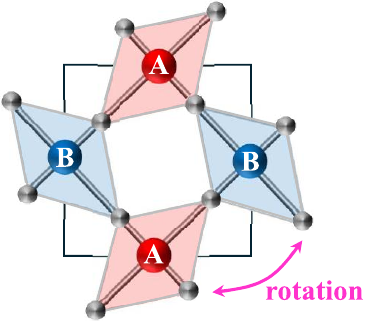}
        \label{fig:Altermagnet}
    }
    \vfill
    \vspace{-0.3cm}
    \setcounter{subfigure}{2} 
    \subfigure[]{
        \includegraphics[height=2.0cm,width=0.9\hsize]{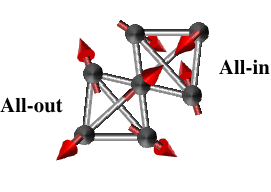}
        \label{fig:AIAO}
    } 
    \setcounter{subfigure}{1} 
    \end{minipage} 
    \begin{minipage}[t]{{0.5\columnwidth}}
        \subfigure[]{
        \includegraphics[height=3.5cm,width=0.8\hsize]{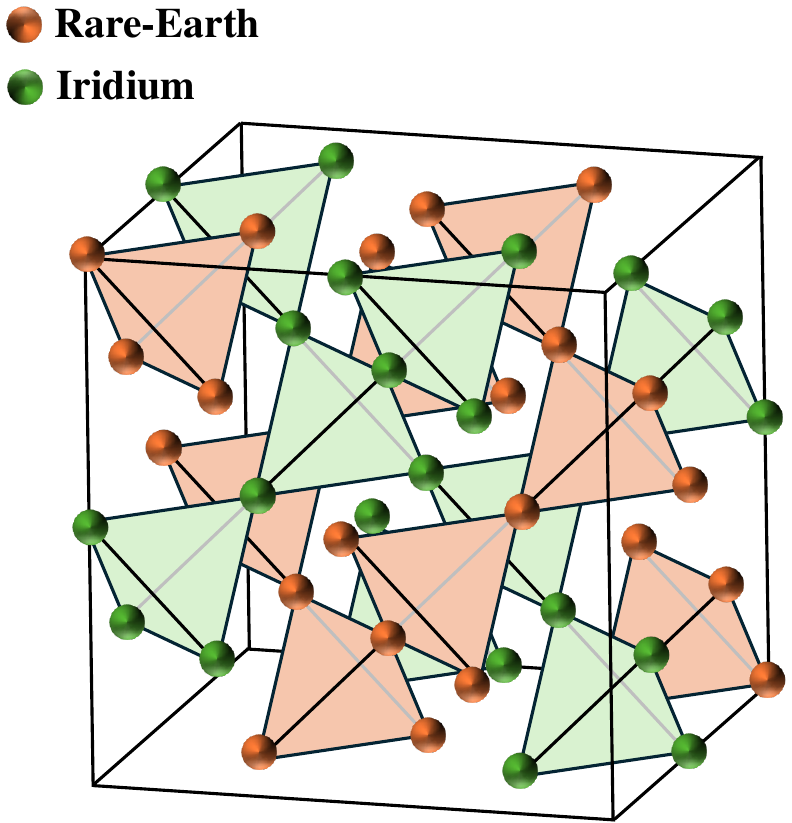}
        \label{fig:Pyrochlore}
    }
    \end{minipage} 
    \setcounter{figure}{0}
    \caption{(a) Illustration of a $d$-wave altermagnet.
             The spins are oriented perpendicular to the plane and point in opposite directions on sublattices A and B.
             The sublattice degrees of freedom arise from the arrangement of the nonmagnetic atoms and break $t_{1/2}\mathcal{T}$ or $\mathcal{P}\mathcal{T}$ symmetries.
             (b) Lattice structure of pyrochlore iridates.
             (c) AIAO magnetic configuration.}
    \label{fig:Altermagnet_Pyrochlore with AIAO}
\end{figure}

%%%%%%%%%%%%%%%%%%%%%%%%%%%%%%%%%%%%%%%%%%%%%%%%%%%%%%%%%%%%%%%%%%%%%%%%%%%%%%%%%%%%%%%%%%%%%%%%%%%%%%%%%%%%%%%%%%%%%%%%%%%%%%%%%%%%%%%%%%%%%%%%%%%%%%%%%%%%%%%%%%%%%%%%%%%%%%%%%%%%%%%%%%%%%%%%
%%%%%%%%%%%%%%%%%%%%%%%%%%%%%%%%%%%%%%%%%%%%%%%%%%%%%%%%%%%%%%%%%%%%%%%%%%%%%%%%%%%%%%%%%%%%%%%%%%%%%%%%%%%%%%%%%%%%%%%%%%%%%%%%%%%%%%%%%%%%%%%%%%%%%%%%%%%%%%%%%%%%%%%%%%%%%%%%%%%%%%%%%%%%%%%%
%%%%%%%%%%%%%%%%%%%%%%%%%%%%%%%%%%%%%%%%%%%%%%%%%%%%%%%%%%%%%%%%%%%%%%%%%%%%%%%%%%%%%%%%%%%%%%%%%%%%%%%%%%%%%%%%%%%%%%%%%%%%%%%%%%%%%%%%%%%%%%%%%%%%%%%%%%%%%%%%%%%%%%%%%%%%%%%%%%%%%%%%%%%%%%%%
%Sec III

\section{Formulation of the NMEE tensor} \label{sec:Formulation}

We first outline the derivation of the NMEE tensor.
Note that here we focus only on spin magnetization and neglect orbital magnetization because spin magnetization usually dominates the total magnetization~\cite{Meyer1961-tf,reck1969orbital}.
The nonequilibrium magnetization induced by an electric field $\bm{E}$ is given by 
\begin{align}
    \label{eq:nonequilibrium_magnetization}
    \bm{M}= \sum_{n,m} \int_{\bm{k}} \bm{s}_{nm}(\bm{k})\rho_{mn}(\bm{k}),
\end{align}
where $\bm{s}_{nm}(\bm{k})$ and $\rho_{nm}(\bm{k})$ are the matrix representations of the spin and density operators in a band basis $\ket{u_n(\bm{k})}$, respectively. 
The eigenstates satisfy
\begin{align}
    H_0(\bm{k})\ket{u_n(\bm{k})}=\varepsilon_n(\bm{k})\ket{u_n(\bm{k})},
\end{align}
where $H_0(\bm{k})$ is an unperturbed Hamiltonian, and $\varepsilon_n(\bm{k})$ is the eigenvalue labeled by crystal momentum $\bm{k}$ and a band index $n$ in the first Brillouin zone (BZ).
For simplicity, we denote $\int_{\mathrm{BZ}} d\bm{k}/(2\pi)^{d}$ as $\int_{\bm{k}}$, where $d$ is the dimension of the system, and we will omit the $\bm{k}$-index of the operators in the following.
From Eq.~\eqref{eq:nonequilibrium_magnetization}, the second-order nonequilibrium magnetization can be calculated by
\begin{align}
    \label{eq:second_order_magnetization}
    M_i^{(2)}= \sum_{n,m} \int_{\bm{k}} s^i_{nm} \rho^{(2)}_{mn},
\end{align}
where $\rho_{nm}$ is expanded in powers of $\bm{E}$: $\rho=\sum_\ell \rho^{(\ell)}$ with $\rho^{(\ell)}=\mathcal{O}(|\bm{E}|^\ell)$.
Therefore, determining the second-order density matrix $\rho^{(2)}_{nm}$ enables us to derive the NMEE tensor.

%%%%%%%%%%%%%%%%%%%%%%%%%%%%%%%%%%%%%%%%%%%%%%%%%%%%%%%%%%%%%%%%%%%%%%%%%%%%%%%%%%%%%%%%%%%%%%%%%%%%%%%%%%%%%%%%%%%%%%%%%%%%%%%%%%%%%%%%%%%%%%%%%%%%%%%%%%%%%%%%%%%%%%%%%%%%%%%%%%%%%%%%%%%%%%%%

The $\ell$-th order density matrix $\rho^{(\ell)}_{nm}$ is obtained by solving the von Neumann equation, 
\begin{align}
    \label{eq:von_Neumann_t}
    (i\hbar\partial_t-\varepsilon_{nm})\rho^{(\ell)}_{nm}(t)=[H_{\mathrm{E}}(t),\rho^{(\ell-1)}(t)]_{nm},
\end{align}
where $\hbar$ is the Planck constant, $\partial_t=\partial/\partial t$, $\varepsilon_{nm}=\varepsilon_n-\varepsilon_m$, and $[A,B]_{nm}=\sum_l (A_{nl}B_{lm}-B_{nl}A_{lm})$.
The perturbed Hamiltonian $H_{\mathrm{E}}(t)$ is given by
\begin{align}
    H_{\mathrm{E}}(t)=e\bm{r}\cdot \bm{E}(t),
\end{align}
where $e=|e|$ is the charge of electrons, and $\bm{r}$ is the position operator.
The position operator breaks translation symmetry, making it difficult to deal with the Hamiltonian in band theory. 
In the infinite volume limit, however, the position operator is written as a derivative by the crystal momentum $\partial_{\bm{k}}=\partial/\partial \bm{k}$ and the $k$-space Berry connection $\bm{\mathcal{A}}_{nm}=i\braket{u_n|\partial_{\bm{k}} u_m}$~\cite{Adams1959-tp,Blount1962-ku}:
\begin{align}
    \bm{r}_{nm}=i\partial_{\bm{k}}\delta_{nm}+\bm{\mathcal{A}}_{nm}.
\end{align}
Before solving Eq.~\eqref{eq:von_Neumann_t}, we introduce a phenomenological treatment of the scattering rate $\eta$~\cite{Cheng2015-qr,Passos2018-tl,Watanabe2022-qp,Das2023-nm}:
\begin{align}
    \label{eq:von_Neumann_t_relaxation}
    (i\hbar\partial_t-\varepsilon_{nm})\rho^{(\ell)}_{nm}(t)&=e\bm{E}(t)\cdot[\bm{r},\rho^{(\ell-1)}(t)]_{nm} \notag \\
    &\quad -i\ell\eta\rho^{(\ell)}_{nm}(t).
\end{align}
Finally, by performing the Fourier transformation to the frequency domain $\omega$ and taking the limit $\omega\rightarrow0$, the $\ell$-th order density matrix is obtained as
\begin{align}  
    \label{eq:rho_l}
    \rho^{(\ell)}_{nm}=e \frac{[\bm{r},\rho^{(\ell-1)}]_{nm}}{\varepsilon_{mn}+i\ell\eta} \cdot \bm{E}.
\end{align}
Note that the zeroth-order density matrix $\rho^{(0)}_{nm}$ is assumed to be $\rho^{(0)}_{nm}=\delta_{nm}f_n$, where $f_n=[1+e^{(\varepsilon_n-\mu)/k_{\mathrm{B}}T}]^{-1}$ is the Fermi distribution function, and $\mu$, $k_{\mathrm{B}}$, and $T$ are the chemical potential, Boltzmann constant, and temperature, respectively.  

%%%%%%%%%%%%%%%%%%%%%%%%%%%%%%%%%%%%%%%%%%%%%%%%%%%%%%%%%%%%%%%%%%%%%%%%%%%%%%%%%%%%%%%%%%%%%%%%%%%%%%%%%%%%%%%%%%%%%%%%%%%%%%%%%%%%%%%%%%%%%%%%%%%%%%%%%%%%%%%%%%%%%%%%%%%%%%%%%%%%%%%%%%%%%%%%

The NMEE tensor can be split into different parts using the action of $\mathcal{T}$: $\zeta^{\mathrm{odd}}_{i;jk}$ and $\zeta^{\mathrm{even}}_{i;jk}$.
The $\mathcal{T}$-odd response ($\zeta^{\mathrm{odd}}_{i;jk}$) is finite only in magnets, and the $\mathcal{T}$-even response ($\zeta^{\mathrm{even}}_{i;jk}$) is finite in both magnetic and nonmagnetic systems.
This difference appears when the NMEE tensor is separated according to the order of the relaxation time $\tau=\hbar/\eta$: 
\begin{align}
    \label{eq:symmetry_tau}
    \zeta^{(2)}_{i;jk}=\zeta^{\tau^2}_{i;jk}+\zeta_{i;jk}^{\tau^1}+\zeta_{i;jk}^{\tau^0},
\end{align}
where $\zeta^{\tau^\ell}_{i;jk}=\mathcal{O}(\tau^{\ell})$.
Specifically, $\mathcal{T}$-odd and $\mathcal{T}$-even responses are proportional to even and odd powers of $\tau$, respectively~\cite{Yatsushiro2021-af}; thus,
\begin{align}
 \zeta^{\mathrm{odd}}_{i;jk}=\zeta^{\tau^2}_{i;jk}+\zeta^{\tau^0}_{i;jk}, \hspace{0.5cm} \zeta^{\mathrm{even}}_{i;jk}=\zeta^{\tau^1}_{i;jk}.
\end{align}
In this study, we assume an applied electric field in the $x$-$y$ plane and thus focus only on the $\mathcal{T}$-odd responses.
This is because the $C_{4z}\mathcal{T}$ symmetry of $d$-wave altermagnets and the symmetry of pyrochlore lattices with AIAO magnetic order prohibit any components in the $\mathcal{T}$-even response~\cite{Xiao2023-yu}.

%%%%%%%%%%%%%%%%%%%%%%%%%%%%%%%%%%%%%%%%%%%%%%%%%%%%%%%%%%%%%%%%%%%%%%%%%%%%%%%%%%%%%%%%%%%%%%%%%%%%%%%%%%%%%%%%%%%%%%%%%%%%%%%%%%%%%%%%%%%%%%%%%%%%%%%%%%%%%%%%%%%%%%%%%%%%%%%%%%%%%%%%%%%%%%%%

From Eqs.~\eqref{eq:second_order_magnetization} and~\eqref{eq:rho_l}, $\zeta^{\tau^2}_{i;jk}$ and $\zeta^{\tau^0}_{i;jk}$ are expressed as 
\begin{align}
    \zeta^{\tau^2}_{i;jk}&=\frac{ e^2}{2\hbar^2} \tau^2 \sum_n \int_{\bm{k}}s^i_{nn}\partial_{k_j}\partial_{k_k}f_n, \label{eq:tau2} \\
    \zeta^{\tau^0}_{i;jk}&=-\frac{e^2}{2}\sum_n \int_{\bm{k}}\Bigl[ \partial_{h_i} G^{jk}_n-2\Bigl(\partial_{k_j} \mathfrak{G}^{ik}_n+\partial_{k_k}\mathfrak{G}^{ij}_n \Bigr)\Bigr]f_n, \label{eq:tau0}
\end{align}
where the details of the derivation are given in Appendix~\ref{app:A}.
The $\tau^2$-term is analogous to the higher-order Drude conductivity~\cite{Das2023-nm,Watanabe2020-bs,Kaplan2023-jt}, and the $\tau^0$-term involves two geometric quantities: $G^{ij}_n$ and $\mathfrak{G}^{ij}_n$.
The quantity $G^{ij}_n$ is the $k$-space Berry connection polarizability (BCP)~\cite{Gao2014-kx} and is related to the $k$-space quantum metric~\cite{Provost1980-qb,Resta2011-da},
\begin{align}
    g^{ij}_n&=\mathrm{Re}\braket{\partial_{k_i}u_n|(1-\ket{u_n}\bra{u_n})|\partial_{k_j}u_n} \notag \\
    &=\sum_{m(\neq n)}\mathrm{Re}\braket{\partial_{k_i}u_n|u_m}\braket{u_m|\partial_{k_j}u_n} \eqqcolon \sum_{m(\neq n)}g^{ij}_{nm}.
\end{align}
Specifically, $G^{ij}_n$ is written as 
\begin{align}
    \label{eq:G_k}
    G^{ij}_n=2\sum_{m(\neq n)}\frac{g^{ij}_{nm}}{\varepsilon_{nm}}=2 \hbar^2 \sum_{m(\neq n)} \mathrm{Re}\biggl[ \frac{v^i_{nm} v^j_{mn}}{\varepsilon^3_{nm}} \biggr],
\end{align}
where $v^{i}_{nm}=\hbar^{-1}\braket{u_n|\partial_{k_i}H_0|u_m}$ is the matrix representation of the velocity operator.
Here, the second equality results from an identity for $n \neq m$, 
\begin{align}
    \label{eq:HF_Theorem}
    \braket{\partial_{k_i}u_n|u_m}=-\braket{u_n|\partial_{k_i}u_m}=\hbar \frac{v^i_{nm}}{\varepsilon_{nm}}.
\end{align}
On the other hand, $\mathfrak{G}^{ij}_n$ is related to the $h$-$k$ space quantum metric~\cite{Feng2024-gj},
\begin{align}
    \mathfrak{g}^{ij}_n&=\mathrm{Re}\braket{\partial_{h_i}u_n|(1-\ket{u_n}\bra{u_n})|\partial_{k_j}u_n} \notag \\
    &=\sum_{m(\neq n)}\mathrm{Re}\braket{\partial_{h_i}u_n|u_m}\braket{u_m|\partial_{k_j}u_n} \eqqcolon \sum_{m(\neq n)}\mathfrak{g}^{ij}_{nm},
\end{align}
which is defined in the extended parameter space spanned by the momentum $\bm{k}$ and a magnetic field $\bm{h}$.
This magnetic field $\bm{h}$ couples to spin $\bm{s}$, which is akin to a vector potential $\bm{A}$ coupling to current $\bm{j}$:
\begin{align}
    \label{eq:delh}
    \bm{s}=-\frac{\delta\mathcal{H}}{\delta \bm{h}} \hspace{0.25cm} \biggl( \Leftrightarrow \bm{j}=-\frac{\delta \mathcal{H}}{\delta \bm{A}} \biggr),
\end{align}
where $\mathcal{H}$ is a general Hamiltonian. 
Based on this analogy, $\mathfrak{G}^{ij}_n$ is referred to as the $h$-space BCP~\cite{Xiao2022-xr} and is written as
\begin{align}
    \label{eq:G_h}
    \mathfrak{G}^{ij}_n=2\sum_{m(\neq n)}\frac{\mathfrak{g}^{ij}_{nm}}{\varepsilon_{nm}}=-2 \hbar \sum_{m(\neq n)} \mathrm{Re}\biggl[ \frac{s^i_{nm} v^j_{mn}}{\varepsilon^3_{nm}} \biggr].
\end{align}
Here, the second equality results from Eq.~\eqref{eq:HF_Theorem} and an identity for $n \neq m$, 
\begin{align}
    \label{eq:HF_Theorem_s}
    \braket{\partial_{h_i}u_n|u_m}=-\braket{u_n|\partial_{h_i}u_m}=-\frac{s^i_{nm}}{\varepsilon_{nm}}.
\end{align}
This identity is derived by performing the following steps on Eq.~\eqref{eq:HF_Theorem}: first, use a relation $\partial_{\bm{k}}=(\hbar/e)\partial_{\bm{A}}$, and then replace $\partial_{\bm{A}}$ with $-\partial_{\bm{h}}$.
These steps are based on minimal coupling and Eq.~\eqref{eq:delh}, respectively.
On the other hand, by taking the opposite steps for Eq.~\eqref{eq:tau0}, we can reproduce the intrinsic nonlinear conductivity~\cite{Das2023-nm},
\begin{align}
    \sigma^{\tau^0}_{i;jk}=-\frac{e^3}{2\hbar}\sum_n \int_{\bm{k}}\Bigl[ \partial_{k_i} G^{jk}_n-2\Bigl(\partial_{k_j} G^{ik}_n+\partial_{k_k} G^{ij}_n \Bigr)\Bigr]f_n. 
\end{align}
Specifically, this is achieved by replacing $\partial_{\bm{h}}$ with $\partial_{\bm{k}}$ and multiplying a factor of $e/\hbar$.

%%%%%%%%%%%%%%%%%%%%%%%%%%%%%%%%%%%%%%%%%%%%%%%%%%%%%%%%%%%%%%%%%%%%%%%%%%%%%%%%%%%%%%%%%%%%%%%%%%%%%%%%%%%%%%%%%%%%%%%%%%%%%%%%%%%%%%%%%%%%%%%%%%%%%%%%%%%%%%%%%%%%%%%%%%%%%%%%%%%%%%%%%%%%%%%%

The intrinsic NMEE exhibits two distinctive properties, as indicated by Eq.~\eqref{eq:tau0}.
First, it is related to the quantum metric, which measures the distance between quantum states in parameter space~\cite{Provost1980-qb,Resta2011-da}.
Thus, band structures with a large distance between neighboring quantum states, such as Weyl points, can lead to a large intrinsic NMEE. 
Second, the intrinsic NMEE can occur even in insulators because of the presence of a Fermi sea term.
This property is absent in the TNHE, essentially a transport phenomenon, rendering the NMEE more ideal for magnetic octupole responses.

%%%%%%%%%%%%%%%%%%%%%%%%%%%%%%%%%%%%%%%%%%%%%%%%%%%%%%%%%%%%%%%%%%%%%%%%%%%%%%%%%%%%%%%%%%%%%%%%%%%%%%%%%%%%%%%%%%%%%%%%%%%%%%%%%%%%%%%%%%%%%%%%%%%%%%%%%%%%%%%%%%%%%%%%%%%%%%%%%%%%%%%%%%%%%%%%
%%%%%%%%%%%%%%%%%%%%%%%%%%%%%%%%%%%%%%%%%%%%%%%%%%%%%%%%%%%%%%%%%%%%%%%%%%%%%%%%%%%%%%%%%%%%%%%%%%%%%%%%%%%%%%%%%%%%%%%%%%%%%%%%%%%%%%%%%%%%%%%%%%%%%%%%%%%%%%%%%%%%%%%%%%%%%%%%%%%%%%%%%%%%%%%%
%%%%%%%%%%%%%%%%%%%%%%%%%%%%%%%%%%%%%%%%%%%%%%%%%%%%%%%%%%%%%%%%%%%%%%%%%%%%%%%%%%%%%%%%%%%%%%%%%%%%%%%%%%%%%%%%%%%%%%%%%%%%%%%%%%%%%%%%%%%%%%%%%%%%%%%%%%%%%%%%%%%%%%%%%%%%%%%%%%%%%%%%%%%%%%%%
%Sec IV

\section{Model calculation for a $d$-wave altermagnet} \label{sec:Model_calculation_altermagnet}

%%%%%%%%%%%%%%%%%%%%%%%%%%%%%%%%%%%%%%%%%%%%%%%%%%%%%%%%%%%%%%%%%%%%%%%%%%%%%%%%%%%%%%%%%%%%%%%%%%%%%%%%%%%%%%%%%%%%%%%%%%%%%%%%%%%%%%%%%%%%%%%%%%%%%%%%%%%%%%%%%%%%%%%%%%%%%%%%%%%%%%%%%%%%%%%%

\begin{figure*}[t]
  \setcounter{figure}{2}
    \subfigure[]{
        \includegraphics[height=3.6cm,width=0.28\hsize]{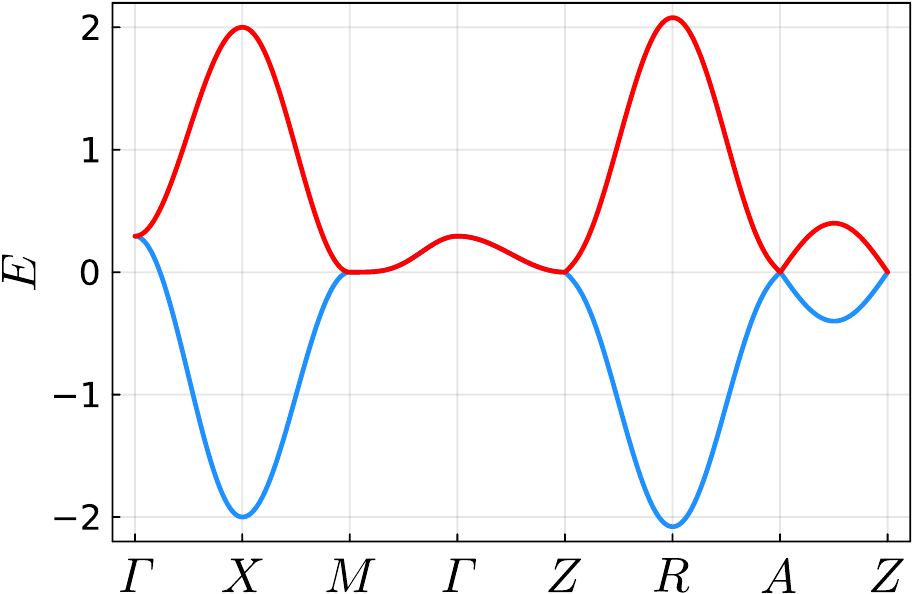}
        \label{fig:Energy_3D}
    } 
    \hfill
    \subfigure[]{
        \includegraphics[height=4.0cm,width=0.31\hsize]{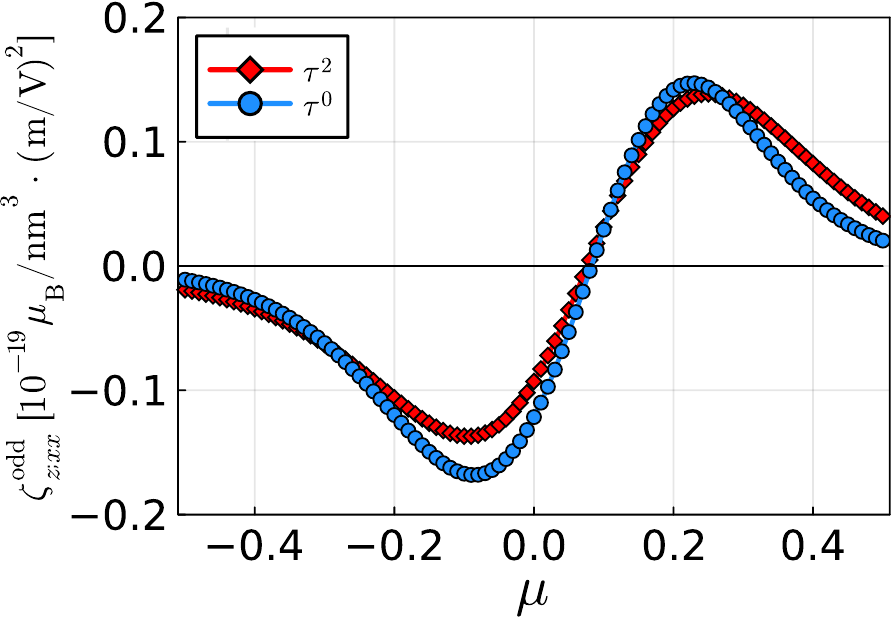}
        \label{fig:zxx}
    }
    \hfill
    \subfigure[]{
        \includegraphics[height=4.0cm,width=0.31\hsize]{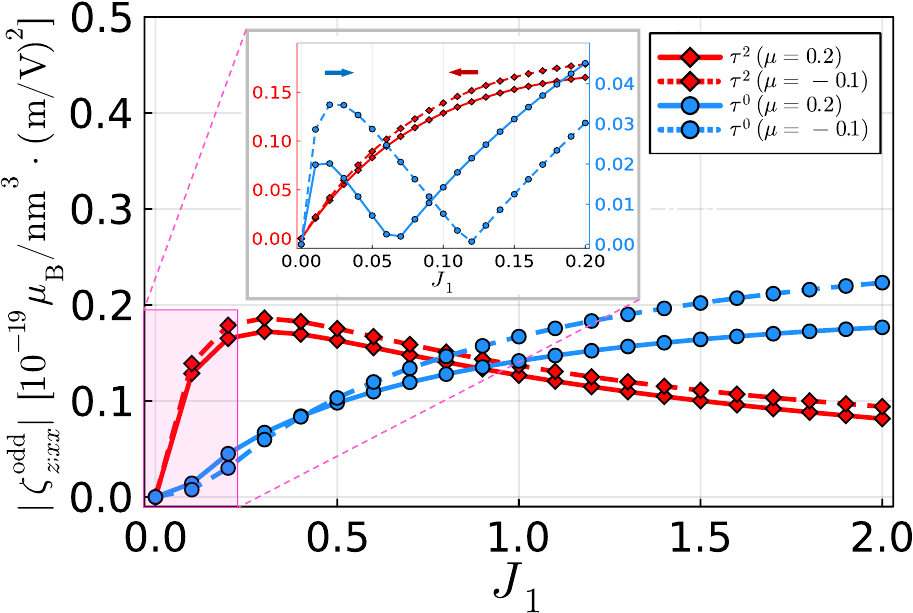}
        \label{fig:zxx_order}
    }
  \setcounter{figure}{1}
  \caption{(a) Band structure of Eq.~\eqref{eq:Altermagnet_Effective_Hamiltonian} for $(t,\lambda,J_0,J_1)=(1.0,0.4,1.7,1.0)$. 
           (b) Chemical potential dependence of $\zeta^{\mathrm{odd}}_{z;xx}$ separated into the $\tau^2$ (red color) and $\tau^0$ (blue color) components.
           (c) Altermagnetic order parameter ($J_1$) dependence of $|\zeta^{\mathrm{odd}}_{z;xx}|$ for different chemical potentials.
           We select two chemical potentials: $\mu=0.2$ (solid line) and $\mu=-0.1$ (dashed line), at which the NMEE takes a local maximum when $J_1=1.0$, as shown in~(b).
           The inset is a magnified view of the region highlighted in the main panel.} 
  \label{fig:Energy_dispersion_zxx}
\end{figure*}

%%%%%%%%%%%%%%%%%%%%%%%%%%%%%%%%%%%%%%%%%%%%%%%%%%%%%%%%%%%%%%%%%%%%%%%%%%%%%%%%%%%%%%%%%%%%%%%%%%%%%%%%%%%%%%%%%%%%%%%%%%%%%%%%%%%%%%%%%%%%%%%%%%%%%%%%%%%%%%%%%%%%%%%%%%%%%%%%%%%%%%%%%%%%%%%%

\subsubsection{Model}

We first introduce a four-band model of a $d$-wave altermagnet.
The four bands consist of the basis with two spins on each of two sublattices, A and B [Fig.~\ref{fig:Altermagnet}].
The Hamiltonian reads~\cite{Smejkal2020-zu,Fang2023-xl}
\begin{align}
    \label{eq:Altermagnet_Hamiltonian}
    &H(\bm{k})=t \cos \Bigl( \frac{k_x}{2} \Bigr) \cos \Bigl( \frac{k_y}{2} \Bigr) \cos \Bigl( \frac{k_z}{2} \Bigr) \sigma^0 \tau^x \notag \\
    &+\lambda \Bigl(\sin \Bigl( \frac{k_x+k_y}{2} \Bigr)\sigma^x+\sin \Bigl( \frac{k_y-k_x}{2} \Bigr)\sigma^y \Bigr)\sin\Bigl(\frac{k_z}{2} \Bigr)\tau^x \notag \\
    &+J_0 \sigma^z \tau^z +J_1(\cos k_x- \cos k_y)\sigma^z \tau^0,
\end{align}
where $\sigma^0$ and $\tau^0$ are the identity matrices, and $\bm{\sigma}=(\sigma^x,\sigma^y,\sigma^z)$ and $\bm{\tau}=(\tau^x,\tau^y,\tau^z)$ are the Pauli matrices of the spin and sublattice, respectively.
Specifically, $t$ describes the nearest-neighbor (NN) hopping, $\lambda$ represents the SOC, and $J_0$ and $J_1$ denote an antiferromagnetic molecular field and the $d$-wave altermagnetic order parameter, respectively.

%%%%%%%%%%%%%%%%%%%%%%%%%%%%%%%%%%%%%%%%%%%%%%%%%%%%%%%%%%%%%%%%%%%%%%%%%%%%%%%%%%%%%%%%%%%%%%%%%%%%%%%%%%%%%%%%%%%%%%%%%%%%%%%%%%%%%%%%%%%%%%%%%%%%%%%%%%%%%%%%%%%%%%%%%%%%%%%%%%%%%%%%%%%%%%%%

We then derive an effective two-band model for the calculation from the four-band model by assuming that $J_0$ dominates Eq.~\eqref{eq:Altermagnet_Hamiltonian}.
The four bands can be divided into two groups, each of which consists of the basis with the opposite spins on the different sublattices~\cite{Fang2023-xl}: $\{\ket{\mathrm{A},\uparrow},\ket{\mathrm{B},\downarrow}\}$ and $\{\ket{\mathrm{A},\downarrow},\ket{\mathrm{B},\uparrow}\}$.
The effective two-band Hamiltonian takes the form,
\begin{align}
  \label{eq:Altermagnet_Effective_Hamiltonian}
  &H_{\mathrm{eff}}(\bm{k}) 
  =\frac{t^2}{2J_0}\cos^2 \Bigl( \frac{k_x}{2} \Bigr)\cos^2\Bigl( \frac{k_y}{2}\Bigr)\cos^2\Bigl(\frac{k_z}{2} \Bigr) \sigma^0  \notag \\
  &+\lambda \Bigl(\sin \Bigl( \frac{k_x+k_y}{2} \Bigr)\sigma^x+\sin \Bigl( \frac{k_y-k_x}{2} \Bigr)\sigma^y \Bigr)\sin\Bigl(\frac{k_z}{2} \Bigr) \notag \\
  &+J_1(\cos k_x- \cos k_y)\sigma^z, 
\end{align}
where we shift the effective two bands by $-J_0$.
In the absence of the SOC, the energy eigenvalues are given by 
\begin{align}
    \varepsilon_{\pm}&=\frac{t^2}{2J_0}\cos^2 \Bigl( \frac{k_x}{2} \Bigr)\cos^2\Bigl( \frac{k_y}{2}\Bigr)\cos^2\Bigl(\frac{k_z}{2} \Bigr) \notag \\
    &\quad \pm J_1(\cos k_x-\cos k_y), \label{eq:eigen_altermagnet}
\end{align}
where ``$\pm$" indicates the upper ($+$) and lower ($-$) bands.
Equation~\eqref{eq:eigen_altermagnet} generates spin-splittings without the SOC and nodal lines along $k_x=\pm k_y$, which correspond to $d$-wave altermagnetism. 
Figure~\ref{fig:Energy_3D} shows the band structure of this effective Hamiltonian, which exhibits such spin-splittings along the $\varGamma$-$X$-$M$ and $Z$-$R$-$A$ lines.
The symmetry leading to these spin-splittings can be identified by the spin groups~\cite{LITVIN1974538,Litvin1977-kw}, which describe the symmetry of magnets without SOC.
In the presence of the SOC, however, the spin group symmetry breaks down, reducing the symmetry of the Hamiltonian to $4'/mm'm$.
Furthermore, gaps open on the spin-group protected nodal lines, such as the $Z$-$A$ line, except for Dirac nodes at $Z=(0,0,\pi)$ and $A=(\pi,\pi,\pi)$.

%%%%%%%%%%%%%%%%%%%%%%%%%%%%%%%%%%%%%%%%%%%%%%%%%%%%%%%%%%%%%%%%%%%%%%%%%%%%%%%%%%%%%%%%%%%%%%%%%%%%%%%%%%%%%%%%%%%%%%%%%%%%%%%%%%%%%%%%%%%%%%%%%%%%%%%%%%%%%%%%%%%%%%%%%%%%%%%%%%%%%%%%%%%%%%%%

The presence or absence of SOC is also an important factor for observing the NMEE.
Collinear magnets without SOC do not exhibit a finite NMEE even if they break $\mathcal{T}$-symmetry because they always preserve effective $\mathcal{T}$-symmetry~\cite{Gosalbez-Martinez2015-ui}.  
Effective $\mathcal{T}$-symmetry combines $\mathcal{T}$-symmetry with a spin rotation and prohibits the manifestation of responses that have identical transformation properties as spin, such as the AHE and NMEE~\cite{Suzuki2017-ve}.
However, SOC breaks effective $\mathcal{T}$-symmetry, activating the NMEE in altermagnets with collinearity.
Indeed, the SOC reduces the symmetry to $4'/mm'm$, which is an MPG allowing the NMEE [see Table~\ref{tab:Classification}].
Note that a complete absence of SOC is not a realistic scenario, and altermagnets typically exhibit, at least, a weak SOC.

%%%%%%%%%%%%%%%%%%%%%%%%%%%%%%%%%%%%%%%%%%%%%%%%%%%%%%%%%%%%%%%%%%%%%%%%%%%%%%%%%%%%%%%%%%%%%%%%%%%%%%%%%%%%%%%%%%%%%%%%%%%%%%%%%%%%%%%%%%%%%%%%%%%%%%%%%%%%%%%%%%%%%%%%%%%%%%%%%%%%%%%%%%%%%%%%

\subsubsection{Results}

First, we calculate the chemical potential dependence of the NMEE tensor by assuming an applied electric field in the $x$-$y$ plane.
Under this assumption, a generator $C_{4z}\mathcal{T}$ of $4'/mm'm$ leaves the following nonvanishing components of $\zeta^{\mathrm{odd}}_{i;jk}$~\cite{Xiao2022-xr}: $\zeta^{\mathrm{odd}}_{z;xx}=-\zeta^{\mathrm{odd}}_{z;yy}$ and $\zeta^{\mathrm{odd}}_{z;xy}$.
Furthermore, mirror symmetries, $\mathcal{M}_{xy}$ and $\mathcal{M}_{x} \mathcal{T}$, which are the other generators, prohibit $\zeta^{\mathrm{odd}}_{z;xy}$ and only leave $\zeta^{\mathrm{odd}}_{z;xx}=-\zeta^{\mathrm{odd}}_{z;yy}$.
Figure~\ref{fig:zxx} shows the chemical potential dependence of $\zeta^{\mathrm{odd}}_{z;xx}$ for $(t,\lambda,J_0,J_1)=(1.0,0.4,1.7,1.0)$, $k_{\mathrm{B}}T=0.1$, $\hbar/\tau=0.1$, and $\hbar=e=1$. 
Note that we make a replacement called smearing~\cite{Ibanez-Azpiroz2018-mm} in Eq.~\eqref{eq:tau0} to avoid divergences at crossing points: $1/\varepsilon_{nm} \rightarrow \varepsilon_{nm}/(\varepsilon^2_{nm}+\gamma^2)$, and set $\gamma=0.005$ in the calculation.
Both the $\tau^2$- and $\tau^0$-responses take finite values without any bias fields, which directly demonstrates the effectiveness of the NMEE for $d$-wave altermagnets.

%%%%%%%%%%%%%%%%%%%%%%%%%%%%%%%%%%%%%%%%%%%%%%%%%%%%%%%%%%%%%%%%%%%%%%%%%%%%%%%%%%%%%%%%%%%%%%%%%%%%%%%%%%%%%%%%%%%%%%%%%%%%%%%%%%%%%%%%%%%%%%%%%%%%%%%%%%%%%%%%%%%%%%%%%%%%%%%%%%%%%%%%%%%%%%%%

Then, we examine the relationship between the NMEE and the magnetic octupole.
Here, we assume that the $d$-wave altermagnetic order parameter $J_1$ reflects the strength of the magnetic octupole. 
Figure~\ref{fig:zxx_order} shows the $J_1$ dependence of $|\zeta^{\mathrm{odd}}_{z;xx}|$ for two chemical potentials and $(t,\lambda,J_0)=(1.0,0.4,1.7)$, $k_{\mathrm{B}}T=0.1$, $\hbar/\tau=0.1$, $\gamma=0.005$, and $\hbar=e=1$. 
For small order parameters, the $\tau^2$- and $\tau^0$-responses show an approximately linear relationship to $J_1$ (see the inset).
As the order parameter increases, however, this linear relationship breaks down.
Specifically, the $\tau^2$-response starts to decrease beyond a certain point ($J_1 \approx 0.3$), while the $\tau^0$-response decreases rapidly before rising again, indicating a sign change. 
Still, the NMEE (combining the $\tau^2$- and $\tau^0$-responses) is finite when $J_1\neq 0$ and zero when $J_1=0$, which satisfies the conditions as a magnetic octupole response.

%%%%%%%%%%%%%%%%%%%%%%%%%%%%%%%%%%%%%%%%%%%%%%%%%%%%%%%%%%%%%%%%%%%%%%%%%%%%%%%%%%%%%%%%%%%%%%%%%%%%%%%%%%%%%%%%%%%%%%%%%%%%%%%%%%%%%%%%%%%%%%%%%%%%%%%%%%%%%%%%%%%%%%%%%%%%%%%%%%%%%%%%%%%%%%%%
%%%%%%%%%%%%%%%%%%%%%%%%%%%%%%%%%%%%%%%%%%%%%%%%%%%%%%%%%%%%%%%%%%%%%%%%%%%%%%%%%%%%%%%%%%%%%%%%%%%%%%%%%%%%%%%%%%%%%%%%%%%%%%%%%%%%%%%%%%%%%%%%%%%%%%%%%%%%%%%%%%%%%%%%%%%%%%%%%%%%%%%%%%%%%%%%
%%%%%%%%%%%%%%%%%%%%%%%%%%%%%%%%%%%%%%%%%%%%%%%%%%%%%%%%%%%%%%%%%%%%%%%%%%%%%%%%%%%%%%%%%%%%%%%%%%%%%%%%%%%%%%%%%%%%%%%%%%%%%%%%%%%%%%%%%%%%%%%%%%%%%%%%%%%%%%%%%%%%%%%%%%%%%%%%%%%%%%%%%%%%%%%%
%Sec V

%%%%%%%%%%%%%%%%%%%%%%%%%%%%%%%%%%%%%%%%%%%%%%%%%%%%%%%%%%%%%%%%%%%%%%%%%%%%%%%%%%%%%%%%%%%%%%%%%%%%%%%%%%%%%%%%%%%%%%%%%%%%%%%%%%%%%%%%%%%%%%%%%%%%%%%%%%%%%%%%%%%%%%%%%%%%%%%%%%%%%%%%%%%%%%%%

\section{Model calculation for a pyrochlore lattice with AIAO magnetic order} \label{sec:Model_calculation_pyrochlore}

\begin{figure}[t]
    \setcounter{figure}{3}
    \begin{minipage}[t]{{0.32\columnwidth}} 
        \subfigure[]{
            \includegraphics[height=2.5cm,width=1.1\hsize]{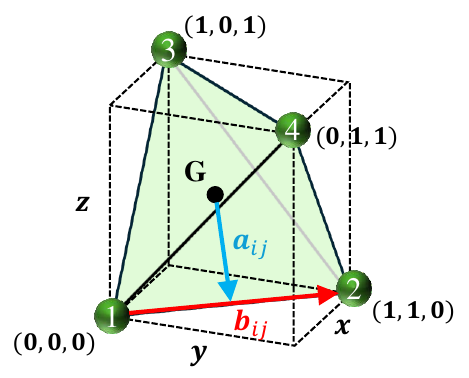}
            \label{fig:unit}
        }
        \vfill
        \vspace{-0.2cm}
        \subfigure[]{
            \includegraphics[height=2.5cm,width=1.0\hsize]{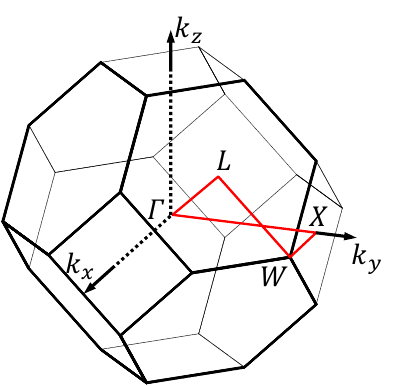}
            \label{fig:1stBZ}
        } 
        \vfill
        \vspace{-0.3cm}
        \setcounter{subfigure}{3}
        \subfigure[]{
            \includegraphics[height=2.5cm,width=1.0\hsize]{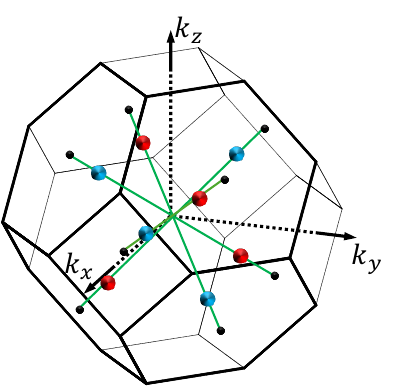}
            \label{fig:pyrochlore_Weyl}
        } 
        \setcounter{subfigure}{2}
    \end{minipage}
    \hfill
    \begin{minipage}[t]{{0.6\columnwidth}} 
        \subfigure[]{
            \includegraphics[height=3.9cm,width=0.98\hsize]{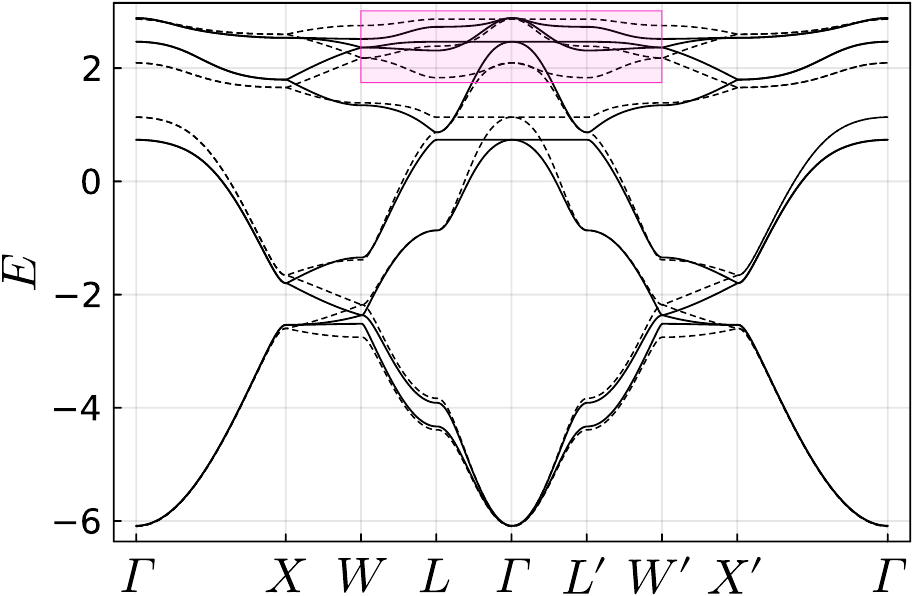}
            \label{fig:Energy_dispersion_1}
        }
        \setcounter{subfigure}{4}
        \subfigure[]{
            \includegraphics[height=3.9cm,width=1.0\hsize]{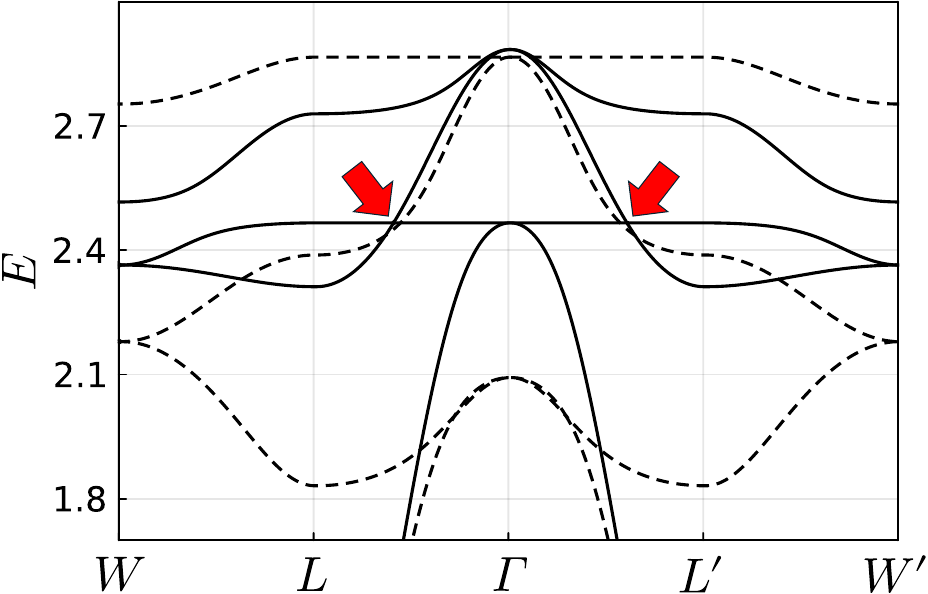}
            \label{fig:Energy_dispersion_2}
        }
        \setcounter{subfigure}{3}
    \end{minipage} 
    \setcounter{figure}{2}
    \caption{(a) Unit cell of a pyrochlore lattice.
             The blue and red arrows correspond to $\bm{a}_{ij}$ in Eq.~\eqref{eq:a_ij} and $\bm{b}_{ij}$ in Eq.~\eqref{eq:b_ij}, respectively. 
             (b) First BZ of a pyrochlore lattice (face-centered cubic lattice) with marked high-symmetry lines.
           (c) Band structure of Eq.~\eqref{eq:Pyrochlore_Hamiltonian} for $(t,m/\sqrt{3})=(1.0,0.5)$ with $\lambda=0.1$ (solid line) and $\lambda=0.0$ (dashed line). 
           The points $L'$, $W'$, and $X'$ are obtained by applying a transformation $(k_x,k_y,k_z)\rightarrow(k_x,-k_y,-k_z)$ to the points $L$, $W$, and $X$.
           (d) Locations of eight Weyl points (red and blue dots) in an AIAO magnetic phase.
           The green lines connect two Weyl points related by $\mathcal{P}$-symmetry.
           (e) Magnified view of the band structure highlighted in (c). }

    \label{fig:Energy_dispersion}
\end{figure}

\begin{figure*}[t]
    \setcounter{figure}{4}
    \subfigure[]{
        \includegraphics[height=4.0cm,width=0.3\hsize]{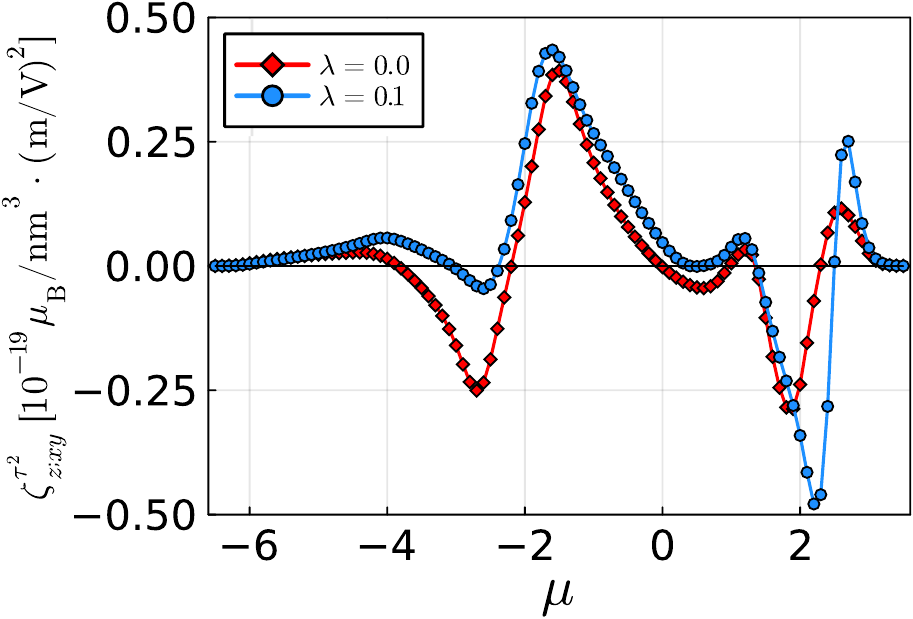}
        \label{fig:zxy_tau2}
    } 
    \hfill
    \subfigure[]{
        \includegraphics[height=4.0cm,width=0.3\hsize]{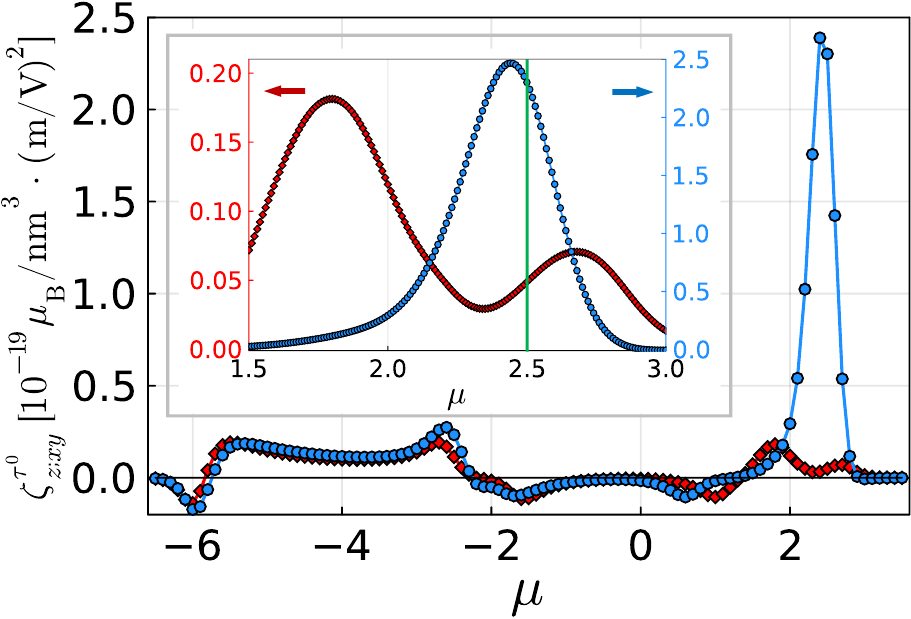}
        \label{fig:zxy_tau0}
    }
    \hfill
    \subfigure[]{
        \includegraphics[height=4.0cm,width=0.3\hsize]{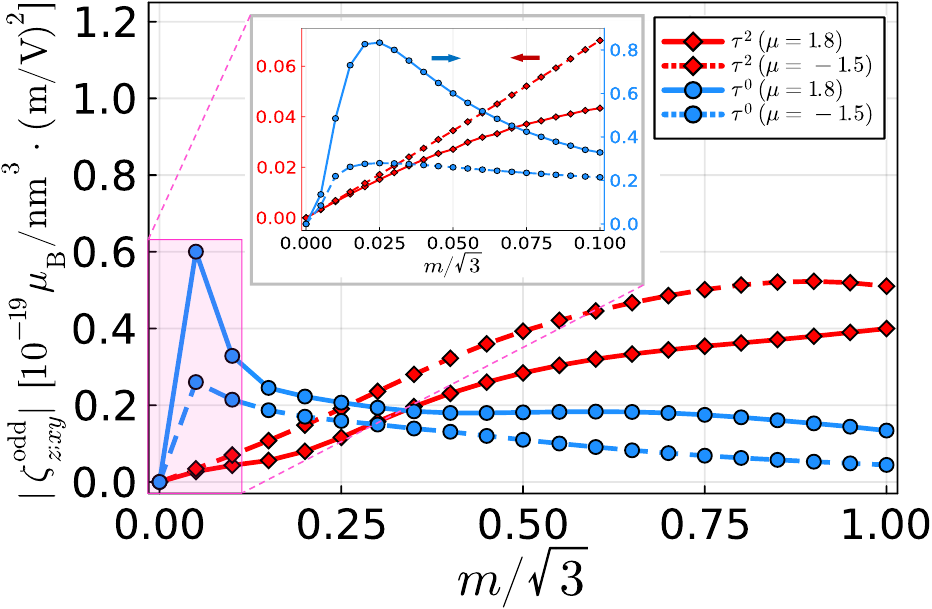}
        \label{fig:zxy_order}
    }
    \setcounter{figure}{3}
    \caption{ (a),(b) Chemical potential dependences of the NMEE tensors.
          Panels~(a) and~(b) correspond to the $\tau^2$- and~$\tau^0$-responses, respectively.
          Furthermore, we examine the SOC dependence: $\lambda=0.0$ (red color) and $\lambda=0.1$ (blue color).
          The inset of (b) is a magnified view of the region of $1.5 \leq \mu \leq 3.0$, and the green line corresponds to $\mu=2.5$.
          (c) AIAO magnetic order parameter ($m/\sqrt{3}$) dependence of $|\zeta^{\mathrm{odd}}_{z;xy}|$ for different chemical potentials and $\lambda=0.0$.
           The red and blue colors denote the $\tau^2$- and $\tau^0$-responses, respectively.  
           We select two chemical potentials: $\mu=1.8$ (solid line) and $\mu=-1.5$ (dashed line), at which the NMEE takes a local maximum when $m/\sqrt{3}=0.5$, as shown in (a) and (b).
           The inset is a magnified view of the region highlighted in the main panel.
           }
    \label{fig:zxy_result}
\end{figure*}

%%%%%%%%%%%%%%%%%%%%%%%%%%%%%%%%%%%%%%%%%%%%%%%%%%%%%%%%%%%%%%%%%%%%%%%%%%%%%%%%%%%%%%%%%%%%%%%%%%%%%%%%%%%%%%%%%%%%%%%%%%%%%%%%%%%%%%%%%%%%%%%%%%%%%%%%%%%%%%%%%%%%%%%%%%%%%%%%%%%%%%%%%%%%%%%%

In this section, we calculate the NMEE for a pyrochlore lattice with AIAO magnetic order.
In Sec.~\ref{subsec:Numerical_calculation}, we numerically show that the intrinsic response is enhanced near band crossings, which are proven to be Weyl points.
In Sec.~\ref{subsec:Analytical_calculation}, we analytically demonstrate that the Weyl points are indeed responsible for the large intrinsic response by analyzing an effective Weyl Hamiltonian.

%%%%%%%%%%%%%%%%%%%%%%%%%%%%%%%%%%%%%%%%%%%%%%%%%%%%%%%%%%%%%%%%%%%%%%%%%%%%%%%%%%%%%%%%%%%%%%%%%%%%%%%%%%%%%%%%%%%%%%%%%%%%%%%%%%%%%%%%%%%%%%%%%%%%%%%%%%%%%%%%%%%%%%%%%%%%%%%%%%%%%%%%%%%%%%%%
%%%%%%%%%%%%%%%%%%%%%%%%%%%%%%%%%%%%%%%%%%%%%%%%%%%%%%%%%%%%%%%%%%%%%%%%%%%%%%%%%%%%%%%%%%%%%%%%%%%%%%%%%%%%%%%%%%%%%%%%%%%%%%%%%%%%%%%%%%%%%%%%%%%%%%%%%%%%%%%%%%%%%%%%%%%%%%%%%%%%%%%%%%%%%%%%

\subsection{Numerical calculation} \label{subsec:Numerical_calculation}

\subsubsection{Model}

The general Hamiltonian composed of NN hopping on a pyrochlore lattice  reads~\cite{Guo2009-nw,Witczak-Krempa2013-bv,Yamaji2014-fh,Ueda2017-jr,Oh2018-xs,Varnava2018-ub,Hayami2019-zk}
\begin{align}
    H_0=-t\sum_{\langle i,j \rangle, \alpha} c^{\dagger}_{i\alpha}c_{j\alpha}+i\lambda \sum_{\langle i,j \rangle, \alpha\beta} c^{\dagger}_{i\alpha} (\bm{d}_{ij}\cdot \bm{\sigma})_{\alpha \beta} c_{j\beta},
\end{align}
where $c^{\dag}_{i \alpha}$ and $c_{i \alpha}$ are creation and annihilation operators of electrons with the spin {$\alpha=\{\uparrow,\downarrow\}$ at a site $i$, $\bm{\sigma}=(\sigma^x,\sigma^y,\sigma^z)$ are the Pauli matrices, and $\sum_{\braket{ij}}$ is the sum over the NN sites.
The first term is the NN hopping with hopping strength $t$.
The second term is the effective SOC with coupling strength $\lambda$ acting 
between the NN bond.
The vector $\bm{d}_{ij}$ is composed of two vectors, $\bm{a}_{ij}$ and $\bm{b}_{ij}$ [Fig.~\ref{fig:unit}]:
\begin{align}
    \bm{d}_{ij}&=2\bm{a}_{ij}\times\bm{b}_{ij}, \\
    \bm{a}_{ij}&=\frac{1}{2}(\bm{x}_i+\bm{x}_j)-\bm{x}_{\mathrm{G}}, \label{eq:a_ij} \\
    \bm{b}_{ij}&=\bm{x}_j-\bm{x}_i, \label{eq:b_ij}
\end{align}
where $\bm{b}_{ij}$ points from the $j$-th site $\bm{x}_j$ to the $i$-th site $\bm{x}_i$, and $\bm{a}_{ij}$ points from the center of the unit tetrahedron, $\bm{x}_{\mathrm{G}}=(1,1,1)/2$, to the midpoint of the $\langle i,j \rangle$ bond.

%%%%%%%%%%%%%%%%%%%%%%%%%%%%%%%%%%%%%%%%%%%%%%%%%%%%%%%%%%%%%%%%%%%%%%%%%%%%%%%%%%%%%%%%%%%%%%%%%%%%%%%%%%%%%%%%%%%%%%%%%%%%%%%%%%%%%%%%%%%%%%%%%%%%%%%%%%%%%%%%%%%%%%%%%%%%%%%%%%%%%%%%%%%%%%%%

The Hamiltonian in momentum space is given by
\begin{align}
    \label{eq:pyrochlore_nonmag}
    H_0&=\sum_{\bm{k}}\Psi^{\dagger}_{\bm{k}}H_0(\bm{k})\Psi_{\bm{k}}, \\
    [H_0(\bm{k})]_{\mu\nu}&=2(-t+i\lambda(\bm{d}_{\mu \nu}\cdot \bm{\sigma}))\cos(\bm{b}_{\mu \nu}\cdot\bm{k}),
\end{align}
where $\Psi_{\bm{k}}=(c_{\bm{k}1\uparrow},c_{\bm{k}2\uparrow},c_{\bm{k}3\uparrow},c_{\bm{k}4\uparrow},c_{\bm{k}1\downarrow},c_{\bm{k}2\downarrow},c_{\bm{k}3\downarrow},c_{\bm{k}4\downarrow})$ is the basis with the momentum $\bm{k}$ and the spin $\{\uparrow,\downarrow\}$} on four sublattices, $\mu=1,2,3,4$ [Fig.~\ref{fig:unit}].
The final Hamiltonian $H$ is constructed by adding an AIAO magnetic order term $H_m$ with strength $m/\sqrt{3}$ to Eq.~\eqref{eq:pyrochlore_nonmag}~\cite{Yamaji2014-fh,Hayami2019-zk}:
\begin{align}
    H&=\sum_{\bm{k}}\Psi^{\dagger}_{\bm{k}}(H_0(\bm{k})+\frac{m}{\sqrt{3}}H_m)\Psi_{\bm{k}}, \label{eq:Pyrochlore_Hamiltonian} \\
    H_m&=\mathrm{diag}(1,-1,-1,1)\sigma^x    \notag \\
    &\quad +\mathrm{diag}(1,-1,1,-1)\sigma^y \notag \\
    &\quad +\mathrm{diag}(1,1,-1,-1)\sigma^z,
\end{align}  
where ``diag" denotes a diagonal matrix. 
This AIAO magnetic order term reduces the symmetry to $m\bar{3}m'$, allowing lowest-rank magnetic octupole order [see Table~\ref{tab:Classification}].
Figure~\ref{fig:1stBZ} shows the first BZ of a pyrochlore lattice, and Fig.~\ref{fig:Energy_dispersion_1} shows the band structure of this Hamiltonian along the high-symmetry lines depicted in it. 
This band structure retains a resemblance to the fourfold degenerate quadratic band crossings at the $\varGamma$ point.
However, AIAO magnetic order splits such a band crossing into four pairs of Weyl points, and each pair is connected by $\mathcal{P}$-symmetry.
Figure~\ref{fig:pyrochlore_Weyl} shows that the Weyl points are located along the [111] direction or the other three equivalent directions~\cite{Witczak-Krempa2012-ro,Witczak-Krempa2013-bv,Yamaji2014-fh,Varnava2018-ub,Ueda2017-jr,Ueda2018-wq,Oh2018-xs}.
Indeed, Fig.~\ref{fig:Energy_dispersion_2}, which magnifies the band structure, suggests the presence of Weyl points at the points designated by the red arrows when the SOC is turned on.

%%%%%%%%%%%%%%%%%%%%%%%%%%%%%%%%%%%%%%%%%%%%%%%%%%%%%%%%%%%%%%%%%%%%%%%%%%%%%%%%%%%%%%%%%%%%%%%%%%%%%%%%%%%%%%%%%%%%%%%%%%%%%%%%%%%%%%%%%%%%%%%%%%%%%%%%%%%%%%%%%%%%%%%%%%%%%%%%%%%%%%%%%%%%%%%%

\subsubsection{Results}

First, we calculate the chemical potential dependence of the NMEE tensor by assuming an applied electric field in the $x$-$y$ plane.
Under this assumption, the generators of $m\bar{3}m'$ leave only one nonvanishing component $\zeta^{\mathrm{odd}}_{z;xy}$.
Figures~\ref{fig:zxy_tau2} and \ref{fig:zxy_tau0} show the chemical potential dependences of $\zeta^{\tau^2}_{z;xy}$ and $\zeta^{\tau^0}_{z;xy}$, respectively, for $(t,m/\sqrt{3})=(1.0,0.5)$, $k_{\mathrm{B}}T=0.1$, $\hbar/\tau=0.1$, and $\hbar=e=1$.
Note that we use a smearing value of $\gamma=0.005$ in the calculation for the same reason as in the calculation for the $d$-wave altermagnet.
Both the $\tau^2$- and $\tau^0$-responses take finite values; in particular, the $\tau^0$-response is strongly enhanced around $\mu=2.5$ when the SOC is turned on.
Furthermore, the system does not require the SOC to activate the NMEE, unlike the $d$-wave altermagnet.
This is because the noncollinearity of the AIAO configuration breaks effective $\mathcal{T}$-symmetry regardless of SOC.

%%%%%%%%%%%%%%%%%%%%%%%%%%%%%%%%%%%%%%%%%%%%%%%%%%%%%%%%%%%%%%%%%%%%%%%%%%%%%%%%%%%%%%%%%%%%%%%%%%%%%%%%%%%%%%%%%%%%%%%%%%%%%%%%%%%%%%%%%%%%%%%%%%%%%%%%%%%%%%%%%%%%%%%%%%%%%%%%%%%%%%%%%%%%%%%%

Then, we examine the relationship between the NMEE and the magnetic octupole.
As in the $d$-wave altermagnet, we assume that the AIAO magnetic order parameter $m/\sqrt{3}$ reflects the strength of the magnetic octupole. 
Figure~\ref{fig:zxy_order} shows the $m/\sqrt{3}$ dependence of $|\zeta^{\mathrm{odd}}_{z;xy}|$ for two chemical potentials and $(t,\lambda)=(1.0,0.0)$, $k_{\mathrm{B}}T=0.1$, $\hbar/\tau=0.1$, $\gamma=0.005$, and $\hbar=e=1$.
Note that we only focus on the case where the SOC is absent ($\lambda=0$) to exclude the effect of the enhanced response. 
The approximately linear relationship in the small order-parameter region [see the inset] and its absence in the large order-parameter region are analogous to the behavior in the $d$-wave altermagnet.
Furthermore, the NMEE is finite only when the order parameter is finite, which further supports its validity as a magnetic octupole response.

%%%%%%%%%%%%%%%%%%%%%%%%%%%%%%%%%%%%%%%%%%%%%%%%%%%%%%%%%%%%%%%%%%%%%%%%%%%%%%%%%%%%%%%%%%%%%%%%%%%%%%%%%%%%%%%%%%%%%%%%%%%%%%%%%%%%%%%%%%%%%%%%%%%%%%%%%%%%%%%%%%%%%%%%%%%%%%%%%%%%%%%%%%%%%%%%

\subsubsection{Discussion}

%%%%%%%%%%%%%%%%%%%%%%%%%%%%%%%%%%%%%%%%%%%%%%%%%%%%%%%%%%%%%%%%%%%%%%%%%%%%%%%%%%%%%%%%%%%%%%%%%%%%%%%%%%%%%%%%%%%%%%%%%%%%%%%%%%%%%%%%%%%%%%%%%%%%%%%%%%%%%%%%%%%%%%%%%%%%%%%%%%%%%%%%%%%%%%%%

\begin{figure}[b]
    \centering
    \includegraphics[height=4.0cm,width=0.65\hsize]{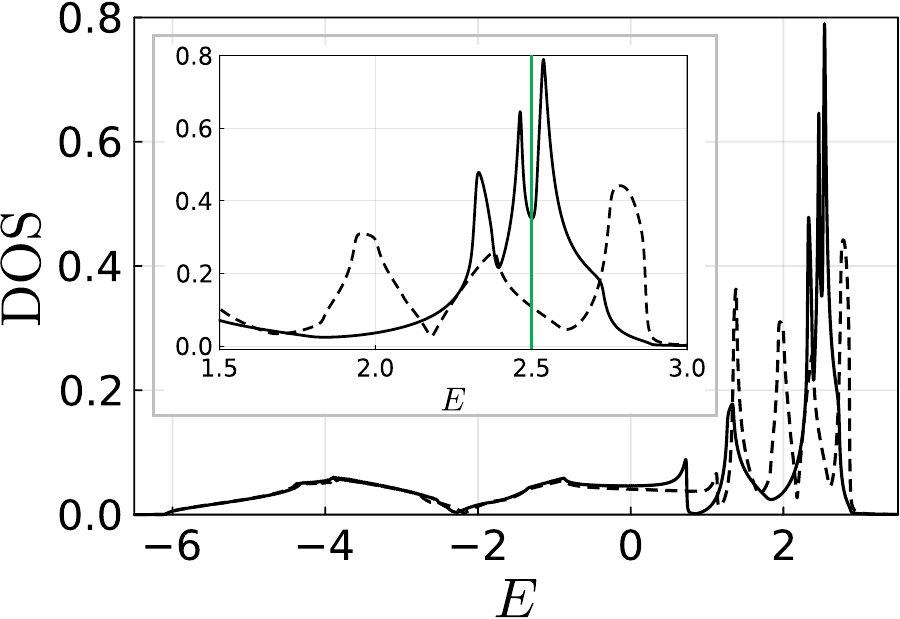}
    \caption{ DOS of the model.
          The solid and dashed lines show the DOS for $\lambda=0.1$ and~$\lambda=0.0$, respectively.
          The inset is a magnified view of the region of $1.5 \leq E \leq 3.0$, and the green line corresponds to $E=2.5$.}
    \label{fig:DOS}
\end{figure}

%%%%%%%%%%%%%%%%%%%%%%%%%%%%%%%%%%%%%%%%%%%%%%%%%%%%%%%%%%%%%%%%%%%%%%%%%%%%%%%%%%%%%%%%%%%%%%%%%%%%%%%%%%%%%%%%%%%%%%%%%%%%%%%%%%%%%%%%%%%%%%%%%%%%%%%%%%%%%%%%%%%%%%%%%%%%%%%%%%%%%%%%%%%%%%%%

Here, we discuss the origin of the enhanced response in Fig.~\ref{fig:zxy_tau0}.
This enhancement may be attributed to three possibilities.
The first possibility is the presence of a flat band around $E=2.5$ [see Fig.~\ref{fig:Energy_dispersion_2}].
Flat bands result in a large density of states (DOS), which can lead to an enhanced response.
Therefore, we show the DOS of the model in Fig.~\ref{fig:DOS}. 
It is clear, however, that a large DOS does not necessarily lead to a large NMEE response.
For example, although the peak in the DOS above $E=2.5$ is higher than below $E=2.5$, the intrinsic response at the corresponding chemical potential does not show an enhancement [compare the insets of Figs.~\ref{fig:zxy_tau0} and~\ref{fig:DOS}].
Thus, the flat band is unlikely the cause of the enhanced response.
The second possibility is the presence of band crossings around $E=2.5$ [see Fig.~\ref{fig:Energy_dispersion_2}].
Band crossings result in a small band gap of $\Delta\varepsilon$, which can enhance the intrinsic NMEE by a factor of $1/(\Delta \varepsilon)^3$, as seen from Eqs.~\eqref{eq:G_k} and~\eqref{eq:G_h}.
However, mere band crossings cannot explain the enhanced response because some band crossings do not lead to an enhanced response. 
For example, the band crossings around $E=-2.0$ do not lead to an enhanced response at the corresponding chemical potential [see Figs.~\ref{fig:Energy_dispersion_1} and~\ref{fig:zxy_tau0}].
This suggests that the band crossings around $E=2.5$ are different from the others; that is, they may be Weyl points, which are hot spots for the quantum metric. 
Thus, the most promising possibility is the presence of Weyl points.

%%%%%%%%%%%%%%%%%%%%%%%%%%%%%%%%%%%%%%%%%%%%%%%%%%%%%%%%%%%%%%%%%%%%%%%%%%%%%%%%%%%%%%%%%%%%%%%%%%%%%%%%%%%%%%%%%%%%%%%%%%%%%%%%%%%%%%%%%%%%%%%%%%%%%%%%%%%%%%%%%%%%%%%%%%%%%%%%%%%%%%%%%%%%%%%%

To confirm that the band crossings around $E=2.5$ are Weyl points, we calculate the Berry curvature $\bm{\Omega}_n(\bm{k})$~\cite{simon1983holonomy,Berry1984-fn,Resta2011-da} and the Chern number $\mathrm{Ch}_n$~\cite{Thouless1982-zh,Kohmoto1985-zj,Resta2011-da}:
\begin{align}
    \Omega^i_{n}(\bm{k})&=\frac{1}{2}\varepsilon_{ijk}\Omega^{jk}_n(\bm{k}),  \\
    \mathrm{Ch}_n&=\frac{1}{2\pi}\int_{S}\bm{\Omega}_n(\bm{k})\cdot d\bm{k}. \label{eq:Chern}
\end{align}
Here, $n$ is the band index of the bands in Fig.~\ref{fig:Energy_dispersion_1}, $S$ is a closed manifold in the first BZ, and $\Omega^{jk}_n(\bm{k})$ is given by
\begin{align}
    \label{eq:Berry_curvature}
    \Omega^{jk}_n=-2\hbar^2 \sum_{m(\neq n)} \mathrm{Im} \biggl[ \frac{v^j_{nm}v^k_{mn} }{\varepsilon^2_{nm}} \biggr].
\end{align}
We note that the band crossings around $E=2.5$ comprise the bands with indices $n=6$ and $n=7$.
Weyl points act as a source or drain of Berry curvature flux.
Therefore, if a closed manifold contains Weyl points, the integral of the Berry curvature is quantized in units of $2\pi$.
Consequently, an integer value of the Chern number in Eq.~\eqref{eq:Chern} demonstrates the presence of Weyl points.

%%%%%%%%%%%%%%%%%%%%%%%%%%%%%%%%%%%%%%%%%%%%%%%%%%%%%%%%%%%%%%%%%%%%%%%%%%%%%%%%%%%%%%%%%%%%%%%%%%%%%%%%%%%%%%%%%%%%%%%%%%%%%%%%%%%%%%%%%%%%%%%%%%%%%%%%%%%%%%%%%%%%%%%%%%%%%%%%%%%%%%%%%%%%%%%%

\begin{figure}[t]
  \centering
  \includegraphics[height=4.5cm,width=1.0\hsize]{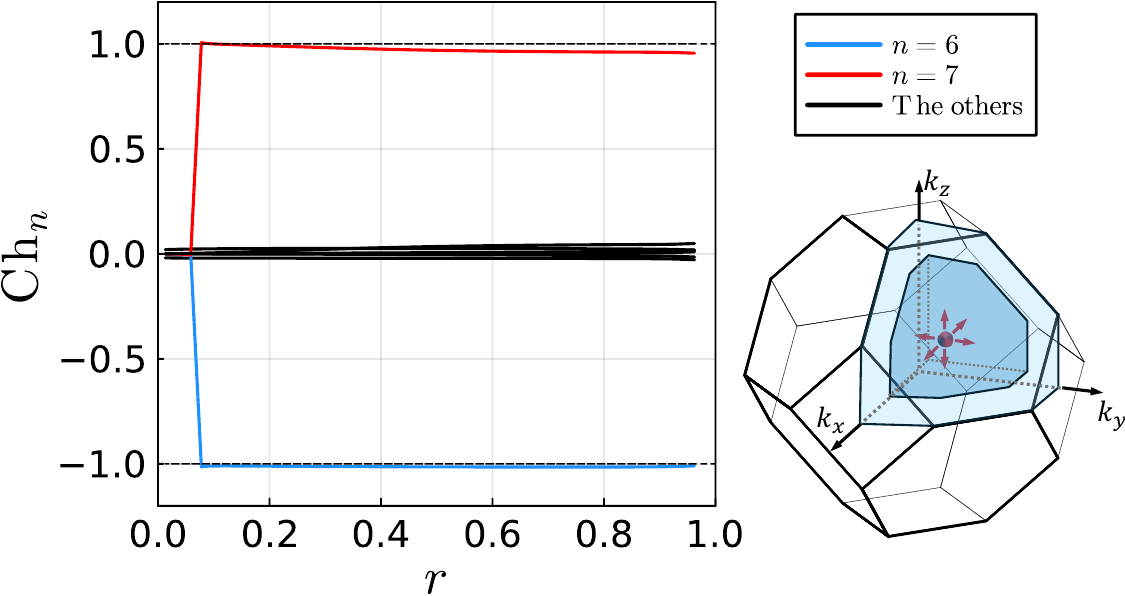}
  \caption{
           Chern number of each band calculated by varying the size of a closed manifold for the integration, which is characterized by the volume ratio $r$ regarding the initial manifold.
           The lightly shaded area in the inset represents the initial manifold ($r=1$), and the darkly shaded area represents a shrunk manifold ($r<1$).
           Note that we do not perform any smearing when calculating the Berry curvature in Eq.~\eqref{eq:Berry_curvature}.}
  \label{fig:Chern}
\end{figure}

We calculate the Chern number by selecting an arbitrary quadrant and focusing on the expected Weyl point within it [see Fig.~\ref{fig:pyrochlore_Weyl}].
Selecting any quadrant does not make any difference because the eight expected Weyl points are related by the symmetry of the system.
Specifically, the inversion symmetry $\mathcal{P}$ and three mirror symmetries $\mathcal{M}_x,\mathcal{M}_y,\mathcal{M}_z$ act on the Berry curvature as~\cite{Smejkal2020-zu} 
\begin{align}
  \mathcal{P}\bm{\Omega}_n(\bm{k})&=\bm{\Omega}_n(-\bm{k}), \\
  \mathcal{M}_x\bm{\Omega}_n(\bm{k})&=(\Omega^x_n,-\Omega^y_n,-\Omega^z_n)(-k_x,k_y,k_z),\\
  \mathcal{M}_y\bm{\Omega}_n(\bm{k})&=(-\Omega^x_n,\Omega^y_n,-\Omega^z_n)(k_x,-k_y,k_z),\\
  \mathcal{M}_z\bm{\Omega}_n(\bm{k})&=(-\Omega^x_n,-\Omega^y_n,\Omega^z_n)(k_x,k_y,-k_z).
\end{align}
Therefore, finding one Weyl point necessarily reveals the presence of the other seven.
Figure~\ref{fig:Chern} shows the Chern number of each band, which is calculated by varying the size of a closed manifold [inset of Fig.~\ref{fig:Chern}].
The manifold used in the calculation is characterized by a variable $r$, which is the volume ratio between the initial manifold and the manifold.
Notably, the bands comprising the band crossings only yield non-zero Chern numbers for a certain range of $r$.
Moreover, the values are nearly integers, which provides evidence that the band crossings are Weyl points.
Note that the sudden change in the Chern numbers near $r=0$ is due to the shrunk manifold no longer containing the Weyl point.

%%%%%%%%%%%%%%%%%%%%%%%%%%%%%%%%%%%%%%%%%%%%%%%%%%%%%%%%%%%%%%%%%%%%%%%%%%%%%%%%%%%%%%%%%%%%%%%%%%%%%%%%%%%%%%%%%%%%%%%%%%%%%%%%%%%%%%%%%%%%%%%%%%%%%%%%%%%%%%%%%%%%%%%%%%%%%%%%%%%%%%%%%%%%%%%%

%%%%%%%%%%%%%%%%%%%%%%%%%%%%%%%%%%%%%%%%%%%%%%%%%%%%%%%%%%%%%%%%%%%%%%%%%%%%%%%%%%%%%%%%%%%%%%%%%%%%%%%%%%%%%%%%%%%%%%%%%%%%%%%%%%%%%%%%%%%%%%%%%%%%%%%%%%%%%%%%%%%%%%%%%%%%%%%%%%%%%%%%%%%%%%%%%%%%%%%
%%%%%%%%%%%%%%%%%%%%%%%%%%%%%%%%%%%%%%%%%%%%%%%%%%%%%%%%%%%%%%%%%%%%%%%%%%%%%%%%%%%%%%%%%%%%%%%%%%%%%%%%%%%%%%%%%%%%%%%%%%%%%%%%%%%%%%%%%%%%%%%%%%%%%%%%%%%%%%%%%%%%%%%%%%%%%%%%%%%%%%%%%%%%%%%%

\subsection{Analytical calculation} \label{subsec:Analytical_calculation}

\subsubsection{Model}

We consider an effective Weyl Hamiltonian for any one of the eight Weyl points, which are located at either $\bm{k}_{\mathrm{Weyl}}=\sqrt{m/2t}(1,1,1)$ ($t,m>0$) or its equivalent positions~\cite{Yamaji2014-fh}.
For concreteness, we focus on $\bm{k}_{\mathrm{Weyl}}=\sqrt{m/2t}(1,1,1)$, for which the effective Weyl Hamiltonian reads~\cite{Yamaji2014-fh}
\begin{align}
    \label{eq:Hamiltonian_Weyl_1}
    &H_{\mathrm{Weyl}}(\bm{k}) \notag \\
    &=-\frac{2t}{3}k^2\sigma^0+\frac{t}{3}(k^2_x+k^2_y-2k^2_z)\sigma^x-\frac{t}{\sqrt{3}}(k^2_x-k^2_y)\sigma^y \notag \\
    & \quad -\frac{2t}{3}(k_xk_y+k_yk_z+k_zk_x)\sigma^z+m\sigma^z.
\end{align}
Here, $\sigma^0$ is the identity matrix, $\bm{\sigma}=(\sigma^x,\sigma^y,\sigma^z)$ are the Pauli matrices, $k=|\bm{k}|$, and the momentum $\bm{k}$ is taken around the $\varGamma$ point. 
Note that this Hamiltonian is derived from the Luttinger Hamiltonian that describes the low-energy physics of pyrochlore lattices~\cite{Moon2013-mr,Yamaji2014-fh,Oh2018-xs}.
Figure~\ref{fig:Energy_Weyl} shows the band structure of this Hamiltonian, which reproduces the band structure of the $\varGamma$-$L$ line in Fig.~\ref{fig:Energy_dispersion_2}, except for an energy shift.
Furthermore, we rewrite Eq.~\eqref{eq:Hamiltonian_Weyl_1} by redefining the momentum $\bm{k}$ around the Weyl point and introducing new coordinates $\bm{q}$ via a transformation, 
\begin{align}
    q_x&=k_x+k_y-2k_z, \label{eq:kx->qx} \\
    q_y&=-\sqrt{3}(k_x-k_y), \label{eq:ky->qy}  \\
    q_z&=-2(k_x+k_y+k_z). \label{eq:kz->qz} 
\end{align}
Specifically, the new effective Weyl Hamiltonian is given by $H_{\mathrm{Weyl}}(\bm{q})=g_0(\bm{q})\sigma^0+\bm{g}(\bm{q})\cdot\bm{\sigma}$ with 
\begin{align}
    g_0(\bm{q})&=-m+a_tq_z+\mathcal{O}(q^2), \label{eq:g0_q} \\
    g_x(\bm{q})&=a_tq_x-\frac{t}{18}(q^2_x-q^2_y+2q_zq_x), \label{eq:gx_q} \\
    g_y(\bm{q})&=a_tq_y-\frac{t}{9}q_y(q_z-q_x), \label{eq:gy_q} \\
    g_z(\bm{q})&=a_tq_z+\frac{t}{18}(q^2_x+q^2_y-q^2_z), \label{eq:gz_q}
\end{align}
where $a_t=(2t/3)\sqrt{m/2t}$.
This Hamiltonian is an example of the tilted Weyl Hamiltonian, which reads~\cite{Soluyanov2015-zh} 
\begin{align}
    \label{eq:tilted_Weyl_Hamiltonian}
    H(\bm{k})=Ck_z\sigma^0+C_0 \bm{k}\cdot \bm{\sigma},
\end{align}
where $C<C_0$ describes type-I Weyl semimetals, and $C>C_0$ describes type-II Weyl semimetals.
Specifically, up to the linear order terms in the momentum, $H_{\mathrm{Weyl}}(\bm{q})$ corresponds to $C=C_0$ in Eq.~\eqref{eq:tilted_Weyl_Hamiltonian}, and thus hosts a flat band, as shown in Fig.~\ref{fig:Energy_Weyl}.
The quadratic order terms describe the anisotropy of the Fermi surface.

%%%%%%%%%%%%%%%%%%%%%%%%%%%%%%%%%%%%%%%%%%%%%%%%%%%%%%%%%%%%%%%%%%%%%%%%%%%%%%%%%%%%%%%%%%%%%%%%%%%%%%%%%%%%%%%%%%%%%%%%%%%%%%%%%%%%%%%%%%%%%%%%%%%%%%%%%%%%%%%%%%%%%%%%%%%%%%%%%%%%%%%%%%%%%%%%

\begin{figure}[t]
    \centering
    \includegraphics[height=3.6cm,width=0.6\hsize]{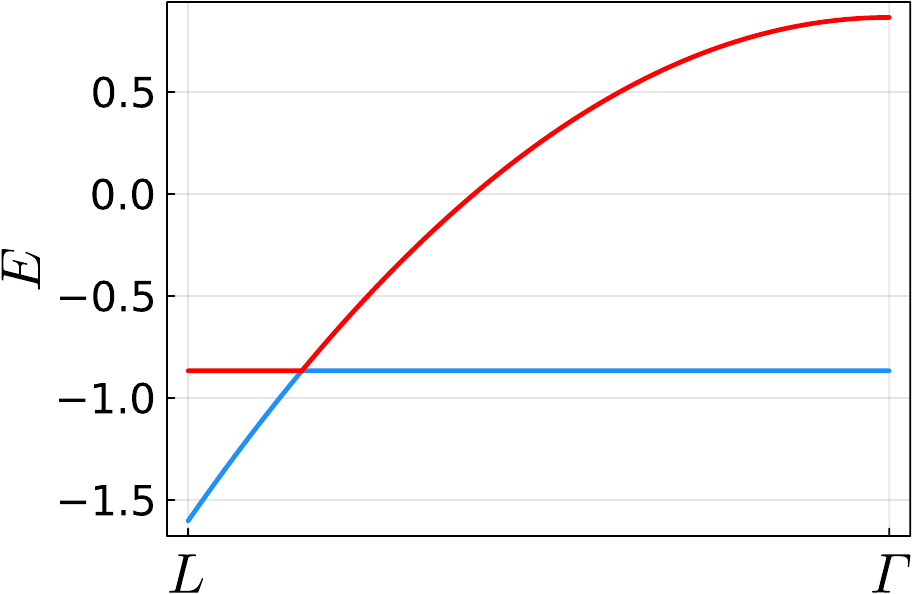}
    \caption{Band structure of Eq.~\eqref{eq:Hamiltonian_Weyl_1} for $(t,m/\sqrt{3})=(1.0,0.5)$. }
    \label{fig:Energy_Weyl}
\end{figure}

%%%%%%%%%%%%%%%%%%%%%%%%%%%%%%%%%%%%%%%%%%%%%%%%%%%%%%%%%%%%%%%%%%%%%%%%%%%%%%%%%%%%%%%%%%%%%%%%%%%%%%%%%%%%%%%%%%%%%%%%%%%%%%%%%%%%%%%%%%%%%%%%%%%%%%%%%%%%%%%%%%%%%%%%%%%%%%%%%%%%%%%%%%%%%%%%

\subsubsection{Results}

First, we introduce the analytical expression of the NMEE tensor in two-level systems, $H(\bm{k})=g_0(\bm{k})\sigma^0+\bm{g}(\bm{k})\cdot\bm{\sigma}$.
The band-resolved NMEE tensor is given by
\begin{align}
    \zeta^{\tau^2,\pm}_{i;jk}&=\pm\frac{e^2 \tau^2}{4\hbar}\int_{\bm{k}}g^{(2)}_{i;jk}(\bm{k};1,3,1)f_{\pm}, \label{eq:tau2_twolevel} \\
    \zeta^{\tau^0,\pm}_{i;jk}&=\mp\frac{e^2 \hbar}{4}\int_{\bm{k}}g^{(0)}_{i;jk}(\bm{k};4,15,7)f_{\pm} \label{eq:tau0_twolevel},
\end{align}
which are derived in Appendix~\ref{app:B}.
Here, $\zeta^{\tau^n,\pm}_{i;jk}$ and $f_{\pm}$ are the NMEE tensor and the Fermi distribution function for the upper ($+$) and lower ($-$) bands, respectively, and $g^{(n)}_{i;jk}(\bm{k};\alpha,\beta,\gamma)$ is given by
\begin{align}
    \label{eq:gn}
    &g^{(n)}_{i;jk}(\bm{k};\alpha,\beta,\gamma)  \notag \\
    &=\frac{1}{|\bm{g}|^{3-n}} \biggl( \partial_{k_j}\partial_{k_k}-\frac{\bm{g}\cdot\partial_{k_j}\partial_{k_k}\bm{g}}{|\bm{g}|^2}\biggr)g_i  \notag \\
    & \quad -\frac{1}{\alpha|\bm{g}|^{5-n}} \biggl[ g_i \biggl( \partial_{k_j}\bm{g}\cdot\partial_{k_k}\bm{g} -\frac{\beta(\bm{g}\cdot\partial_{k_j}\bm{g} )(\bm{g}\cdot\partial_{k_k}\bm{g})}{|\bm{g}|^2} \biggr) \notag \\
    & \quad +\gamma\biggl( \partial_{k_j} g_i (\bm{g}\cdot\partial_{k_k}\bm{g})+\partial_{k_k} g_i (\bm{g}\cdot\partial_{k_j}\bm{g}) \biggr)\biggr].
\end{align}

%%%%%%%%%%%%%%%%%%%%%%%%%%%%%%%%%%%%%%%%%%%%%%%%%%%%%%%%%%%%%%%%%%%%%%%%%%%%%%%%%%%%%%%%%%%%%%%%%%%%%%%%%%%%%%%%%%%%%%%%%%%%%%%%%%%%%%%%%%%%%%%%%%%%%%%%%%%%%%%%%%%%%%%%%%%%%%%%%%%%%%%%%%%%%%%%

Then, we derive the analytical expression of the NMEE tensor in the effective Weyl Hamiltonian $H_{\mathrm{Weyl}}(\bm{k})$ by performing the momentum integrals in Eqs.~\eqref{eq:tau2_twolevel} and~\eqref{eq:tau0_twolevel}.
Here, we assume zero temperature ($T=0$) and set the chemical potential $\mu$ to cross the lower band.
Note that the origin of the chemical potential is defined as the energy of the Weyl point.
Furthermore, changing the coordinates from $\bm{k}$ to $\bm{q}$ by Eqs.~\eqref{eq:kx->qx}$\sim$\eqref{eq:kz->qz} and taking the limit $q \ll m/t$, we expand the NMEE tensor in powers of $t^0/a^2_t$.
After the integration, the leading order term ($\sim t^0/a^2_t$) cancels out, leaving the next leading order term ($\sim t/a^3_t$) to determine the NMEE near the Weyl point.
The expression is given by
\begin{align}
    \zeta^{\tau^2}_{z;xy}&=\frac{e^2 \tau^2}{8c\hbar} \frac{t}{a^3_t} (\Lambda^2_{a_t}-|\mu|^2), \label{eq:tau2_Weyl} \\
    \zeta^{\tau^0}_{z;xy}&=\frac{e^2 \hbar}{4c} \frac{t}{a^3_t} \ln (\Lambda_{a_t}/|\mu|), \label{eq:tau0_Weyl}
\end{align}
where $c=648\sqrt{3}\pi^2$, and $\Lambda_{a_t}=2a_t \Lambda$.
Note that we introduce a cutoff $\Lambda$ because the radial integral, $\int q^2 dq$, diverges for the flat band.  
The details of the derivation are given in Appendix~\ref{app:D}.

%%%%%%%%%%%%%%%%%%%%%%%%%%%%%%%%%%%%%%%%%%%%%%%%%%%%%%%%%%%%%%%%%%%%%%%%%%%%%%%%%%%%%%%%%%%%%%%%%%%%%%%%%%%%%%%%%%%%%%%%%%%%%%%%%%%%%%%%%%%%%%%%%%%%%%%%%%%%%%%%%%%%%%%%%%%%%%%%%%%%%%%%%%%%%%%%

\subsubsection{Discussion}

Taking the limit $\mu \rightarrow 0$ in these expressions, we can qualitatively explain the behavior of the NMEE tensors near the Weyl points.
Equation~\eqref{eq:tau2_Weyl} does not exhibit divergences but takes a local maximum.
Indeed, Fig.~\ref{fig:zxy_tau2} shows such extreme points around the chemical potential corresponding to the Weyl points.
Notably, Eq.~\eqref{eq:tau0_Weyl} logarithmically diverges, which explains the large peak in Fig.~\ref{fig:zxy_tau0}.
Therefore, from Eq.~\eqref{eq:tau0_Weyl} and the discussion in Sec.~\ref{subsec:Numerical_calculation}, we conclude that the Weyl points strongly enhance the intrinsic NMEE.

%%%%%%%%%%%%%%%%%%%%%%%%%%%%%%%%%%%%%%%%%%%%%%%%%%%%%%%%%%%%%%%%%%%%%%%%%%%%%%%%%%%%%%%%%%%%%%%%%%%%%%%%%%%%%%%%%%%%%%%%%%%%%%%%%%%%%%%%%%%%%%%%%%%%%%%%%%%%%%%%%%%%%%%%%%%%%%%%%%%%%%%%%%%%%%%%
%%%%%%%%%%%%%%%%%%%%%%%%%%%%%%%%%%%%%%%%%%%%%%%%%%%%%%%%%%%%%%%%%%%%%%%%%%%%%%%%%%%%%%%%%%%%%%%%%%%%%%%%%%%%%%%%%%%%%%%%%%%%%%%%%%%%%%%%%%%%%%%%%%%%%%%%%%%%%%%%%%%%%%%%%%%%%%%%%%%%%%%%%%%%%%%%
%%%%%%%%%%%%%%%%%%%%%%%%%%%%%%%%%%%%%%%%%%%%%%%%%%%%%%%%%%%%%%%%%%%%%%%%%%%%%%%%%%%%%%%%%%%%%%%%%%%%%%%%%%%%%%%%%%%%%%%%%%%%%%%%%%%%%%%%%%%%%%%%%%%%%%%%%%%%%%%%%%%%%%%%%%%%%%%%%%%%%%%%%%%%%%%%
%Sec VI

\section{Conclusions and outlook} \label{sec:Conclusion}

In this paper, we have proposed the NMEE as a magnetic octupole response.
First, we have classified the MPGs [Table~\ref{tab:Classification}] and have found many candidates with lowest-rank magnetic octupole order [Table~\ref{tab:candidates}]. 
Then, we have derived the NMEE tensor and have confirmed the effectiveness of the NMEE through model calculations for a $d$-wave altermagnet and a pyrochlore lattice with AIAO magnetic order.
Notably, the intrinsic NMEE exhibits a large response in magnetic Weyl semimetal phases because its response tensor involves the quantum metric, which is enhanced near Weyl points.
These results demonstrate that the NMEE is capable of detecting and controlling lowest-rank magnetic octupole order, which cannot be achieved by conventional methods such as the AHE.
Furthermore, the NMEE can be one of the most promising octupole responses because of its electrical controllability and possible effectiveness in insulators.
With these unique properties, the NMEE will give a new direction for antiferromagnetic spintronics based on the perspective of magnetic octupoles. 

%%%%%%%%%%%%%%%%%%%%%%%%%%%%%%%%%%%%%%%%%%%%%%%%%%%%%%%%%%%%%%%%%%%%%%%%%%%%%%%%%%%%%%%%%%%%%%%%%%%%%%%%%%%%%%%%%%%%%%%%%%%%%%%%%%%%%%%%%%%%%%%%%%%%%%%%%%%%%%%%%%%%%%%%%%%%%%%%%%%%%%%%%%%%%%%%

Finally, we comment on the experimental realization of the NMEE in a $d$-wave altermagnet, \ce{RuO2}, and pyrochlore iridates, \ce{\textit{R}2Ir2Al20}.
As for \ce{RuO2}, its high N\'{e}el temperature ($T_{\mathrm{N}}$), which exceeds 300~K~\cite{Berlijn2017-yg,Zhu2019-fx}, allows the measurement at room temperature.
Pyrochlore iridates forms AIAO magnetic order below $T_{\mathrm{N}}$, with $T_{\mathrm{N}}$ increasing monotonically from 30~K to 150~K as the atomic number of \textit{R} elements increases~\cite{Matsuhira2007-rt,Matsuhira2011-ca,Witczak-Krempa2014-rt}.
However, its magnetic Weyl semimetal phase can only be realized within a narrow temperature window just below $T_{\mathrm{N}}$.
This is because a charge gap can easily appear due to the pair annihilation of the Weyl points. 
Nevertheless, Ref.~\cite{Ueda2018-wq} has observed this phase at a $\sim$ 4~K width in the \textit{R}=Nd compound, which suggests the possibility of observing an enhanced NMEE there.

%%%%%%%%%%%%%%%%%%%%%%%%%%%%%%%%%%%%%%%%%%%%%%%%%%%%%%%%%%%%%%%%%%%%%%%%%%%%%%%%%%%%%%%%%%%%%%%%%%%%%%%%%%%%%%%%%%%%%%%%%%%%%%%%%%%%%%%%%%%%%%%%%%%%%%%%%%%%%%%%%%%%%%%%%%%%%%%%%%%%%%%%%%%%%%%%

In addition, we estimate the magnitude of the spin density generated by the NMEE from Figs.~\ref{fig:zxx},~\ref{fig:zxy_tau2}, and~\ref{fig:zxy_tau0} by considering various magnitudes of applied electric fields.
Note that we assume room temperature ($T=300$~K) and take the relaxation time $\tau$ as 10~fs in the calculations.
For example, with a driving electric field of $E=10^5$~V/m, which is feasible in experiments~\cite{Chernyshov2009-yj}, the value is on the order of $10^{-10}$ $\sim$ $10^{-9}$~$\mu_{\mathrm{B}}/\mathrm{nm}^3$.
Furthermore, if one can apply a terahertz electric field with an intensity exceeding $E=10^7$~V/m~\cite{Olejnik2018-ix}, the value can reach $10^{-7} \sim 10^{-6}$~$\mu_{\mathrm{B}}/\mathrm{nm}^3$.  
On the other hand, spin density with a magnitude of $10^{-9} \sim 10^{-8}$~$\mu_{\mathrm{B}}/\mathrm{nm}^3$ has already been measured by using the magneto-optical Kerr effect~\cite{Stern2006-ny}.
Therefore, spin density generated by the NMEE can be detected by the same method.

%%%%%%%%%%%%%%%%%%%%%%%%%%%%%%%%%%%%%%%%%%%%%%%%%%%%%%%%%%%%%%%%%%%%%%%%%%%%%%%%%%%%%%%%%%%%%%%%%%%%%%%%%%%%%%%%%%%%%%%%%%%%%%%%%%%%%%%%%%%%%%%%%%%%%%%%%%%%%%%%%%%%%%%%%%%%%%%%%%%%%%%%%%%%%%%%
%%%%%%%%%%%%%%%%%%%%%%%%%%%%%%%%%%%%%%%%%%%%%%%%%%%%%%%%%%%%%%%%%%%%%%%%%%%%%%%%%%%%%%%%%%%%%%%%%%%%%%%%%%%%%%%%%%%%%%%%%%%%%%%%%%%%%%%%%%%%%%%%%%%%%%%%%%%%%%%%%%%%%%%%%%%%%%%%%%%%%%%%%%%%%%%%
%%%%%%%%%%%%%%%%%%%%%%%%%%%%%%%%%%%%%%%%%%%%%%%%%%%%%%%%%%%%%%%%%%%%%%%%%%%%%%%%%%%%%%%%%%%%%%%%%%%%%%%%%%%%%%%%%%%%%%%%%%%%%%%%%%%%%%%%%%%%%%%%%%%%%%%%%%%%%%%%%%%%%%%%%%%%%%%%%%%%%%%%%%%%%%%%
%Sec Acknowledgments

\section*{Acknowledgments} \label{sec:Acknowledgments} 
J.O. thanks Tomoya Higo, Kensuke Kobayashi, and Shingo Yonezawa for valuable comments from their experimental perspectives.
K.S. acknowledges support as a JSPS research fellow and is supported by JSPS KAKENHI, Grant No.22J23393 and No.22KJ2008.
R.P. is supported by JSPS KAKENHI No.~JP23K03300. 
Parts of the numerical simulations in this work have been done using the facilities of the Supercomputer Center at the Institute for Solid State Physics, the University of Tokyo.

%%%%%%%%%%%%%%%%%%%%%%%%%%%%%%%%%%%%%%%%%%%%%%%%%%%%%%%%%%%%%%%%%%%%%%%%%%%%%%%%%%%%%%%%%%%%%%%%%%%%%%%%%%%%%%%%%%%%%%%%%%%%%%%%%%%%%%%%%%%%%%%%%%%%%%%%%%%%%%%%%%%%%%%%%%%%%%%%%%%%%%%%%%%%%%%%
%%%%%%%%%%%%%%%%%%%%%%%%%%%%%%%%%%%%%%%%%%%%%%%%%%%%%%%%%%%%%%%%%%%%%%%%%%%%%%%%%%%%%%%%%%%%%%%%%%%%%%%%%%%%%%%%%%%%%%%%%%%%%%%%%%%%%%%%%%%%%%%%%%%%%%%%%%%%%%%%%%%%%%%%%%%%%%%%%%%%%%%%%%%%%%%%
%%%%%%%%%%%%%%%%%%%%%%%%%%%%%%%%%%%%%%%%%%%%%%%%%%%%%%%%%%%%%%%%%%%%%%%%%%%%%%%%%%%%%%%%%%%%%%%%%%%%%%%%%%%%%%%%%%%%%%%%%%%%%%%%%%%%%%%%%%%%%%%%%%%%%%%%%%%%%%%%%%%%%%%%%%%%%%%%%%%%%%%%%%%%%%%%

\begin{widetext}

\tabcolsep = 0.6mm
\renewcommand\arraystretch{1.3}
\setlength\LTleft{-40pt}    
\setlength\LTright{-40pt}           
\setlength\LTcapwidth{\linewidth}
\begin{longtable}{cccccc}  
\caption{Candidate AFMs that are classified as a type-III MPG in Table~\ref{tab:Classification}. 
        Compounds and their structures and N\'{e}el temperatures $T_{\mathrm{N}}$ are collected from the MAGNDATA database~\cite{MAGNDATA}.
        Collinear AFMs that allow a spin-split band are noted by a symbol * next to their names~\cite{Guo2023-yl}.}  \label{tab:candidates} \\ \hline \hline 
    MPGs & Compounds & Structure & $T_{\mathrm{N}}$ [K]  & References  \\ \hline
    \endfirsthead
    \caption*{TABLE II:~(\textit{Continued.})} \\
    \hline \hline 
    MPGs & Compounds & Structure & $T_{\mathrm{N}}$ [K]  & References  \\ \hline
    \endhead
    \hline \hline
    \endfoot
    \multicolumn{1}{c|}{\multirow{23}{*}{$mmm$}} & \ce{Mn2GeO4} & $Pnma$ (\#62) & 17 & \cite{White2012-zh} \\
    \multicolumn{1}{c|}{} & $\alpha$-\ce{Mn2O3} & $Pbca$ (\#61) & 80 & \cite{Cockayne2013-pm} \\
    \multicolumn{1}{c|}{} & \ce{CoSO4} & $Pnma$ (\#62) & - & \cite{brown1963magnetic} \\
    \multicolumn{1}{c|}{} & \ce{CoSO4} & $Pbnm$ (\#62)& 12 & \cite{Bertaut1963-oq} \\
    \multicolumn{1}{c|}{} & \ce{Rb2Fe2O(AsO4)2} & $Pnma$ (\#62) & 25 & \cite{Garlea2014-pv}  \\
    \multicolumn{1}{c|}{} & \ce{\textit{X} Fe2F6} & $Pnma$ (\#62)& 19 (\textit{X}=\ce{NH4}), 16 (\textit{X}=Rb) & \cite{ferey1985frustrated,Kim2012-hn} \\
    \multicolumn{1}{c|}{} & \ce{\textit{X}2SiO4} & $Pnma$ (\#62) & 65.3 (\textit{X}=Fe), 49 $\sim$ 49.5 (\textit{X}=Co) & \cite{Lottermoser1986-fq,Lottermoser1988-gl,Sazonov2009-dy} \\
    \multicolumn{1}{c|}{} & \ce{\textit{X} FePO5} & $Pnma$ (\#62) & 250 (\textit{X}=Fe$^*$), 178 (\textit{X}=Ni$^*$), 195 (\textit{X}=Cu$^*$) & \cite{Diffraction1992-zi,El_Khayati2001-rw} \\
    \multicolumn{1}{c|}{} & \ce{NiTe2O5} & $Pnma$ (\#62) & 30.5 & \cite{Lee2019-bn} \\
    \multicolumn{1}{c|}{} & \multirow{2}{*}{\ce{\textit{X} CrO3}} & \multirow{2}{*}{$Pnma$ (\#62)}  & 73 (\textit{X}=Sc$^*$), 93 (\textit{X}=In$^*$), 89 (\textit{X}=Tl$^*$) & \cite{Ding2017-ct} \\
    \multicolumn{1}{c|}{} & & & 290 (\textit{X}=La$^*$), 10 (\textit{X}=Sm) & \cite{Martinelli2011-vk,Tripathi2017-bu} \\
    \multicolumn{1}{c|}{} & La$_{0.75}$Bi$_{0.25}$Fe$_{1-x}$Cr$_{x}$O$_3$ & $Pnma$ (\#62) & 350 ($x$=0.5) & \cite{Ivanov2017-er} \\
    \multicolumn{1}{c|}{} & Fe$_{3-x}$Mn$_{x}$BO$_5$ & $Pbam$ (\#55) & 100 ($x$=1.5) & \cite{Damay2020-gz}\\
    \multicolumn{1}{c|}{} & \ce{Ca2RuO4} & $Pbca$ (\#61)& 110 & \cite{Porter2018-tf} \\
    \multicolumn{1}{c|}{} & \ce{MnSe2} & $Pa\bar{3}$ (\#205) & 49 & \cite{Corliss1958-vk} \\
    \multicolumn{1}{c|}{} & Nd$_{1-x}$Sr$_{x}$CrO$_3$ & $Pbnm$ (\#62) & - ($0.05 \leq x \leq 0.15$) & \cite{Chakraborty2014-zs} \\
    \multicolumn{1}{c|}{} & \ce{\textit{X} FeO3} & $Pbnm$ (\#62) & 220 (\textit{X}=Ce), 4 or 73 (\textit{X}=Dy) & \cite{Ritter2021-be,Ritter2022-cq} \\
    \multicolumn{1}{c|}{} & \ce{NdCoO3} & $Pbnm$ (\#62) & 1.20 & \cite{plaza1997neutron} \\
    \multicolumn{1}{c|}{} & \ce{\textit{X} CrO4} & $Cmcm$ (\#63) & - (\textit{X}=Co,Ni) & \cite{Pernet1969-vq} \\
    \multicolumn{1}{c|}{} & \ce{LaErO3} & $Pnma$ (\#62) & 2.4 & \cite{Martinelli2016-tv} \\
    \multicolumn{1}{c|}{} & \ce{Na2Mn(H2C3O4)2(H2O)2} & $Pbca$ (\#61) & 8 & \cite{Rousse2016-iq} \\
    \multicolumn{1}{c|}{} & \ce{TmVO3} & $Pnma$ (\#62) & 20 & \cite{Ritter2016-ye} \\
    \multicolumn{1}{c|}{} & \ce{MnTe} & $P6_3/mmc$ (\#194) & 323 & \cite{Kunitomi1964-vz} \\
    \hline
    \multicolumn{1}{c|}{\multirow{2}{*}{$4/mmm$}} & \ce{CdYb2\textit{X}4} & $Fd\bar{3}m$ (\#227) & 1.92 (\textit{X}=S), 1.75 (\textit{X}=Se) & \cite{Dalmas_de_Reotier2017-wv}  \\
    \multicolumn{1}{c|}{} & \ce{KMnF3}$^*$ & $I4/mcm$ (\#140) & 86.8 & \cite{Knight2020-rn} \\
    \hline
    \multicolumn{1}{c|}{\multirow{9}{*}{$4'/mm'm$}} & \ce{\textit{X} F2} & $P4_2/mnm$ (\#136) & 67 (\textit{X}=Mn$^*$), 39 (\textit{X}=Co$^*$) & \cite{Yamani2010-hj,Jauch2004-yr} \\
    \multicolumn{1}{c|}{} & \ce{Er2\textit{X}2O7} & $Fd\bar{3}m$ (\#227) & 1.173 (\textit{X}=Ti), 90 (\textit{X}=Ru), 0.1 (\textit{X}=Sn), 0.38 (\textit{X}=Pt) & \cite{Poole2007-bt,Taira2003-ye,Petit2017-bi,Hallas2017-kk} \\ 
    \multicolumn{1}{c|}{} & \ce{Gd2\textit{X}2O7} & $Fd\bar{3}m$ (\#227) & 1.0 (\textit{X}=Sn), 1.6 (\textit{X}=Pt) & \cite{Wills2006-oi,Welch2022-gb} \\
    \multicolumn{1}{c|}{} & \ce{\textit{X} Mn2Ge4O12} & $P4/nbm$ (\#125) & 8 (\textit{X}=Zr$^*$), 8 (\textit{X}=Ce) & \cite{Xu2017-ee,Xu2017-zt} \\
    \multicolumn{1}{c|}{} & \ce{LiFe2F6}$^*$ & $P4_2/mnm$ (\#136) & 105 & \cite{shachar1972neutron}\\
    \multicolumn{1}{c|}{} & \multirow{2}{*}{Sr$_{0.7}$\textit{X}$_{0.3}$CoO$_{3-x}$} & \multirow{2}{*}{$I4/mmm$ (\#139)} & 300 (\textit{X}=Tb, $x=0.1$), 300 (\textit{X}=Ho, $x$=0.3) & \multirow{2}{*}{\cite{Cascos2020-at}} \\
    \multicolumn{1}{c|}{} & & & 290 (\textit{X}=Er, $x$=0.2) &  \\
    \multicolumn{1}{c|}{} & \ce{RuO2}$^*$ & $P4_2/mnm$ (\#136) & > 300 & \cite{Berlijn2017-yg} \\
    \multicolumn{1}{c|}{} & \ce{CaFe4Al8} & $I4/mmm$ (\#139) & 180 & \cite{Gvozdetskyi2018-gx} \\
    \hline 
    \multicolumn{1}{c|}{\multirow{7}{*}{$\bar{3}m$}} & \ce{FeCO3}$^*$ & $R\bar{3}c$ (\#167) & 38 & \cite{ALIKHANOV1959} \\
    \multicolumn{1}{c|}{} & Mn$_3$Cu$_{1-x}$Ge$_{x}$N & $Pm\bar{3}m$ (\#221) & 380 ($x$=0.5) & \cite{Iikubo2008-ye} \\
    \multicolumn{1}{c|}{} & \ce{Mn3\textit{X} N} & $Pm\bar{3}m$ (\#221) & 183 (\textit{X}=Zn), 298 (\textit{X}=Ga) & \cite{Kren1971-bg,Fruchart1978-es} \\
    \multicolumn{1}{c|}{} & \ce{CoF3}$^*$ & $R\bar{3}c$ (\#167) & 460 & \cite{Lee2018-os} \\
    \multicolumn{1}{c|}{} & \ce{LaCrO3} & $R\bar{3}c$ (\#167) & 380 & \cite{Zhou2011-vw} \\
    \multicolumn{1}{c|}{} & \ce{Li2MnTeO6} & $P\bar{3}1c$ (\#163) & 8.5 & \cite{Zvereva2020-my} \\
    \multicolumn{1}{c|}{} & K$_{2-x}$Fe$_4$O$_{7-x}$(OH)$_{x}$ & $P\bar{3}1c$ (\#163) & - ($x$=0.38) & \cite{Albrecht2019-nh} \\
    \hline
    \multicolumn{1}{c|}{\multirow{9}{*}{$6'/m'mm'$}} & \ce{Ba5Co5ClO13} & $P6_3/mmc$ (\#194) & 110 & \cite{Mentre2008-ks} \\
    \multicolumn{1}{c|}{} & \ce{CsCoCl3} & $P6_3/mmc$ (\#194) & 20.82 $\sim$ 21.5 & \cite{Melamud1974-tq,Mekata1978-sz} \\
    \multicolumn{1}{c|}{} & Mn$_{3-x}$Ga$_{1+x}$ & $P6_3/mmc$ (\#194) & 460 ($x$=0.15) & \cite{Kren1970-np} \\
    \multicolumn{1}{c|}{} & \ce{RbCoBr3} & $P6_3/mmc$ (\#194) & 36 & \cite{Minkiewicz1971-ac} \\
    \multicolumn{1}{c|}{} & CrSb$^*$ & $P6_3/mmc$ (\#194) & > 600 & \cite{Yuan2020-nx} \\
    \multicolumn{1}{c|}{} & \ce{BaMnO3} & $P6_3/mmc$ (\#194) & 2.3 & \cite{Christensen1972-pv} \\
    \multicolumn{1}{c|}{} & \ce{CrNb4S8}$^*$ & $P6_3/mmc$ (\#194) & - & \cite{Van_Laar_undated-rm} \\
    \multicolumn{1}{c|}{} & \ce{Ba3NiRu2O9}$^*$ & $P6_3/mmc$ (\#194) & 95 & \cite{Lightfoot_undated-jw} \\
    \multicolumn{1}{c|}{} & Fe$_{0.25}$NbS$_2$ & $P6_3/mmc$ (\#194) & 150 & \cite{Lawrence2023-ky} \\
    \hline
    \multicolumn{1}{c|}{\multirow{5}{*}{$m\bar{3}$}} & \ce{NiS2} & $Pa\bar{3}$ (\#205) & 39 & \cite{Yano2016-hm} \\ 
    \multicolumn{1}{c|}{} & \ce{MnTe2} & $Pa\bar{3}$ (\#205) & 86.5 & \cite{Burlet1997-da} \\
    \multicolumn{1}{c|}{} & \ce{Na3Co(CO3)2Cl} & $Fd\bar{3}$ (\#203) & 1.5 & \cite{Fu2013-gs} \\
    \multicolumn{1}{c|}{} & \ce{Er2O3} & $Ia\bar{3}$ (\#206) & 3.4 & \cite{moon1968magnetic} \\
    \multicolumn{1}{c|}{}& (La$_{0.5}$Er$_{0.5}$)$_2$O$_3$ & $Ia\bar{3}$ (\#206) & - & \cite{Martinelli2016-tv} \\
    \hline
    \multicolumn{1}{c|}{\multirow{14}{*}{$m\bar{3}m'$}} & \ce{UO2} & $Fm\bar{3}m$ (\#225) & 30.8 & \cite{Frazer1965-iw} \\
    \multicolumn{1}{c|}{} & Fe$_{1-x}$Mn$_{x}$ & $Fm\bar{3}m$ (\#225) & - ($x$=0.3) & \cite{Umebayashi1966-qj} \\
    \multicolumn{1}{c|}{} & \ce{DyCu} & $Pm\bar{3}m$ (\#221) & 64 & \cite{Wintenberger1971-mx} \\
    \multicolumn{1}{c|}{} & \ce{NpBi} & $Fm\bar{3}m$ (\#225) & 192.5 & \cite{Burlet1992-qe} \\
    \multicolumn{1}{c|}{} & \ce{USb} & $Fm\bar{3}m$ (\#225) & 213 & \cite{Lander1995-yi} \\
    \multicolumn{1}{c|}{} & \ce{Cd2Os2O7} & $Fd\bar{3}m$ (\#227) & 225 $\sim$ 227 & \cite{Yamaura2012-zd,Calder2016-sl} \\
    \multicolumn{1}{c|}{} & \ce{Nd2\textit{X}2O7} & $Fd\bar{3}m$ (\#227) & 0.4 (\textit{X}=Zr), 0.91 (\textit{X}=Sn), 0.55 (\textit{X}=Hf) & \cite{Xu2015-dw,Bertin2015-iz,Anand2015-bv} \\
    \multicolumn{1}{c|}{} & \ce{Sm2Ti2O7} & $Fd\bar{3}m$ (\#227) & 0.35 & \cite{Mauws2018-hw} \\
    \multicolumn{1}{c|}{} & \multirow{2}{*}{\ce{\textit{X}3Ga5O12}} & \multirow{2}{*}{$Ia\bar{3}d$ (\#230)} & 0.24 $\sim$ 0.25 (\textit{X}=Tb), 0.15 (\textit{X}=Ho) & \cite{Hammann1977-sw,Wawrzynczak2019-zi} \\
    \multicolumn{1}{c|}{} & & & 0.37 (\textit{X}=Dy), 0.8 (\textit{X}=Er) & \cite{Kibalin2020-fl,Cai2019-ip} \\
    \multicolumn{1}{c|}{} & \ce{\textit{X}3Al5O12} & $Ia\bar{3}d$ (\#230) & 1.35 (\textit{X}=Tb), 0.85 (\textit{X}=Ho), 2.49 (\textit{X}=Dy) & \cite{noauthor_undated-ww,Hastings1965-fy} \\
    \multicolumn{1}{c|}{} & \ce{Nd2ScNbO7} & $Fd\bar{3}m$ (\#227) & 0.371 & \cite{Scheie2021-sh} \\
    \multicolumn{1}{c|}{} & \ce{\textit{X}2Ir2O7} & $Fd\bar{3}m$ (\#227) & 30 (\textit{X}=Nd), 120 (\textit{X}=Eu), 150 (\textit{X}=Yb) & \cite{Guo2016-is,Das2022-ms,Jacobsen2020-ey} \\
    \multicolumn{1}{c|}{} & \ce{TmGa3} & $Pm\bar{3}m$ (\#221) & 4.26 & \cite{Morin_1987-wy} \\
    \hline 
    \multicolumn{1}{c|}{\multirow{1}{*}{$6'$}} & \ce{YMnO3} & $P6_3cm$ (\#185) & 66 & \cite{Brown2006-ie} \\
    \hline
    \multicolumn{1}{c|}{\multirow{4}{*}{$6'mm'$}} & \ce{\textit{X} MnO3} & $P6_3cm$ (\#185) & 70 (\textit{X}=Ho), 85 (\textit{X}=Yb)  & \cite{Brown2006-ie,Fabreges2008-pw,Chattopadhyay2018-at} \\
    \multicolumn{1}{c|}{} & HoMn$_{1-x}$Fe$_x$O$_3$ & $P6_3cm$ (\#185) & 72 (0.0 $\leq$ $x$ $\leq$ 0.25) & \cite{Prakash2021-cp} \\
    \multicolumn{1}{c|}{} & \ce{\textit{X}_2Mn3O8} & $P6_3mc$ (\#186) & 59 (\textit{X}=Fe), 39 $\sim$ 42 (\textit{X}=Co) & \cite{Bertrand1975-dv,Tang2019-ab} \\
    \multicolumn{1}{c|}{} & Co$_6$(OH)$_3$(TeO$_3$)$_4$(OH)$_{0.9}$H$_{20}$ & $P6_3mc$ (\#186) & 75.5 & \cite{Poupon2019-kh} \\
    \hline
    \multicolumn{1}{c|}{\multirow{1}{*}{$\bar{6}m2$}} & \ce{Ba3CoSb2O9} & $P6_3/mmc$ (\#194) & 3.8 & \cite{Ma2016-la} \\
    \hline 
    \multicolumn{1}{c|}{\multirow{1}{*}{$4'32'$}} & \ce{BaCuTe2O6} & $P4_132$ (\#213) & 6.3 & \cite{Samartzis2021-as} \\
\end{longtable}
\end{widetext}

%%%%%%%%%%%%%%%%%%%%%%%%%%%%%%%%%%%%%%%%%%%%%%%%%%%%%%%%%%%%%%%%%%%%%%%%%%%%%%%%%%%%%%%%%%%%%%%%%%%%%%%%%%%%%%%%%%%%%%%%%%%%%%%%%%%%%%%%%%%%%%%%%%%%%%%%%%%%%%%%%%%%%%%%%%%%%%%%%%%%%%%%%%%%%%%%
%%%%%%%%%%%%%%%%%%%%%%%%%%%%%%%%%%%%%%%%%%%%%%%%%%%%%%%%%%%%%%%%%%%%%%%%%%%%%%%%%%%%%%%%%%%%%%%%%%%%%%%%%%%%%%%%%%%%%%%%%%%%%%%%%%%%%%%%%%%%%%%%%%%%%%%%%%%%%%%%%%%%%%%%%%%%%%%%%%%%%%%%%%%%%%%%
%%%%%%%%%%%%%%%%%%%%%%%%%%%%%%%%%%%%%%%%%%%%%%%%%%%%%%%%%%%%%%%%%%%%%%%%%%%%%%%%%%%%%%%%%%%%%%%%%%%%%%%%%%%%%%%%%%%%%%%%%%%%%%%%%%%%%%%%%%%%%%%%%%%%%%%%%%%%%%%%%%%%%%%%%%%%%%%%%%%%%%%%%%%%%%%%
\appendix

\section{REVIEW OF PREVIOUS AHE MEASUREMENTS} \label{app:F}

In this section, we review previous AHE measurements in some centrosymmetric AFMs.

%%%%%%%%%%%%%%%%%%%%%%%%%%%%%%%%%%%%%%%%%%%%%%%%%%%%%%%%%%%%%%%%%%%%%%%%%%%%%%%%%%%%%%%%%%%%%%%%%%%%%%%%%%%%%%%%%%%%%%%%%%%%%%%%%%%%%%%%%%%%%%%%%%%%%%%%%%%%%%%%%%%%%%%%%%%%%%%%%%%%%%%%%%%%%%%%
%%%%%%%%%%%%%%%%%%%%%%%%%%%%%%%%%%%%%%%%%%%%%%%%%%%%%%%%%%%%%%%%%%%%%%%%%%%%%%%%%%%%%%%%%%%%%%%%%%%%%%%%%%%%%%%%%%%%%%%%%%%%%%%%%%%%%%%%%%%%%%%%%%%%%%%%%%%%%%%%%%%%%%%%%%%%%%%%%%%%%%%%%%%%%%%%

\subsection{Altermagnets: \ce{RuO2} and \ce{MnTe}} \label{subsec:Altermagnet}

Altermagnets, \ce{RuO2} and \ce{MnTe}, do not exhibit the AHE in the ground state, and different approaches are employed to induce the AHE.
Tetragonal \ce{RuO2} belongs to a space group $P4_2/mnm$ (\#136), with the collinear \ce{Ru} spin along the [001] magnetic easy axis~\cite{Berlijn2017-yg,Zhu2019-fx}.
The magnetic structure belongs to a type-III MPG $4'/mm'm$, which excludes the AHE.
Therefore, Ref.~\cite{Feng2022-jm} applied a magnetic field to reduce the symmetry and induce the AHE.
Specifically, the magnetic field along the [110] direction induces a continuous rotation of the N\'{e}el vector towards the [110] direction [Fig.~\ref{fig:AHA1}].
This reduces the symmetry from $4'/mm'm$ to a type-I MPG $m'm'm$~\cite{Feng2022-jm,Smejkal2020-zu}, activating the AHE.
On the other hand, hexagonal \ce{MnTe} belongs to a space group $P6_3/mmc$ (\#194), and the N\'{e}el vector aligns along the [2$\bar{1}\bar{1}$0] magnetic easy axis~\cite{Kunitomi1964-vz}.
The magnetic structure belongs to a type-III MPG $mmm$, which excludes the AHE.
Therefore, Ref.~\cite{Gonzalez_Betancourt2023-qp} fabricated a thin film to reduce the symmetry and induce the AHE.
Specifically, the thin film changes the magnetic anisotropy: the magnetic easy axis is modulated from the [2$\bar{1}\bar{1}$0] direction to the [1$\bar{1}$00] direction.
This reduces the symmetry from $mmm$ to a type-I MPG $m'm'm$.
Thus, unlike \ce{RuO2}, thin-film \ce{MnTe} does not require any bias fields to induce the AHE. 

%%%%%%%%%%%%%%%%%%%%%%%%%%%%%%%%%%%%%%%%%%%%%%%%%%%%%%%%%%%%%%%%%%%%%%%%%%%%%%%%%%%%%%%%%%%%%%%%%%%%%%%%%%%%%%%%%%%%%%%%%%%%%%%%%%%%%%%%%%%%%%%%%%%%%%%%%%%%%%%%%%%%%%%%%%%%%%%%%%%%%%%%%%%%%%%%

\subsection{Noncollinear AFM : \ce{Mn3Sn}} \label{subsec:Mn3Sn}

A noncollinear AFM, \ce{Mn3Sn}, exhibits a distinctive AHE in the ground state.
This AFM has a hexagonal Ni$_3$Sn-type structure with a space group $P6_3/mmc$ (\#194) and forms an inverse triangular spin texture of the Mn magnetic moments [Fig.~\ref{fig:AHA2}]. 
The magnetic structure belongs to a type-I MPG $m'm'm$, which allows the AHE.
Furthermore, angle-resolved photoemission spectroscopy measurements have revealed the presence of Weyl points near the Fermi level~\cite{Kuroda2017-oa}.
Together with its high N\'{e}el temperature, $T_{\mathrm{N}}=420~\mathrm{K}$~\cite{Brown1990-qs}, this leads to a giant AHE comparable to ferromagnets at room temperature~\cite{Nakatsuji2015-ll}.
Therefore, \ce{Mn3Sn} is a strong candidate for antiferromagnetic spintronics~\cite{Jungwirth2016-td,Baltz2018-or}.

%%%%%%%%%%%%%%%%%%%%%%%%%%%%%%%%%%%%%%%%%%%%%%%%%%%%%%%%%%%%%%%%%%%%%%%%%%%%%%%%%%%%%%%%%%%%%%%%%%%%%%%%%%%%%%%%%%%%%%%%%%%%%%%%%%%%%%%%%%%%%%%%%%%%%%%%%%%%%%%%%%%%%%%%%%%%%%%%%%%%%%%%%%%%%%%%

\begin{figure}[t]
    \setcounter{figure}{8} 
    \begin{minipage}[t]{{0.25\columnwidth}}
        \subfigure[]{
            \includegraphics[height=2.3cm,width=1.4\hsize]{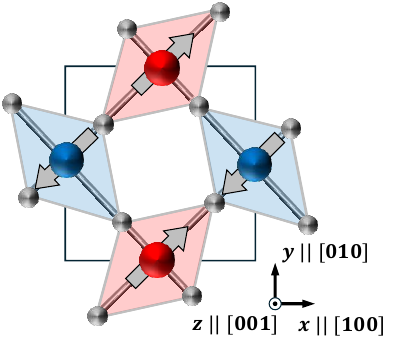}
            \label{fig:AHA1}
        }
        \vfill
        \vspace{-0.3cm}
        \setcounter{subfigure}{1} 
        \subfigure[]{
        \includegraphics[height=1.8cm,width=0.9\hsize]{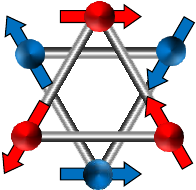}
        \label{fig:AHA2}
        }
        \setcounter{subfigure}{0} 
    \end{minipage}
    \hfill
    \begin{minipage}[t]{{0.63\columnwidth}}
        \setcounter{subfigure}{2} 
        \subfigure[]{
            \includegraphics[height=4.7cm,width=1.0\hsize]{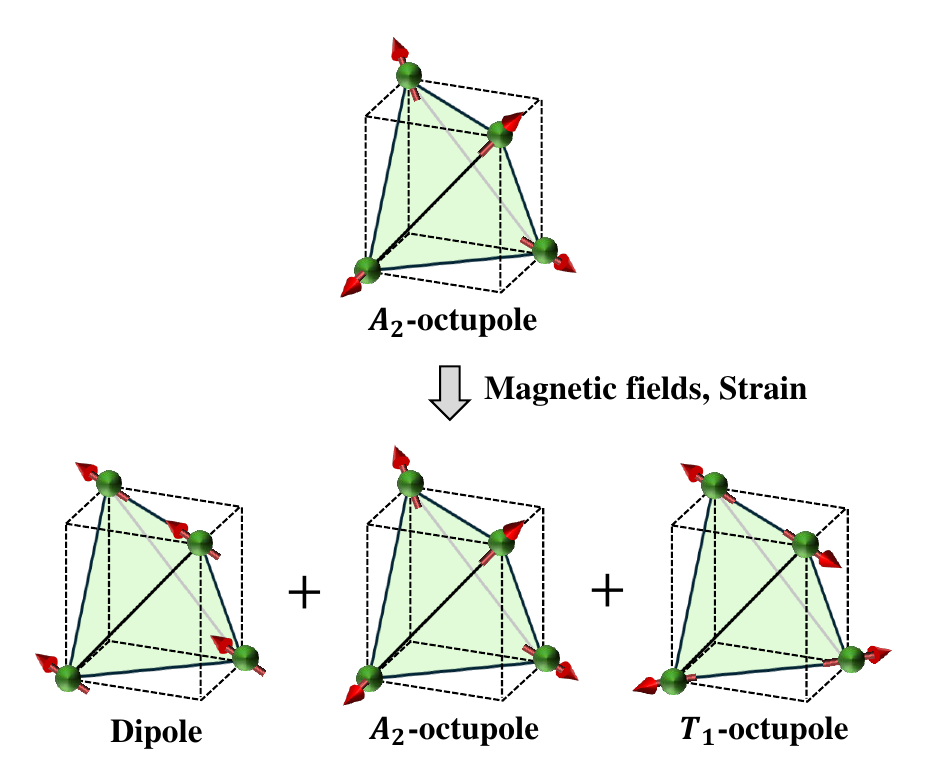}
            \label{fig:AHA3}
        } 
        \setcounter{subfigure}{1} 
    \end{minipage}
    \setcounter{figure}{7}
    \caption{(a) Crystal structure of \ce{RuO2} with the N\'{e}el vector along the [110] direction.
             The spins in the $x$-$y$ plane break the $C_{4z}\mathcal{T}$ symmetry, allowing the AHE.
             (b) Spin texture of \ce{Mn3Sn}.
             (c) Magnetic moments of \ce{\textit{R}2Ir2Al20}.
             The magnetic ground state is usually described only by an $A_2$-octupole.
             Magnetic fields or strain cause the Ir spin to reorient, leading to a superposition of a dipole, $A_2$-octupole, and $T_1$-octupole and allowing the AHE.
    }
    \label{fig:AHA}
\end{figure}

%%%%%%%%%%%%%%%%%%%%%%%%%%%%%%%%%%%%%%%%%%%%%%%%%%%%%%%%%%%%%%%%%%%%%%%%%%%%%%%%%%%%%%%%%%%%%%%%%%%%%%%%%%%%%%%%%%%%%%%%%%%%%%%%%%%%%%%%%%%%%%%%%%%%%%%%%%%%%%%%%%%%%%%%%%%%%%%%%%%%%%%%%%%%%%%%

\subsection{Noncoplanar AFM : \ce{\textit{R}2Ir2Al20} (\textit{R}=Nd, Eu) \label{subsec:Pyrochlore}}

A noncoplanar AFM, pyrochlore iridates, \ce{\textit{R}2Ir2Al20} (\textit{R}=Nd,Eu), does not exhibit the AHE in the ground state and requires further symmetry reduction for a finite AHE. 
This AFM belongs to a space group $Fd\bar{3}m$ (\#227).
In particular, the paramagnetic states of \textit{R}=Nd and \textit{R}=Eu exhibit a metallic behavior above $T_{\mathrm{N}}=33~\mathrm{K}$ for \textit{R}=Nd and $T_{\mathrm{N}}=120~\mathrm{K}$ for \textit{R}=Eu, below which AIAO magnetic order stabilizes.
The magnetic structure belongs to a type-III MPG $m\bar{3}m'$, which forbids the AHE.
Therefore, Refs.~\cite{Ueda2017-jr,Ueda2018-wq} applied a magnetic field to reduce the symmetry and induce the AHE.
This symmetry reduction and the resulting AHE can be understood from the following argument~\cite{Oh2018-xs,Kim2020-er}.
In general, AIAO magnetic order is characterized by an $A_2$-octupole, which forms the basis of the $A_2$-representation of pyrochlore iridates.
Meanwhile, it intrinsically contains a magnetic dipole and a $T_1$-octupole, which is an antiferromagnetic order distinct from AIAO magnetic order.
This $T_1$-octupole shares the irreducible representation of the magnetic dipole, which satisfies the condition to activate the AHE~\cite{Chen2014-ld,Suzuki2017-ve}.
The magnetic field induces a spin rearrangement and activates a $T_1$-octupole [Fig.~\ref{fig:AHA3}].
In the same way, Refs.~\cite{Kim2020-er,Li2021-zv,Ghosh2023-hl} applied a strain to a thin film to induce a $T_1$-octupole.

%%%%%%%%%%%%%%%%%%%%%%%%%%%%%%%%%%%%%%%%%%%%%%%%%%%%%%%%%%%%%%%%%%%%%%%%%%%%%%%%%%%%%%%%%%%%%%%%%%%%%%%%%%%%%%%%%%%%%%%%%%%%%%%%%%%%%%%%%%%%%%%%%%%%%%%%%%%%%%%%%%%%%%%%%%%%%%%%%%%%%%%%%%%%%%%%

\subsection{Chiral spin liquid : \ce{\textit{R}2Ir2Al20} (\textit{R}=Pr) \label{subsec:Pyrochlore}}

The \textit{R}=Pr compound of pyrochlore iridates exhibits an unconventional AHE, unlike the other pyrochlore iridates.
This compound uniquely remains in a paramagnetic metal state down to an extremely low temperature~\cite{Nakatsuji2006-ml}. 
In particular, the state in the temperature range of 0.3 $\leq$ $T$ $\leq$ 1.5 is a chiral spin liquid, which orders the scalar spin chirality.
This ordering is composed of a superposition of the above three multipoles (dipole, $A_2$-octupole, and $T_1$-octupole)~\cite{Kim2020-er} and symmetrically allows for the AHE~\cite{Machida2010-uf}. 

%%%%%%%%%%%%%%%%%%%%%%%%%%%%%%%%%%%%%%%%%%%%%%%%%%%%%%%%%%%%%%%%%%%%%%%%%%%%%%%%%%%%%%%%%%%%%%%%%%%%%%%%%%%%%%%%%%%%%%%%%%%%%%%%%%%%%%%%%%%%%%%%%%%%%%%%%%%%%%%%%%%%%%%%%%%%%%%%%%%%%%%%%%%%%%%%
%%%%%%%%%%%%%%%%%%%%%%%%%%%%%%%%%%%%%%%%%%%%%%%%%%%%%%%%%%%%%%%%%%%%%%%%%%%%%%%%%%%%%%%%%%%%%%%%%%%%%%%%%%%%%%%%%%%%%%%%%%%%%%%%%%%%%%%%%%%%%%%%%%%%%%%%%%%%%%%%%%%%%%%%%%%%%%%%%%%%%%%%%%%%%%%%
%%%%%%%%%%%%%%%%%%%%%%%%%%%%%%%%%%%%%%%%%%%%%%%%%%%%%%%%%%%%%%%%%%%%%%%%%%%%%%%%%%%%%%%%%%%%%%%%%%%%%%%%%%%%%%%%%%%%%%%%%%%%%%%%%%%%%%%%%%%%%%%%%%%%%%%%%%%%%%%%%%%%%%%%%%%%%%%%%%%%%%%%%%%%%%%%
%app B

\section{DERIVATION OF THE NMEE TENSOR WITH QUANTUM KINETIC THEORY} \label{app:A}

In this section, we present the details of the derivation of the NMEE tensor, which is originally defined as
\begin{align}
    \label{eq:NMEE_tensor_app}
    \zeta^{(2)}_{i;jk}= \sum_{n,m} \int_{\bm{k}} s^i_{\bm{k}nm} \rho^{(2)}_{\bm{k}mn}/E_jE_k.
\end{align}
Specifically, we first review the dynamics of the density operator in Sec.~\ref{subapp:density} and then discuss the derivation of the NMEE tensor in Sec.~\ref{subapp:tensor}.

%%%%%%%%%%%%%%%%%%%%%%%%%%%%%%%%%%%%%%%%%%%%%%%%%%%%%%%%%%%%%%%%%%%%%%%%%%%%%%%%%%%%%%%%%%%%%%%%%%%%%%%%%%%%%%%%%%%%%%%%%%%%%%%%%%%%%%%%%%%%%%%%%%%%%%%%%%%%%%%%%%%%%%%%%%%%%%%%%%%%%%%%%%%%%%%%
%%%%%%%%%%%%%%%%%%%%%%%%%%%%%%%%%%%%%%%%%%%%%%%%%%%%%%%%%%%%%%%%%%%%%%%%%%%%%%%%%%%%%%%%%%%%%%%%%%%%%%%%%%%%%%%%%%%%%%%%%%%%%%%%%%%%%%%%%%%%%%%%%%%%%%%%%%%%%%%%%%%%%%%%%%%%%%%%%%%%%%%%%%%%%%%%

\subsection{Dynamics of the density operator} \label{subapp:density}

The dynamics of the density operator $\rho_{\bm{k}}(t)$ are derived from the fact that the full density operator $\rho(t)$ obeys the von Neumann equation,
\begin{align}
    \label{eq:vN}
    i\hbar \partial_t \rho(t)=[H(t),\rho(t)].
\end{align}
Here, $\hbar$ is the Planck constant, $\partial_t=\partial/\partial t$, $[A,B]=AB-BA$, and $H(t)$ is a Hamiltonian. 
The operator $\rho_{\bm{k}}(t)$ is defined within a subspace labeled by crystal momentum $\bm{k}$, and $\rho(t)$ is written as a tensor product of $\rho_{\bm{k}}(t)$: $\rho(t)=\prod_{\bm{k}} \otimes \rho_{\bm{k}}(t)$.
The matrix representation of $\rho_{\bm{k}}(t)$ is defined as
\begin{align}
    \rho_{\bm{k}nm}(t)=\mathrm{Tr}\bigl[ \rho(t) c_{\bm{k}m}^\dagger c_{\bm{k}n}\bigr],
\end{align}
and from Eq.~\eqref{eq:vN}, its equation of motion is written as
\begin{align}
    \label{eq:eqm_RDM_app}
    i\hbar \partial_t \rho_{\bm{k}nm}(t)
    =\mathrm{Tr}\bigl[\rho(t)[c_{\bm{k}m}^\dagger c_{\bm{k}n},H(t)] \bigr].
\end{align}
Here, $c_{\bm{k}n}^\dagger$ and $c_{\bm{k}n}$ are fermionic creation and annihilation operators of a Bloch state $\ket{u_{\bm{k}n}}$ labeled by the momentum $\bm{k}$ and a band $n$.
In general, $H(t)$ is given by $H(t)=H_0+V(t)$, where $H_0$ is an unperturbed Hamiltonian, and $V(t)$ is a perturbation by an external field $\bm{F}(t)$.
These Hamiltonians are described as
\begin{align} 
    H_0&=\sum_n \int_{\bm{k}} \varepsilon_{\bm{k}n} c_{\bm{k}n}^\dagger c_{\bm{k}n}, \\
    V(t)&=\sum_{n,m} \int_{\bm{k}} c_{\bm{k}n}^\dagger V_{\bm{k}nm}(t) c_{\bm{k}m},
\end{align}
where $\int_{\bm{k}}=\int_{\mathrm{BZ}} d\bm{k}/(2\pi)^{d}$, $d$ is the dimension of the system, $\varepsilon_{\bm{k}n}$ is the eigenvalue of $H_0$, and $V_{\bm{k}nm}(t)$ is the matrix representation of $V(t)$.
Under these expressions, Eq.~\eqref{eq:eqm_RDM_app} becomes
\begin{align}
    \label{eq:eqm_RDM_perturb_app}
    (i\hbar \partial_t-\varepsilon_{\bm{k}nm}) \rho_{\bm{k}nm}(t)=[V_{\bm{k}}(t),\rho_{\bm{k}}(t)]_{nm},
\end{align}
where $\varepsilon_{\bm{k}nm}=\varepsilon_{\bm{k}n}-\varepsilon_{\bm{k}m}$, $[A_{\bm{k}},B_{\bm{k}}]_{nm}=\sum_l (A_{\bm{k}nl}B_{\bm{k}lm}-B_{\bm{k}nl}A_{\bm{k}lm})$, and we use anticommutation relations,
\begin{align}
   \{c_{\bm{k}n}, c_{\bm{k}'m}\}&=\{c_{\bm{k}n}^\dagger, c_{\bm{k}'m}^\dagger\}=0, \\
   \{c_{\bm{k}n}, c_{\bm{k}'m}^\dagger\}&=(2\pi)^{d}\delta_{nm}\delta(\bm{k}-\bm{k}').
\end{align}
In particular, we focus on the equation for the $\ell$-th order density matrix $\rho^{(\ell)}_{\bm{k}nm}(t)$,
\begin{align}
    \label{eq:eqm_RDM_l_app}
    (i\hbar \partial_t-\varepsilon_{\bm{k}nm}) \rho_{\bm{k}nm}^{(\ell)}(t)=\sum_{\lambda=0}^{\ell}[V^{(\ell-\lambda)}_{\bm{k}}(t),\rho_{\bm{k}}^{(\lambda)}(t)]_{nm},
\end{align}
where $\rho_{\bm{k}}(t)$ and $V_{\bm{k}}(t)$ are expanded in powers of $\bm{F}(t)$ as $\rho_{\bm{k}}(t)=\sum_{\ell} \rho^{(\ell)}_{\bm{k}}(t)$ and $V_{\bm{k}}(t)=\sum_{\ell} V^{(\ell)}_{\bm{k}}(t)$ with $\rho^{(\ell)}_{\bm{k}},V^{(\ell)}_{\bm{k}}=\mathcal{O}(|\bm{F}|^\ell)$.
Furthermore, we phenomenologically introduce a scattering term into Eq.~\eqref{eq:eqm_RDM_l_app}~\cite{Cheng2015-qr,Passos2018-tl,Watanabe2022-qp,Das2023-nm}:
\begin{align}
    \label{eq:eqm_RDM_scat_app}
    (i\hbar \partial_t-\varepsilon_{\bm{k}nm}) \rho_{\bm{k}nm}^{(\ell)}(t)&=\sum_{\lambda=0}^{\ell}[V^{(\ell-\lambda)}_{\bm{k}}(t),\rho_{\bm{k}}^{(\lambda)}(t)]_{nm} \notag \\
    & \quad -i\ell \eta \rho^{(\ell)}_{\bm{k}nm}(t), 
\end{align}
where $\eta$ is the scattering rate.

%%%%%%%%%%%%%%%%%%%%%%%%%%%%%%%%%%%%%%%%%%%%%%%%%%%%%%%%%%%%%%%%%%%%%%%%%%%%%%%%%%%%%%%%%%%%%%%%%%%%%%%%%%%%%%%%%%%%%%%%%%%%%%%%%%%%%%%%%%%%%%%%%%%%%%%%%%%%%%%%%%%%%%%%%%%%%%%%%%%%%%%%%%%%%%%%

To solve this equation, we first take the Fourier transformation to the frequency domain $\omega$:
\begin{align}
    \label{eq:eqm_RDM_omega_app}
    &(\hbar \omega -\varepsilon_{\bm{k}nm}+i \ell \eta) \rho^{(\ell)}_{\bm{k}nm}(\omega) \notag \\
    &= \int_{-\infty}^{\infty} d\omega_1d\omega_2 \sum_{\lambda=0}^{\ell}[V_{\bm{k}}^{(\ell-\lambda)}(\omega_1),\rho_{\bm{k}}^{(\lambda)}(\omega_2)]_{nm} \notag \\
    &\quad \times \delta(\omega,\omega_1+\omega_2).
\end{align}
Then, taking the limit $\omega_1,\omega_2 \rightarrow 0$, we can obtain the $\ell$-th order density matrix as 
\begin{align}
    \label{eq:rho_l_app_general}
    \rho^{(\ell)}_{\bm{k}nm} = \sum_{\lambda=0}^{\ell} \frac{[V_{\bm{k}}^{(\ell-\lambda)},\rho_{\bm{k}}^{(\lambda)}]_{nm}}{\varepsilon_{\bm{k}mn}+i \ell \eta}.
\end{align}
In this study, we set $V_{\bm{k}}=e\bm{r}_{\bm{k}}\cdot \bm{E}$, resulting in
\begin{align}  
    \label{eq:rho_l_app}
    \rho^{(\ell)}_{\bm{k}nm}=e \frac{[\bm{r}_{\bm{k}},\rho_{\bm{k}}^{(\ell-1)}]_{nm}}{\varepsilon_{\bm{k}mn}+i\ell\eta} \cdot \bm{E},
\end{align}
where $e=|e|$ is the charge of electrons, $\bm{r}_{\bm{k}}$ is the position operator, and $\bm{E}$ is an electric field.
In this context, the matrix presentation of $\bm{r}_{\bm{k}}$ is given by $\bm{r}_{\bm{k}nm}=i\partial_{\bm{k}}\delta_{nm}+\bm{\mathcal{A}}_{\bm{k}nm}$~\cite{Adams1959-tp,Blount1962-ku}, where $\partial_{\bm{k}}=\partial/\partial \bm{k}$, and $\bm{\mathcal{A}}_{\bm{k}nm}=i\braket{u_{\bm{k}n}|\partial_{\bm{k}} u_{\bm{k}m}}$ is the $k$-space Berry connection.

%%%%%%%%%%%%%%%%%%%%%%%%%%%%%%%%%%%%%%%%%%%%%%%%%%%%%%%%%%%%%%%%%%%%%%%%%%%%%%%%%%%%%%%%%%%%%%%%%%%%%%%%%%%%%%%%%%%%%%%%%%%%%%%%%%%%%%%%%%%%%%%%%%%%%%%%%%%%%%%%%%%%%%%%%%%%%%%%%%%%%%%%%%%%%%%%

Each order of the density matrix is obtained as follows:
Note that we will omit the $\bm{k}$-index for simplicity in the following.
The zeroth-order density matrix is defined as $\rho^{(0)}_{nm}=\delta_{nm}f_{n}$, where $f_{n}$ is the Fermi distribution function.
The first-order density matrix is given by
\begin{align}
    \rho^{(1)}_{nm}&=e \frac{[r^j,\rho^{(0)}]_{nm}}{\varepsilon_{mn}+i\eta}E_j \notag \\
    &=e\sum_l\frac{r^j_{nl}\rho^{(0)}_{lm}-\rho^{(0)}_{nl}r^j_{lm}}{\varepsilon_{mn}+i\eta}E_j \notag \\
    &=e\sum_l\frac{(i\partial_{k_j}\delta_{nl}+\mathcal{A}^j_{nl})\rho^{(0)}_{lm}-\rho^{(0)}_{nl}\mathcal{A}^j_{lm}}{\varepsilon_{mn}+i\eta}E_j \notag \\
    &=\frac{e}{\eta}\delta_{nm}\partial_{k_j}f_mE_j-\frac{e \mathcal{A}^j_{nm}f_{nm}}{\varepsilon_{mn}+i\eta}E_j,
\end{align}
where $f_{nm}=f_n-f_m$.
The first and second terms in the last line are intraband (i) and interband (e) effects, respectively.
Therefore, we symbolically describe each term as
\begin{align}
    \rho^{(1\mathrm{i})}_{nm}&=\frac{e}{\eta}\delta_{nm}\partial_{k_j}f_mE_j, \\
    \rho^{(1\mathrm{e})}_{nm}&=-\frac{e \mathcal{A}^j_{nm}f_{nm}}{\varepsilon_{mn}+i\eta}E_j.
\end{align}
Similarly, the second-order density matrix is further decomposed into intraband and interband effects as
\begin{align}
    \rho^{(2\mathrm{i})}_{nm}&=e \frac{[r^j,\rho^{(1\mathrm{i})}]_{nm}}{\varepsilon_{mn}+2i\eta}E_j=\rho^{(2\mathrm{ii})}_{nm}+\rho^{(2\mathrm{ie})}_{nm}, \\
    \rho^{(2\mathrm{e})}_{nm}&=e \frac{[r^j,\rho^{(1\mathrm{e})}]_{nm}}{\varepsilon_{mn}+2i\eta}E_j=\rho^{(2\mathrm{ei})}_{nm}+\rho^{(2\mathrm{ee})}_{nm},
\end{align}
where each term is given by
\begin{align}
    \rho^{(2\mathrm{ii})}_{nm}&=\frac{e^2}{2\eta^2}\delta_{nm}\partial_{k_j}\partial_{k_k}f_mE_jE_k, \label{eq:ii} \\
    \rho^{(2\mathrm{ie})}_{nm}&=-\frac{e^2 \mathcal{A}^j_{nm}\partial_{k_k}f_{nm} }{\eta(\varepsilon_{mn}+2i\eta)}E_jE_k, \label{eq:ie} \\
    \rho^{(2\mathrm{ei})}_{nm}&=\frac{-ie^2}{\varepsilon_{mn}+2i\eta}\partial_{k_j}\biggl( \frac{\mathcal{A}^k_{nm}f_{nm}}{\varepsilon_{mn}+i\eta} \biggr) E_jE_k, \label{eq:ei} \\
    \rho^{(2\mathrm{ee})}_{nm}&=\sum_l\frac{-e^2}{\varepsilon_{mn}+2i\eta} \notag \\
    &\quad \times\biggl(\frac{\mathcal{A}^j_{nl}\mathcal{A}^k_{lm}f_{lm}}{\varepsilon_{ml}+i\eta}-\frac{\mathcal{A}^j_{lm}\mathcal{A}^k_{nl}f_{nl}}{\varepsilon_{ln}+i\eta} \biggr)E_jE_k. \label{eq:ee}
\end{align}

%%%%%%%%%%%%%%%%%%%%%%%%%%%%%%%%%%%%%%%%%%%%%%%%%%%%%%%%%%%%%%%%%%%%%%%%%%%%%%%%%%%%%%%%%%%%%%%%%%%%%%%%%%%%%%%%%%%%%%%%%%%%%%%%%%%%%%%%%%%%%%%%%%%%%%%%%%%%%%%%%%%%%%%%%%%%%%%%%%%%%%%%%%%%%%%%
%%%%%%%%%%%%%%%%%%%%%%%%%%%%%%%%%%%%%%%%%%%%%%%%%%%%%%%%%%%%%%%%%%%%%%%%%%%%%%%%%%%%%%%%%%%%%%%%%%%%%%%%%%%%%%%%%%%%%%%%%%%%%%%%%%%%%%%%%%%%%%%%%%%%%%%%%%%%%%%%%%%%%%%%%%%%%%%%%%%%%%%%%%%%%%%%

\subsection{Derivation of the NMEE tensor} \label{subapp:tensor}

From Eq.~\eqref{eq:NMEE_tensor_app} and Eqs.~\eqref{eq:ii}$\sim$\eqref{eq:ee}, the corresponding NMEE tensors are given by 
\begin{align}
    \zeta^{(2\mathrm{ii})}_{i;jk}&=\frac{e^2}{2\eta^2} \sum_n \int_{\bm{k}}s^i_{nn}\partial_{k_j}\partial_{k_k}f_n, \\
    \zeta^{(2\mathrm{ie})}_{i;jk}&=\frac{e^2}{2\eta} \sum_{n,m} \int_{\bm{k}} \frac{s^i_{nm}\mathcal{A}^j_{mn}\partial_{k_k}f_{nm} }{\varepsilon_{nm}+2i\eta}+(j \leftrightarrow k), \\
    \zeta^{(2\mathrm{ei})}_{i;jk}&=\frac{ie^2}{2} \sum_{n,m} \int_{\bm{k}} \frac{s^i_{nm}}{\varepsilon_{nm}+2i\eta}\partial_{k_j}\biggl( \frac{\mathcal{A}^k_{mn}f_{nm}}{\varepsilon_{nm}+i\eta} \biggr)+(j \leftrightarrow k), \\
    \zeta^{(2\mathrm{ee})}_{i;jk}&=\frac{e^2}{2} \sum_{n,m,l} \int_{\bm{k} }\frac{s^i_{nm}}{\varepsilon_{nm}+2i\eta} \notag \\
    &\quad \times\biggl(\frac{\mathcal{A}^j_{ml}\mathcal{A}^k_{ln}f_{nl}}{\varepsilon_{nl}+i\eta}- \frac{\mathcal{A}^j_{ln}\mathcal{A}^k_{ml}f_{lm}}{\varepsilon_{lm}+i\eta} \biggr)+(j \leftrightarrow k),
\end{align}
where $\bm{s}_{nm}$ is the matrix representation of the spin operator.
Here, we symmetrize the indices of the electric fields, $(j,k)$, to explicitly indicate that their permutation does not affect the result.
In particular, the (2ee) term is further separated into the (2eed) term and the (2eeo) term, which correspond to the diagonal and off-diagonal parts of the spin operator, respectively:
\begin{align}   
        \zeta^{(2\mathrm{eed})}_{i;jk}&=\frac{e^2}{2} \sum_{n,m} \int_{\bm{k} }\frac{s^i_{nn}}{2i\eta}  \notag \\
        & \quad \times \biggl(\frac{\mathcal{A}^j_{nm}\mathcal{A}^k_{mn}f_{nm}}{\varepsilon_{nm}+i\eta}- \frac{\mathcal{A}^j_{mn}\mathcal{A}^k_{nm}f_{mn}}{\varepsilon_{mn}+i\eta} \biggr)+(j \leftrightarrow k) \notag \\
        &= \frac{e^2}{4i\eta} \sum_{n,m} \int_{\bm{k} } \frac{(s^i_{nn}-s^i_{mm})\mathcal{A}^j_{nm}\mathcal{A}^k_{mn}f_{nm}}{\varepsilon_{nm}+i\eta} \notag \\
        &\quad +(j \leftrightarrow k), \\
        \zeta^{(2\mathrm{eeo})}_{i;jk}&=\frac{e^2}{2} \sum_{n,m(\neq n),l} \int_{\bm{k} }\frac{s^i_{nm}}{\varepsilon_{nm}+2i\eta} \notag \\
        &\quad \times\biggl(\frac{\mathcal{A}^j_{ml}\mathcal{A}^k_{ln}f_{nl}}{\varepsilon_{nl}+i\eta}- \frac{\mathcal{A}^j_{ln}\mathcal{A}^k_{ml}f_{lm}}{\varepsilon_{lm}+i\eta} \biggr)+(j \leftrightarrow k).
\end{align}
In the following, we expand the above terms in powers of $\eta$ up to $\mathcal{O}(\eta)$ and derive each term of Eq.~\eqref{eq:symmetry_tau} by replacing $\eta$ with the relaxation time $\tau$ based on $\tau=\hbar/\eta$.
Obviously, the (2ii) term determines the $\tau^2$-response,
\begin{align}
    \zeta^{\tau^2}_{i;jk}=\frac{e^2}{2\hbar^2}\tau^2 \sum_n \int_{\bm{k}}s^i_{nn}\partial_{k_j}\partial_{k_k}f_n. 
\end{align}

%%%%%%%%%%%%%%%%%%%%%%%%%%%%%%%%%%%%%%%%%%%%%%%%%%%%%%%%%%%%%%%%%%%%%%%%%%%%%%%%%%%%%%%%%%%%%%%%%%%%%%%%%%%%%%%%%%%%%%%%%%%%%%%%%%%%%%%%%%%%%%%%%%%%%%%%%%%%%%%%%%%%%%%%%%%%%%%%%%%%%%%%%%%%%%%%

\subsubsection{2ie term}

The (2ie) term results in 
\begin{align}
    \label{eq:2ie}
    \zeta^{(2\mathrm{ie})}_{i;jk} &=\frac{e^2}{2\eta} \sum_{n,m} \int_{\bm{k}} \frac{s^i_{nm}\mathcal{A}^j_{mn}\partial_{k_k}f_{nm} }{\varepsilon_{nm}+2i\eta}+(j \leftrightarrow k) \notag \\
    &\rightarrow \frac{e^2}{2\eta} \sum_{n,m(\neq n)} \int_{\bm{k}} \frac{s^i_{nm} \mathcal{A}^j_{mn}\partial_{k_k}f_{nm} }{\varepsilon_{nm}}\biggl(1-\frac{2i\eta}{\varepsilon_{nm}} \biggr) \notag \\
    &\quad +(j \leftrightarrow k) \notag \\
    &=-\frac{ie^2}{2\eta} \sum_{n,m(\neq n)} \int_{\bm{k}} \mathfrak{A}^i_{nm} \mathcal{A}^j_{mn} \partial_{k_k}f_{nm} \notag \\
    &\quad -e^2 \sum_{n,m(\neq n)} \int_{\bm{k}} \frac{ \mathfrak{A}^i_{nm} \mathcal{A}^j_{mn} }{\varepsilon_{nm}}\partial_{k_k}f_{nm}+(j \leftrightarrow k), 
\end{align}
where we use the relation $\braket{u_n|\partial_{h_i}u_m}=s^{i}_{nm}/\varepsilon_{nm}$ in the third line [see Eq.~\eqref{eq:HF_Theorem_s}] and define the $h$-space Berry connection as $\mathfrak{A}_{nm}=i\braket{u_n|\partial_{\bm{h}}u_m}$.
Furthermore, performing $m \leftrightarrow n$ for the second subscript of $f_{nm}$ and taking a partial integral, we can write Eq.~\eqref{eq:2ie} as 
\begin{align}
    \zeta^{(2\mathrm{ie})}_{i;jk}&=\zeta^{\tau^1}_{i;jk}+\zeta^{\tau^0,\mathrm{A}}_{i;jk}, \label{eq:2ie_geometry} \\
    \zeta^{\tau^1}_{i;jk}&=\frac{e^2}{2\hbar} \tau \sum_n \int_{\bm{k}} \Bigl[ \partial_{k_j} \Upsilon^{ik}_n+\partial_{k_k}\Upsilon^{ij}_n\Bigr] f_n, \label{eq:tau1} \\
    \zeta^{\tau^0,\mathrm{A}}_{i;jk}&=e^2 \sum_n \int_{\bm{k}} \Bigl[ \partial_{k_j} \mathfrak{G}^{ik}_n+\partial_{k_k}\mathfrak{G}^{ij}_n\Bigr] f_n. \label{eq:tau0_A}
\end{align}
Equation~\eqref{eq:tau1} describes the $\tau^1$-response, and Eq.~\eqref{eq:tau0_A} forms a part of the intrinsic response tensor, the rest of which is derived in the following two subsections.
Here, we introduce the anomalous spin polarizability $\Upsilon^{ij}_n$~\cite{Xiao2023-yu}, the $h$-$k$ space quantum metric $\mathfrak{g}^{ij}_n$~\cite{Feng2024-gj}, and the $h$-space BCP $\mathfrak{G}^{ij}_n$~\cite{Xiao2022-xr}:
\begin{align}
    \Upsilon^{ij}_n&=-2\mathrm{Im}\braket{\partial_{h_i}u_n|(1-\ket{u_n}\bra{u_n})|\partial_{k_j}u_n} \notag \\
    &=-2 \sum_{m (\neq n)} \mathrm{Im} [\mathfrak{A}^i_{nm} \mathcal{A}^j_{mn}],  \label{eq:Upsilon} \\
    \mathfrak{g}^{ij}_n&=\mathrm{Re}\braket{\partial_{h_i}u_n|(1-\ket{u_n}\bra{u_n})|\partial_{k_j}u_n} \notag \\
    &=\sum_{m (\neq n)} \mathrm{Re}  [\mathfrak{A}^i_{nm} \mathcal{A}^j_{mn}] \eqqcolon \sum_{m (\neq n)} \mathfrak{g}^{ij}_{nm}, \\
    \mathfrak{G}^{ij}_n&=2\sum_{m(\neq n)}\frac{\mathfrak{g}^{ij}_{nm}}{\varepsilon_{nm}}. \label{eq:Gh}
\end{align}
These geometric quantities correspond to the Berry curvature $\Omega^{ij}_n$~\cite{simon1983holonomy,Berry1984-fn,Resta2011-da}, the $k$-space quantum metric $g^{ij}_n$~\cite{Provost1980-qb,Resta2011-da}, and the $k$-space BCP $G^{ij}_n$~\cite{Gao2014-kx}, respectively, which are given by
\begin{align}
    \Omega^{ij}_n&=-2\mathrm{Im}\braket{\partial_{k_i}u_n|(1-\ket{u_n}\bra{u_n})|\partial_{k_j}u_n} \notag \\
    &=-2 \sum_{m (\neq n)} \mathrm{Im} [\mathcal{A}^i_{nm} \mathcal{A}^j_{mn}], \label{eq:Omega}  \\
    g^{ij}_n&=\mathrm{Re}\braket{\partial_{k_i}u_n|(1-\ket{u_n}\bra{u_n})|\partial_{k_j}u_n} \notag \\
    &= \sum_{m (\neq n)} \mathrm{Re} [\mathcal{A}^i_{nm} \mathcal{A}^j_{mn}] \eqqcolon \sum_{m (\neq n)} g^{ij}_{nm},  \label{eq:gnm} \\
    G^{ij}_n&=2\sum_{m(\neq n)} \frac{g^{ij}_{nm}}{\varepsilon_{nm}}. \label{eq:Gk} 
\end{align}
Therefore, replacing $\Upsilon^{ij}_n$ with $\Omega^{ij}_n$ and multiplying a factor of $e/\hbar$ in Eq.~\eqref{eq:tau1}, we can reproduce the Berry curvature dipole contribution in the nonlinear conductivity~\cite{Watanabe2020-bs,Kaplan2023-jt,Das2023-nm,Sodemann2015-pu}, 
\begin{align}
    \sigma^{\tau^1}_{i;jk}=\frac{e^3}{2\hbar^2} \tau \sum_n \int_{\bm{k}} \Bigl[ \partial_{k_j} \Omega^{ik}_n+\partial_{k_k}\Omega^{ij}_n\Bigr] f_n. 
\end{align}

%%%%%%%%%%%%%%%%%%%%%%%%%%%%%%%%%%%%%%%%%%%%%%%%%%%%%%%%%%%%%%%%%%%%%%%%%%%%%%%%%%%%%%%%%%%%%%%%%%%%%%%%%%%%%%%%%%%%%%%%%%%%%%%%%%%%%%%%%%%%%%%%%%%%%%%%%%%%%%%%%%%%%%%%%%%%%%%%%%%%%%%%%%%%%%%%

\subsubsection{2eed term}

The (2eed) term becomes
\begin{align}   
    \label{eq:eed}
    &\zeta^{(2\mathrm{eed})}_{i;jk} \notag \\
    &= \frac{e^2}{4i\eta} \sum_{n,m} \int_{\bm{k} } \frac{(s^i_{nn}-s^i_{mm})\mathcal{A}^j_{nm}\mathcal{A}^k_{mn}f_{nm}}{\varepsilon_{nm}+i\eta}+(j \leftrightarrow k) \notag \\
    &\rightarrow \frac{e^2 }{4i\eta} \sum_{n,m(\neq n)} \int_{\bm{k} } \frac{(s^i_{nn}-s^i_{mm}) \mathcal{A}^j_{nm}\mathcal{A}^k_{mn}f_{nm}}{\varepsilon_{nm}} \notag \\
    & \quad \times\biggl(1-\frac{i\eta}{\varepsilon_{nm}}\biggr) +(j \leftrightarrow k) \notag \\
    &= \frac{e^2 }{4i\eta} \sum_{n,m(\neq n)} \int_{\bm{k} } \frac{(s^i_{nn}-s^i_{mm}) \mathcal{A}^j_{nm} \mathcal{A}^k_{mn}f_{nm}}{\varepsilon_{nm}} \notag \\
    & \quad -\frac{e^2}{4} \sum_{n,m(\neq n)} \int_{\bm{k} } \frac{(s^i_{nn}-s^i_{mm}) \mathcal{A}^j_{nm} \mathcal{A}^k_{mn}f_{nm}}{\varepsilon^2_{nm}} +(j \leftrightarrow k) \notag \\
    &=- e^2 \sum_{n,m(\neq n)} \int_{\bm{k} } \mathrm{Re}\biggl[\frac{(s^i_{nn}-s^i_{mm}) \mathcal{A}^j_{nm} \mathcal{A}^k_{mn}}{\varepsilon^2_{nm}} \biggr]f_n \notag \\
    &= -e^2 \sum_{n,m(\neq n)} \int_{\bm{k} } \mathrm{Re} [\mathcal{A}^j_{nm} \mathcal{A}^k_{mn}] \partial_{h_i}\biggl( \frac{1}{\varepsilon_{nm}}\biggr) f_n.   
\end{align}
Here, the transformations in the bottom three lines are specifically as follows:
In the third line, $n \leftrightarrow m$ is performed to the $(j \leftrightarrow k)$ terms, resulting in the cancellation of the first term and leaving twice the second term.
This remaining term leads to the fourth line after performing $m \leftrightarrow n$ for the second subscript of $f_{nm}$.
The last line results from a relation derived from Eq.~\eqref{eq:delh}, $\partial_{h_i}(1/\varepsilon_{nm})=(s^i_{nn}-s^i_{mm})/\varepsilon^2_{nm}$. 

%%%%%%%%%%%%%%%%%%%%%%%%%%%%%%%%%%%%%%%%%%%%%%%%%%%%%%%%%%%%%%%%%%%%%%%%%%%%%%%%%%%%%%%%%%%%%%%%%%%%%%%%%%%%%%%%%%%%%%%%%%%%%%%%%%%%%%%%%%%%%%%%%%%%%%%%%%%%%%%%%%%%%%%%%%%%%%%%%%%%%%%%%%%%%%%%

\subsubsection{2ei and 2eeo terms}

The (2ei) and (2eeo) terms are described as 
\begin{widetext}
    \begin{align}
        \label{eq:ei_eeo}
        &\zeta^{(2\mathrm{ei})}_{i;jk}+\zeta^{(2\mathrm{eeo})}_{i;jk} \notag \\
        &=\frac{ie^2}{2} \sum_{n,m} \int_{\bm{k}} \frac{s^i_{nm}}{\varepsilon_{nm}+2i\eta}\partial_{k_j}\biggl( \frac{\mathcal{A}^k_{mn}f_{nm}}{\varepsilon_{nm}+i\eta} \biggr)+\frac{e^2}{2} \sum_{n,m(\neq n),l} \int_{\bm{k} }\frac{s^i_{nm}}{\varepsilon_{nm}+2i\eta}\biggl(\frac{\mathcal{A}^j_{ml}\mathcal{A}^k_{ln}f_{nl}}{\varepsilon_{nl}+i\eta}- \frac{\mathcal{A}^j_{ln}\mathcal{A}^k_{ml}f_{lm}}{\varepsilon_{lm}+i\eta} \biggr)+(j \leftrightarrow k) \notag \\
        &\rightarrow\frac{ie^2}{2} \sum_{n,m(\neq n)} \int_{\bm{k}}\frac{s^i_{nm}}{\varepsilon_{nm}} \biggl\{ \partial_{k_j}\biggl( \frac{\mathcal{A}^k_{mn}f_{nm}}{\varepsilon_{nm}} \biggr) -i \biggl(\sum_{l(\neq n)}\frac{\mathcal{A}^j_{ml}\mathcal{A}^k_{ln}f_{nl}}{\varepsilon_{nl}}- \sum_{l(\neq m)}\frac{\mathcal{A}^j_{ln}\mathcal{A}^k_{ml}f_{lm}}{\varepsilon_{lm}} \biggr) \biggr\}+(j \leftrightarrow k)  \notag \\
        &=\frac{ie^2}{2} \sum_{n,m(\neq n)} \int_{\bm{k}}\frac{s^i_{nm}}{\varepsilon_{nm}} \biggl\{ \Bigl[\partial_{k_j}+i(\mathcal{A}^j_{nn}-\mathcal{A}^j_{mm}) \Bigr]\biggl( \frac{\mathcal{A}^k_{mn}f_{nm}}{\varepsilon_{nm}} \biggr) -i \sum_{l(\neq n,m)} \biggl(\frac{\mathcal{A}^j_{ml}\mathcal{A}^k_{ln}f_{nl}}{\varepsilon_{nl}}- \frac{\mathcal{A}^j_{ln}\mathcal{A}^k_{ml}f_{lm}}{\varepsilon_{lm}} \biggr) \biggr\}+(j \leftrightarrow k) \notag  \\
        &=\frac{ie^2}{2} \sum_{n,m(\neq n)} \int_{\bm{k}}  \biggl( \frac{\mathcal{A}^k_{mn}f_{nm}}{\varepsilon_{nm}} \biggr) \biggl\{ \Bigl[-\partial_{k_j}+i(\mathcal{A}^j_{nn}-\mathcal{A}^j_{mm}) \Bigr]\frac{s^i_{nm}}{\varepsilon_{nm}} -i \sum_{l(\neq n,m)} \biggl(\frac{s^i_{nl} \mathcal{A}^j_{lm}}{\varepsilon_{nl}}- \frac{\mathcal{A}^j_{nl}s^i_{lm}}{\varepsilon_{lm}} \biggr) \biggr\}+(j \leftrightarrow k) \notag \\
        &=\frac{e^2}{2} \sum_{n,m(\neq n)} \int_{\bm{k}}  \biggl( \frac{\mathcal{A}^k_{mn}f_{nm}}{\varepsilon_{nm}} \biggr) \biggl\{ -\bigl[D^j\mathfrak{A}^i\bigr]_{nm} -i \sum_{l(\neq n,m)} \bigl( \mathfrak{A}^i_{nl} \mathcal{A}^j_{lm}- \mathcal{A}^j_{nl} \mathfrak{A}^i_{lm} \bigr) \biggr\}+(j \leftrightarrow k). 
    \end{align}
\end{widetext}
Here, the transformations in the bottom three lines are specifically as follows:
In the third line, we divide it into the two-band terms and the multiband terms.
In the fourth line, we take a partial integral for the first two-band term and perform $l \leftrightarrow m$ and $l \leftrightarrow n$ for the first and second multi-band terms, respectively.
In the last line, we use $\mathfrak{A}^i_{nm}=is^i_{nm}/\varepsilon_{nm}$ and introduce 
the $k$-space covariant derivative $\bm{D}$, which acts on a physical quantity $O$ in Bloch representation as~\cite{Aversa1995-pb,Sipe2000-ne,Watanabe2021-kw}
\begin{align}
    \bigl[D^j O \bigr]_{nm}=\partial_{k_j}O_{nm}-i(\mathcal{A}^j_{nn}-\mathcal{A}^j_{mm})O_{nm}.
\end{align}

%%%%%%%%%%%%%%%%%%%%%%%%%%%%%%%%%%%%%%%%%%%%%%%%%%%%%%%%%%%%%%%%%%%%%%%%%%%%%%%%%%%%%%%%%%%%%%%%%%%%%%%%%%%%%%%%%%%%%%%%%%%%%%%%%%%%%%%%%%%%%%%%%%%%%%%%%%%%%%%%%%%%%%%%%%%%%%%%%%%%%%%%%%%%%%%%

Equation~\eqref{eq:ei_eeo} is further transformed by defining the $h$-space covariant derivative $\bm{\mathfrak{D}}$ as
\begin{align}
    \bigl[\mathfrak{D}^j O \bigr]_{nm}=\partial_{h_j}O_{nm}-i(\mathfrak{A}^j_{nn}-\mathfrak{A}^j_{mm})O_{nm}.
\end{align}
This covariant derivative satisfies the following sum rule,
\begin{align}
    \Bigl[ \mathfrak{D}^i \mathcal{A}^j \Bigr]_{nm}-\Bigl[ \mathcal{D}^j \mathfrak{A}^i \Bigr]_{nm}=i \sum_{l(\neq n,m)} \bigl( \mathfrak{A}^i_{nl} \mathcal{A}^j_{lm}- \mathcal{A}^j_{nl} \mathfrak{A}^i_{lm} \bigr),
\end{align}
which results in 
\begin{align}
    &\zeta^{(2\mathrm{ei})}_{i;jk}+\zeta^{(2\mathrm{eeo})}_{i;jk} \notag \\
    &=-\frac{e^2}{2} \sum_{n,m(\neq n)} \int_{\bm{k}}  \biggl( \frac{\mathcal{A}^k_{mn}f_{nm}}{\varepsilon_{nm}} \biggr) \bigl[\mathfrak{D}^i\mathcal{A}^j\bigr]_{nm} +(j \leftrightarrow k) \notag \\
    &=-\frac{e^2}{2} \sum_{n,m(\neq n)} \int_{\bm{k}}  \biggl( \frac{\mathcal{A}^k_{mn}f_{nm}}{\varepsilon_{nm}} \biggr) \partial_{h_i} \mathcal{A}^j_{nm} \notag \\
    &\quad +\frac{ie^2}{2} \sum_{n,m(\neq n)} \int_{\bm{k}}  \biggl( \frac{\mathcal{A}^k_{mn}f_{nm}}{\varepsilon_{nm}} \biggr) (\mathfrak{A}^i_{nn}-\mathfrak{A}^i_{mm})\mathcal{A}^j_{nm} \notag \\
    &\quad +(j \leftrightarrow k). 
\end{align}
In particular, the second term in the second line cancels with its $(j \leftrightarrow k)$ term after performing $n \leftrightarrow m$, leaving only the first term.
This remaining term leads to
\begin{align}
    \label{eq:ei_eeo_g}
    &\zeta^{(2\mathrm{ei})}_{i;jk}+\zeta^{(2\mathrm{eeo})}_{i;jk} \notag \\
    &=-\frac{e^2}{2} \sum_{n,m(\neq n)} \int_{\bm{k}}  (\mathcal{A}^k_{mn}\partial_{h_i}\mathcal{A}^j_{nm}+\mathcal{A}^j_{nm}\partial_{h_i}\mathcal{A}^k_{mn})\biggl( \frac{f_{nm}}{\varepsilon_{nm}} \biggr) \notag \\
    &=-\frac{e^2}{2} \sum_{n,m(\neq n)} \int_{\bm{k}}  \partial_{h_i}(\mathcal{A}^j_{nm} \mathcal{A}^k_{mn})\biggl( \frac{f_{nm}}{\varepsilon_{nm}} \biggr) \notag \\
    &=-e^2 \sum_{n,m(\neq n)} \int_{\bm{k}}  \partial_{h_i}\mathrm{Re} \Bigl[\mathcal{A}^j_{nm} \mathcal{A}^k_{mn} \Bigr] \biggl( \frac{f_n}{\varepsilon_{nm}} \biggr), 
\end{align}
where the third line results from performing $m \leftrightarrow n$ for the second subscript of $f_{nm}$.
From this expression and Eq.~\eqref{eq:eed}, the rest of the intrinsic response tensor is obtained as 
\begin{align}
    \label{eq:tau0_B}   \zeta^{\tau^0,\mathrm{B}}_{i;jk}&=\zeta^{(2\mathrm{eed})}_{i;jk}+\zeta^{(2\mathrm{ei})}_{i;jk}+\zeta^{(2\mathrm{eeo})}_{i;jk} \notag  \\
    &=-e^2 \sum_{n,m(\neq n)} \int_{\bm{k}}  \partial_{h_i}  \biggl( \frac{\mathrm{Re} [\mathcal{A}^j_{nm} \mathcal{A}^k_{mn} ]}{\varepsilon_{nm}} \biggr) f_n \notag \\
    &= -\frac{e^2}{2} \sum_n \int_{\bm{k}} \partial_{h_i} G^{jk}_n f_n,
\end{align}
where the last line results from Eqs.~\eqref{eq:gnm} and~\eqref{eq:Gk}.

%%%%%%%%%%%%%%%%%%%%%%%%%%%%%%%%%%%%%%%%%%%%%%%%%%%%%%%%%%%%%%%%%%%%%%%%%%%%%%%%%%%%%%%%%%%%%%%%%%%%%%%%%%%%%%%%%%%%%%%%%%%%%%%%%%%%%%%%%%%%%%%%%%%%%%%%%%%%%%%%%%%%%%%%%%%%%%%%%%%%%%%%%%%%%%%%

\subsubsection{Intrinsic response tensor}

The final expression of the intrinsic response tensor is derived from Eqs.~\eqref{eq:tau0_A} and~\eqref{eq:tau0_B}:
\begin{align}
    \zeta^{\tau^0}_{i;jk}&=\zeta^{\tau^0,\mathrm{A}}_{i;jk}+\zeta^{\tau^0,\mathrm{B}}_{i;jk} \notag \\
    &=-\frac{e^2}{2}\sum_n \int_{\bm{k}}\Bigl[ \partial_{h_i} G^{jk}_n-2\Bigl(\partial_{k_j} \mathfrak{G}^{ik}_n+\partial_{k_k}\mathfrak{G}^{ij}_n \Bigr)\Bigr]f_n.
\end{align}
Note that the analytical expression of $\partial_{h_i} G^{jk}_n$ is given by~\cite{Xiao2022-xr}
\begin{align} 
    \label{eq:delh_G}
    \partial_{h_i} G^{jk}_n&=2\hbar^2 \sum_{m(\neq n)}\mathrm{Re}\biggl[\frac{3(s^i_{nn}-s^i_{mm})v^j_{nm}v^k_{mn}}{\varepsilon^4_{nm}} \biggr]  \notag \\
    &\quad -2\hbar^2\sum_{m(\neq n)}\sum_{l(\neq n)}\mathrm{Re}\biggl[ \frac{s^i_{nl}v^j_{lm}v^k_{mn}}{\varepsilon^3_{nm}\varepsilon_{nl}}+(j \leftrightarrow k) \biggr]  \notag \\
    &\quad -2\hbar^2\sum_{m(\neq n)}\sum_{l(\neq m)}\mathrm{Re}\biggl[ \frac{s^i_{ml}v^j_{ln}v^k_{nm}}{\varepsilon^3_{nm}\varepsilon_{ml}}+(j \leftrightarrow k) \biggr].
\end{align}

%%%%%%%%%%%%%%%%%%%%%%%%%%%%%%%%%%%%%%%%%%%%%%%%%%%%%%%%%%%%%%%%%%%%%%%%%%%%%%%%%%%%%%%%%%%%%%%%%%%%%%%%%%%%%%%%%%%%%%%%%%%%%%%%%%%%%%%%%%%%%%%%%%%%%%%%%%%%%%%%%%%%%%%%%%%%%%%%%%%%%%%%%%%%%%%%
%%%%%%%%%%%%%%%%%%%%%%%%%%%%%%%%%%%%%%%%%%%%%%%%%%%%%%%%%%%%%%%%%%%%%%%%%%%%%%%%%%%%%%%%%%%%%%%%%%%%%%%%%%%%%%%%%%%%%%%%%%%%%%%%%%%%%%%%%%%%%%%%%%%%%%%%%%%%%%%%%%%%%%%%%%%%%%%%%%%%%%%%%%%%%%%%
%%%%%%%%%%%%%%%%%%%%%%%%%%%%%%%%%%%%%%%%%%%%%%%%%%%%%%%%%%%%%%%%%%%%%%%%%%%%%%%%%%%%%%%%%%%%%%%%%%%%%%%%%%%%%%%%%%%%%%%%%%%%%%%%%%%%%%%%%%%%%%%%%%%%%%%%%%%%%%%%%%%%%%%%%%%%%%%%%%%%%%%%%%%%%%%%
%app C

\section{ANALYTICAL EXPRESSION OF THE NMEE TENSOR IN TWO-LEVEL SYSTEMS} \label{app:B}

In this section, we derive the analytical expression of the NMEE tensor in two-level systems,
\begin{align}
    \label{eq:twolevels_Hamiltonian}
    H=g_{0}\sigma^0+\bm{g}\cdot \bm{\sigma}=
    \begin{pmatrix}
        g_0+g_z  & g_x- ig_y \\
        g_x+ ig_y & g_0-g_z 
    \end{pmatrix},
\end{align}
where $\sigma^0$ is the identity matrix, and $\bm{\sigma}=(\sigma^x,\sigma^y,\sigma^z)$ are the Pauli matrices.
This derivation is performed by expressing the geometric quantities comprising the NMEE tensor in the basis that diagonalizes Eq.~\eqref{eq:twolevels_Hamiltonian} after converting the Cartesian coordinates ($g_x,g_y,g_z)$ to spherical coordinates $(g,\theta,\phi)$.

%%%%%%%%%%%%%%%%%%%%%%%%%%%%%%%%%%%%%%%%%%%%%%%%%%%%%%%%%%%%%%%%%%%%%%%%%%%%%%%%%%%%%%%%%%%%%%%%%%%%%%%%%%%%%%%%%%%%%%%%%%%%%%%%%%%%%%%%%%%%%%%%%%%%%%%%%%%%%%%%%%%%%%%%%%%%%%%%%%%%%%%%%%%%%%%%

We first derive the analytical expressions of two physical quantities comprising the geometric quantities: the Pauli matrices $\bm{\sigma}$ and a $\bm{k}$-derivative of the Hamiltonian $\partial_{\bm{k}}H$.
Equation~\eqref{eq:twolevels_Hamiltonian} is diagonalized as
\begin{align}
    \label{eq:lambda}
    U^{-1}HU=\mathrm{diag}(\varepsilon_+,\varepsilon_-)=\mathrm{diag}(g_0+g,g_0-g),
\end{align}
where ``diag" denotes a diagonal matrix, $\varepsilon_{\pm}$ is the energy eigenvalue for the upper ($+$) and lower ($-$) bands, $g=|\bm{g}|$, and $U$ is a unitary matrix,
\begin{align}
    U=\frac{1}{\sqrt{2g(g+g_z)}}
    \begin{pmatrix}
        g_z + g & g_x- ig_y \\
        g_x+ ig_y & -g_z-g 
    \end{pmatrix}.
\end{align}
In the spherical coordinates, $(g_x,g_y,g_z)=g(\sin\theta\cos\phi,\sin\theta\sin\phi,\cos\theta)$, the Hamiltonian and unitary matrix become
\begin{align}
    H&=
    \begin{pmatrix}
        g_0+g\cos\theta & ge^{-i\phi}\sin\theta \\
        ge^{ i\phi}\sin\theta & g_0-g\cos\theta
    \end{pmatrix}, \\
    U&=
    \begin{pmatrix}
        \cos\frac{\theta}{2} & e^{-i\phi} \sin\frac{\theta}{2} \\
        e^{i\phi}\sin\frac{\theta}{2} & -\cos\frac{\theta}{2}
    \end{pmatrix}.
\end{align}
The unitary transformation of a physical quantity $A$ by this unitary matrix yields
\begin{align}
    U^{-1}AU&=U^{-1}
    \begin{pmatrix}
        A^{\mathrm{d}}_+ & A^{\mathrm{o}}_+ \\
        A^{\mathrm{o}}_- & A^{\mathrm{d}}_-
    \end{pmatrix}
    U \notag \\
    &=
    \begin{pmatrix}
        (U^{-1}AU)^{\mathrm{d}}_+ & (U^{-1}AU)^{\mathrm{o}}_+ \\
        (U^{-1}AU)^{\mathrm{o}}_- & (U^{-1}AU)^{\mathrm{d}}_-
    \end{pmatrix},
\end{align}
where each component is given by 
\begin{align}
    (U^{-1}AU)^{\mathrm{d}}_{\pm}&=\cos^2\frac{\theta}{2}A^{\mathrm{d}}_{\pm} \pm \frac{1}{2}e^{\pm i\phi}\sin\theta A^{\mathrm{o}}_{\pm} \notag \\
    & \quad \pm \frac{1}{2}e^{\mp i\phi} \sin\theta A^{\mathrm{o}}_{\mp}+\sin^2\frac{\theta}{2}A^{\mathrm{d}}_{\mp}, \\
    (U^{-1}AU)^{\mathrm{o}}_{\pm}&=\pm\frac{1}{2}e^{\mp i\phi}\sin\theta A^{\mathrm{d}}_{\pm} -\cos^2\frac{\theta}{2}A^{\mathrm{o}}_{\pm} \notag \\
    & \quad +e^{\mp 2i\phi}\sin^2\frac{\theta}{2}A^{\mathrm{o}}_{\mp}\mp \frac{1}{2}e^{\mp i\phi}\sin\theta A^{\mathrm{d}}_{\mp}.
\end{align}
Here, we apply the above expressions to the Pauli matrices $\bm{\sigma}$ and the $\bm{k}$-derivative of the Hamiltonian,
\begin{align}
    \partial_{\bm{k}}H^{\mathrm{d}}_{\pm}&=\partial_{\bm{k}}g_0\pm (\cos\theta\partial_{\bm{k}}g-g\sin\theta\partial_{\bm{k}}\theta), \\
    \partial_{\bm{k}}H^{\mathrm{o}}_{\pm}&=e^{\mp i\phi}(\sin\theta\partial_{\bm{k}}g+g\cos\theta\partial_{\bm{k}}\theta\mp ig\sin\theta\partial_{\bm{k}}\phi).
\end{align}
Specifically, the unitary transformation of the Pauli matrices is written as
\begin{align}
    (U^{-1}\sigma^iU)^{\mathrm{d}}_{\pm}&=\pm\frac{g_i}{|\bm{g}|}, \label{eq:U-1sigmaU} \\ 
    (U^{-1}\sigma^xU)^{\mathrm{o}}_{\pm}&=-\cos^2\frac{\theta}{2}+e^{\mp 2i\phi}\sin^2\frac{\theta}{2}, \label{eq:sigma_x} \\
    (U^{-1}\sigma^yU)^{\mathrm{o}}_{\pm}&=\pm i \biggl(\cos^2\frac{\theta}{2}+e^{\mp 2i\phi}\sin^2\frac{\theta}{2} \biggr), \label{eq:sigma_y} \\
    (U^{-1}\sigma^zU)^{\mathrm{o}}_{\pm}&=e^{\mp i\phi}\sin\theta. \label{eq:sigma_z}
\end{align}
Similarly, the unitary transformation of the $\bm{k}$-derivative of the Hamiltonian is given by
\begin{align}
    (U^{-1}\partial_{\bm{k}}HU)^{\mathrm{d}}_{\pm}&=\partial_{\bm{k}}(g_0\pm g), \label{eq:U-1partial_kU} \\
    (U^{-1}\partial_{\bm{k}}H U)^{\mathrm{o}}_{\pm}&=-ge^{\mp i\phi}(\partial_{\bm{k}}\theta\mp i\sin\theta\partial_{\bm{k}}\phi). \label{eq:partial_k}
\end{align}
Equations~\eqref{eq:U-1sigmaU}$\sim$\eqref{eq:partial_k} lead to the analytical expressions of the geometric quantities.
In the following, we first introduce these expressions in Sec.~\ref{subsec:twolevels_geo} and then derive the analytical expression of the NMEE tensor in Sec.~\ref{subsec:twolevels_tensor}.

%%%%%%%%%%%%%%%%%%%%%%%%%%%%%%%%%%%%%%%%%%%%%%%%%%%%%%%%%%%%%%%%%%%%%%%%%%%%%%%%%%%%%%%%%%%%%%%%%%%%%%%%%%%%%%%%%%%%%%%%%%%%%%%%%%%%%%%%%%%%%%%%%%%%%%%%%%%%%%%%%%%%%%%%%%%%%%%%%%%%%%%%%%%%%%%%
%%%%%%%%%%%%%%%%%%%%%%%%%%%%%%%%%%%%%%%%%%%%%%%%%%%%%%%%%%%%%%%%%%%%%%%%%%%%%%%%%%%%%%%%%%%%%%%%%%%%%%%%%%%%%%%%%%%%%%%%%%%%%%%%%%%%%%%%%%%%%%%%%%%%%%%%%%%%%%%%%%%%%%%%%%%%%%%%%%%%%%%%%%%%%%%%

\subsection{Geometric quantities in two-level systems} \label{subsec:twolevels_geo}

The analytical expressions of the geometric quantities are obtained by rewriting Eqs.~\eqref{eq:Upsilon}$\sim$\eqref{eq:Gk} based on identities derived from Eqs.~\eqref{eq:HF_Theorem} and~\eqref{eq:HF_Theorem_s}, $\mathcal{A}^i_{nm}=-i\hbar v^i_{nm}/\varepsilon_{nm}$ and $\mathfrak{A}^i_{nm}=is^i_{nm}/\varepsilon_{nm}$.
The rewritten expressions are given by
\begin{align}
    \Omega^{ij}_n&=-2 \hbar^2 \sum_{m(\neq n)} \mathrm{Im}\biggl[ \frac{v^i_{nm} v^j_{mn}}{\varepsilon^2_{nm}} \biggr], \\
    \Upsilon^{ij}_n&=2 \hbar \sum_{m(\neq n)} \mathrm{Im}\biggl[ \frac{s^i_{nm} v^j_{mn}}{\varepsilon^2_{nm}} \biggr], \\
    G^{ij}_n&=2 \hbar^2 \sum_{m(\neq n)} \mathrm{Re}\biggl[ \frac{v^i_{nm} v^j_{mn}}{\varepsilon^3_{nm}} \biggr],\\
    \mathfrak{G}^{ij}_n&=-2 \hbar \sum_{m(\neq n)} \mathrm{Re}\biggl[ \frac{s^i_{nm} v^j_{mn}}{\varepsilon^3_{nm}} \biggr],
\end{align}
which lead to the analytical expressions in two-level systems,
\begin{align}
    \Omega^{ij}_{\pm}&=-2\mathrm{Im}\biggl[\frac{(U^{-1}\partial_{k_i}HU)^{\mathrm{o}}_{\pm} (U^{-1}\partial_{k_j}HU)^{\mathrm{o}}_{\mp}}{(\varepsilon_{\pm}-\varepsilon_{\mp})^2}\biggr] \notag \\
    &=\mp \frac{1}{2|\bm{g}|^3}\bm{g}\cdot(\partial_{k_i} \bm{g}\times\partial_{k_j} \bm{g}), \\   \Upsilon^{ij}_{\pm}&=\hbar\mathrm{Im}\biggl[\frac{(U^{-1}\sigma^iU)^{\mathrm{o}}_{\pm} (U^{-1}\partial_{k_j}HU)^{\mathrm{o}}_{\mp}}{(\varepsilon_{\pm}-\varepsilon_{\mp})^2}\biggr] \notag \\
    &=\mp\frac{\hbar}{4|\bm{g}|^3}(\bm{g}\times\partial_{k_j}\bm{g})_i, \\   G^{ij}_{\pm}&=2\mathrm{Re}\biggl[\frac{(U^{-1}\partial_{k_i}HU)^{\mathrm{o}}_{\pm} (U^{-1}\partial_{k_j}HU)^{\mathrm{o}}_{\mp}}{(\varepsilon_{\pm}-\varepsilon_{\mp})^3}\biggr] \notag \\
    &=\pm\frac{1}{4|\bm{g}|^3} \biggl( \partial_{k_i}\bm{g}\cdot\partial_{k_j}\bm{g}-\frac{(\bm{g}\cdot\partial_{k_i}\bm{g} )(\bm{g}\cdot\partial_{k_j}\bm{g})}{|\bm{g}|^2}\biggr), \\
    \label{eq:hG_tls}
    \mathfrak{G}^{ij}_{\pm}&=-\hbar\mathrm{Re}\biggl[\frac{(U^{-1}\sigma^iU)^{\mathrm{o}}_{\pm} (U^{-1}\partial_{k_j}HU)^{\mathrm{o}}_{\mp}}{(\varepsilon_{\pm}-\varepsilon_{\mp})^3}\biggr] \notag \\
    &=\mp\frac{\hbar}{8|\bm{g}|^3} \biggl( \partial_{k_j} -\frac{\bm{g}\cdot\partial_{k_j}\bm{g}}{|\bm{g}|^2} \biggr) g_i, 
\end{align}
where $s^i=(\hbar/2)\sigma^i$, and $v^i=\hbar^{-1}\partial_{k_i}H$.
Here, we symbolically denote the quantities for the upper band as ``$+$" and for the lower band as ``$-$", and use the following identities: $\partial_{k_i}g=(\bm{g}\cdot\partial_{k_i}\bm{g})/|\bm{g}|$, 
\begin{align}
    &\partial_{k_i}\bm{g}-\frac{\bm{g}\cdot\partial_{k_i}\bm{g}}{|\bm{g}|^2}\bm{g}=g
    \begin{pmatrix}
        \cos\theta\cos\phi\partial_{k_i}\theta-\sin\theta\sin\phi\partial_{k_i}\phi \\
        \cos\theta\sin\phi\partial_{k_i}\theta+\sin\theta\cos\phi\partial_{k_i}\phi \\
        -\sin\theta\partial_{k_i}\theta
    \end{pmatrix}, \\
    &\bm{g}\times \partial_{k_i}\bm{g}=g^2
    \begin{pmatrix}
        -\sin\phi\partial_{k_i}\theta-\sin\theta\cos\theta\cos\phi\partial_{k_i}\phi \\
        \cos\phi\partial_{k_i}\theta-\sin\theta\cos\theta\sin\phi\partial_{k_i}\phi \\
        \sin^2\theta\partial_{k_i}\phi
    \end{pmatrix}, \\
    &\partial_{k_i}\bm{g}\cdot\partial_{k_j}\bm{g}-\frac{(\bm{g}\cdot\partial_{k_i}\bm{g} )(\bm{g}\cdot\partial_{k_j}\bm{g})}{|\bm{g}|^2} \notag  \\
    &=g^2(\partial_{k_i}\theta\partial_{k_j}\theta+\sin^2\theta\partial_{k_i}\phi\partial_{k_j}\phi), \\
    &\bm{g}\cdot(\partial_{k_i}\bm{g}\times\partial_{k_j}\bm{g})=g^3\sin\theta(\partial_{k_i}\theta\partial_{k_j}\phi-\partial_{k_i}\phi\partial_{k_j}\theta). 
\end{align}

%%%%%%%%%%%%%%%%%%%%%%%%%%%%%%%%%%%%%%%%%%%%%%%%%%%%%%%%%%%%%%%%%%%%%%%%%%%%%%%%%%%%%%%%%%%%%%%%%%%%%%%%%%%%%%%%%%%%%%%%%%%%%%%%%%%%%%%%%%%%%%%%%%%%%%%%%%%%%%%%%%%%%%%%%%%%%%%%%%%%%%%%%%%%%%%%
%%%%%%%%%%%%%%%%%%%%%%%%%%%%%%%%%%%%%%%%%%%%%%%%%%%%%%%%%%%%%%%%%%%%%%%%%%%%%%%%%%%%%%%%%%%%%%%%%%%%%%%%%%%%%%%%%%%%%%%%%%%%%%%%%%%%%%%%%%%%%%%%%%%%%%%%%%%%%%%%%%%%%%%%%%%%%%%%%%%%%%%%%%%%%%%%

\subsection{NMEE tensor in two-level systems} \label{subsec:twolevels_tensor}

The final expression of the NMEE tensor is obtained by deriving the analytical expressions of derivatives of the physical quantities that compose Eqs.~\eqref{eq:tau2} and~\eqref{eq:tau0}.
Note that here we only focus on the $\tau^2$- and $\tau^0$-responses, but a similar expression can be obtained for the $\tau^1$-response.
From Eqs.~\eqref{eq:hG_tls} and~\eqref{eq:delh_G}, the $\bm{k}$-derivative of the $h$-space BCP and the $\bm{h}$-derivative of the $k$-space BCP are given by
\begin{widetext}
\begin{align}
    \partial_{k_j}\mathfrak{G}^{ik}_{\pm}&=
     \pm\hbar\frac{3(\bm{g}\cdot\partial_{k_j}\bm{g})}{8|\bm{g}|^5} \biggl( \partial_{k_k} g_i-\frac{\bm{g}\cdot\partial_{k_k}\bm{g}}{|\bm{g}|^2} g_i \biggr) \notag \\
    &\quad \mp\frac{\hbar}{8|\bm{g}|^3} \biggl( \partial_{k_j} \partial_{k_k} g_i-\frac{\partial_{k_j}\bm{g}\cdot\partial_{k_k}\bm{g}+\bm{g}\cdot\partial_{k_j}\partial_{k_k}\bm{g}}{|\bm{g}|^2} g_i 
    -\frac{\bm{g}\cdot\partial_{k_k}\bm{g}}{|\bm{g}|^2} \partial_{k_j}g_i +\frac{2(\bm{g}\cdot\partial_{k_j}\bm{g} )(\bm{g}\cdot\partial_{k_k}\bm{g})}{|\bm{g}|^4}g_i \biggr) \notag \\
    &= \mp \frac{\hbar}{8|\bm{g}|^3} \biggl( \partial_{k_j}\partial_{k_k}-\frac{\bm{g}\cdot\partial_{k_j}\partial_{k_k}\bm{g}}{|\bm{g}|^2}\biggr)g_i \notag \\
    &\quad \pm \frac{\hbar}{8|\bm{g}|^5} \biggl[ g_i \biggl( \partial_{k_j}\bm{g}\cdot\partial_{k_k}\bm{g} -\frac{5(\bm{g}\cdot\partial_{k_j}\bm{g} )(\bm{g}\cdot\partial_{k_k}\bm{g})}{|\bm{g}|^2} \biggr) + \biggl( \partial_{k_j} g_i (\bm{g}\cdot\partial_{k_k}\bm{g})+3\partial_{k_k} g_i (\bm{g}\cdot\partial_{k_j}\bm{g}) \biggr)\biggr], \\
    \partial_{h_i} G^{jk}_{\pm}&=\hbar\mathrm{Re}\biggl[\frac{3\bigl[(U^{-1}\sigma^iU)^{\mathrm{d}}_{\pm}-(U^{-1}\sigma^iU)^{\mathrm{d}}_{\mp}\bigr](U^{-1}\partial_{k_j}HU)^{\mathrm{o}}_{\pm}(U^{-1}\partial_{k_k}HU)^{\mathrm{o}}_{\mp}}{(\varepsilon_{\pm}-\varepsilon_{\mp})^4} \biggr] \notag \\
    &\quad -\hbar\mathrm{Re}\biggl[ \frac{(U^{-1}\sigma^iU)^{\mathrm{o}}_{\pm}(U^{-1}\partial_{k_j}HU)^{\mathrm{d}}_{\mp}(U^{-1}\partial_{k_k}HU)^{\mathrm{o}}_{\mp}}{(\varepsilon_{\pm}-\varepsilon_{\mp})^4}-\frac{(U^{-1}\sigma^iU)^{\mathrm{o}}_{\mp}(U^{-1}\partial_{k_j}HU)^{\mathrm{d}}_{\pm}(U^{-1}\partial_{k_k}HU)^{\mathrm{o}}_{\pm}}{(\varepsilon_{\pm}-\varepsilon_{\mp})^4}+(j \leftrightarrow k) \biggr] \notag \\
    &= \pm \hbar \frac{6g_i}{16|\bm{g}|^5} \biggl( \partial_{k_j}\bm{g}\cdot\partial_{k_k}\bm{g}-\frac{(\bm{g}\cdot\partial_{k_j}\bm{g} )(\bm{g}\cdot\partial_{k_k}\bm{g})}{|\bm{g}|^2} \biggr)
    \pm\hbar \frac{2}{16|\bm{g}|^4} \biggl[ \frac{\bm{g}\cdot\partial_{k_j}\bm{g}}{|\bm{g}|} \biggl( \partial_{k_k} g_i-\frac{(\bm{g}\cdot\partial_{k_k}\bm{g})}{|\bm{g}|^2} g_i \biggr) +(j \leftrightarrow k) \biggr] \notag \\
    &=\pm \frac{\hbar}{8|\bm{g}|^5} \biggl[ g_i \biggl( 3 \partial_{k_j}\bm{g}\cdot\partial_{k_k}\bm{g} -\frac{5(\bm{g}\cdot\partial_{k_j}\bm{g} )(\bm{g}\cdot\partial_{k_k}\bm{g})}{|\bm{g}|^2} \biggr)+ \biggl( \partial_{k_j} g_i (\bm{g}\cdot\partial_{k_k}\bm{g})+(j \leftrightarrow k) \biggr)\biggr],
\end{align}
\end{widetext}
which result in 
\begin{align}
    &\partial_{h_i} G^{jk}_{\pm}-2\Bigl(\partial_{k_j} \mathfrak{G}^{ik}_{\pm}+\partial_{k_k}\mathfrak{G}^{ij}_{\pm} \Bigr) \notag \\
    &=\pm \frac{\hbar}{2|\bm{g}|^3} \biggl( \partial_{k_j}\partial_{k_k}-\frac{\bm{g}\cdot\partial_{k_j}\partial_{k_k}\bm{g}}{|\bm{g}|^2}\biggr)g_i \notag \\
    &\quad \mp \frac{\hbar}{8|\bm{g}|^5} \biggl[ g_i \biggl( \partial_{k_j}\bm{g}\cdot\partial_{k_k}\bm{g} -\frac{15(\bm{g}\cdot\partial_{k_j}\bm{g} )(\bm{g}\cdot\partial_{k_k}\bm{g})}{|\bm{g}|^2} \biggr) \notag  \\
    &\quad +7\biggl( \partial_{k_j} g_i (\bm{g}\cdot\partial_{k_k}\bm{g})+\partial_{k_k} g_i (\bm{g}\cdot\partial_{k_j}\bm{g}) \biggr)\biggr]. 
\end{align}
Furthermore, the $\bm{k}$-derivative of Eq.~\eqref{eq:U-1sigmaU} is described~as
\begin{align}
    &\partial_{k_j}\partial_{k_k}\biggl( \pm \frac{g_i}{|\bm{g}|} \biggr) \notag \\
    &=\pm\frac{1}{|\bm{g}|} \biggl( \partial_{k_j}\partial_{k_k}-\frac{\bm{g}\cdot\partial_{k_j}\partial_{k_k}\bm{g}}{|\bm{g}|^2}\biggr)g_i \notag \\
    &\quad \mp\frac{1}{|\bm{g}|^3} \biggl[ g_i \biggl( \partial_{k_j}\bm{g}\cdot\partial_{k_k}\bm{g} -\frac{3(\bm{g}\cdot\partial_{k_j}\bm{g} )(\bm{g}\cdot\partial_{k_k}\bm{g})}{|\bm{g}|^2} \biggr) \notag \\
    &\quad +\biggl( \partial_{k_j} g_i (\bm{g}\cdot\partial_{k_k}\bm{g})+\partial_{k_k} g_i (\bm{g}\cdot\partial_{k_j}\bm{g}) \biggr)\biggr]. 
\end{align}
From these expressions, we can derive the analytical expression of the NMEE tensor in two-level systems, as shown in Eqs.~\eqref{eq:tau2_twolevel} and~\eqref{eq:tau0_twolevel}.

%%%%%%%%%%%%%%%%%%%%%%%%%%%%%%%%%%%%%%%%%%%%%%%%%%%%%%%%%%%%%%%%%%%%%%%%%%%%%%%%%%%%%%%%%%%%%%%%%%%%%%%%%%%%%%%%%%%%%%%%%%%%%%%%%%%%%%%%%%%%%%%%%%%%%%%%%%%%%%%%%%%%%%%%%%%%%%%%%%%%%%%%%%%%%%%%
%%%%%%%%%%%%%%%%%%%%%%%%%%%%%%%%%%%%%%%%%%%%%%%%%%%%%%%%%%%%%%%%%%%%%%%%%%%%%%%%%%%%%%%%%%%%%%%%%%%%%%%%%%%%%%%%%%%%%%%%%%%%%%%%%%%%%%%%%%%%%%%%%%%%%%%%%%%%%%%%%%%%%%%%%%%%%%%%%%%%%%%%%%%%%%%%
%%%%%%%%%%%%%%%%%%%%%%%%%%%%%%%%%%%%%%%%%%%%%%%%%%%%%%%%%%%%%%%%%%%%%%%%%%%%%%%%%%%%%%%%%%%%%%%%%%%%%%%%%%%%%%%%%%%%%%%%%%%%%%%%%%%%%%%%%%%%%%%%%%%%%%%%%%%%%%%%%%%%%%%%%%%%%%%%%%%%%%%%%%%%%%%%
%app D

\section{ANALYTICAL EXPRESSION OF THE NMEE TENSOR IN THE EFFECTIVE WEYL HAMILTONIAN} \label{app:D}

In this section, we derive the analytical expression of the NMEE tensor in the effective Weyl Hamiltonian, $H_{\mathrm{Weyl}}(\bm{k})=\bm{g}(\bm{k})\cdot\bm{\sigma}$ with
\begin{align}
  g_x(\bm{k})&=a_t ( k_x+k_y-2k_z)+\frac{t}{3}(k^2_x+k^2_y-2k^2_z), \label{eq:gx_k_redefine} \\
  g_y(\bm{k})&=-\sqrt{3}a_t(k_x-k_y)-\frac{t}{\sqrt{3}}(k^2_x-k^2_y), \label{eq:gy_k_redefine} \\
  g_z(\bm{k})&=-2a_t(k_x+k_y+k_z)-\frac{2t}{3}(k_xk_y+k_yk_z+k_zk_x), \label{eq:gz_k_redefine}
\end{align}
where $a_t=(2t/3)\sqrt{m/2t}$.
Specifically, we first express the $\bm{k}$-resolved NMEE tensor of Eq.~\eqref{eq:gn} in spherical coordinates of the $\bm{q}$-space defined by Eqs.~\eqref{eq:kx->qx}$\sim$\eqref{eq:kz->qz}.
Then, we expand it by taking the limit $ q \ll m/t $ and perform the $\bm{q}$-integral.
The details of each step are presented in Secs.~\ref{subsec:EffWeyl_1} and~\ref{subsec:EffWeyl_2}.
Note that Eqs.~\eqref{eq:gx_k_redefine}$\sim$\eqref{eq:gz_k_redefine} are derived by redefining the momentum around the Weyl point in Eq.~\eqref{eq:Hamiltonian_Weyl_1} after ignoring the $g_0(\bm{k})$ term ($\propto \sigma^0$).

%%%%%%%%%%%%%%%%%%%%%%%%%%%%%%%%%%%%%%%%%%%%%%%%%%%%%%%%%%%%%%%%%%%%%%%%%%%%%%%%%%%%%%%%%%%%%%%%%%%%%%%%%%%%%%%%%%%%%%%%%%%%%%%%%%%%%%%%%%%%%%%%%%%%%%%%%%%%%%%%%%%%%%%%%%%%%%%%%%%%%%%%%%%%%%%%

\subsection{Expression of the momentum-resolved NMEE tensor} \label{subsec:EffWeyl_1}

First, we rewrite $\bm{g}(\bm{k})$ and its $\bm{k}$-derivatives involved in the $\bm{k}$-resolved NMEE tensor in terms of the $\bm{q}$-space.
The $\bm{k}$-derivatives of $\bm{g}(\bm{k})$ are given by
\begin{align}
    \partial_{k_x/k_y}g_x(\bm{k})&=a_t+\frac{2t}{3}k_x/k_y, \\
    \partial_{k_x/k_y}g_y(\bm{k})&=\mp\sqrt{3}a_t \mp \frac{2t}{\sqrt{3}}k_x/k_y, \\
    \partial_{k_x/k_y}g_z(\bm{k})&=-2a_t-\frac{2t}{3}(k_y/k_x+k_z), \\
    \partial_{k_x}\partial_{k_y}g_x(\bm{k})&=\partial_{k_x}\partial_{k_y}g_y(\bm{k})=0, \\
    \partial_{k_x}\partial_{k_y}g_z(\bm{k})&=-\frac{2t}{3}, 
\end{align}
where $\partial_{k_x}$ and $\partial_{k_y}$ take the upper and lower signs of $\mp$, respectively.
The transformation of these expressions to the $\bm{q}$-space results in
\begin{align}
    \partial_{k_x/k_y}g_x(\bm{q})&=a_t+\frac{t}{9}(q_x\mp\sqrt{3}q_y-q_z), \\
    \partial_{k_x/k_y}g_y(\bm{q})&=\mp\sqrt{3}a_t \mp \frac{\sqrt{3}t}{9}(q_x \mp \sqrt{3}q_y-q_z), \\
    \partial_{k_x/k_y}g_z(\bm{q})&=-2a_t+\frac{t}{9}(q_x \mp \sqrt{3}q_y+2q_z), \\
    \partial_{k_x}\partial_{k_y}g_x(\bm{q})&=\partial_{k_x}\partial_{k_y}g_y(\bm{q})=0, \\
    \partial_{k_x}\partial_{k_y}g_z(\bm{q})&=-\frac{2t}{3}, 
\end{align}
where we use relations derived from Eqs.~\eqref{eq:kx->qx}$\sim$\eqref{eq:kz->qz},
\begin{align}
    k_x/k_y&=\frac{1}{6}(q_x \mp \sqrt{3}q_y-q_z), \label{eq:kx/ky->qx/qy_inverse} \\
    k_z&=-\frac{1}{6}(2q_x+q_z).  \label{eq:kz->qz_inverse}
\end{align}
Meanwhile, Eqs.~\eqref{eq:gx_k_redefine}$\sim$\eqref{eq:gz_k_redefine} are rewritten as
\begin{align}
    g_x(\bm{q})&=a_tq_x-\frac{t}{18}(q^2_x-q^2_y+2q_zq_x), \\
    g_y(\bm{q})&=a_tq_y-\frac{t}{9}q_y(q_z-q_x),  \\
    g_z(\bm{q})&=a_tq_z+\frac{t}{18}(q^2_x+q^2_y-q^2_z).
\end{align}
Therefore, each component of the $\bm{k}$-resolved NMEE tensor is given by 
\begin{align}
    \label{eq:gddg}
    &\bm{g}\cdot\partial_{k_x}\partial_{k_y}\bm{g}=-\frac{2a_t t}{3}q_z+\mathcal{O}(q^2_i), \\
    %%%%%%%%%%%%%%%%%%%%%%%%%%%%%%%%%%%%%%%%%%%%%%%%%%%%%%%%%%%%%%%%%%%%%%%%%%%%%%%%%%%%%%%%%%%%%%%%%%%%%%%%%%%%%%%%%%%%%%%%%%%%%%%%%%%%%%%%%%%%%%%%%%%%%%%%%%%%%%%%%%%%%%%%%%%%%%%%%%%%%%%%%%%%%%%
    \label{eq:dgdg}
    &\partial_{k_x}\bm{g} \cdot \partial_{k_y}\bm{g}=2a^2_t-\frac{4a_t t}{9}(2q_x+q_z)+\mathcal{O}(q^2_i), \\
    %%%%%%%%%%%%%%%%%%%%%%%%%%%%%%%%%%%%%%%%%%%%%%%%%%%%%%%%%%%%%%%%%%%%%%%%%%%%%%%%%%%%%%%%%%%%%%%%%%%%%%%%%%%%%%%%%%%%%%%%%%%%%%%%%%%%%%%%%%%%%%%%%%%%%%%%%%%%%%%%%%%%%%%%%%%%%%%%%%%%%%%%%%%%%%%
    \label{eq:gdg_gdg}
    &(\bm{g}\cdot\partial_{k_x}\bm{g})(\bm{g}\cdot\partial_{k_y}\bm{g}) \notag  \\
    &=a^4_t(q^2_x-3q^2_y+4q^2_z-4q_zq_x)+\frac{a^3_t t}{9}(-q^3_x-13q_xq^2_y \notag \\
    &\quad+10q_xq^2_z-4q^2_yq_z-12q^3_z)+\mathcal{O}(q^4_i), \\
    %%%%%%%%%%%%%%%%%%%%%%%%%%%%%%%%%%%%%%%%%%%%%%%%%%%%%%%%%%%%%%%%%%%%%%%%%%%%%%%%%%%%%%%%%%%%%%%%%%%%%%%%%%%%%%%%%%%%%%%%%%%%%%%%%%%%%%%%%%%%%%%%%%%%%%%%%%%%%%%%%%%%%%%%%%%%%%%%%%%%%%%%%%%%%%%
    \label{eq:dg_gdg}
    &\partial_{k_x/k_y}g_z(\bm{g}\cdot\partial_{k_y/k_x}\bm{g}) \notag \\
    &=-2a^3_t(q_x \pm \sqrt{3}q_y-2q_z)+\frac{2 a^2_t t}{9}(q^2_x-4q^2_y-5q^2_z \notag \\
    &\quad \mp 3\sqrt{3}q_xq_y\pm3\sqrt{3}q_yq_z+q_zq_x)+\mathcal{O}(q^3_i), \\
    %%%%%%%%%%%%%%%%%%%%%%%%%%%%%%%%%%%%%%%%%%%%%%%%%%%%%%%%%%%%%%%%%%%%%%%%%%%%%%%%%%%%%%%%%%%%%%%%%%%%%%%%%%%%%%%%%%%%%%%%%%%%%%%%%%%%%%%%%%%%%%%%%%%%%%%%%%%%%%%%%%%%%%%%%%%%%%%%%%%%%%%%%%%%%%%
    \label{eq:gnk}
    &\frac{1}{|\bm{g}|^n}=\biggl[ a^2_t |\bm{q}|^2-\frac{a_t t}{9}(q^3_x-3q_xq^2_y \notag \\
    & \qquad \qquad +q^2_xq_z+q^2_yq_z+q^3_z)+\cdots \biggr]^{-n/2},
\end{align}
where we use the following relation to derive Eqs.~\eqref{eq:gdg_gdg} and~\eqref{eq:dg_gdg}:
\begin{align}
    \bm{g}\cdot\partial_{k_x/k_y}\bm{g}
    &=a_t^2(q_x\mp\sqrt{3}q_y-2q_z)-\frac{a_t t}{18}(q^2_x-5q^2_y-6q^2_z \notag \\
    &\quad \pm 6\sqrt{3}q_xq_y \mp 2\sqrt{3}q_yq_z+2q_zq_x)+\mathcal{O}(q^3_i).
\end{align}

%%%%%%%%%%%%%%%%%%%%%%%%%%%%%%%%%%%%%%%%%%%%%%%%%%%%%%%%%%%%%%%%%%%%%%%%%%%%%%%%%%%%%%%%%%%%%%%%%%%%%%%%%%%%%%%%%%%%%%%%%%%%%%%%%%%%%%%%%%%%%%%%%%%%%%%%%%%%%%%%%%%%%%%%%%%%%%%%%%%%%%%%%%%%%%%%

Then, we rewrite Eqs.~\eqref{eq:gddg}$\sim$\eqref{eq:gnk} in the spherical coordinates, $(q_x,q_y,q_z)=q(\sin\theta\cos\phi,\sin\theta\sin\phi,\cos\theta)$.
Specifically, the expressions are given by
\begin{align}
    \label{eq:gddg_sph}
    &\bm{g}\cdot\partial_{k_x}\partial_{k_y}\bm{g}=-\frac{2a_t t}{3}q\cos\theta+\mathcal{O}(q^2), \\
    &\partial_{k_x}\bm{g} \cdot \partial_{k_y}\bm{g}=2 a^2_t-\frac{4a_t t}{9}q(2\sin\theta\cos\phi+\cos\theta)+\mathcal{O}(q^2), \\
    &(\bm{g}\cdot\partial_{k_x}\bm{g})(\bm{g}\cdot\partial_{k_y}\bm{g}) \notag  \\
    &=a^4_t q^2g_1(\theta,\phi)+\frac{a^3_t t }{9}q^3g_2(\theta,\phi)+\mathcal{O}(q^4),  \\   &\partial_{k_x}g_z(\bm{g}\cdot\partial_{k_y}\bm{g})+\partial_{k_y}g_z(\bm{g}\cdot\partial_{k_x}\bm{g}) \notag \\
    &=-4a^3_tq(\sin\theta\cos\phi-2\cos\theta)+\frac{4a^2_t t}{9}q^2 g_3(\theta,\phi)+\mathcal{O}(q^3),\\
    \label{eq:gnk_sph}
    &\frac{1}{|\bm{g}|^n}=\biggl(\frac{1}{a_t q} \biggr)^n \biggl[1-\frac{t}{9a_t}qf(\theta,\phi)+\cdots \biggr]^{-n/2},
\end{align}
where $g_i(\theta,\phi)\ (i=1,2,3)$ and $f(\theta,\phi)$ are given by 
\begin{align}
    \label{eq:g1}
    g_1(\theta,\phi)&=-1+5\cos^2\theta-4\sin\theta\cos\theta\cos\phi \notag \\
    &\quad +2\sin^2\theta\cos2\phi, \\
    \label{eq:g2}    g_2(\theta,\phi)&=-2\cos\theta(1+5\cos^2\theta)-14\sin^3\theta\cos\phi \notag \\
    &\quad+10\sin\theta\cos\phi+2\sin^2\theta\cos\theta\cos2\phi \notag \\
    &\quad+3\sin^3\theta\cos3\phi,  \\
    \label{eq:g3}
    g_3(\theta,\phi)&=-\frac{1}{2}(3+7\cos^2\theta)+\sin\theta\cos\theta\cos\phi \notag \\
    &\quad+\frac{5}{2}\sin^2\theta\cos2\phi, \\
    \label{eq:f}
    f(\theta,\phi)&=\sin^3\theta\cos3\phi+\cos\theta.
\end{align}
Equations~\eqref{eq:gddg_sph}$\sim$\eqref{eq:gnk_sph} lead to the expression of the $\bm{k}$-resolved NMEE tensor. 
In particular, we focus on a specific expression derived from a relation obtained by taking the limit $ q \ll m/t $ for Eq.~\eqref{eq:gnk_sph}, 
\begin{align}
    \frac{1}{|\bm{g}|^n}\simeq \biggl( \frac{1}{a_t q}\biggr)^n \biggl[ 1+\frac{n}{2} \frac{t}{9 a_t}q f(\theta,\phi)+\cdots \biggr].
\end{align}
In the following, we perform the $\bm{q}$-integral for the specific expression, which is given by
\begin{widetext}
\begin{align}
    \label{eq:g_zxy_k}
    g^{(n)}_{z;xy}(\bm{q};\alpha,\beta,\gamma)
    &=\biggl( \frac{1}{a_t q}\biggr)^{3-n} \biggl(-\frac{2t}{3}+\frac{2a_t t}{3} q \cos\theta \biggl( \frac{1}{a_t q}\biggr)^2 a_t q \cos\theta+\cdots \biggr)  \notag \\
    &\quad -\frac{1}{\alpha}\biggl( \frac{1}{a_t q}\biggr)^{5-n} \biggl( 1+\frac{5-n}{2} \frac{t}{9 a_t}q f(\theta,\phi)\biggr) 
    \biggl[ \biggl( a_tq\cos\theta-\frac{t}{18}q^2(\cos^2\theta-\sin^2\theta) \biggr) \notag  \\
    &\quad \times \biggl\{ 2 a^2_t-\frac{4a_t t}{9}q(2\sin\theta\cos\phi+\cos\theta)-\beta\biggl( \frac{1}{a_t q}\biggr)^2 \biggl( 1+\frac{t}{9 a_t}q f(\theta,\phi)\biggr) \biggl( a^4_t q^2g_1(\theta,\phi)+\frac{a^3_t t }{9}q^3g_2(\theta,\phi) \biggr) \biggr\} \notag \\
    &\quad -4\gamma a^3_tq(\sin\theta\cos\phi-2\cos\theta)+\frac{4 \gamma a^2_t t}{9}q^2 g_3(\theta,\phi) \biggr]+\cdots \notag \\
    &=-\frac{1}{\alpha} \frac{1}{a^{2-n}_t q^{4-n}}  \Psi_{\beta \gamma}(\theta,\phi) -\frac{1}{18\alpha} \frac{t}{a^{3-n}_t q^{3-n}}  \Bigl\{ 12\alpha(1-\cos^2\theta)-8\cos\theta(2\sin\theta\cos\phi+\cos\theta) \notag \\
    &\quad -2\beta \cos\theta\Bigl( f(\theta,\phi)g_1(\theta,\phi)+g_2(\theta,\phi) \Bigr)+8\gamma g_3(\theta,\phi)+(1-2\cos^2\theta)(2-\beta g_1(\theta,\phi)) \notag \\
    &\quad +(5-n)f(\theta,\phi) \Psi_{\beta\gamma}(\theta,\phi)\Bigr\}+\mathcal{O}(t^2/a^{4-n}_tq^{2-n}), 
\end{align}
where $\Psi_{\beta \gamma}(\theta,\phi)$ is written as
\begin{align}
    \label{eq:Psi}
    \Psi_{\beta \gamma}(\theta,\phi)=\cos\theta \Bigl(2-\beta g_1(\theta,\phi) \Bigr)-4\gamma(\sin\theta\cos\phi-2\cos\theta).
\end{align}

%%%%%%%%%%%%%%%%%%%%%%%%%%%%%%%%%%%%%%%%%%%%%%%%%%%%%%%%%%%%%%%%%%%%%%%%%%%%%%%%%%%%%%%%%%%%%%%%%%%%%%%%%%%%%%%%%%%%%%%%%%%%%%%%%%%%%%%%%%%%%%%%%%%%%%%%%%%%%%%%%%%%%%%%%%%%%%%%%%%%%%%%%%%%%%%%

\subsection{Integration of the momentum-resolved NMEE tensor} \label{subsec:EffWeyl_2}

Equation~\eqref{eq:g_zxy_k} is integrated over the volume unit in the spherical coordinates, $\int q^2dq\int^{\pi}_{0}\sin\theta d\theta\int^{2\pi}_{0}d\phi$.
First, we perform the integral over the surface unit, $\int^{\pi}_{0}\sin\theta d\theta\int^{2\pi}_{0}d\phi$, by using the following relations,
\begin{align}
    \int^{2\pi}_{0}\sin n\phi d\phi=\int^{2\pi}_{0} \cos n\phi d\phi=0 \hspace{0.4cm} (n=\mathrm{integer}), \qquad
    \int^{\pi}_0\sin\theta \cos^n\theta d\theta=\left\{
        \begin{array}{ll}
        2/(n+1) & (n=\mathrm{even}) \\
        0 & (n=\mathrm{odd})
        \end{array}
        \right. .
\end{align}
Explicitly including the terms that are finite after the surface integration, we can write Eq.~\eqref{eq:g_zxy_k} as
\begin{align}
    \label{eq:g_zxy_k_integral}
    g^{(n)}_{z;xy}(\bm{q};\alpha,\beta,\gamma)&=-\frac{1}{18\alpha}\frac{t}{a^{3-n}_tq^{3-n}}\biggl\{12\alpha(1-\cos^2\theta)-8\cos^2\theta -2\beta \cos\theta \Bigl(\cos\theta(-1+5\cos^2\theta) -2\cos\theta(1+5\cos^2\theta)\Bigr) \notag \\
    &\quad -4\gamma(3+7\cos^2\theta)+(1-2\cos^2\theta)\Bigl(2-\beta(-1+5\cos^2\theta)\Bigr) \notag  \\
    &\quad +(5-n)\cos\theta\Bigl( \cos\theta \bigl(2-\beta(-1+5\cos^2\theta)\bigr)+8\gamma\cos\theta\Bigr) \biggr\}+\mathcal{O}(t^2/a^{4-n}_tq^{2-n})  \notag \\
    &=-\frac{1}{18\alpha}\frac{t}{a^{3-n}_tq^{3-n}}\biggl\{\Bigl(2+12\alpha+\beta-12\gamma\Bigr)+\Bigl(-2(1+n)-12\alpha+(4-n)\beta+(12-8n)\gamma \Bigr)\cos^2\theta \notag \\
    &\quad -5(1-n)\beta\cos^4\theta \biggr\}+\mathcal{O}(t^2/a^{4-n}_tq^{2-n})  \notag \\
    &\rightarrow -\frac{2\pi}{27\alpha}\frac{t}{a^{3-n}_tq^{3-n}}\Bigl\{2(2-n)+24\alpha+2(2+n)\beta-8(3+n)\gamma\Bigr\}+\mathcal{O}(t^2/a^{4-n}_tq^{2-n}), 
\end{align}
where the surface integration is performed in the last line.
Note that the leading order term vanishes after the surface integration.
Then, the radial integration in the next leading order term of Eq.~\eqref{eq:g_zxy_k_integral} results in
\begin{align}
    \int_{\bm{k}}g^{(n)}_{z;xy}(\bm{k};\alpha,\beta,\gamma)
    &=\frac{1}{(2\pi)^3}\int_{q_0}^{\Lambda} q^2dq\int_{0}^{\pi} \sin\theta d\theta \int_{0}^{2\pi} d\phi |J(q_x,q_y,q_z)| g^{(n)}_{z;xy}(\bm{q};\alpha,\beta,\gamma) \notag \\
    &=\frac{t}{648\sqrt{3}\pi^2 a_t}
        \begin{dcases}
    -2\biggl( \Lambda^2-\biggl(\frac{|\mu|}{2a_t} \biggr)^2 \biggr) & \mathrm{for}~(n,\alpha,\beta,\gamma)=(2,1,3,1) \\
    \frac{1}{a^2_t}\ln \biggl(\frac{\Lambda}{|\mu|/2a_t} \biggr) & \mathrm{for}~(n,\alpha,\beta,\gamma)=(0,4,15,7)
    \end{dcases},
\end{align}
where $q_0=|\mu|/2a_t$~\cite{Michishita2022-ag}, and $|J(q_x,q_y,q_z)|=|\partial(k_x,k_y,k_z)/\partial(q_x,q_y,q_z)|=1/12\sqrt{3}$ is the Jacobian determinant for the change of the coordinates from $\bm{k}$ to $\bm{q}$.
Note that we introduce a cutoff $\Lambda$ to avoid the divergence of the radial integral, $\int q^2 dq$.
From this expression, we can obtain the analytical expression of the NMEE tensor in the effective Weyl Hamitonian, as shown in Eqs.~\eqref{eq:tau2_Weyl} and~\eqref{eq:tau0_Weyl}.

\end{widetext}

%%%%%%%%%%%%%%%%%%%%%%%%%%%%%%%%%%%%%%%%%%%%%%%%%%%%%%%%%%%%%%%%%%%%%%%%%%%%%%%%%%%%%%%%%%%%%%%%%%%%%%%%%%%%%%%%%%%%%%%%%%%%%%%%%%%%%%%%%%%%%%%%%%%%%%%%%%%%%%%%%%%%%%%%%%%%%%%%%%%%%%%%%%%%%%%%
%%%%%%%%%%%%%%%%%%%%%%%%%%%%%%%%%%%%%%%%%%%%%%%%%%%%%%%%%%%%%%%%%%%%%%%%%%%%%%%%%%%%%%%%%%%%%%%%%%%%%%%%%%%%%%%%%%%%%%%%%%%%%%%%%%%%%%%%%%%%%%%%%%%%%%%%%%%%%%%%%%%%%%%%%%%%%%%%%%%%%%%%%%%%%%%%
%%%%%%%%%%%%%%%%%%%%%%%%%%%%%%%%%%%%%%%%%%%%%%%%%%%%%%%%%%%%%%%%%%%%%%%%%%%%%%%%%%%%%%%%%%%%%%%%%%%%%%%%%%%%%%%%%%%%%%%%%%%%%%%%%%%%%%%%%%%%%%%%%%%%%%%%%%%%%%%%%%%%%%%%%%%%%%%%%%%%%%%%%%%%%%%%
%bib

\bibliography{main.bib}% Produces the bibliography via BibTeX.

\providecommand{\noopsort}[1]{}\providecommand{\singleletter}[1]{#1}%
\begin{thebibliography}{215}%
\makeatletter
\providecommand \@ifxundefined [1]{%
 \@ifx{#1\undefined}
}%
\providecommand \@ifnum [1]{%
 \ifnum #1\expandafter \@firstoftwo
 \else \expandafter \@secondoftwo
 \fi
}%
\providecommand \@ifx [1]{%
 \ifx #1\expandafter \@firstoftwo
 \else \expandafter \@secondoftwo
 \fi
}%
\providecommand \natexlab [1]{#1}%
\providecommand \enquote  [1]{``#1''}%
\providecommand \bibnamefont  [1]{#1}%
\providecommand \bibfnamefont [1]{#1}%
\providecommand \citenamefont [1]{#1}%
\providecommand \href@noop [0]{\@secondoftwo}%
\providecommand \href [0]{\begingroup \@sanitize@url \@href}%
\providecommand \@href[1]{\@@startlink{#1}\@@href}%
\providecommand \@@href[1]{\endgroup#1\@@endlink}%
\providecommand \@sanitize@url [0]{\catcode `\\12\catcode `\$12\catcode
  `\&12\catcode `\#12\catcode `\^12\catcode `\_12\catcode `\%12\relax}%
\providecommand \@@startlink[1]{}%
\providecommand \@@endlink[0]{}%
\providecommand \url  [0]{\begingroup\@sanitize@url \@url }%
\providecommand \@url [1]{\endgroup\@href {#1}{\urlprefix }}%
\providecommand \urlprefix  [0]{URL }%
\providecommand \Eprint [0]{\href }%
\providecommand \doibase [0]{https://doi.org/}%
\providecommand \selectlanguage [0]{\@gobble}%
\providecommand \bibinfo  [0]{\@secondoftwo}%
\providecommand \bibfield  [0]{\@secondoftwo}%
\providecommand \translation [1]{[#1]}%
\providecommand \BibitemOpen [0]{}%
\providecommand \bibitemStop [0]{}%
\providecommand \bibitemNoStop [0]{.\EOS\space}%
\providecommand \EOS [0]{\spacefactor3000\relax}%
\providecommand \BibitemShut  [1]{\csname bibitem#1\endcsname}%
\let\auto@bib@innerbib\@empty
%</preamble>
\bibitem [{\citenamefont {Kuramoto}\ \emph {et~al.}(2009)\citenamefont
  {Kuramoto}, \citenamefont {Kusunose},\ and\ \citenamefont
  {Kiss}}]{Kuramoto2009-ln}%
  \BibitemOpen
  \bibfield  {author} {\bibinfo {author} {\bibfnamefont {Y.}~\bibnamefont
  {Kuramoto}}, \bibinfo {author} {\bibfnamefont {H.}~\bibnamefont {Kusunose}},\
  and\ \bibinfo {author} {\bibfnamefont {A.}~\bibnamefont {Kiss}},\ }\bibfield
  {title} {\bibinfo {title} {{Multipole Orders and Fluctuations in Strongly
  Correlated Electron Systems}},\ }\href
  {https://doi.org/10.1143/JPSJ.78.072001} {\bibfield  {journal} {\bibinfo
  {journal} {J. Phys. Soc. Jpn.}\ }\textbf {\bibinfo {volume} {78}},\ \bibinfo
  {pages} {072001} (\bibinfo {year} {2009})}\BibitemShut {NoStop}%
\bibitem [{\citenamefont {Santini}\ \emph {et~al.}(2009)\citenamefont
  {Santini}, \citenamefont {Carretta}, \citenamefont {Amoretti}, \citenamefont
  {Caciuffo}, \citenamefont {Magnani},\ and\ \citenamefont
  {Lander}}]{Santini2009-er}%
  \BibitemOpen
  \bibfield  {author} {\bibinfo {author} {\bibfnamefont {P.}~\bibnamefont
  {Santini}}, \bibinfo {author} {\bibfnamefont {S.}~\bibnamefont {Carretta}},
  \bibinfo {author} {\bibfnamefont {G.}~\bibnamefont {Amoretti}}, \bibinfo
  {author} {\bibfnamefont {R.}~\bibnamefont {Caciuffo}}, \bibinfo {author}
  {\bibfnamefont {N.}~\bibnamefont {Magnani}},\ and\ \bibinfo {author}
  {\bibfnamefont {G.~H.}\ \bibnamefont {Lander}},\ }\bibfield  {title}
  {\bibinfo {title} {{Multipolar interactions in $f$-electron systems: The
  paradigm of actinide dioxides}},\ }\href
  {https://doi.org/10.1103/RevModPhys.81.807} {\bibfield  {journal} {\bibinfo
  {journal} {Rev. Mod. Phys.}\ }\textbf {\bibinfo {volume} {81}},\ \bibinfo
  {pages} {807} (\bibinfo {year} {2009})}\BibitemShut {NoStop}%
\bibitem [{\citenamefont {Mydosh}\ \emph {et~al.}(2020)\citenamefont {Mydosh},
  \citenamefont {Oppeneer},\ and\ \citenamefont {Riseborough}}]{Mydosh2020-gm}%
  \BibitemOpen
  \bibfield  {author} {\bibinfo {author} {\bibfnamefont {J.~A.}\ \bibnamefont
  {Mydosh}}, \bibinfo {author} {\bibfnamefont {P.~M.}\ \bibnamefont
  {Oppeneer}},\ and\ \bibinfo {author} {\bibfnamefont {P.~S.}\ \bibnamefont
  {Riseborough}},\ }\bibfield  {title} {\bibinfo {title} {{Hidden order and
  beyond: an experimental-theoretical overview of the multifaceted behavior of
  {URu$_2$Si$_2$}}},\ }\href {https://doi.org/10.1088/1361-648X/ab5eba}
  {\bibfield  {journal} {\bibinfo  {journal} {J. Phys. Condens. Matter}\
  }\textbf {\bibinfo {volume} {32}},\ \bibinfo {pages} {143002} (\bibinfo
  {year} {2020})}\BibitemShut {NoStop}%
\bibitem [{\citenamefont {Bhowal}\ and\ \citenamefont
  {Spaldin}(2024)}]{Bhowal2024-vj}%
  \BibitemOpen
  \bibfield  {author} {\bibinfo {author} {\bibfnamefont {S.}~\bibnamefont
  {Bhowal}}\ and\ \bibinfo {author} {\bibfnamefont {N.~A.}\ \bibnamefont
  {Spaldin}},\ }\bibfield  {title} {\bibinfo {title} {{Ferroically Ordered
  Magnetic Octupoles in {$d$-Wave} Altermagnets}},\ }\href
  {https://doi.org/10.1103/PhysRevX.14.011019} {\bibfield  {journal} {\bibinfo
  {journal} {Phys. Rev. X}\ }\textbf {\bibinfo {volume} {14}},\ \bibinfo
  {pages} {011019} (\bibinfo {year} {2024})}\BibitemShut {NoStop}%
\bibitem [{\citenamefont {McClarty}\ and\ \citenamefont
  {Rau}(2024)}]{McClarty2024-eo}%
  \BibitemOpen
  \bibfield  {author} {\bibinfo {author} {\bibfnamefont {P.~A.}\ \bibnamefont
  {McClarty}}\ and\ \bibinfo {author} {\bibfnamefont {J.~G.}\ \bibnamefont
  {Rau}},\ }\bibfield  {title} {\bibinfo {title} {{Landau Theory of
  Altermagnetism}},\ }\href {https://doi.org/10.1103/PhysRevLett.132.176702}
  {\bibfield  {journal} {\bibinfo  {journal} {Phys. Rev. Lett.}\ }\textbf
  {\bibinfo {volume} {132}},\ \bibinfo {pages} {176702} (\bibinfo {year}
  {2024})}\BibitemShut {NoStop}%
\bibitem [{\citenamefont {Noda}\ \emph {et~al.}(2016)\citenamefont {Noda},
  \citenamefont {Ohno},\ and\ \citenamefont {Nakamura}}]{Noda2016-wt}%
  \BibitemOpen
  \bibfield  {author} {\bibinfo {author} {\bibfnamefont {Y.}~\bibnamefont
  {Noda}}, \bibinfo {author} {\bibfnamefont {K.}~\bibnamefont {Ohno}},\ and\
  \bibinfo {author} {\bibfnamefont {S.}~\bibnamefont {Nakamura}},\ }\bibfield
  {title} {\bibinfo {title} {{Momentum-dependent band spin splitting in
  semiconducting {MnO$_2$}: a density functional calculation}},\ }\href
  {https://doi.org/10.1039/c5cp07806g} {\bibfield  {journal} {\bibinfo
  {journal} {Phys. Chem. Chem. Phys.}\ }\textbf {\bibinfo {volume} {18}},\
  \bibinfo {pages} {13294} (\bibinfo {year} {2016})}\BibitemShut {NoStop}%
\bibitem [{\citenamefont {Naka}\ \emph {et~al.}(2019)\citenamefont {Naka},
  \citenamefont {Hayami}, \citenamefont {Kusunose}, \citenamefont {Yanagi},
  \citenamefont {Motome},\ and\ \citenamefont {Seo}}]{Naka2019-uk}%
  \BibitemOpen
  \bibfield  {author} {\bibinfo {author} {\bibfnamefont {M.}~\bibnamefont
  {Naka}}, \bibinfo {author} {\bibfnamefont {S.}~\bibnamefont {Hayami}},
  \bibinfo {author} {\bibfnamefont {H.}~\bibnamefont {Kusunose}}, \bibinfo
  {author} {\bibfnamefont {Y.}~\bibnamefont {Yanagi}}, \bibinfo {author}
  {\bibfnamefont {Y.}~\bibnamefont {Motome}},\ and\ \bibinfo {author}
  {\bibfnamefont {H.}~\bibnamefont {Seo}},\ }\bibfield  {title} {\bibinfo
  {title} {{Spin current generation in organic antiferromagnets}},\ }\href
  {https://doi.org/10.1038/s41467-019-12229-y} {\bibfield  {journal} {\bibinfo
  {journal} {Nat. Commun.}\ }\textbf {\bibinfo {volume} {10}},\ \bibinfo
  {pages} {4305} (\bibinfo {year} {2019})}\BibitemShut {NoStop}%
\bibitem [{\citenamefont {Naka}\ \emph {et~al.}(2020)\citenamefont {Naka},
  \citenamefont {Hayami}, \citenamefont {Kusunose}, \citenamefont {Yanagi},
  \citenamefont {Motome},\ and\ \citenamefont {Seo}}]{Naka2020-pm}%
  \BibitemOpen
  \bibfield  {author} {\bibinfo {author} {\bibfnamefont {M.}~\bibnamefont
  {Naka}}, \bibinfo {author} {\bibfnamefont {S.}~\bibnamefont {Hayami}},
  \bibinfo {author} {\bibfnamefont {H.}~\bibnamefont {Kusunose}}, \bibinfo
  {author} {\bibfnamefont {Y.}~\bibnamefont {Yanagi}}, \bibinfo {author}
  {\bibfnamefont {Y.}~\bibnamefont {Motome}},\ and\ \bibinfo {author}
  {\bibfnamefont {H.}~\bibnamefont {Seo}},\ }\bibfield  {title} {\bibinfo
  {title} {{Anomalous Hall effect in $\kappa$-type organic antiferromagnets}},\
  }\href {https://doi.org/10.1103/PhysRevB.102.075112} {\bibfield  {journal}
  {\bibinfo  {journal} {Phys. Rev. B}\ }\textbf {\bibinfo {volume} {102}},\
  \bibinfo {pages} {075112} (\bibinfo {year} {2020})}\BibitemShut {NoStop}%
\bibitem [{\citenamefont {Hayami}\ \emph {et~al.}(2019)\citenamefont {Hayami},
  \citenamefont {Yanagi},\ and\ \citenamefont {Kusunose}}]{Hayami2019-zk}%
  \BibitemOpen
  \bibfield  {author} {\bibinfo {author} {\bibfnamefont {S.}~\bibnamefont
  {Hayami}}, \bibinfo {author} {\bibfnamefont {Y.}~\bibnamefont {Yanagi}},\
  and\ \bibinfo {author} {\bibfnamefont {H.}~\bibnamefont {Kusunose}},\
  }\bibfield  {title} {\bibinfo {title} {{Momentum-Dependent Spin Splitting by
  Collinear Antiferromagnetic Ordering}},\ }\href
  {https://doi.org/10.7566/JPSJ.88.123702} {\bibfield  {journal} {\bibinfo
  {journal} {J. Phys. Soc. Jpn.}\ }\textbf {\bibinfo {volume} {88}},\ \bibinfo
  {pages} {123702} (\bibinfo {year} {2019})}\BibitemShut {NoStop}%
\bibitem [{\citenamefont {Ahn}\ \emph {et~al.}(2019)\citenamefont {Ahn},
  \citenamefont {Hariki}, \citenamefont {Lee},\ and\ \citenamefont {Kune{\v
  s}}}]{Ahn2019-cy}%
  \BibitemOpen
  \bibfield  {author} {\bibinfo {author} {\bibfnamefont {K.-H.}\ \bibnamefont
  {Ahn}}, \bibinfo {author} {\bibfnamefont {A.}~\bibnamefont {Hariki}},
  \bibinfo {author} {\bibfnamefont {K.-W.}\ \bibnamefont {Lee}},\ and\ \bibinfo
  {author} {\bibfnamefont {J.}~\bibnamefont {Kune{\v s}}},\ }\bibfield  {title}
  {\bibinfo {title} {{Antiferromagnetism in {RuO$_{2}$} as d-wave Pomeranchuk
  instability}},\ }\href {https://doi.org/10.1103/PhysRevB.99.184432}
  {\bibfield  {journal} {\bibinfo  {journal} {Phys. Rev. B}\ }\textbf {\bibinfo
  {volume} {99}},\ \bibinfo {pages} {184432} (\bibinfo {year}
  {2019})}\BibitemShut {NoStop}%
\bibitem [{\citenamefont {Yuan}\ \emph
  {et~al.}(2020{\natexlab{a}})\citenamefont {Yuan}, \citenamefont {Wang},
  \citenamefont {Luo}, \citenamefont {Rashba},\ and\ \citenamefont
  {Zunger}}]{Yuan2020-al}%
  \BibitemOpen
  \bibfield  {author} {\bibinfo {author} {\bibfnamefont {L.-D.}\ \bibnamefont
  {Yuan}}, \bibinfo {author} {\bibfnamefont {Z.}~\bibnamefont {Wang}}, \bibinfo
  {author} {\bibfnamefont {J.-W.}\ \bibnamefont {Luo}}, \bibinfo {author}
  {\bibfnamefont {E.~I.}\ \bibnamefont {Rashba}},\ and\ \bibinfo {author}
  {\bibfnamefont {A.}~\bibnamefont {Zunger}},\ }\bibfield  {title} {\bibinfo
  {title} {{Giant momentum-dependent spin splitting in centrosymmetric
  {low-$Z$} antiferromagnets}},\ }\href
  {https://doi.org/10.1103/PhysRevB.102.014422} {\bibfield  {journal} {\bibinfo
   {journal} {Phys. Rev. B}\ }\textbf {\bibinfo {volume} {102}},\ \bibinfo
  {pages} {014422} (\bibinfo {year} {2020}{\natexlab{a}})}\BibitemShut
  {NoStop}%
\bibitem [{\citenamefont {{\v S}mejkal}\ \emph {et~al.}(2020)\citenamefont {{\v
  S}mejkal}, \citenamefont {Gonz{\'a}lez-Hern{\'a}ndez}, \citenamefont
  {Jungwirth},\ and\ \citenamefont {Sinova}}]{Smejkal2020-zu}%
  \BibitemOpen
  \bibfield  {author} {\bibinfo {author} {\bibfnamefont {L.}~\bibnamefont {{\v
  S}mejkal}}, \bibinfo {author} {\bibfnamefont {R.}~\bibnamefont
  {Gonz{\'a}lez-Hern{\'a}ndez}}, \bibinfo {author} {\bibfnamefont
  {T.}~\bibnamefont {Jungwirth}},\ and\ \bibinfo {author} {\bibfnamefont
  {J.}~\bibnamefont {Sinova}},\ }\bibfield  {title} {\bibinfo {title} {{Crystal
  time-reversal symmetry breaking and spontaneous Hall effect in collinear
  antiferromagnets}},\ }\href {https://doi.org/10.1126/sciadv.aaz8809}
  {\bibfield  {journal} {\bibinfo  {journal} {Sci Adv}\ }\textbf {\bibinfo
  {volume} {6}},\ \bibinfo {pages} {eaaz8809} (\bibinfo {year}
  {2020})}\BibitemShut {NoStop}%
\bibitem [{\citenamefont {Yuan}\ \emph {et~al.}(2021)\citenamefont {Yuan},
  \citenamefont {Wang}, \citenamefont {Luo},\ and\ \citenamefont
  {Zunger}}]{Yuan2021-xz}%
  \BibitemOpen
  \bibfield  {author} {\bibinfo {author} {\bibfnamefont {L.-D.}\ \bibnamefont
  {Yuan}}, \bibinfo {author} {\bibfnamefont {Z.}~\bibnamefont {Wang}}, \bibinfo
  {author} {\bibfnamefont {J.-W.}\ \bibnamefont {Luo}},\ and\ \bibinfo {author}
  {\bibfnamefont {A.}~\bibnamefont {Zunger}},\ }\bibfield  {title} {\bibinfo
  {title} {{Prediction of {low-$Z$} collinear and noncollinear
  antiferromagnetic compounds having momentum-dependent spin splitting even
  without spin-orbit coupling}},\ }\href
  {https://doi.org/10.1103/PhysRevMaterials.5.014409} {\bibfield  {journal}
  {\bibinfo  {journal} {Phys. Rev. Mater.}\ }\textbf {\bibinfo {volume} {5}},\
  \bibinfo {pages} {014409} (\bibinfo {year} {2021})}\BibitemShut {NoStop}%
\bibitem [{\citenamefont {Mazin}\ \emph {et~al.}(2021)\citenamefont {Mazin},
  \citenamefont {Koepernik}, \citenamefont {Johannes}, \citenamefont
  {Gonz{\'a}lez-Hern{\'a}ndez},\ and\ \citenamefont {{\v
  S}mejkal}}]{Mazin2021-ea}%
  \BibitemOpen
  \bibfield  {author} {\bibinfo {author} {\bibfnamefont {I.~I.}\ \bibnamefont
  {Mazin}}, \bibinfo {author} {\bibfnamefont {K.}~\bibnamefont {Koepernik}},
  \bibinfo {author} {\bibfnamefont {M.~D.}\ \bibnamefont {Johannes}}, \bibinfo
  {author} {\bibfnamefont {R.}~\bibnamefont {Gonz{\'a}lez-Hern{\'a}ndez}},\
  and\ \bibinfo {author} {\bibfnamefont {L.}~\bibnamefont {{\v S}mejkal}},\
  }\bibfield  {title} {\bibinfo {title} {{Prediction of unconventional
  magnetism in doped {FeSb$_2$}}},\ }\href
  {https://doi.org/10.1073/pnas.2108924118} {\bibfield  {journal} {\bibinfo
  {journal} {Proc. Natl. Acad. Sci. U. S. A.}\ }\textbf {\bibinfo {volume}
  {118}},\ \bibinfo {pages} {e2108924118} (\bibinfo {year} {2021})}\BibitemShut
  {NoStop}%
\bibitem [{\citenamefont {Egorov}\ and\ \citenamefont
  {Evarestov}(2021)}]{Egorov2021-uo}%
  \BibitemOpen
  \bibfield  {author} {\bibinfo {author} {\bibfnamefont {S.~A.}\ \bibnamefont
  {Egorov}}\ and\ \bibinfo {author} {\bibfnamefont {R.~A.}\ \bibnamefont
  {Evarestov}},\ }\bibfield  {title} {\bibinfo {title} {{Colossal Spin
  Splitting in the Monolayer of the Collinear Antiferromagnet {MnF2}}},\ }\href
  {https://doi.org/10.1021/acs.jpclett.1c00282} {\bibfield  {journal} {\bibinfo
   {journal} {J. Phys. Chem. Lett.}\ }\textbf {\bibinfo {volume} {12}},\
  \bibinfo {pages} {2363} (\bibinfo {year} {2021})}\BibitemShut {NoStop}%
\bibitem [{\citenamefont {{\v S}mejkal}\ \emph
  {et~al.}(2022{\natexlab{a}})\citenamefont {{\v S}mejkal}, \citenamefont
  {Sinova},\ and\ \citenamefont {Jungwirth}}]{Smejkal2022-xr}%
  \BibitemOpen
  \bibfield  {author} {\bibinfo {author} {\bibfnamefont {L.}~\bibnamefont {{\v
  S}mejkal}}, \bibinfo {author} {\bibfnamefont {J.}~\bibnamefont {Sinova}},\
  and\ \bibinfo {author} {\bibfnamefont {T.}~\bibnamefont {Jungwirth}},\
  }\bibfield  {title} {\bibinfo {title} {{Beyond Conventional Ferromagnetism
  and Antiferromagnetism: A Phase with Nonrelativistic Spin and Crystal
  Rotation Symmetry}},\ }\href {https://doi.org/10.1103/PhysRevX.12.031042}
  {\bibfield  {journal} {\bibinfo  {journal} {Phys. Rev. X}\ }\textbf {\bibinfo
  {volume} {12}},\ \bibinfo {pages} {031042} (\bibinfo {year}
  {2022}{\natexlab{a}})}\BibitemShut {NoStop}%
\bibitem [{\citenamefont {{\v S}mejkal}\ \emph
  {et~al.}(2022{\natexlab{b}})\citenamefont {{\v S}mejkal}, \citenamefont
  {Sinova},\ and\ \citenamefont {Jungwirth}}]{Smejkal2022-xx}%
  \BibitemOpen
  \bibfield  {author} {\bibinfo {author} {\bibfnamefont {L.}~\bibnamefont {{\v
  S}mejkal}}, \bibinfo {author} {\bibfnamefont {J.}~\bibnamefont {Sinova}},\
  and\ \bibinfo {author} {\bibfnamefont {T.}~\bibnamefont {Jungwirth}},\
  }\bibfield  {title} {\bibinfo {title} {{Emerging Research Landscape of
  Altermagnetism}},\ }\href {https://doi.org/10.1103/PhysRevX.12.040501}
  {\bibfield  {journal} {\bibinfo  {journal} {Phys. Rev. X}\ }\textbf {\bibinfo
  {volume} {12}},\ \bibinfo {pages} {040501} (\bibinfo {year}
  {2022}{\natexlab{b}})}\BibitemShut {NoStop}%
\bibitem [{\citenamefont {Mazin}(2022)}]{Mazin2022-nf}%
  \BibitemOpen
  \bibfield  {author} {\bibinfo {author} {\bibfnamefont {I.}~\bibnamefont
  {Mazin}},\ }\bibfield  {title} {\bibinfo {title} {{Editorial:
  {Altermagnetism---A} New Punch Line of Fundamental Magnetism}},\ }\href
  {https://doi.org/10.1103/PhysRevX.12.040002} {\bibfield  {journal} {\bibinfo
  {journal} {Phys. Rev. X}\ }\textbf {\bibinfo {volume} {12}},\ \bibinfo
  {pages} {040002} (\bibinfo {year} {2022})}\BibitemShut {NoStop}%
\bibitem [{\citenamefont {Bose}\ \emph {et~al.}(2022)\citenamefont {Bose},
  \citenamefont {Schreiber}, \citenamefont {Jain}, \citenamefont {Shao},
  \citenamefont {Nair}, \citenamefont {Sun}, \citenamefont {Zhang},
  \citenamefont {Muller}, \citenamefont {Tsymbal}, \citenamefont {Schlom},\
  and\ \citenamefont {Ralph}}]{Bose2022-bp}%
  \BibitemOpen
  \bibfield  {author} {\bibinfo {author} {\bibfnamefont {A.}~\bibnamefont
  {Bose}}, \bibinfo {author} {\bibfnamefont {N.~J.}\ \bibnamefont {Schreiber}},
  \bibinfo {author} {\bibfnamefont {R.}~\bibnamefont {Jain}}, \bibinfo {author}
  {\bibfnamefont {D.-F.}\ \bibnamefont {Shao}}, \bibinfo {author}
  {\bibfnamefont {H.~P.}\ \bibnamefont {Nair}}, \bibinfo {author}
  {\bibfnamefont {J.}~\bibnamefont {Sun}}, \bibinfo {author} {\bibfnamefont
  {X.~S.}\ \bibnamefont {Zhang}}, \bibinfo {author} {\bibfnamefont {D.~A.}\
  \bibnamefont {Muller}}, \bibinfo {author} {\bibfnamefont {E.~Y.}\
  \bibnamefont {Tsymbal}}, \bibinfo {author} {\bibfnamefont {D.~G.}\
  \bibnamefont {Schlom}},\ and\ \bibinfo {author} {\bibfnamefont {D.~C.}\
  \bibnamefont {Ralph}},\ }\bibfield  {title} {\bibinfo {title} {{Tilted spin
  current generated by the collinear antiferromagnet ruthenium dioxide}},\
  }\href {https://doi.org/10.1038/s41928-022-00744-8} {\bibfield  {journal}
  {\bibinfo  {journal} {Nature Electronics}\ }\textbf {\bibinfo {volume} {5}},\
  \bibinfo {pages} {267} (\bibinfo {year} {2022})}\BibitemShut {NoStop}%
\bibitem [{\citenamefont {Bai}\ \emph {et~al.}(2022)\citenamefont {Bai},
  \citenamefont {Han}, \citenamefont {Feng}, \citenamefont {Zhou},
  \citenamefont {Su}, \citenamefont {Wang}, \citenamefont {Liao}, \citenamefont
  {Zhu}, \citenamefont {Chen}, \citenamefont {Pan}, \citenamefont {Fan},\ and\
  \citenamefont {Song}}]{Bai2022-lx}%
  \BibitemOpen
  \bibfield  {author} {\bibinfo {author} {\bibfnamefont {H.}~\bibnamefont
  {Bai}}, \bibinfo {author} {\bibfnamefont {L.}~\bibnamefont {Han}}, \bibinfo
  {author} {\bibfnamefont {X.~Y.}\ \bibnamefont {Feng}}, \bibinfo {author}
  {\bibfnamefont {Y.~J.}\ \bibnamefont {Zhou}}, \bibinfo {author}
  {\bibfnamefont {R.~X.}\ \bibnamefont {Su}}, \bibinfo {author} {\bibfnamefont
  {Q.}~\bibnamefont {Wang}}, \bibinfo {author} {\bibfnamefont {L.~Y.}\
  \bibnamefont {Liao}}, \bibinfo {author} {\bibfnamefont {W.~X.}\ \bibnamefont
  {Zhu}}, \bibinfo {author} {\bibfnamefont {X.~Z.}\ \bibnamefont {Chen}},
  \bibinfo {author} {\bibfnamefont {F.}~\bibnamefont {Pan}}, \bibinfo {author}
  {\bibfnamefont {X.~L.}\ \bibnamefont {Fan}},\ and\ \bibinfo {author}
  {\bibfnamefont {C.}~\bibnamefont {Song}},\ }\bibfield  {title} {\bibinfo
  {title} {{Observation of Spin Splitting Torque in a Collinear Antiferromagnet
  {RuO$_2$}}},\ }\href {https://doi.org/10.1103/PhysRevLett.128.197202}
  {\bibfield  {journal} {\bibinfo  {journal} {Phys. Rev. Lett.}\ }\textbf
  {\bibinfo {volume} {128}},\ \bibinfo {pages} {197202} (\bibinfo {year}
  {2022})}\BibitemShut {NoStop}%
\bibitem [{\citenamefont {Karube}\ \emph {et~al.}(2022)\citenamefont {Karube},
  \citenamefont {Tanaka}, \citenamefont {Sugawara}, \citenamefont {Kadoguchi},
  \citenamefont {Kohda},\ and\ \citenamefont {Nitta}}]{Karube2022-mg}%
  \BibitemOpen
  \bibfield  {author} {\bibinfo {author} {\bibfnamefont {S.}~\bibnamefont
  {Karube}}, \bibinfo {author} {\bibfnamefont {T.}~\bibnamefont {Tanaka}},
  \bibinfo {author} {\bibfnamefont {D.}~\bibnamefont {Sugawara}}, \bibinfo
  {author} {\bibfnamefont {N.}~\bibnamefont {Kadoguchi}}, \bibinfo {author}
  {\bibfnamefont {M.}~\bibnamefont {Kohda}},\ and\ \bibinfo {author}
  {\bibfnamefont {J.}~\bibnamefont {Nitta}},\ }\bibfield  {title} {\bibinfo
  {title} {{Observation of {Spin-Splitter} Torque in Collinear
  Antiferromagnetic {RuO$_2$}}},\ }\href
  {https://doi.org/10.1103/PhysRevLett.129.137201} {\bibfield  {journal}
  {\bibinfo  {journal} {Phys. Rev. Lett.}\ }\textbf {\bibinfo {volume} {129}},\
  \bibinfo {pages} {137201} (\bibinfo {year} {2022})}\BibitemShut {NoStop}%
\bibitem [{\citenamefont {Feng}\ \emph {et~al.}(2022)\citenamefont {Feng},
  \citenamefont {Zhou}, \citenamefont {{\v S}mejkal}, \citenamefont {Wu},
  \citenamefont {Zhu}, \citenamefont {Guo}, \citenamefont
  {Gonz{\'a}lez-Hern{\'a}ndez}, \citenamefont {Wang}, \citenamefont {Yan},
  \citenamefont {Qin}, \citenamefont {Zhang}, \citenamefont {Wu}, \citenamefont
  {Chen}, \citenamefont {Meng}, \citenamefont {Liu}, \citenamefont {Xia},
  \citenamefont {Sinova}, \citenamefont {Jungwirth},\ and\ \citenamefont
  {Liu}}]{Feng2022-jm}%
  \BibitemOpen
  \bibfield  {author} {\bibinfo {author} {\bibfnamefont {Z.}~\bibnamefont
  {Feng}}, \bibinfo {author} {\bibfnamefont {X.}~\bibnamefont {Zhou}}, \bibinfo
  {author} {\bibfnamefont {L.}~\bibnamefont {{\v S}mejkal}}, \bibinfo {author}
  {\bibfnamefont {L.}~\bibnamefont {Wu}}, \bibinfo {author} {\bibfnamefont
  {Z.}~\bibnamefont {Zhu}}, \bibinfo {author} {\bibfnamefont {H.}~\bibnamefont
  {Guo}}, \bibinfo {author} {\bibfnamefont {R.}~\bibnamefont
  {Gonz{\'a}lez-Hern{\'a}ndez}}, \bibinfo {author} {\bibfnamefont
  {X.}~\bibnamefont {Wang}}, \bibinfo {author} {\bibfnamefont {H.}~\bibnamefont
  {Yan}}, \bibinfo {author} {\bibfnamefont {P.}~\bibnamefont {Qin}}, \bibinfo
  {author} {\bibfnamefont {X.}~\bibnamefont {Zhang}}, \bibinfo {author}
  {\bibfnamefont {H.}~\bibnamefont {Wu}}, \bibinfo {author} {\bibfnamefont
  {H.}~\bibnamefont {Chen}}, \bibinfo {author} {\bibfnamefont {Z.}~\bibnamefont
  {Meng}}, \bibinfo {author} {\bibfnamefont {L.}~\bibnamefont {Liu}}, \bibinfo
  {author} {\bibfnamefont {Z.}~\bibnamefont {Xia}}, \bibinfo {author}
  {\bibfnamefont {J.}~\bibnamefont {Sinova}}, \bibinfo {author} {\bibfnamefont
  {T.}~\bibnamefont {Jungwirth}},\ and\ \bibinfo {author} {\bibfnamefont
  {Z.}~\bibnamefont {Liu}},\ }\bibfield  {title} {\bibinfo {title} {{An
  anomalous Hall effect in altermagnetic ruthenium dioxide}},\ }\href
  {https://doi.org/10.1038/s41928-022-00866-z} {\bibfield  {journal} {\bibinfo
  {journal} {Nature Electronics}\ }\textbf {\bibinfo {volume} {5}},\ \bibinfo
  {pages} {735} (\bibinfo {year} {2022})}\BibitemShut {NoStop}%
\bibitem [{\citenamefont {Fedchenko}\ \emph {et~al.}(2024)\citenamefont
  {Fedchenko}, \citenamefont {Min{\'a}r}, \citenamefont {Akashdeep},
  \citenamefont {D'Souza}, \citenamefont {Vasilyev}, \citenamefont {Tkach},
  \citenamefont {Odenbreit}, \citenamefont {Nguyen}, \citenamefont
  {Kutnyakhov}, \citenamefont {Wind}, \citenamefont {Wenthaus}, \citenamefont
  {Scholz}, \citenamefont {Rossnagel}, \citenamefont {Hoesch}, \citenamefont
  {Aeschlimann}, \citenamefont {Stadtm{\"u}ller}, \citenamefont {Kl{\"a}ui},
  \citenamefont {Sch{\"o}nhense}, \citenamefont {Jungwirth}, \citenamefont
  {Hellenes}, \citenamefont {Jakob}, \citenamefont {{\v S}mejkal},
  \citenamefont {Sinova},\ and\ \citenamefont {Elmers}}]{Fedchenko2024-lw}%
  \BibitemOpen
  \bibfield  {author} {\bibinfo {author} {\bibfnamefont {O.}~\bibnamefont
  {Fedchenko}}, \bibinfo {author} {\bibfnamefont {J.}~\bibnamefont
  {Min{\'a}r}}, \bibinfo {author} {\bibfnamefont {A.}~\bibnamefont
  {Akashdeep}}, \bibinfo {author} {\bibfnamefont {S.~W.}\ \bibnamefont
  {D'Souza}}, \bibinfo {author} {\bibfnamefont {D.}~\bibnamefont {Vasilyev}},
  \bibinfo {author} {\bibfnamefont {O.}~\bibnamefont {Tkach}}, \bibinfo
  {author} {\bibfnamefont {L.}~\bibnamefont {Odenbreit}}, \bibinfo {author}
  {\bibfnamefont {Q.}~\bibnamefont {Nguyen}}, \bibinfo {author} {\bibfnamefont
  {D.}~\bibnamefont {Kutnyakhov}}, \bibinfo {author} {\bibfnamefont
  {N.}~\bibnamefont {Wind}}, \bibinfo {author} {\bibfnamefont {L.}~\bibnamefont
  {Wenthaus}}, \bibinfo {author} {\bibfnamefont {M.}~\bibnamefont {Scholz}},
  \bibinfo {author} {\bibfnamefont {K.}~\bibnamefont {Rossnagel}}, \bibinfo
  {author} {\bibfnamefont {M.}~\bibnamefont {Hoesch}}, \bibinfo {author}
  {\bibfnamefont {M.}~\bibnamefont {Aeschlimann}}, \bibinfo {author}
  {\bibfnamefont {B.}~\bibnamefont {Stadtm{\"u}ller}}, \bibinfo {author}
  {\bibfnamefont {M.}~\bibnamefont {Kl{\"a}ui}}, \bibinfo {author}
  {\bibfnamefont {G.}~\bibnamefont {Sch{\"o}nhense}}, \bibinfo {author}
  {\bibfnamefont {T.}~\bibnamefont {Jungwirth}}, \bibinfo {author}
  {\bibfnamefont {A.~B.}\ \bibnamefont {Hellenes}}, \bibinfo {author}
  {\bibfnamefont {G.}~\bibnamefont {Jakob}}, \bibinfo {author} {\bibfnamefont
  {L.}~\bibnamefont {{\v S}mejkal}}, \bibinfo {author} {\bibfnamefont
  {J.}~\bibnamefont {Sinova}},\ and\ \bibinfo {author} {\bibfnamefont {H.-J.}\
  \bibnamefont {Elmers}},\ }\bibfield  {title} {\bibinfo {title} {{Observation
  of time-reversal symmetry breaking in the band structure of altermagnetic
  {RuO$_2$}}},\ }\href {https://doi.org/10.1126/sciadv.adj4883} {\bibfield
  {journal} {\bibinfo  {journal} {Sci Adv}\ }\textbf {\bibinfo {volume} {10}},\
  \bibinfo {pages} {eadj4883} (\bibinfo {year} {2024})}\BibitemShut {NoStop}%
\bibitem [{\citenamefont {Lin}\ \emph {et~al.}()\citenamefont {Lin},
  \citenamefont {Chen}, \citenamefont {Lu}, \citenamefont {Liang},
  \citenamefont {Feng}, \citenamefont {Yamagami}, \citenamefont {Osiecki},
  \citenamefont {Leandersson}, \citenamefont {Thiagarajan}, \citenamefont
  {Liu}, \citenamefont {Felser},\ and\ \citenamefont {Ma}}]{Lin2024-qk}%
  \BibitemOpen
  \bibfield  {author} {\bibinfo {author} {\bibfnamefont {Z.}~\bibnamefont
  {Lin}}, \bibinfo {author} {\bibfnamefont {D.}~\bibnamefont {Chen}}, \bibinfo
  {author} {\bibfnamefont {W.}~\bibnamefont {Lu}}, \bibinfo {author}
  {\bibfnamefont {X.}~\bibnamefont {Liang}}, \bibinfo {author} {\bibfnamefont
  {S.}~\bibnamefont {Feng}}, \bibinfo {author} {\bibfnamefont {K.}~\bibnamefont
  {Yamagami}}, \bibinfo {author} {\bibfnamefont {J.}~\bibnamefont {Osiecki}},
  \bibinfo {author} {\bibfnamefont {M.}~\bibnamefont {Leandersson}}, \bibinfo
  {author} {\bibfnamefont {B.}~\bibnamefont {Thiagarajan}}, \bibinfo {author}
  {\bibfnamefont {J.}~\bibnamefont {Liu}}, \bibinfo {author} {\bibfnamefont
  {C.}~\bibnamefont {Felser}},\ and\ \bibinfo {author} {\bibfnamefont
  {J.}~\bibnamefont {Ma}},\ }\bibfield  {title} {\bibinfo {title} {{Observation
  of Giant Spin Splitting and d-wave Spin Texture in Room Temperature
  Altermagnet {RuO$_2$}}},\ }\Eprint {https://arxiv.org/abs/2402.04995}
  {arXiv:2402.04995} \BibitemShut {NoStop}%
\bibitem [{\citenamefont {Hayami}\ \emph {et~al.}(2016)\citenamefont {Hayami},
  \citenamefont {Kusunose},\ and\ \citenamefont {Motome}}]{Hayami2016-xs}%
  \BibitemOpen
  \bibfield  {author} {\bibinfo {author} {\bibfnamefont {S.}~\bibnamefont
  {Hayami}}, \bibinfo {author} {\bibfnamefont {H.}~\bibnamefont {Kusunose}},\
  and\ \bibinfo {author} {\bibfnamefont {Y.}~\bibnamefont {Motome}},\
  }\bibfield  {title} {\bibinfo {title} {{Emergent spin-valley-orbital physics
  by spontaneous parity breaking}},\ }\href
  {https://doi.org/10.1088/0953-8984/28/39/395601} {\bibfield  {journal}
  {\bibinfo  {journal} {J. Phys. Condens. Matter}\ }\textbf {\bibinfo {volume}
  {28}},\ \bibinfo {pages} {395601} (\bibinfo {year} {2016})}\BibitemShut
  {NoStop}%
\bibitem [{\citenamefont {Suzuki}\ \emph {et~al.}(2017)\citenamefont {Suzuki},
  \citenamefont {Koretsune}, \citenamefont {Ochi},\ and\ \citenamefont
  {Arita}}]{Suzuki2017-ve}%
  \BibitemOpen
  \bibfield  {author} {\bibinfo {author} {\bibfnamefont {M.-T.}\ \bibnamefont
  {Suzuki}}, \bibinfo {author} {\bibfnamefont {T.}~\bibnamefont {Koretsune}},
  \bibinfo {author} {\bibfnamefont {M.}~\bibnamefont {Ochi}},\ and\ \bibinfo
  {author} {\bibfnamefont {R.}~\bibnamefont {Arita}},\ }\bibfield  {title}
  {\bibinfo {title} {{Cluster multipole theory for anomalous Hall effect in
  antiferromagnets}},\ }\href {https://doi.org/10.1103/PhysRevB.95.094406}
  {\bibfield  {journal} {\bibinfo  {journal} {Phys. Rev. B}\ }\textbf {\bibinfo
  {volume} {95}},\ \bibinfo {pages} {094406} (\bibinfo {year}
  {2017})}\BibitemShut {NoStop}%
\bibitem [{\citenamefont {Suzuki}\ \emph {et~al.}(2018)\citenamefont {Suzuki},
  \citenamefont {Ikeda},\ and\ \citenamefont {Oppeneer}}]{Suzuki2018-kz}%
  \BibitemOpen
  \bibfield  {author} {\bibinfo {author} {\bibfnamefont {M.-T.}\ \bibnamefont
  {Suzuki}}, \bibinfo {author} {\bibfnamefont {H.}~\bibnamefont {Ikeda}},\ and\
  \bibinfo {author} {\bibfnamefont {P.~M.}\ \bibnamefont {Oppeneer}},\
  }\bibfield  {title} {\bibinfo {title} {{First-principles Theory of Magnetic
  Multipoles in Condensed Matter Systems}},\ }\href
  {https://doi.org/10.7566/JPSJ.87.041008} {\bibfield  {journal} {\bibinfo
  {journal} {J. Phys. Soc. Jpn.}\ }\textbf {\bibinfo {volume} {87}},\ \bibinfo
  {pages} {041008} (\bibinfo {year} {2018})}\BibitemShut {NoStop}%
\bibitem [{\citenamefont {Suzuki}\ \emph {et~al.}(2019)\citenamefont {Suzuki},
  \citenamefont {Nomoto}, \citenamefont {Arita}, \citenamefont {Yanagi},
  \citenamefont {Hayami},\ and\ \citenamefont {Kusunose}}]{Suzuki2019-ek}%
  \BibitemOpen
  \bibfield  {author} {\bibinfo {author} {\bibfnamefont {M.-T.}\ \bibnamefont
  {Suzuki}}, \bibinfo {author} {\bibfnamefont {T.}~\bibnamefont {Nomoto}},
  \bibinfo {author} {\bibfnamefont {R.}~\bibnamefont {Arita}}, \bibinfo
  {author} {\bibfnamefont {Y.}~\bibnamefont {Yanagi}}, \bibinfo {author}
  {\bibfnamefont {S.}~\bibnamefont {Hayami}},\ and\ \bibinfo {author}
  {\bibfnamefont {H.}~\bibnamefont {Kusunose}},\ }\bibfield  {title} {\bibinfo
  {title} {{Multipole expansion for magnetic structures: A generation scheme
  for a symmetry-adapted orthonormal basis set in the crystallographic point
  group}},\ }\href {https://doi.org/10.1103/PhysRevB.99.174407} {\bibfield
  {journal} {\bibinfo  {journal} {Phys. Rev. B}\ }\textbf {\bibinfo {volume}
  {99}},\ \bibinfo {pages} {174407} (\bibinfo {year} {2019})}\BibitemShut
  {NoStop}%
\bibitem [{\citenamefont {Huebsch}\ \emph {et~al.}(2021)\citenamefont
  {Huebsch}, \citenamefont {Nomoto}, \citenamefont {Suzuki},\ and\
  \citenamefont {Arita}}]{Huebsch2021-vx}%
  \BibitemOpen
  \bibfield  {author} {\bibinfo {author} {\bibfnamefont {M.-T.}\ \bibnamefont
  {Huebsch}}, \bibinfo {author} {\bibfnamefont {T.}~\bibnamefont {Nomoto}},
  \bibinfo {author} {\bibfnamefont {M.-T.}\ \bibnamefont {Suzuki}},\ and\
  \bibinfo {author} {\bibfnamefont {R.}~\bibnamefont {Arita}},\ }\bibfield
  {title} {\bibinfo {title} {{Benchmark for Ab Initio Prediction of Magnetic
  Structures Based on {Cluster-Multipole} Theory}},\ }\href
  {https://doi.org/10.1103/PhysRevX.11.011031} {\bibfield  {journal} {\bibinfo
  {journal} {Phys. Rev. X}\ }\textbf {\bibinfo {volume} {11}},\ \bibinfo
  {pages} {011031} (\bibinfo {year} {2021})}\BibitemShut {NoStop}%
\bibitem [{\citenamefont {Nakatsuji}\ \emph {et~al.}(2015)\citenamefont
  {Nakatsuji}, \citenamefont {Kiyohara},\ and\ \citenamefont
  {Higo}}]{Nakatsuji2015-ll}%
  \BibitemOpen
  \bibfield  {author} {\bibinfo {author} {\bibfnamefont {S.}~\bibnamefont
  {Nakatsuji}}, \bibinfo {author} {\bibfnamefont {N.}~\bibnamefont
  {Kiyohara}},\ and\ \bibinfo {author} {\bibfnamefont {T.}~\bibnamefont
  {Higo}},\ }\bibfield  {title} {\bibinfo {title} {{Large anomalous Hall effect
  in a non-collinear antiferromagnet at room temperature}},\ }\href
  {https://doi.org/10.1038/nature15723} {\bibfield  {journal} {\bibinfo
  {journal} {Nature}\ }\textbf {\bibinfo {volume} {527}},\ \bibinfo {pages}
  {212} (\bibinfo {year} {2015})}\BibitemShut {NoStop}%
\bibitem [{\citenamefont {Kiyohara}\ \emph {et~al.}(2016)\citenamefont
  {Kiyohara}, \citenamefont {Tomita},\ and\ \citenamefont
  {Nakatsuji}}]{Kiyohara2016-wl}%
  \BibitemOpen
  \bibfield  {author} {\bibinfo {author} {\bibfnamefont {N.}~\bibnamefont
  {Kiyohara}}, \bibinfo {author} {\bibfnamefont {T.}~\bibnamefont {Tomita}},\
  and\ \bibinfo {author} {\bibfnamefont {S.}~\bibnamefont {Nakatsuji}},\
  }\bibfield  {title} {\bibinfo {title} {{Giant Anomalous Hall Effect in the
  Chiral Antiferromagnet {Mn$_{3}$Ge}}},\ }\href
  {https://doi.org/10.1103/PhysRevApplied.5.064009} {\bibfield  {journal}
  {\bibinfo  {journal} {Phys. Rev. Appl.}\ }\textbf {\bibinfo {volume} {5}},\
  \bibinfo {pages} {064009} (\bibinfo {year} {2016})}\BibitemShut {NoStop}%
\bibitem [{\citenamefont {Nayak}\ \emph {et~al.}(2016)\citenamefont {Nayak},
  \citenamefont {Fischer}, \citenamefont {Sun}, \citenamefont {Yan},
  \citenamefont {Karel}, \citenamefont {Komarek}, \citenamefont {Shekhar},
  \citenamefont {Kumar}, \citenamefont {Schnelle}, \citenamefont {K{\"u}bler},
  \citenamefont {Felser},\ and\ \citenamefont {Parkin}}]{Nayak2016-tl}%
  \BibitemOpen
  \bibfield  {author} {\bibinfo {author} {\bibfnamefont {A.~K.}\ \bibnamefont
  {Nayak}}, \bibinfo {author} {\bibfnamefont {J.~E.}\ \bibnamefont {Fischer}},
  \bibinfo {author} {\bibfnamefont {Y.}~\bibnamefont {Sun}}, \bibinfo {author}
  {\bibfnamefont {B.}~\bibnamefont {Yan}}, \bibinfo {author} {\bibfnamefont
  {J.}~\bibnamefont {Karel}}, \bibinfo {author} {\bibfnamefont {A.~C.}\
  \bibnamefont {Komarek}}, \bibinfo {author} {\bibfnamefont {C.}~\bibnamefont
  {Shekhar}}, \bibinfo {author} {\bibfnamefont {N.}~\bibnamefont {Kumar}},
  \bibinfo {author} {\bibfnamefont {W.}~\bibnamefont {Schnelle}}, \bibinfo
  {author} {\bibfnamefont {J.}~\bibnamefont {K{\"u}bler}}, \bibinfo {author}
  {\bibfnamefont {C.}~\bibnamefont {Felser}},\ and\ \bibinfo {author}
  {\bibfnamefont {S.~S.~P.}\ \bibnamefont {Parkin}},\ }\bibfield  {title}
  {\bibinfo {title} {{Large anomalous Hall effect driven by a nonvanishing
  Berry curvature in the noncolinear antiferromagnet {Mn$_3$Ge}}},\ }\href
  {https://doi.org/10.1126/sciadv.1501870} {\bibfield  {journal} {\bibinfo
  {journal} {Sci Adv}\ }\textbf {\bibinfo {volume} {2}},\ \bibinfo {pages}
  {e1501870} (\bibinfo {year} {2016})}\BibitemShut {NoStop}%
\bibitem [{\citenamefont {Zhao}\ \emph {et~al.}(2011)\citenamefont {Zhao},
  \citenamefont {Mackie}, \citenamefont {MacLaughlin}, \citenamefont {Bernal},
  \citenamefont {Ishikawa}, \citenamefont {Ohta},\ and\ \citenamefont
  {Nakatsuji}}]{Zhao2011-uc}%
  \BibitemOpen
  \bibfield  {author} {\bibinfo {author} {\bibfnamefont {S.}~\bibnamefont
  {Zhao}}, \bibinfo {author} {\bibfnamefont {J.~M.}\ \bibnamefont {Mackie}},
  \bibinfo {author} {\bibfnamefont {D.~E.}\ \bibnamefont {MacLaughlin}},
  \bibinfo {author} {\bibfnamefont {O.~O.}\ \bibnamefont {Bernal}}, \bibinfo
  {author} {\bibfnamefont {J.~J.}\ \bibnamefont {Ishikawa}}, \bibinfo {author}
  {\bibfnamefont {Y.}~\bibnamefont {Ohta}},\ and\ \bibinfo {author}
  {\bibfnamefont {S.}~\bibnamefont {Nakatsuji}},\ }\bibfield  {title} {\bibinfo
  {title} {{Magnetic transition, long-range order, and moment fluctuations in
  the pyrochlore iridate {Eu$_{2}$Ir$_{2}$O$_{7}$}}},\ }\href
  {https://doi.org/10.1103/PhysRevB.83.180402} {\bibfield  {journal} {\bibinfo
  {journal} {Phys. Rev. B}\ }\textbf {\bibinfo {volume} {83}},\ \bibinfo
  {pages} {180402(R)} (\bibinfo {year} {2011})}\BibitemShut {NoStop}%
\bibitem [{\citenamefont {Ishikawa}\ \emph {et~al.}(2012)\citenamefont
  {Ishikawa}, \citenamefont {O'Farrell},\ and\ \citenamefont
  {Nakatsuji}}]{Ishikawa2012-mf}%
  \BibitemOpen
  \bibfield  {author} {\bibinfo {author} {\bibfnamefont {J.~J.}\ \bibnamefont
  {Ishikawa}}, \bibinfo {author} {\bibfnamefont {E.~C.~T.}\ \bibnamefont
  {O'Farrell}},\ and\ \bibinfo {author} {\bibfnamefont {S.}~\bibnamefont
  {Nakatsuji}},\ }\bibfield  {title} {\bibinfo {title} {{Continuous transition
  between antiferromagnetic insulator and paramagnetic metal in the pyrochlore
  iridate {Eu$_{2}$Ir$_{2}$O$_{7}$}}},\ }\href
  {https://doi.org/10.1103/PhysRevB.85.245109} {\bibfield  {journal} {\bibinfo
  {journal} {Phys. Rev. B}\ }\textbf {\bibinfo {volume} {85}},\ \bibinfo
  {pages} {245109} (\bibinfo {year} {2012})}\BibitemShut {NoStop}%
\bibitem [{\citenamefont {Sagayama}\ \emph {et~al.}(2013)\citenamefont
  {Sagayama}, \citenamefont {Uematsu}, \citenamefont {Arima}, \citenamefont
  {Sugimoto}, \citenamefont {Ishikawa}, \citenamefont {O'Farrell},\ and\
  \citenamefont {Nakatsuji}}]{Sagayama2013-xe}%
  \BibitemOpen
  \bibfield  {author} {\bibinfo {author} {\bibfnamefont {H.}~\bibnamefont
  {Sagayama}}, \bibinfo {author} {\bibfnamefont {D.}~\bibnamefont {Uematsu}},
  \bibinfo {author} {\bibfnamefont {T.}~\bibnamefont {Arima}}, \bibinfo
  {author} {\bibfnamefont {K.}~\bibnamefont {Sugimoto}}, \bibinfo {author}
  {\bibfnamefont {J.~J.}\ \bibnamefont {Ishikawa}}, \bibinfo {author}
  {\bibfnamefont {E.}~\bibnamefont {O'Farrell}},\ and\ \bibinfo {author}
  {\bibfnamefont {S.}~\bibnamefont {Nakatsuji}},\ }\bibfield  {title} {\bibinfo
  {title} {{Determination of long-range all-in-all-out ordering of {Ir$^{4+}$}
  moments in a pyrochlore iridate {Eu$_{2}$Ir$_{2}$O$_{7}$} by resonant x-ray
  diffraction}},\ }\href {https://doi.org/10.1103/PhysRevB.87.100403}
  {\bibfield  {journal} {\bibinfo  {journal} {Phys. Rev. B}\ }\textbf {\bibinfo
  {volume} {87}},\ \bibinfo {pages} {100403(R)} (\bibinfo {year}
  {2013})}\BibitemShut {NoStop}%
\bibitem [{\citenamefont {Tomiyasu}\ \emph {et~al.}(2012)\citenamefont
  {Tomiyasu}, \citenamefont {Matsuhira}, \citenamefont {Iwasa}, \citenamefont
  {Watahiki}, \citenamefont {Takagi}, \citenamefont {Wakeshima}, \citenamefont
  {Hinatsu}, \citenamefont {Yokoyama}, \citenamefont {Ohoyama},\ and\
  \citenamefont {Yamada}}]{Tomiyasu2012-cy}%
  \BibitemOpen
  \bibfield  {author} {\bibinfo {author} {\bibfnamefont {K.}~\bibnamefont
  {Tomiyasu}}, \bibinfo {author} {\bibfnamefont {K.}~\bibnamefont {Matsuhira}},
  \bibinfo {author} {\bibfnamefont {K.}~\bibnamefont {Iwasa}}, \bibinfo
  {author} {\bibfnamefont {M.}~\bibnamefont {Watahiki}}, \bibinfo {author}
  {\bibfnamefont {S.}~\bibnamefont {Takagi}}, \bibinfo {author} {\bibfnamefont
  {M.}~\bibnamefont {Wakeshima}}, \bibinfo {author} {\bibfnamefont
  {Y.}~\bibnamefont {Hinatsu}}, \bibinfo {author} {\bibfnamefont
  {M.}~\bibnamefont {Yokoyama}}, \bibinfo {author} {\bibfnamefont
  {K.}~\bibnamefont {Ohoyama}},\ and\ \bibinfo {author} {\bibfnamefont
  {K.}~\bibnamefont {Yamada}},\ }\bibfield  {title} {\bibinfo {title}
  {{Emergence of Magnetic Long-range Order in Frustrated Pyrochlore
  {Nd$_2$Ir$_2$O$_7$} with {Metal--Insulator} Transition}},\ }\href
  {https://doi.org/10.1143/JPSJ.81.034709} {\bibfield  {journal} {\bibinfo
  {journal} {J. Phys. Soc. Jpn.}\ }\textbf {\bibinfo {volume} {81}},\ \bibinfo
  {pages} {034709} (\bibinfo {year} {2012})}\BibitemShut {NoStop}%
\bibitem [{\citenamefont {Matsuhira}\ \emph {et~al.}(2007)\citenamefont
  {Matsuhira}, \citenamefont {Wakeshima}, \citenamefont {Nakanishi},
  \citenamefont {Yamada}, \citenamefont {Nakamura}, \citenamefont {Kawano},
  \citenamefont {Takagi},\ and\ \citenamefont {Hinatsu}}]{Matsuhira2007-rt}%
  \BibitemOpen
  \bibfield  {author} {\bibinfo {author} {\bibfnamefont {K.}~\bibnamefont
  {Matsuhira}}, \bibinfo {author} {\bibfnamefont {M.}~\bibnamefont
  {Wakeshima}}, \bibinfo {author} {\bibfnamefont {R.}~\bibnamefont
  {Nakanishi}}, \bibinfo {author} {\bibfnamefont {T.}~\bibnamefont {Yamada}},
  \bibinfo {author} {\bibfnamefont {A.}~\bibnamefont {Nakamura}}, \bibinfo
  {author} {\bibfnamefont {W.}~\bibnamefont {Kawano}}, \bibinfo {author}
  {\bibfnamefont {S.}~\bibnamefont {Takagi}},\ and\ \bibinfo {author}
  {\bibfnamefont {Y.}~\bibnamefont {Hinatsu}},\ }\bibfield  {title} {\bibinfo
  {title} {{{Metal--Insulator} Transition in Pyrochlore Iridates
  {\textit{Ln}$_2$Ir$_2$O$_7$} (\textit{Ln} = Nd, Sm, and Eu)}},\ }\href
  {https://doi.org/10.1143/JPSJ.76.043706} {\bibfield  {journal} {\bibinfo
  {journal} {J. Phys. Soc. Jpn.}\ }\textbf {\bibinfo {volume} {76}},\ \bibinfo
  {pages} {043706} (\bibinfo {year} {2007})}\BibitemShut {NoStop}%
\bibitem [{\citenamefont {Matsuhira}\ \emph {et~al.}(2011)\citenamefont
  {Matsuhira}, \citenamefont {Wakeshima}, \citenamefont {Hinatsu},\ and\
  \citenamefont {Takagi}}]{Matsuhira2011-ca}%
  \BibitemOpen
  \bibfield  {author} {\bibinfo {author} {\bibfnamefont {K.}~\bibnamefont
  {Matsuhira}}, \bibinfo {author} {\bibfnamefont {M.}~\bibnamefont
  {Wakeshima}}, \bibinfo {author} {\bibfnamefont {Y.}~\bibnamefont {Hinatsu}},\
  and\ \bibinfo {author} {\bibfnamefont {S.}~\bibnamefont {Takagi}},\
  }\bibfield  {title} {\bibinfo {title} {{{Metal--Insulator} Transitions in
  Pyrochlore Oxides {\textit{Ln}$_2$Ir$_2$O$_7$}}},\ }\href
  {https://doi.org/10.1143/JPSJ.80.094701} {\bibfield  {journal} {\bibinfo
  {journal} {J. Phys. Soc. Jpn.}\ }\textbf {\bibinfo {volume} {80}},\ \bibinfo
  {pages} {094701} (\bibinfo {year} {2011})}\BibitemShut {NoStop}%
\bibitem [{\citenamefont {Witczak-Krempa}\ \emph {et~al.}(2014)\citenamefont
  {Witczak-Krempa}, \citenamefont {Chen}, \citenamefont {Kim},\ and\
  \citenamefont {Balents}}]{Witczak-Krempa2014-rt}%
  \BibitemOpen
  \bibfield  {author} {\bibinfo {author} {\bibfnamefont {W.}~\bibnamefont
  {Witczak-Krempa}}, \bibinfo {author} {\bibfnamefont {G.}~\bibnamefont
  {Chen}}, \bibinfo {author} {\bibfnamefont {Y.~B.}\ \bibnamefont {Kim}},\ and\
  \bibinfo {author} {\bibfnamefont {L.}~\bibnamefont {Balents}},\ }\bibfield
  {title} {\bibinfo {title} {{Correlated Quantum Phenomena in the Strong
  {Spin-Orbit} Regime}},\ }\href
  {https://doi.org/10.1146/annurev-conmatphys-020911-125138} {\bibfield
  {journal} {\bibinfo  {journal} {Annu. Rev. Condens. Matter Phys.}\ }\textbf
  {\bibinfo {volume} {5}},\ \bibinfo {pages} {57} (\bibinfo {year}
  {2014})}\BibitemShut {NoStop}%
\bibitem [{\citenamefont {Arima}(2013)}]{Arima2013-mp}%
  \BibitemOpen
  \bibfield  {author} {\bibinfo {author} {\bibfnamefont {T.-H.}\ \bibnamefont
  {Arima}},\ }\bibfield  {title} {\bibinfo {title} {{Time-Reversal Symmetry
  Breaking and Consequent Physical Responses Induced by All-In-All-Out Type
  Magnetic Order on the Pyrochlore Lattice}},\ }\href
  {https://doi.org/10.7566/JPSJ.82.013705} {\bibfield  {journal} {\bibinfo
  {journal} {J. Phys. Soc. Jpn.}\ }\textbf {\bibinfo {volume} {82}},\ \bibinfo
  {pages} {013705} (\bibinfo {year} {2013})}\BibitemShut {NoStop}%
\bibitem [{\citenamefont {Ueda}\ \emph {et~al.}(2017)\citenamefont {Ueda},
  \citenamefont {Oh}, \citenamefont {Yang}, \citenamefont {Kaneko},
  \citenamefont {Fujioka}, \citenamefont {Nagaosa},\ and\ \citenamefont
  {Tokura}}]{Ueda2017-jr}%
  \BibitemOpen
  \bibfield  {author} {\bibinfo {author} {\bibfnamefont {K.}~\bibnamefont
  {Ueda}}, \bibinfo {author} {\bibfnamefont {T.}~\bibnamefont {Oh}}, \bibinfo
  {author} {\bibfnamefont {B.-J.}\ \bibnamefont {Yang}}, \bibinfo {author}
  {\bibfnamefont {R.}~\bibnamefont {Kaneko}}, \bibinfo {author} {\bibfnamefont
  {J.}~\bibnamefont {Fujioka}}, \bibinfo {author} {\bibfnamefont
  {N.}~\bibnamefont {Nagaosa}},\ and\ \bibinfo {author} {\bibfnamefont
  {Y.}~\bibnamefont {Tokura}},\ }\bibfield  {title} {\bibinfo {title}
  {{Magnetic-field induced multiple topological phases in pyrochlore iridates
  with Mott criticality}},\ }\href {https://doi.org/10.1038/ncomms15515}
  {\bibfield  {journal} {\bibinfo  {journal} {Nat. Commun.}\ }\textbf {\bibinfo
  {volume} {8}},\ \bibinfo {pages} {15515} (\bibinfo {year}
  {2017})}\BibitemShut {NoStop}%
\bibitem [{\citenamefont {Ueda}\ \emph {et~al.}(2018)\citenamefont {Ueda},
  \citenamefont {Kaneko}, \citenamefont {Ishizuka}, \citenamefont {Fujioka},
  \citenamefont {Nagaosa},\ and\ \citenamefont {Tokura}}]{Ueda2018-wq}%
  \BibitemOpen
  \bibfield  {author} {\bibinfo {author} {\bibfnamefont {K.}~\bibnamefont
  {Ueda}}, \bibinfo {author} {\bibfnamefont {R.}~\bibnamefont {Kaneko}},
  \bibinfo {author} {\bibfnamefont {H.}~\bibnamefont {Ishizuka}}, \bibinfo
  {author} {\bibfnamefont {J.}~\bibnamefont {Fujioka}}, \bibinfo {author}
  {\bibfnamefont {N.}~\bibnamefont {Nagaosa}},\ and\ \bibinfo {author}
  {\bibfnamefont {Y.}~\bibnamefont {Tokura}},\ }\bibfield  {title} {\bibinfo
  {title} {{Spontaneous Hall effect in the Weyl semimetal candidate of all-in
  all-out pyrochlore iridate}},\ }\href
  {https://doi.org/10.1038/s41467-018-05530-9} {\bibfield  {journal} {\bibinfo
  {journal} {Nat. Commun.}\ }\textbf {\bibinfo {volume} {9}},\ \bibinfo {pages}
  {3032} (\bibinfo {year} {2018})}\BibitemShut {NoStop}%
\bibitem [{\citenamefont {Kim}\ \emph {et~al.}(2020)\citenamefont {Kim},
  \citenamefont {Oh}, \citenamefont {Song}, \citenamefont {Ko}, \citenamefont
  {Li}, \citenamefont {Mun}, \citenamefont {Kim}, \citenamefont {Son},
  \citenamefont {Yang}, \citenamefont {Kohama}, \citenamefont {Kim},
  \citenamefont {Yang},\ and\ \citenamefont {Noh}}]{Kim2020-er}%
  \BibitemOpen
  \bibfield  {author} {\bibinfo {author} {\bibfnamefont {W.~J.}\ \bibnamefont
  {Kim}}, \bibinfo {author} {\bibfnamefont {T.}~\bibnamefont {Oh}}, \bibinfo
  {author} {\bibfnamefont {J.}~\bibnamefont {Song}}, \bibinfo {author}
  {\bibfnamefont {E.~K.}\ \bibnamefont {Ko}}, \bibinfo {author} {\bibfnamefont
  {Y.}~\bibnamefont {Li}}, \bibinfo {author} {\bibfnamefont {J.}~\bibnamefont
  {Mun}}, \bibinfo {author} {\bibfnamefont {B.}~\bibnamefont {Kim}}, \bibinfo
  {author} {\bibfnamefont {J.}~\bibnamefont {Son}}, \bibinfo {author}
  {\bibfnamefont {Z.}~\bibnamefont {Yang}}, \bibinfo {author} {\bibfnamefont
  {Y.}~\bibnamefont {Kohama}}, \bibinfo {author} {\bibfnamefont
  {M.}~\bibnamefont {Kim}}, \bibinfo {author} {\bibfnamefont {B.-J.}\
  \bibnamefont {Yang}},\ and\ \bibinfo {author} {\bibfnamefont {T.~W.}\
  \bibnamefont {Noh}},\ }\bibfield  {title} {\bibinfo {title} {{Strain
  engineering of the magnetic multipole moments and anomalous Hall effect in
  pyrochlore iridate thin films}},\ }\href
  {https://doi.org/10.1126/sciadv.abb1539} {\bibfield  {journal} {\bibinfo
  {journal} {Sci. Adv.}\ }\textbf {\bibinfo {volume} {6}},\ \bibinfo {pages}
  {eabb1539} (\bibinfo {year} {2020})}\BibitemShut {NoStop}%
\bibitem [{\citenamefont {Li}\ \emph {et~al.}(2021)\citenamefont {Li},
  \citenamefont {Oh}, \citenamefont {Son}, \citenamefont {Song}, \citenamefont
  {Kim}, \citenamefont {Song}, \citenamefont {Kim}, \citenamefont {Chang},
  \citenamefont {Kim}, \citenamefont {Yang},\ and\ \citenamefont
  {Noh}}]{Li2021-zv}%
  \BibitemOpen
  \bibfield  {author} {\bibinfo {author} {\bibfnamefont {Y.}~\bibnamefont
  {Li}}, \bibinfo {author} {\bibfnamefont {T.}~\bibnamefont {Oh}}, \bibinfo
  {author} {\bibfnamefont {J.}~\bibnamefont {Son}}, \bibinfo {author}
  {\bibfnamefont {J.}~\bibnamefont {Song}}, \bibinfo {author} {\bibfnamefont
  {M.~K.}\ \bibnamefont {Kim}}, \bibinfo {author} {\bibfnamefont
  {D.}~\bibnamefont {Song}}, \bibinfo {author} {\bibfnamefont {S.}~\bibnamefont
  {Kim}}, \bibinfo {author} {\bibfnamefont {S.~H.}\ \bibnamefont {Chang}},
  \bibinfo {author} {\bibfnamefont {C.}~\bibnamefont {Kim}}, \bibinfo {author}
  {\bibfnamefont {B.-J.}\ \bibnamefont {Yang}},\ and\ \bibinfo {author}
  {\bibfnamefont {T.~W.}\ \bibnamefont {Noh}},\ }\bibfield  {title} {\bibinfo
  {title} {{Correlated Magnetic Weyl Semimetal State in Strained
  Pr$_2$Ir$_2$O$_7$}},\ }\href {https://doi.org/10.1002/adma.202008528}
  {\bibfield  {journal} {\bibinfo  {journal} {Adv. Mater.}\ }\textbf {\bibinfo
  {volume} {33}},\ \bibinfo {pages} {e2008528} (\bibinfo {year}
  {2021})}\BibitemShut {NoStop}%
\bibitem [{\citenamefont {Ghosh}\ \emph {et~al.}(2023)\citenamefont {Ghosh},
  \citenamefont {Samal},\ and\ \citenamefont {Kumar}}]{Ghosh2023-hl}%
  \BibitemOpen
  \bibfield  {author} {\bibinfo {author} {\bibfnamefont {M.}~\bibnamefont
  {Ghosh}}, \bibinfo {author} {\bibfnamefont {D.}~\bibnamefont {Samal}},\ and\
  \bibinfo {author} {\bibfnamefont {P.~S.~A.}\ \bibnamefont {Kumar}},\
  }\bibfield  {title} {\bibinfo {title} {{Spontaneous Hall effect in the
  magnetic Weyl semimetallic {Eu$_2$Ir$_2$O$_7$} (111) thin films}},\ }\href
  {https://doi.org/10.1063/5.0172127} {\bibfield  {journal} {\bibinfo
  {journal} {Appl. Phys. Lett.}\ }\textbf {\bibinfo {volume} {123}},\ \bibinfo
  {pages} {213101} (\bibinfo {year} {2023})}\BibitemShut {NoStop}%
\bibitem [{\citenamefont {Nagaosa}\ \emph {et~al.}(2010)\citenamefont
  {Nagaosa}, \citenamefont {Sinova}, \citenamefont {Onoda}, \citenamefont
  {MacDonald},\ and\ \citenamefont {Ong}}]{Nagaosa2010-we}%
  \BibitemOpen
  \bibfield  {author} {\bibinfo {author} {\bibfnamefont {N.}~\bibnamefont
  {Nagaosa}}, \bibinfo {author} {\bibfnamefont {J.}~\bibnamefont {Sinova}},
  \bibinfo {author} {\bibfnamefont {S.}~\bibnamefont {Onoda}}, \bibinfo
  {author} {\bibfnamefont {A.~H.}\ \bibnamefont {MacDonald}},\ and\ \bibinfo
  {author} {\bibfnamefont {N.~P.}\ \bibnamefont {Ong}},\ }\bibfield  {title}
  {\bibinfo {title} {{Anomalous Hall effect}},\ }\href
  {https://doi.org/10.1103/RevModPhys.82.1539} {\bibfield  {journal} {\bibinfo
  {journal} {Rev. Mod. Phys.}\ }\textbf {\bibinfo {volume} {82}},\ \bibinfo
  {pages} {1539} (\bibinfo {year} {2010})}\BibitemShut {NoStop}%
\bibitem [{\citenamefont {Jungwirth}\ \emph {et~al.}(2016)\citenamefont
  {Jungwirth}, \citenamefont {Marti}, \citenamefont {Wadley},\ and\
  \citenamefont {Wunderlich}}]{Jungwirth2016-td}%
  \BibitemOpen
  \bibfield  {author} {\bibinfo {author} {\bibfnamefont {T.}~\bibnamefont
  {Jungwirth}}, \bibinfo {author} {\bibfnamefont {X.}~\bibnamefont {Marti}},
  \bibinfo {author} {\bibfnamefont {P.}~\bibnamefont {Wadley}},\ and\ \bibinfo
  {author} {\bibfnamefont {J.}~\bibnamefont {Wunderlich}},\ }\bibfield  {title}
  {\bibinfo {title} {Antiferromagnetic spintronics},\ }\href
  {https://doi.org/10.1038/nnano.2016.18} {\bibfield  {journal} {\bibinfo
  {journal} {Nat. Nanotechnol.}\ }\textbf {\bibinfo {volume} {11}},\ \bibinfo
  {pages} {231} (\bibinfo {year} {2016})}\BibitemShut {NoStop}%
\bibitem [{\citenamefont {Baltz}\ \emph {et~al.}(2018)\citenamefont {Baltz},
  \citenamefont {Manchon}, \citenamefont {Tsoi}, \citenamefont {Moriyama},
  \citenamefont {Ono},\ and\ \citenamefont {Tserkovnyak}}]{Baltz2018-or}%
  \BibitemOpen
  \bibfield  {author} {\bibinfo {author} {\bibfnamefont {V.}~\bibnamefont
  {Baltz}}, \bibinfo {author} {\bibfnamefont {A.}~\bibnamefont {Manchon}},
  \bibinfo {author} {\bibfnamefont {M.}~\bibnamefont {Tsoi}}, \bibinfo {author}
  {\bibfnamefont {T.}~\bibnamefont {Moriyama}}, \bibinfo {author}
  {\bibfnamefont {T.}~\bibnamefont {Ono}},\ and\ \bibinfo {author}
  {\bibfnamefont {Y.}~\bibnamefont {Tserkovnyak}},\ }\bibfield  {title}
  {\bibinfo {title} {Antiferromagnetic spintronics},\ }\href
  {https://doi.org/10.1103/RevModPhys.90.015005} {\bibfield  {journal}
  {\bibinfo  {journal} {Rev. Mod. Phys.}\ }\textbf {\bibinfo {volume} {90}},\
  \bibinfo {pages} {015005} (\bibinfo {year} {2018})}\BibitemShut {NoStop}%
\bibitem [{\citenamefont {Xiao}\ \emph {et~al.}(2022)\citenamefont {Xiao},
  \citenamefont {Liu}, \citenamefont {Wu}, \citenamefont {Wang}, \citenamefont
  {Niu},\ and\ \citenamefont {Yang}}]{Xiao2022-xr}%
  \BibitemOpen
  \bibfield  {author} {\bibinfo {author} {\bibfnamefont {C.}~\bibnamefont
  {Xiao}}, \bibinfo {author} {\bibfnamefont {H.}~\bibnamefont {Liu}}, \bibinfo
  {author} {\bibfnamefont {W.}~\bibnamefont {Wu}}, \bibinfo {author}
  {\bibfnamefont {H.}~\bibnamefont {Wang}}, \bibinfo {author} {\bibfnamefont
  {Q.}~\bibnamefont {Niu}},\ and\ \bibinfo {author} {\bibfnamefont {S.~A.}\
  \bibnamefont {Yang}},\ }\bibfield  {title} {\bibinfo {title} {{Intrinsic
  Nonlinear Electric Spin Generation in Centrosymmetric Magnets}},\ }\href
  {https://doi.org/10.1103/PhysRevLett.129.086602} {\bibfield  {journal}
  {\bibinfo  {journal} {Phys. Rev. Lett.}\ }\textbf {\bibinfo {volume} {129}},\
  \bibinfo {pages} {086602} (\bibinfo {year} {2022})}\BibitemShut {NoStop}%
\bibitem [{\citenamefont {Xiao}\ \emph {et~al.}(2023)\citenamefont {Xiao},
  \citenamefont {Wu}, \citenamefont {Wang}, \citenamefont {Huang},
  \citenamefont {Feng}, \citenamefont {Liu}, \citenamefont {Guo}, \citenamefont
  {Niu},\ and\ \citenamefont {Yang}}]{Xiao2023-yu}%
  \BibitemOpen
  \bibfield  {author} {\bibinfo {author} {\bibfnamefont {C.}~\bibnamefont
  {Xiao}}, \bibinfo {author} {\bibfnamefont {W.}~\bibnamefont {Wu}}, \bibinfo
  {author} {\bibfnamefont {H.}~\bibnamefont {Wang}}, \bibinfo {author}
  {\bibfnamefont {Y.-X.}\ \bibnamefont {Huang}}, \bibinfo {author}
  {\bibfnamefont {X.}~\bibnamefont {Feng}}, \bibinfo {author} {\bibfnamefont
  {H.}~\bibnamefont {Liu}}, \bibinfo {author} {\bibfnamefont {G.-Y.}\
  \bibnamefont {Guo}}, \bibinfo {author} {\bibfnamefont {Q.}~\bibnamefont
  {Niu}},\ and\ \bibinfo {author} {\bibfnamefont {S.~A.}\ \bibnamefont
  {Yang}},\ }\bibfield  {title} {\bibinfo {title} {{Time-Reversal-Even
  Nonlinear Current Induced Spin Polarization}},\ }\href
  {https://doi.org/10.1103/PhysRevLett.130.166302} {\bibfield  {journal}
  {\bibinfo  {journal} {Phys. Rev. Lett.}\ }\textbf {\bibinfo {volume} {130}},\
  \bibinfo {pages} {166302} (\bibinfo {year} {2023})}\BibitemShut {NoStop}%
\bibitem [{\citenamefont {Baek}\ \emph {et~al.}()\citenamefont {Baek},
  \citenamefont {Han}, \citenamefont {Cheon},\ and\ \citenamefont
  {Lee}}]{Baek2023-aq}%
  \BibitemOpen
  \bibfield  {author} {\bibinfo {author} {\bibfnamefont {I.}~\bibnamefont
  {Baek}}, \bibinfo {author} {\bibfnamefont {S.}~\bibnamefont {Han}}, \bibinfo
  {author} {\bibfnamefont {S.}~\bibnamefont {Cheon}},\ and\ \bibinfo {author}
  {\bibfnamefont {H.-W.}\ \bibnamefont {Lee}},\ }\bibfield  {title} {\bibinfo
  {title} {Nonlinear orbital and spin edelstein effect in centrosymmetric
  metals},\ }\Eprint {https://arxiv.org/abs/2310.05113} {arXiv:2310.05113}
  \BibitemShut {NoStop}%
\bibitem [{\citenamefont {Feng}\ \emph {et~al.}()\citenamefont {Feng},
  \citenamefont {Wu}, \citenamefont {Wang}, \citenamefont {Gao}, \citenamefont
  {Ang}, \citenamefont {Zhao}, \citenamefont {Xiao},\ and\ \citenamefont
  {Yang}}]{Feng2024-gj}%
  \BibitemOpen
  \bibfield  {author} {\bibinfo {author} {\bibfnamefont {X.}~\bibnamefont
  {Feng}}, \bibinfo {author} {\bibfnamefont {W.}~\bibnamefont {Wu}}, \bibinfo
  {author} {\bibfnamefont {H.}~\bibnamefont {Wang}}, \bibinfo {author}
  {\bibfnamefont {W.}~\bibnamefont {Gao}}, \bibinfo {author} {\bibfnamefont
  {L.~K.}\ \bibnamefont {Ang}}, \bibinfo {author} {\bibfnamefont {Y.~X.}\
  \bibnamefont {Zhao}}, \bibinfo {author} {\bibfnamefont {C.}~\bibnamefont
  {Xiao}},\ and\ \bibinfo {author} {\bibfnamefont {S.~A.}\ \bibnamefont
  {Yang}},\ }\bibfield  {title} {\bibinfo {title} {{Quantum Metric Nonlinear
  Spin-Orbit Torque Enhanced by Topological Bands}},\ }\Eprint
  {https://arxiv.org/abs/2402.00532} {arXiv:2402.00532} \BibitemShut {NoStop}%
\bibitem [{\citenamefont {Guo}\ \emph {et~al.}(2024)\citenamefont {Guo},
  \citenamefont {Huang}, \citenamefont {Yang}, \citenamefont {Liu},
  \citenamefont {Xiao},\ and\ \citenamefont {Yuan}}]{Guo2024-vh}%
  \BibitemOpen
  \bibfield  {author} {\bibinfo {author} {\bibfnamefont {R.}~\bibnamefont
  {Guo}}, \bibinfo {author} {\bibfnamefont {Y.-X.}\ \bibnamefont {Huang}},
  \bibinfo {author} {\bibfnamefont {X.}~\bibnamefont {Yang}}, \bibinfo {author}
  {\bibfnamefont {Y.}~\bibnamefont {Liu}}, \bibinfo {author} {\bibfnamefont
  {C.}~\bibnamefont {Xiao}},\ and\ \bibinfo {author} {\bibfnamefont
  {Z.}~\bibnamefont {Yuan}},\ }\bibfield  {title} {\bibinfo {title} {{Extrinsic
  contribution to nonlinear current induced spin polarization}},\ }\href
  {https://doi.org/10.1103/PhysRevB.109.235413} {\bibfield  {journal} {\bibinfo
   {journal} {Phys. Rev. B}\ }\textbf {\bibinfo {volume} {109}},\ \bibinfo
  {pages} {235413} (\bibinfo {year} {2024})}\BibitemShut {NoStop}%
\bibitem [{\citenamefont {{\=O}ik{\'e}}\ and\ \citenamefont
  {Peters}()}]{Oike2024-ms}%
  \BibitemOpen
  \bibfield  {author} {\bibinfo {author} {\bibfnamefont {J.}~\bibnamefont
  {{\=O}ik{\'e}}}\ and\ \bibinfo {author} {\bibfnamefont {R.}~\bibnamefont
  {Peters}},\ }\bibfield  {title} {\bibinfo {title} {{Nonlinear Edelstein
  Effect in Strongly Correlated Electron Systems}},\ }\Eprint
  {https://arxiv.org/abs/2403.17189} {arXiv:2403.17189} \BibitemShut {NoStop}%
\bibitem [{\citenamefont {Hu}\ \emph {et~al.}()\citenamefont {Hu},
  \citenamefont {Matsyshyn},\ and\ \citenamefont {Song}}]{Hu2024-gc}%
  \BibitemOpen
  \bibfield  {author} {\bibinfo {author} {\bibfnamefont {J.-X.}\ \bibnamefont
  {Hu}}, \bibinfo {author} {\bibfnamefont {O.}~\bibnamefont {Matsyshyn}},\ and\
  \bibinfo {author} {\bibfnamefont {J.~C.~W.}\ \bibnamefont {Song}},\
  }\bibfield  {title} {\bibinfo {title} {{Nonlinear Superconducting
  Magnetoelectric Effect}},\ }\Eprint {https://arxiv.org/abs/2404.18616}
  {arXiv:2404.18616} \BibitemShut {NoStop}%
\bibitem [{\citenamefont {Urru}\ and\ \citenamefont
  {Spaldin}(2022)}]{Urru2022-ww}%
  \BibitemOpen
  \bibfield  {author} {\bibinfo {author} {\bibfnamefont {A.}~\bibnamefont
  {Urru}}\ and\ \bibinfo {author} {\bibfnamefont {N.~A.}\ \bibnamefont
  {Spaldin}},\ }\bibfield  {title} {\bibinfo {title} {{Magnetic octupole tensor
  decomposition and second-order magnetoelectric effect}},\ }\href
  {https://doi.org/10.1016/j.aop.2022.168964} {\bibfield  {journal} {\bibinfo
  {journal} {Ann. Phys.}\ }\textbf {\bibinfo {volume} {447}},\ \bibinfo {pages}
  {168964} (\bibinfo {year} {2022})}\BibitemShut {NoStop}%
\bibitem [{\citenamefont {Elcoro}\ \emph {et~al.}(2019)\citenamefont {Elcoro},
  \citenamefont {Etxebarria}, \citenamefont {Gallego}, \citenamefont
  {Perez-Mato},\ and\ \citenamefont {Tasci}}]{Elcoro2019-yl}%
  \BibitemOpen
  \bibfield  {author} {\bibinfo {author} {\bibfnamefont {L.}~\bibnamefont
  {Elcoro}}, \bibinfo {author} {\bibfnamefont {J.}~\bibnamefont {Etxebarria}},
  \bibinfo {author} {\bibfnamefont {S.~V.}\ \bibnamefont {Gallego}}, \bibinfo
  {author} {\bibfnamefont {J.~M.}\ \bibnamefont {Perez-Mato}},\ and\ \bibinfo
  {author} {\bibfnamefont {E.~S.}\ \bibnamefont {Tasci}},\ }\bibfield  {title}
  {\bibinfo {title} {{Automatic calculation of symmetry-adapted tensors in
  magnetic and non-magnetic materials: a new tool of the Bilbao
  Crystallographic Server}},\ }\href
  {https://doi.org/10.1107/S2053273319001748} {\bibfield  {journal} {\bibinfo
  {journal} {Acta Cryst. A}\ }\textbf {\bibinfo {volume} {75}},\ \bibinfo
  {pages} {438} (\bibinfo {year} {2019})}\BibitemShut {NoStop}%
\bibitem [{\citenamefont {Wan}\ \emph {et~al.}(2011)\citenamefont {Wan},
  \citenamefont {Turner}, \citenamefont {Vishwanath},\ and\ \citenamefont
  {Savrasov}}]{Wan2011-sf}%
  \BibitemOpen
  \bibfield  {author} {\bibinfo {author} {\bibfnamefont {X.}~\bibnamefont
  {Wan}}, \bibinfo {author} {\bibfnamefont {A.~M.}\ \bibnamefont {Turner}},
  \bibinfo {author} {\bibfnamefont {A.}~\bibnamefont {Vishwanath}},\ and\
  \bibinfo {author} {\bibfnamefont {S.~Y.}\ \bibnamefont {Savrasov}},\
  }\bibfield  {title} {\bibinfo {title} {{Topological semimetal and Fermi-arc
  surface states in the electronic structure of pyrochlore iridates}},\ }\href
  {https://doi.org/10.1103/PhysRevB.83.205101} {\bibfield  {journal} {\bibinfo
  {journal} {Phys. Rev. B}\ }\textbf {\bibinfo {volume} {83}},\ \bibinfo
  {pages} {205101} (\bibinfo {year} {2011})}\BibitemShut {NoStop}%
\bibitem [{\citenamefont {Witczak-Krempa}\ and\ \citenamefont
  {Kim}(2012)}]{Witczak-Krempa2012-ro}%
  \BibitemOpen
  \bibfield  {author} {\bibinfo {author} {\bibfnamefont {W.}~\bibnamefont
  {Witczak-Krempa}}\ and\ \bibinfo {author} {\bibfnamefont {Y.~B.}\
  \bibnamefont {Kim}},\ }\bibfield  {title} {\bibinfo {title} {{Topological and
  magnetic phases of interacting electrons in the pyrochlore iridates}},\
  }\href {https://doi.org/10.1103/PhysRevB.85.045124} {\bibfield  {journal}
  {\bibinfo  {journal} {Phys. Rev. B}\ }\textbf {\bibinfo {volume} {85}},\
  \bibinfo {pages} {045124} (\bibinfo {year} {2012})}\BibitemShut {NoStop}%
\bibitem [{\citenamefont {Witczak-Krempa}\ \emph {et~al.}(2013)\citenamefont
  {Witczak-Krempa}, \citenamefont {Go},\ and\ \citenamefont
  {Kim}}]{Witczak-Krempa2013-bv}%
  \BibitemOpen
  \bibfield  {author} {\bibinfo {author} {\bibfnamefont {W.}~\bibnamefont
  {Witczak-Krempa}}, \bibinfo {author} {\bibfnamefont {A.}~\bibnamefont {Go}},\
  and\ \bibinfo {author} {\bibfnamefont {Y.~B.}\ \bibnamefont {Kim}},\
  }\bibfield  {title} {\bibinfo {title} {{Pyrochlore electrons under pressure,
  heat, and field: Shedding light on the iridates}},\ }\href
  {https://doi.org/10.1103/PhysRevB.87.155101} {\bibfield  {journal} {\bibinfo
  {journal} {Phys. Rev. B}\ }\textbf {\bibinfo {volume} {87}},\ \bibinfo
  {pages} {155101} (\bibinfo {year} {2013})}\BibitemShut {NoStop}%
\bibitem [{\citenamefont {Yamaji}\ and\ \citenamefont
  {Imada}(2014)}]{Yamaji2014-fh}%
  \BibitemOpen
  \bibfield  {author} {\bibinfo {author} {\bibfnamefont {Y.}~\bibnamefont
  {Yamaji}}\ and\ \bibinfo {author} {\bibfnamefont {M.}~\bibnamefont {Imada}},\
  }\bibfield  {title} {\bibinfo {title} {{Metallic Interface Emerging at
  Magnetic Domain Wall of Antiferromagnetic Insulator: Fate of Extinct Weyl
  Electrons}},\ }\href {https://doi.org/10.1103/PhysRevX.4.021035} {\bibfield
  {journal} {\bibinfo  {journal} {Phys. Rev. X}\ }\textbf {\bibinfo {volume}
  {4}},\ \bibinfo {pages} {021035} (\bibinfo {year} {2014})}\BibitemShut
  {NoStop}%
\bibitem [{\citenamefont {Varnava}\ and\ \citenamefont
  {Vanderbilt}(2018)}]{Varnava2018-ub}%
  \BibitemOpen
  \bibfield  {author} {\bibinfo {author} {\bibfnamefont {N.}~\bibnamefont
  {Varnava}}\ and\ \bibinfo {author} {\bibfnamefont {D.}~\bibnamefont
  {Vanderbilt}},\ }\bibfield  {title} {\bibinfo {title} {{Surfaces of axion
  insulators}},\ }\href {https://doi.org/10.1103/PhysRevB.98.245117} {\bibfield
   {journal} {\bibinfo  {journal} {Phys. Rev. B}\ }\textbf {\bibinfo {volume}
  {98}},\ \bibinfo {pages} {245117} (\bibinfo {year} {2018})}\BibitemShut
  {NoStop}%
\bibitem [{\citenamefont {Spaldin}\ \emph {et~al.}(2008)\citenamefont
  {Spaldin}, \citenamefont {Fiebig},\ and\ \citenamefont
  {Mostovoy}}]{Spaldin2008-uv}%
  \BibitemOpen
  \bibfield  {author} {\bibinfo {author} {\bibfnamefont {N.~A.}\ \bibnamefont
  {Spaldin}}, \bibinfo {author} {\bibfnamefont {M.}~\bibnamefont {Fiebig}},\
  and\ \bibinfo {author} {\bibfnamefont {M.}~\bibnamefont {Mostovoy}},\
  }\bibfield  {title} {\bibinfo {title} {{The toroidal moment in
  condensed-matter physics and its relation to the magnetoelectric effect*}},\
  }\href {https://doi.org/10.1088/0953-8984/20/43/434203} {\bibfield  {journal}
  {\bibinfo  {journal} {J. Phys. Condens. Matter}\ }\textbf {\bibinfo {volume}
  {20}},\ \bibinfo {pages} {434203} (\bibinfo {year} {2008})}\BibitemShut
  {NoStop}%
\bibitem [{\citenamefont {Spaldin}\ \emph {et~al.}(2013)\citenamefont
  {Spaldin}, \citenamefont {Fechner}, \citenamefont {Bousquet}, \citenamefont
  {Balatsky},\ and\ \citenamefont {Nordstr{\"o}m}}]{Spaldin2013-vj}%
  \BibitemOpen
  \bibfield  {author} {\bibinfo {author} {\bibfnamefont {N.~A.}\ \bibnamefont
  {Spaldin}}, \bibinfo {author} {\bibfnamefont {M.}~\bibnamefont {Fechner}},
  \bibinfo {author} {\bibfnamefont {E.}~\bibnamefont {Bousquet}}, \bibinfo
  {author} {\bibfnamefont {A.}~\bibnamefont {Balatsky}},\ and\ \bibinfo
  {author} {\bibfnamefont {L.}~\bibnamefont {Nordstr{\"o}m}},\ }\bibfield
  {title} {\bibinfo {title} {{Monopole-based formalism for the diagonal
  magnetoelectric response}},\ }\href
  {https://doi.org/10.1103/PhysRevB.88.094429} {\bibfield  {journal} {\bibinfo
  {journal} {Phys. Rev. B}\ }\textbf {\bibinfo {volume} {88}},\ \bibinfo
  {pages} {094429} (\bibinfo {year} {2013})}\BibitemShut {NoStop}%
\bibitem [{\citenamefont {Yatsushiro}\ \emph {et~al.}(2021)\citenamefont
  {Yatsushiro}, \citenamefont {Kusunose},\ and\ \citenamefont
  {Hayami}}]{Yatsushiro2021-af}%
  \BibitemOpen
  \bibfield  {author} {\bibinfo {author} {\bibfnamefont {M.}~\bibnamefont
  {Yatsushiro}}, \bibinfo {author} {\bibfnamefont {H.}~\bibnamefont
  {Kusunose}},\ and\ \bibinfo {author} {\bibfnamefont {S.}~\bibnamefont
  {Hayami}},\ }\bibfield  {title} {\bibinfo {title} {{Multipole classification
  in 122 magnetic point groups for unified understanding of multiferroic
  responses and transport phenomena}},\ }\href
  {https://doi.org/10.1103/PhysRevB.104.054412} {\bibfield  {journal} {\bibinfo
   {journal} {Phys. Rev. B}\ }\textbf {\bibinfo {volume} {104}},\ \bibinfo
  {pages} {054412} (\bibinfo {year} {2021})}\BibitemShut {NoStop}%
\bibitem [{\citenamefont {Hayami}\ \emph {et~al.}(2018)\citenamefont {Hayami},
  \citenamefont {Yatsushiro}, \citenamefont {Yanagi},\ and\ \citenamefont
  {Kusunose}}]{Hayami2018-ip}%
  \BibitemOpen
  \bibfield  {author} {\bibinfo {author} {\bibfnamefont {S.}~\bibnamefont
  {Hayami}}, \bibinfo {author} {\bibfnamefont {M.}~\bibnamefont {Yatsushiro}},
  \bibinfo {author} {\bibfnamefont {Y.}~\bibnamefont {Yanagi}},\ and\ \bibinfo
  {author} {\bibfnamefont {H.}~\bibnamefont {Kusunose}},\ }\bibfield  {title}
  {\bibinfo {title} {{Classification of atomic-scale multipoles under
  crystallographic point groups and application to linear response tensors}},\
  }\href {https://doi.org/10.1103/PhysRevB.98.165110} {\bibfield  {journal}
  {\bibinfo  {journal} {Phys. Rev. B}\ }\textbf {\bibinfo {volume} {98}},\
  \bibinfo {pages} {165110} (\bibinfo {year} {2018})}\BibitemShut {NoStop}%
\bibitem [{\citenamefont {Watanabe}\ and\ \citenamefont
  {Yanase}(2018)}]{Watanabe2018-xc}%
  \BibitemOpen
  \bibfield  {author} {\bibinfo {author} {\bibfnamefont {H.}~\bibnamefont
  {Watanabe}}\ and\ \bibinfo {author} {\bibfnamefont {Y.}~\bibnamefont
  {Yanase}},\ }\bibfield  {title} {\bibinfo {title} {{Group-theoretical
  classification of multipole order: Emergent responses and candidate
  materials}},\ }\href {https://doi.org/10.1103/PhysRevB.98.245129} {\bibfield
  {journal} {\bibinfo  {journal} {Phys. Rev. B}\ }\textbf {\bibinfo {volume}
  {98}},\ \bibinfo {pages} {245129} (\bibinfo {year} {2018})}\BibitemShut
  {NoStop}%
\bibitem [{\citenamefont {Kirikoshi}\ and\ \citenamefont
  {Hayami}(2023)}]{Kirikoshi2023-ii}%
  \BibitemOpen
  \bibfield  {author} {\bibinfo {author} {\bibfnamefont {A.}~\bibnamefont
  {Kirikoshi}}\ and\ \bibinfo {author} {\bibfnamefont {S.}~\bibnamefont
  {Hayami}},\ }\bibfield  {title} {\bibinfo {title} {{Microscopic mechanism for
  intrinsic nonlinear anomalous Hall conductivity in noncollinear
  antiferromagnetic metals}},\ }\href
  {https://doi.org/10.1103/PhysRevB.107.155109} {\bibfield  {journal} {\bibinfo
   {journal} {Phys. Rev. B}\ }\textbf {\bibinfo {volume} {107}},\ \bibinfo
  {pages} {155109} (\bibinfo {year} {2023})}\BibitemShut {NoStop}%
\bibitem [{\citenamefont {Aoyama}\ and\ \citenamefont
  {Ohgushi}(2024)}]{Aoyama2024-nz}%
  \BibitemOpen
  \bibfield  {author} {\bibinfo {author} {\bibfnamefont {T.}~\bibnamefont
  {Aoyama}}\ and\ \bibinfo {author} {\bibfnamefont {K.}~\bibnamefont
  {Ohgushi}},\ }\bibfield  {title} {\bibinfo {title} {{Piezomagnetic properties
  in altermagnetic {MnTe}}},\ }\href
  {https://doi.org/10.1103/PhysRevMaterials.8.L041402} {\bibfield  {journal}
  {\bibinfo  {journal} {Phys. Rev. Mater.}\ }\textbf {\bibinfo {volume} {8}},\
  \bibinfo {pages} {L041402} (\bibinfo {year} {2024})}\BibitemShut {NoStop}%
\bibitem [{\citenamefont {Fang}\ \emph {et~al.}()\citenamefont {Fang},
  \citenamefont {Cano},\ and\ \citenamefont {Ghorashi}}]{Fang2023-xl}%
  \BibitemOpen
  \bibfield  {author} {\bibinfo {author} {\bibfnamefont {Y.}~\bibnamefont
  {Fang}}, \bibinfo {author} {\bibfnamefont {J.}~\bibnamefont {Cano}},\ and\
  \bibinfo {author} {\bibfnamefont {S.~A.~A.}\ \bibnamefont {Ghorashi}},\
  }\bibfield  {title} {\bibinfo {title} {{Quantum geometry induced nonlinear
  transport in altermagnets}},\ }\Eprint {https://arxiv.org/abs/2310.11489}
  {arXiv:2310.11489} \BibitemShut {NoStop}%
\bibitem [{\citenamefont {Farajollahpour}\ \emph {et~al.}()\citenamefont
  {Farajollahpour}, \citenamefont {Ganesh},\ and\ \citenamefont
  {Samokhin}}]{Farajollahpour2024-yf}%
  \BibitemOpen
  \bibfield  {author} {\bibinfo {author} {\bibfnamefont {T.}~\bibnamefont
  {Farajollahpour}}, \bibinfo {author} {\bibfnamefont {R.}~\bibnamefont
  {Ganesh}},\ and\ \bibinfo {author} {\bibfnamefont {K.}~\bibnamefont
  {Samokhin}},\ }\bibfield  {title} {\bibinfo {title} {{Light-induced Hall
  currents in altermagnets}},\ }\Eprint {https://arxiv.org/abs/2405.03779}
  {arXiv:2405.03779} \BibitemShut {NoStop}%
\bibitem [{\citenamefont {Sorn}\ and\ \citenamefont {Patri}()}]{Sorn2023-be}%
  \BibitemOpen
  \bibfield  {author} {\bibinfo {author} {\bibfnamefont {S.}~\bibnamefont
  {Sorn}}\ and\ \bibinfo {author} {\bibfnamefont {A.~S.}\ \bibnamefont
  {Patri}},\ }\bibfield  {title} {\bibinfo {title} {{Signatures of hidden
  octupolar order from nonlinear Hall effects}},\ }\Eprint
  {https://arxiv.org/abs/2311.03435} {arXiv:2311.03435} \BibitemShut {NoStop}%
\bibitem [{\citenamefont {{\v S}mejkal}\ \emph
  {et~al.}(2022{\natexlab{c}})\citenamefont {{\v S}mejkal}, \citenamefont
  {MacDonald}, \citenamefont {Sinova}, \citenamefont {Nakatsuji},\ and\
  \citenamefont {Jungwirth}}]{Smejkal2022-el}%
  \BibitemOpen
  \bibfield  {author} {\bibinfo {author} {\bibfnamefont {L.}~\bibnamefont {{\v
  S}mejkal}}, \bibinfo {author} {\bibfnamefont {A.~H.}\ \bibnamefont
  {MacDonald}}, \bibinfo {author} {\bibfnamefont {J.}~\bibnamefont {Sinova}},
  \bibinfo {author} {\bibfnamefont {S.}~\bibnamefont {Nakatsuji}},\ and\
  \bibinfo {author} {\bibfnamefont {T.}~\bibnamefont {Jungwirth}},\ }\bibfield
  {title} {\bibinfo {title} {{Anomalous Hall antiferromagnets}},\ }\href
  {https://doi.org/10.1038/s41578-022-00430-3} {\bibfield  {journal} {\bibinfo
  {journal} {Nat. Rev. Mater.}\ }\textbf {\bibinfo {volume} {7}},\ \bibinfo
  {pages} {482} (\bibinfo {year} {2022}{\natexlab{c}})}\BibitemShut {NoStop}%
\bibitem [{\citenamefont {Yang}\ and\ \citenamefont {Nagaosa}()}]{Yang2014-iz}%
  \BibitemOpen
  \bibfield  {author} {\bibinfo {author} {\bibfnamefont {B.-J.}\ \bibnamefont
  {Yang}}\ and\ \bibinfo {author} {\bibfnamefont {N.}~\bibnamefont {Nagaosa}},\
  }\bibfield  {title} {\bibinfo {title} {{Emergent topological phenomena in
  thin films of pyrochlore iridates}},\ }\href
  {https://doi.org/10.1103/PhysRevLett.112.246402} {\bibfield  {journal}
  {\bibinfo  {journal} {Phys. Rev. Lett.}\ }\textbf {\bibinfo {volume} {112}},\
  \bibinfo {pages} {246402}}\BibitemShut {NoStop}%
\bibitem [{\citenamefont {Yamaura}\ \emph {et~al.}(2012)\citenamefont
  {Yamaura}, \citenamefont {Ohgushi}, \citenamefont {Ohsumi}, \citenamefont
  {Hasegawa}, \citenamefont {Yamauchi}, \citenamefont {Sugimoto}, \citenamefont
  {Takeshita}, \citenamefont {Tokuda}, \citenamefont {Takata}, \citenamefont
  {Udagawa}, \citenamefont {Takigawa}, \citenamefont {Harima}, \citenamefont
  {Arima},\ and\ \citenamefont {Hiroi}}]{Yamaura2012-zd}%
  \BibitemOpen
  \bibfield  {author} {\bibinfo {author} {\bibfnamefont {J.}~\bibnamefont
  {Yamaura}}, \bibinfo {author} {\bibfnamefont {K.}~\bibnamefont {Ohgushi}},
  \bibinfo {author} {\bibfnamefont {H.}~\bibnamefont {Ohsumi}}, \bibinfo
  {author} {\bibfnamefont {T.}~\bibnamefont {Hasegawa}}, \bibinfo {author}
  {\bibfnamefont {I.}~\bibnamefont {Yamauchi}}, \bibinfo {author}
  {\bibfnamefont {K.}~\bibnamefont {Sugimoto}}, \bibinfo {author}
  {\bibfnamefont {S.}~\bibnamefont {Takeshita}}, \bibinfo {author}
  {\bibfnamefont {A.}~\bibnamefont {Tokuda}}, \bibinfo {author} {\bibfnamefont
  {M.}~\bibnamefont {Takata}}, \bibinfo {author} {\bibfnamefont
  {M.}~\bibnamefont {Udagawa}}, \bibinfo {author} {\bibfnamefont
  {M.}~\bibnamefont {Takigawa}}, \bibinfo {author} {\bibfnamefont
  {H.}~\bibnamefont {Harima}}, \bibinfo {author} {\bibfnamefont
  {T.}~\bibnamefont {Arima}},\ and\ \bibinfo {author} {\bibfnamefont
  {Z.}~\bibnamefont {Hiroi}},\ }\bibfield  {title} {\bibinfo {title}
  {{Tetrahedral magnetic order and the metal-insulator transition in the
  pyrochlore lattice of {Cd$_2$Os$_2$O$_7$}}},\ }\href
  {https://doi.org/10.1103/PhysRevLett.108.247205} {\bibfield  {journal}
  {\bibinfo  {journal} {Phys. Rev. Lett.}\ }\textbf {\bibinfo {volume} {108}},\
  \bibinfo {pages} {247205} (\bibinfo {year} {2012})}\BibitemShut {NoStop}%
\bibitem [{\citenamefont {Moon}\ \emph {et~al.}(2013)\citenamefont {Moon},
  \citenamefont {Xu}, \citenamefont {Kim},\ and\ \citenamefont
  {Balents}}]{Moon2013-mr}%
  \BibitemOpen
  \bibfield  {author} {\bibinfo {author} {\bibfnamefont {E.-G.}\ \bibnamefont
  {Moon}}, \bibinfo {author} {\bibfnamefont {C.}~\bibnamefont {Xu}}, \bibinfo
  {author} {\bibfnamefont {Y.~B.}\ \bibnamefont {Kim}},\ and\ \bibinfo {author}
  {\bibfnamefont {L.}~\bibnamefont {Balents}},\ }\bibfield  {title} {\bibinfo
  {title} {{Non-Fermi-liquid and topological states with strong spin-orbit
  coupling}},\ }\href {https://doi.org/10.1103/PhysRevLett.111.206401}
  {\bibfield  {journal} {\bibinfo  {journal} {Phys. Rev. Lett.}\ }\textbf
  {\bibinfo {volume} {111}},\ \bibinfo {pages} {206401} (\bibinfo {year}
  {2013})}\BibitemShut {NoStop}%
\bibitem [{\citenamefont {Kondo}\ \emph {et~al.}(2015)\citenamefont {Kondo},
  \citenamefont {Nakayama}, \citenamefont {Chen}, \citenamefont {Ishikawa},
  \citenamefont {Moon}, \citenamefont {Yamamoto}, \citenamefont {Ota},
  \citenamefont {Malaeb}, \citenamefont {Kanai}, \citenamefont {Nakashima},
  \citenamefont {Ishida}, \citenamefont {Yoshida}, \citenamefont {Yamamoto},
  \citenamefont {Matsunami}, \citenamefont {Kimura}, \citenamefont {Inami},
  \citenamefont {Ono}, \citenamefont {Kumigashira}, \citenamefont {Nakatsuji},
  \citenamefont {Balents},\ and\ \citenamefont {Shin}}]{Kondo2015-dr}%
  \BibitemOpen
  \bibfield  {author} {\bibinfo {author} {\bibfnamefont {T.}~\bibnamefont
  {Kondo}}, \bibinfo {author} {\bibfnamefont {M.}~\bibnamefont {Nakayama}},
  \bibinfo {author} {\bibfnamefont {R.}~\bibnamefont {Chen}}, \bibinfo {author}
  {\bibfnamefont {J.~J.}\ \bibnamefont {Ishikawa}}, \bibinfo {author}
  {\bibfnamefont {E.-G.}\ \bibnamefont {Moon}}, \bibinfo {author}
  {\bibfnamefont {T.}~\bibnamefont {Yamamoto}}, \bibinfo {author}
  {\bibfnamefont {Y.}~\bibnamefont {Ota}}, \bibinfo {author} {\bibfnamefont
  {W.}~\bibnamefont {Malaeb}}, \bibinfo {author} {\bibfnamefont
  {H.}~\bibnamefont {Kanai}}, \bibinfo {author} {\bibfnamefont
  {Y.}~\bibnamefont {Nakashima}}, \bibinfo {author} {\bibfnamefont
  {Y.}~\bibnamefont {Ishida}}, \bibinfo {author} {\bibfnamefont
  {R.}~\bibnamefont {Yoshida}}, \bibinfo {author} {\bibfnamefont
  {H.}~\bibnamefont {Yamamoto}}, \bibinfo {author} {\bibfnamefont
  {M.}~\bibnamefont {Matsunami}}, \bibinfo {author} {\bibfnamefont
  {S.}~\bibnamefont {Kimura}}, \bibinfo {author} {\bibfnamefont
  {N.}~\bibnamefont {Inami}}, \bibinfo {author} {\bibfnamefont
  {K.}~\bibnamefont {Ono}}, \bibinfo {author} {\bibfnamefont {H.}~\bibnamefont
  {Kumigashira}}, \bibinfo {author} {\bibfnamefont {S.}~\bibnamefont
  {Nakatsuji}}, \bibinfo {author} {\bibfnamefont {L.}~\bibnamefont {Balents}},\
  and\ \bibinfo {author} {\bibfnamefont {S.}~\bibnamefont {Shin}},\ }\bibfield
  {title} {\bibinfo {title} {{Quadratic Fermi node in a 3D strongly correlated
  semimetal}},\ }\href {https://doi.org/10.1038/ncomms10042} {\bibfield
  {journal} {\bibinfo  {journal} {Nat. Commun.}\ }\textbf {\bibinfo {volume}
  {6}},\ \bibinfo {pages} {10042} (\bibinfo {year} {2015})}\BibitemShut
  {NoStop}%
\bibitem [{\citenamefont {Meyer}\ and\ \citenamefont
  {Asch}(1961)}]{Meyer1961-tf}%
  \BibitemOpen
  \bibfield  {author} {\bibinfo {author} {\bibfnamefont {A.~J.~P.}\
  \bibnamefont {Meyer}}\ and\ \bibinfo {author} {\bibfnamefont
  {G.}~\bibnamefont {Asch}},\ }\bibfield  {title} {\bibinfo {title}
  {Experimental $g'$ and $g$ {V}alues of {Fe, Co, Ni}, and {Their Alloys}},\
  }\href {https://doi.org/10.1063/1.2000457} {\bibfield  {journal} {\bibinfo
  {journal} {J. Appl. Phys.}\ }\textbf {\bibinfo {volume} {32}},\ \bibinfo
  {pages} {S330} (\bibinfo {year} {1961})}\BibitemShut {NoStop}%
\bibitem [{\citenamefont {Reck}\ and\ \citenamefont
  {Fry}(1969)}]{reck1969orbital}%
  \BibitemOpen
  \bibfield  {author} {\bibinfo {author} {\bibfnamefont {R.~A.}\ \bibnamefont
  {Reck}}\ and\ \bibinfo {author} {\bibfnamefont {D.~L.}\ \bibnamefont {Fry}},\
  }\bibfield  {title} {\bibinfo {title} {Orbital and {S}pin {M}agnetization in
  {Fe-Co, Fe-Ni, and Ni-Co}},\ }\href {https://doi.org/10.1103/PhysRev.184.492}
  {\bibfield  {journal} {\bibinfo  {journal} {Phys. Rev.}\ }\textbf {\bibinfo
  {volume} {184}},\ \bibinfo {pages} {492} (\bibinfo {year}
  {1969})}\BibitemShut {NoStop}%
\bibitem [{\citenamefont {Adams}\ and\ \citenamefont
  {Blount}(1959)}]{Adams1959-tp}%
  \BibitemOpen
  \bibfield  {author} {\bibinfo {author} {\bibfnamefont {E.~N.}\ \bibnamefont
  {Adams}}\ and\ \bibinfo {author} {\bibfnamefont {E.~I.}\ \bibnamefont
  {Blount}},\ }\bibfield  {title} {\bibinfo {title} {{Energy bands in the
  presence of an external force {field---II}: Anomalous velocities}},\ }\href
  {https://doi.org/10.1016/0022-3697(59)90004-6} {\bibfield  {journal}
  {\bibinfo  {journal} {J. Phys. Chem. Solids}\ }\textbf {\bibinfo {volume}
  {10}},\ \bibinfo {pages} {286} (\bibinfo {year} {1959})}\BibitemShut
  {NoStop}%
\bibitem [{\citenamefont {Blount}(1962)}]{Blount1962-ku}%
  \BibitemOpen
  \bibfield  {author} {\bibinfo {author} {\bibfnamefont {E.~I.}\ \bibnamefont
  {Blount}},\ }\bibfield  {title} {\bibinfo {title} {{Bloch Electrons in a
  Magnetic Field}},\ }\href {https://doi.org/10.1103/PhysRev.126.1636}
  {\bibfield  {journal} {\bibinfo  {journal} {Phys. Rev.}\ }\textbf {\bibinfo
  {volume} {126}},\ \bibinfo {pages} {1636} (\bibinfo {year}
  {1962})}\BibitemShut {NoStop}%
\bibitem [{\citenamefont {Cheng}\ \emph {et~al.}(2015)\citenamefont {Cheng},
  \citenamefont {Vermeulen},\ and\ \citenamefont {Sipe}}]{Cheng2015-qr}%
  \BibitemOpen
  \bibfield  {author} {\bibinfo {author} {\bibfnamefont {J.~L.}\ \bibnamefont
  {Cheng}}, \bibinfo {author} {\bibfnamefont {N.}~\bibnamefont {Vermeulen}},\
  and\ \bibinfo {author} {\bibfnamefont {J.~E.}\ \bibnamefont {Sipe}},\
  }\bibfield  {title} {\bibinfo {title} {{Third-order nonlinearity of graphene:
  Effects of phenomenological relaxation and finite temperature}},\ }\href
  {https://doi.org/10.1103/PhysRevB.91.235320} {\bibfield  {journal} {\bibinfo
  {journal} {Phys. Rev. B}\ }\textbf {\bibinfo {volume} {91}},\ \bibinfo
  {pages} {235320} (\bibinfo {year} {2015})}\BibitemShut {NoStop}%
\bibitem [{\citenamefont {Passos}\ \emph {et~al.}(2018)\citenamefont {Passos},
  \citenamefont {Ventura}, \citenamefont {Lopes}, \citenamefont {Santos},\ and\
  \citenamefont {Peres}}]{Passos2018-tl}%
  \BibitemOpen
  \bibfield  {author} {\bibinfo {author} {\bibfnamefont {D.~J.}\ \bibnamefont
  {Passos}}, \bibinfo {author} {\bibfnamefont {G.~B.}\ \bibnamefont {Ventura}},
  \bibinfo {author} {\bibfnamefont {J.~M. V.~P.}\ \bibnamefont {Lopes}},
  \bibinfo {author} {\bibfnamefont {J.~M. B. L.~d.}\ \bibnamefont {Santos}},\
  and\ \bibinfo {author} {\bibfnamefont {N.~M.~R.}\ \bibnamefont {Peres}},\
  }\bibfield  {title} {\bibinfo {title} {{Nonlinear optical responses of
  crystalline systems: Results from a velocity gauge analysis}},\ }\href
  {https://doi.org/10.1103/PhysRevB.97.235446} {\bibfield  {journal} {\bibinfo
  {journal} {Phys. Rev. B}\ }\textbf {\bibinfo {volume} {97}},\ \bibinfo
  {pages} {235446} (\bibinfo {year} {2018})}\BibitemShut {NoStop}%
\bibitem [{\citenamefont {Watanabe}\ \emph {et~al.}(2022)\citenamefont
  {Watanabe}, \citenamefont {Daido},\ and\ \citenamefont
  {Yanase}}]{Watanabe2022-qp}%
  \BibitemOpen
  \bibfield  {author} {\bibinfo {author} {\bibfnamefont {H.}~\bibnamefont
  {Watanabe}}, \bibinfo {author} {\bibfnamefont {A.}~\bibnamefont {Daido}},\
  and\ \bibinfo {author} {\bibfnamefont {Y.}~\bibnamefont {Yanase}},\
  }\bibfield  {title} {\bibinfo {title} {{Nonreciprocal optical response in
  parity-breaking superconductors}},\ }\href
  {https://doi.org/10.1103/PhysRevB.105.024308} {\bibfield  {journal} {\bibinfo
   {journal} {Phys. Rev. B}\ }\textbf {\bibinfo {volume} {105}},\ \bibinfo
  {pages} {024308} (\bibinfo {year} {2022})}\BibitemShut {NoStop}%
\bibitem [{\citenamefont {Das}\ \emph {et~al.}(2023)\citenamefont {Das},
  \citenamefont {Lahiri}, \citenamefont {Atencia}, \citenamefont {Culcer},\
  and\ \citenamefont {Agarwal}}]{Das2023-nm}%
  \BibitemOpen
  \bibfield  {author} {\bibinfo {author} {\bibfnamefont {K.}~\bibnamefont
  {Das}}, \bibinfo {author} {\bibfnamefont {S.}~\bibnamefont {Lahiri}},
  \bibinfo {author} {\bibfnamefont {R.~B.}\ \bibnamefont {Atencia}}, \bibinfo
  {author} {\bibfnamefont {D.}~\bibnamefont {Culcer}},\ and\ \bibinfo {author}
  {\bibfnamefont {A.}~\bibnamefont {Agarwal}},\ }\bibfield  {title} {\bibinfo
  {title} {{Intrinsic nonlinear conductivities induced by the quantum
  metric}},\ }\href {https://doi.org/10.1103/PhysRevB.108.L201405} {\bibfield
  {journal} {\bibinfo  {journal} {Phys. Rev. B}\ }\textbf {\bibinfo {volume}
  {108}},\ \bibinfo {pages} {L201405} (\bibinfo {year} {2023})}\BibitemShut
  {NoStop}%
\bibitem [{\citenamefont {Watanabe}\ and\ \citenamefont
  {Yanase}(2020)}]{Watanabe2020-bs}%
  \BibitemOpen
  \bibfield  {author} {\bibinfo {author} {\bibfnamefont {H.}~\bibnamefont
  {Watanabe}}\ and\ \bibinfo {author} {\bibfnamefont {Y.}~\bibnamefont
  {Yanase}},\ }\bibfield  {title} {\bibinfo {title} {{Nonlinear electric
  transport in odd-parity magnetic multipole systems: Application to Mn-based
  compounds}},\ }\href {https://doi.org/10.1103/PhysRevResearch.2.043081}
  {\bibfield  {journal} {\bibinfo  {journal} {Phys. Rev. Research}\ }\textbf
  {\bibinfo {volume} {2}},\ \bibinfo {pages} {043081} (\bibinfo {year}
  {2020})}\BibitemShut {NoStop}%
\bibitem [{\citenamefont {Kaplan}\ \emph {et~al.}(2023)\citenamefont {Kaplan},
  \citenamefont {Holder},\ and\ \citenamefont {Yan}}]{Kaplan2023-jt}%
  \BibitemOpen
  \bibfield  {author} {\bibinfo {author} {\bibfnamefont {D.}~\bibnamefont
  {Kaplan}}, \bibinfo {author} {\bibfnamefont {T.}~\bibnamefont {Holder}},\
  and\ \bibinfo {author} {\bibfnamefont {B.}~\bibnamefont {Yan}},\ }\bibfield
  {title} {\bibinfo {title} {{Unifying semiclassics and quantum perturbation
  theory at nonlinear order}},\ }\href
  {https://doi.org/10.21468/scipostphys.14.4.082} {\bibfield  {journal}
  {\bibinfo  {journal} {SciPost Phys.}\ }\textbf {\bibinfo {volume} {14}},\
  \bibinfo {pages} {082} (\bibinfo {year} {2023})}\BibitemShut {NoStop}%
\bibitem [{\citenamefont {Gao}\ \emph {et~al.}(2014)\citenamefont {Gao},
  \citenamefont {Yang},\ and\ \citenamefont {Niu}}]{Gao2014-kx}%
  \BibitemOpen
  \bibfield  {author} {\bibinfo {author} {\bibfnamefont {Y.}~\bibnamefont
  {Gao}}, \bibinfo {author} {\bibfnamefont {S.~A.}\ \bibnamefont {Yang}},\ and\
  \bibinfo {author} {\bibfnamefont {Q.}~\bibnamefont {Niu}},\ }\bibfield
  {title} {\bibinfo {title} {{Field induced positional shift of Bloch electrons
  and its dynamical implications}},\ }\href
  {https://doi.org/10.1103/PhysRevLett.112.166601} {\bibfield  {journal}
  {\bibinfo  {journal} {Phys. Rev. Lett.}\ }\textbf {\bibinfo {volume} {112}},\
  \bibinfo {pages} {166601} (\bibinfo {year} {2014})}\BibitemShut {NoStop}%
\bibitem [{\citenamefont {Provost}\ and\ \citenamefont
  {Vallee}(1980)}]{Provost1980-qb}%
  \BibitemOpen
  \bibfield  {author} {\bibinfo {author} {\bibfnamefont {J.~P.}\ \bibnamefont
  {Provost}}\ and\ \bibinfo {author} {\bibfnamefont {G.}~\bibnamefont
  {Vallee}},\ }\bibfield  {title} {\bibinfo {title} {{Riemannian structure on
  manifolds of quantum states}},\ }\href {https://doi.org/10.1007/bf02193559}
  {\bibfield  {journal} {\bibinfo  {journal} {Commun. Math. Phys.}\ }\textbf
  {\bibinfo {volume} {76}},\ \bibinfo {pages} {289} (\bibinfo {year}
  {1980})}\BibitemShut {NoStop}%
\bibitem [{\citenamefont {Resta}(2011)}]{Resta2011-da}%
  \BibitemOpen
  \bibfield  {author} {\bibinfo {author} {\bibfnamefont {R.}~\bibnamefont
  {Resta}},\ }\bibfield  {title} {\bibinfo {title} {{The insulating state of
  matter: a geometrical theory}},\ }\href
  {https://doi.org/10.1140/epjb/e2010-10874-4} {\bibfield  {journal} {\bibinfo
  {journal} {Eur. Phys. J. B}\ }\textbf {\bibinfo {volume} {79}},\ \bibinfo
  {pages} {121} (\bibinfo {year} {2011})}\BibitemShut {NoStop}%
\bibitem [{\citenamefont {Litvin}\ and\ \citenamefont
  {Opechowski}(1974)}]{LITVIN1974538}%
  \BibitemOpen
  \bibfield  {author} {\bibinfo {author} {\bibfnamefont {D.}~\bibnamefont
  {Litvin}}\ and\ \bibinfo {author} {\bibfnamefont {W.}~\bibnamefont
  {Opechowski}},\ }\bibfield  {title} {\bibinfo {title} {{Spin groups}},\
  }\href {https://doi.org/10.1016/0031-8914(74)90157-8} {\bibfield  {journal}
  {\bibinfo  {journal} {Physica}\ }\textbf {\bibinfo {volume} {76}},\ \bibinfo
  {pages} {538} (\bibinfo {year} {1974})}\BibitemShut {NoStop}%
\bibitem [{\citenamefont {Litvin}(1977)}]{Litvin1977-kw}%
  \BibitemOpen
  \bibfield  {author} {\bibinfo {author} {\bibfnamefont {D.~B.}\ \bibnamefont
  {Litvin}},\ }\bibfield  {title} {\bibinfo {title} {{Spin point groups}},\
  }\href {https://doi.org/10.1107/S0567739477000709} {\bibfield  {journal}
  {\bibinfo  {journal} {Acta Crystallogr. A}\ }\textbf {\bibinfo {volume}
  {33}},\ \bibinfo {pages} {279} (\bibinfo {year} {1977})}\BibitemShut
  {NoStop}%
\bibitem [{\citenamefont {Gos{\'a}lbez-Mart{\'\i}nez}\ \emph
  {et~al.}(2015)\citenamefont {Gos{\'a}lbez-Mart{\'\i}nez}, \citenamefont
  {Souza},\ and\ \citenamefont {Vanderbilt}}]{Gosalbez-Martinez2015-ui}%
  \BibitemOpen
  \bibfield  {author} {\bibinfo {author} {\bibfnamefont {D.}~\bibnamefont
  {Gos{\'a}lbez-Mart{\'\i}nez}}, \bibinfo {author} {\bibfnamefont
  {I.}~\bibnamefont {Souza}},\ and\ \bibinfo {author} {\bibfnamefont
  {D.}~\bibnamefont {Vanderbilt}},\ }\bibfield  {title} {\bibinfo {title}
  {{Chiral degeneracies and Fermi-surface Chern numbers in bcc Fe}},\ }\href
  {https://doi.org/10.1103/PhysRevB.92.085138} {\bibfield  {journal} {\bibinfo
  {journal} {Phys. Rev. B}\ }\textbf {\bibinfo {volume} {92}},\ \bibinfo
  {pages} {085138} (\bibinfo {year} {2015})}\BibitemShut {NoStop}%
\bibitem [{\citenamefont {Iba{\~n}ez-Azpiroz}\ \emph
  {et~al.}(2018)\citenamefont {Iba{\~n}ez-Azpiroz}, \citenamefont {Tsirkin},\
  and\ \citenamefont {Souza}}]{Ibanez-Azpiroz2018-mm}%
  \BibitemOpen
  \bibfield  {author} {\bibinfo {author} {\bibfnamefont {J.}~\bibnamefont
  {Iba{\~n}ez-Azpiroz}}, \bibinfo {author} {\bibfnamefont {S.~S.}\ \bibnamefont
  {Tsirkin}},\ and\ \bibinfo {author} {\bibfnamefont {I.}~\bibnamefont
  {Souza}},\ }\bibfield  {title} {\bibinfo {title} {{Ab initio calculation of
  the shift photocurrent by Wannier interpolation}},\ }\href
  {https://doi.org/10.1103/PhysRevB.97.245143} {\bibfield  {journal} {\bibinfo
  {journal} {Phys. Rev. B}\ }\textbf {\bibinfo {volume} {97}},\ \bibinfo
  {pages} {245143} (\bibinfo {year} {2018})}\BibitemShut {NoStop}%
\bibitem [{\citenamefont {Guo}\ and\ \citenamefont {Franz}(2009)}]{Guo2009-nw}%
  \BibitemOpen
  \bibfield  {author} {\bibinfo {author} {\bibfnamefont {H.-M.}\ \bibnamefont
  {Guo}}\ and\ \bibinfo {author} {\bibfnamefont {M.}~\bibnamefont {Franz}},\
  }\bibfield  {title} {\bibinfo {title} {{Three-dimensional topological
  insulators on the pyrochlore lattice}},\ }\href
  {https://doi.org/10.1103/PhysRevLett.103.206805} {\bibfield  {journal}
  {\bibinfo  {journal} {Phys. Rev. Lett.}\ }\textbf {\bibinfo {volume} {103}},\
  \bibinfo {pages} {206805} (\bibinfo {year} {2009})}\BibitemShut {NoStop}%
\bibitem [{\citenamefont {Oh}\ \emph {et~al.}(2018)\citenamefont {Oh},
  \citenamefont {Ishizuka},\ and\ \citenamefont {Yang}}]{Oh2018-xs}%
  \BibitemOpen
  \bibfield  {author} {\bibinfo {author} {\bibfnamefont {T.}~\bibnamefont
  {Oh}}, \bibinfo {author} {\bibfnamefont {H.}~\bibnamefont {Ishizuka}},\ and\
  \bibinfo {author} {\bibfnamefont {B.-J.}\ \bibnamefont {Yang}},\ }\bibfield
  {title} {\bibinfo {title} {{Magnetic field induced topological semimetals
  near the quantum critical point of pyrochlore iridates}},\ }\href
  {https://doi.org/10.1103/PhysRevB.98.144409} {\bibfield  {journal} {\bibinfo
  {journal} {Phys. Rev. B}\ }\textbf {\bibinfo {volume} {98}},\ \bibinfo
  {pages} {144409} (\bibinfo {year} {2018})}\BibitemShut {NoStop}%
\bibitem [{\citenamefont {Simon}(1983)}]{simon1983holonomy}%
  \BibitemOpen
  \bibfield  {author} {\bibinfo {author} {\bibfnamefont {B.}~\bibnamefont
  {Simon}},\ }\bibfield  {title} {\bibinfo {title} {{Holonomy, the Quantum
  Adiabatic Theorem, and Berry's Phase}},\ }\href
  {https://doi.org/10.1103/PhysRevLett.51.2167} {\bibfield  {journal} {\bibinfo
   {journal} {Phys. Rev. Lett.}\ }\textbf {\bibinfo {volume} {51}},\ \bibinfo
  {pages} {2167} (\bibinfo {year} {1983})}\BibitemShut {NoStop}%
\bibitem [{\citenamefont {Berry}(1984)}]{Berry1984-fn}%
  \BibitemOpen
  \bibfield  {author} {\bibinfo {author} {\bibfnamefont {M.~V.}\ \bibnamefont
  {Berry}},\ }\bibfield  {title} {\bibinfo {title} {{Quantal phase factors
  accompanying adiabatic changes}},\ }\href
  {https://doi.org/10.1098/rspa.1984.0023} {\bibfield  {journal} {\bibinfo
  {journal} {Proc. R. Soc. Lond. A}\ }\textbf {\bibinfo {volume} {392}},\
  \bibinfo {pages} {45} (\bibinfo {year} {1984})}\BibitemShut {NoStop}%
\bibitem [{\citenamefont {Thouless}\ \emph {et~al.}(1982)\citenamefont
  {Thouless}, \citenamefont {Kohmoto}, \citenamefont {Nightingale},\ and\
  \citenamefont {den Nijs}}]{Thouless1982-zh}%
  \BibitemOpen
  \bibfield  {author} {\bibinfo {author} {\bibfnamefont {D.~J.}\ \bibnamefont
  {Thouless}}, \bibinfo {author} {\bibfnamefont {M.}~\bibnamefont {Kohmoto}},
  \bibinfo {author} {\bibfnamefont {M.~P.}\ \bibnamefont {Nightingale}},\ and\
  \bibinfo {author} {\bibfnamefont {M.}~\bibnamefont {den Nijs}},\ }\bibfield
  {title} {\bibinfo {title} {{Quantized Hall Conductance in a {Two-Dimensional}
  Periodic Potential}},\ }\href {https://doi.org/10.1103/PhysRevLett.49.405}
  {\bibfield  {journal} {\bibinfo  {journal} {Phys. Rev. Lett.}\ }\textbf
  {\bibinfo {volume} {49}},\ \bibinfo {pages} {405} (\bibinfo {year}
  {1982})}\BibitemShut {NoStop}%
\bibitem [{\citenamefont {Kohmoto}(1985)}]{Kohmoto1985-zj}%
  \BibitemOpen
  \bibfield  {author} {\bibinfo {author} {\bibfnamefont {M.}~\bibnamefont
  {Kohmoto}},\ }\bibfield  {title} {\bibinfo {title} {{Topological invariant
  and the quantization of the Hall conductance}},\ }\href
  {https://doi.org/10.1016/0003-4916(85)90148-4} {\bibfield  {journal}
  {\bibinfo  {journal} {Ann. Phys.}\ }\textbf {\bibinfo {volume} {160}},\
  \bibinfo {pages} {343} (\bibinfo {year} {1985})}\BibitemShut {NoStop}%
\bibitem [{\citenamefont {Soluyanov}\ \emph {et~al.}(2015)\citenamefont
  {Soluyanov}, \citenamefont {Gresch}, \citenamefont {Wang}, \citenamefont
  {Wu}, \citenamefont {Troyer}, \citenamefont {Dai},\ and\ \citenamefont
  {Bernevig}}]{Soluyanov2015-zh}%
  \BibitemOpen
  \bibfield  {author} {\bibinfo {author} {\bibfnamefont {A.~A.}\ \bibnamefont
  {Soluyanov}}, \bibinfo {author} {\bibfnamefont {D.}~\bibnamefont {Gresch}},
  \bibinfo {author} {\bibfnamefont {Z.}~\bibnamefont {Wang}}, \bibinfo {author}
  {\bibfnamefont {Q.}~\bibnamefont {Wu}}, \bibinfo {author} {\bibfnamefont
  {M.}~\bibnamefont {Troyer}}, \bibinfo {author} {\bibfnamefont
  {X.}~\bibnamefont {Dai}},\ and\ \bibinfo {author} {\bibfnamefont {B.~A.}\
  \bibnamefont {Bernevig}},\ }\bibfield  {title} {\bibinfo {title} {{{Type-II}
  Weyl semimetals}},\ }\href {https://doi.org/10.1038/nature15768} {\bibfield
  {journal} {\bibinfo  {journal} {Nature}\ }\textbf {\bibinfo {volume} {527}},\
  \bibinfo {pages} {495} (\bibinfo {year} {2015})}\BibitemShut {NoStop}%
\bibitem [{\citenamefont {Berlijn}\ \emph {et~al.}(2017)\citenamefont
  {Berlijn}, \citenamefont {Snijders}, \citenamefont {Delaire}, \citenamefont
  {Zhou}, \citenamefont {Maier}, \citenamefont {Cao}, \citenamefont {Chi},
  \citenamefont {Matsuda}, \citenamefont {Wang}, \citenamefont {Koehler},
  \citenamefont {Kent},\ and\ \citenamefont {Weitering}}]{Berlijn2017-yg}%
  \BibitemOpen
  \bibfield  {author} {\bibinfo {author} {\bibfnamefont {T.}~\bibnamefont
  {Berlijn}}, \bibinfo {author} {\bibfnamefont {P.~C.}\ \bibnamefont
  {Snijders}}, \bibinfo {author} {\bibfnamefont {O.}~\bibnamefont {Delaire}},
  \bibinfo {author} {\bibfnamefont {H.-D.}\ \bibnamefont {Zhou}}, \bibinfo
  {author} {\bibfnamefont {T.~A.}\ \bibnamefont {Maier}}, \bibinfo {author}
  {\bibfnamefont {H.-B.}\ \bibnamefont {Cao}}, \bibinfo {author} {\bibfnamefont
  {S.-X.}\ \bibnamefont {Chi}}, \bibinfo {author} {\bibfnamefont
  {M.}~\bibnamefont {Matsuda}}, \bibinfo {author} {\bibfnamefont
  {Y.}~\bibnamefont {Wang}}, \bibinfo {author} {\bibfnamefont {M.~R.}\
  \bibnamefont {Koehler}}, \bibinfo {author} {\bibfnamefont {P.~R.~C.}\
  \bibnamefont {Kent}},\ and\ \bibinfo {author} {\bibfnamefont {H.~H.}\
  \bibnamefont {Weitering}},\ }\bibfield  {title} {\bibinfo {title} {{Itinerant
  Antiferromagnetism in {RuO$_2$}}},\ }\href
  {https://doi.org/10.1103/PhysRevLett.118.077201} {\bibfield  {journal}
  {\bibinfo  {journal} {Phys. Rev. Lett.}\ }\textbf {\bibinfo {volume} {118}},\
  \bibinfo {pages} {077201} (\bibinfo {year} {2017})}\BibitemShut {NoStop}%
\bibitem [{\citenamefont {Zhu}\ \emph {et~al.}(2019)\citenamefont {Zhu},
  \citenamefont {Strempfer}, \citenamefont {Rao}, \citenamefont {Occhialini},
  \citenamefont {Pelliciari}, \citenamefont {Choi}, \citenamefont {Kawaguchi},
  \citenamefont {You}, \citenamefont {Mitchell}, \citenamefont {Shao-Horn},\
  and\ \citenamefont {Comin}}]{Zhu2019-fx}%
  \BibitemOpen
  \bibfield  {author} {\bibinfo {author} {\bibfnamefont {Z.~H.}\ \bibnamefont
  {Zhu}}, \bibinfo {author} {\bibfnamefont {J.}~\bibnamefont {Strempfer}},
  \bibinfo {author} {\bibfnamefont {R.~R.}\ \bibnamefont {Rao}}, \bibinfo
  {author} {\bibfnamefont {C.~A.}\ \bibnamefont {Occhialini}}, \bibinfo
  {author} {\bibfnamefont {J.}~\bibnamefont {Pelliciari}}, \bibinfo {author}
  {\bibfnamefont {Y.}~\bibnamefont {Choi}}, \bibinfo {author} {\bibfnamefont
  {T.}~\bibnamefont {Kawaguchi}}, \bibinfo {author} {\bibfnamefont
  {H.}~\bibnamefont {You}}, \bibinfo {author} {\bibfnamefont {J.~F.}\
  \bibnamefont {Mitchell}}, \bibinfo {author} {\bibfnamefont {Y.}~\bibnamefont
  {Shao-Horn}},\ and\ \bibinfo {author} {\bibfnamefont {R.}~\bibnamefont
  {Comin}},\ }\bibfield  {title} {\bibinfo {title} {{Anomalous
  Antiferromagnetism in Metallic {RuO$_2$} Determined by Resonant X-ray
  Scattering}},\ }\href {https://doi.org/10.1103/PhysRevLett.122.017202}
  {\bibfield  {journal} {\bibinfo  {journal} {Phys. Rev. Lett.}\ }\textbf
  {\bibinfo {volume} {122}},\ \bibinfo {pages} {017202} (\bibinfo {year}
  {2019})}\BibitemShut {NoStop}%
\bibitem [{\citenamefont {Chernyshov}\ \emph {et~al.}(2009)\citenamefont
  {Chernyshov}, \citenamefont {Overby}, \citenamefont {Liu}, \citenamefont
  {Furdyna}, \citenamefont {Lyanda-Geller},\ and\ \citenamefont
  {Rokhinson}}]{Chernyshov2009-yj}%
  \BibitemOpen
  \bibfield  {author} {\bibinfo {author} {\bibfnamefont {A.}~\bibnamefont
  {Chernyshov}}, \bibinfo {author} {\bibfnamefont {M.}~\bibnamefont {Overby}},
  \bibinfo {author} {\bibfnamefont {X.}~\bibnamefont {Liu}}, \bibinfo {author}
  {\bibfnamefont {J.~K.}\ \bibnamefont {Furdyna}}, \bibinfo {author}
  {\bibfnamefont {Y.}~\bibnamefont {Lyanda-Geller}},\ and\ \bibinfo {author}
  {\bibfnamefont {L.~P.}\ \bibnamefont {Rokhinson}},\ }\bibfield  {title}
  {\bibinfo {title} {{Evidence for reversible control of magnetization in a
  ferromagnetic material by means of spin--orbit magnetic field}},\ }\href
  {https://doi.org/10.1038/nphys1362} {\bibfield  {journal} {\bibinfo
  {journal} {Nat. Phys.}\ }\textbf {\bibinfo {volume} {5}},\ \bibinfo {pages}
  {656} (\bibinfo {year} {2009})}\BibitemShut {NoStop}%
\bibitem [{\citenamefont {Olejn{\'\i}k}\ \emph {et~al.}(2018)\citenamefont
  {Olejn{\'\i}k}, \citenamefont {Seifert}, \citenamefont {Ka{\v s}par},
  \citenamefont {Nov{\'a}k}, \citenamefont {Wadley}, \citenamefont {Campion},
  \citenamefont {Baumgartner}, \citenamefont {Gambardella}, \citenamefont {N{\v
  e}mec}, \citenamefont {Wunderlich}, \citenamefont {Sinova}, \citenamefont
  {Ku{\v z}el}, \citenamefont {M{\"u}ller}, \citenamefont {Kampfrath},\ and\
  \citenamefont {Jungwirth}}]{Olejnik2018-ix}%
  \BibitemOpen
  \bibfield  {author} {\bibinfo {author} {\bibfnamefont {K.}~\bibnamefont
  {Olejn{\'\i}k}}, \bibinfo {author} {\bibfnamefont {T.}~\bibnamefont
  {Seifert}}, \bibinfo {author} {\bibfnamefont {Z.}~\bibnamefont {Ka{\v
  s}par}}, \bibinfo {author} {\bibfnamefont {V.}~\bibnamefont {Nov{\'a}k}},
  \bibinfo {author} {\bibfnamefont {P.}~\bibnamefont {Wadley}}, \bibinfo
  {author} {\bibfnamefont {R.~P.}\ \bibnamefont {Campion}}, \bibinfo {author}
  {\bibfnamefont {M.}~\bibnamefont {Baumgartner}}, \bibinfo {author}
  {\bibfnamefont {P.}~\bibnamefont {Gambardella}}, \bibinfo {author}
  {\bibfnamefont {P.}~\bibnamefont {N{\v e}mec}}, \bibinfo {author}
  {\bibfnamefont {J.}~\bibnamefont {Wunderlich}}, \bibinfo {author}
  {\bibfnamefont {J.}~\bibnamefont {Sinova}}, \bibinfo {author} {\bibfnamefont
  {P.}~\bibnamefont {Ku{\v z}el}}, \bibinfo {author} {\bibfnamefont
  {M.}~\bibnamefont {M{\"u}ller}}, \bibinfo {author} {\bibfnamefont
  {T.}~\bibnamefont {Kampfrath}},\ and\ \bibinfo {author} {\bibfnamefont
  {T.}~\bibnamefont {Jungwirth}},\ }\bibfield  {title} {\bibinfo {title}
  {{Terahertz electrical writing speed in an antiferromagnetic memory}},\
  }\href {https://doi.org/10.1126/sciadv.aar3566} {\bibfield  {journal}
  {\bibinfo  {journal} {Sci Adv}\ }\textbf {\bibinfo {volume} {4}},\ \bibinfo
  {pages} {eaar3566} (\bibinfo {year} {2018})}\BibitemShut {NoStop}%
\bibitem [{\citenamefont {Stern}\ \emph {et~al.}(2006)\citenamefont {Stern},
  \citenamefont {Ghosh}, \citenamefont {Xiang}, \citenamefont {Zhu},
  \citenamefont {Samarth},\ and\ \citenamefont {Awschalom}}]{Stern2006-ny}%
  \BibitemOpen
  \bibfield  {author} {\bibinfo {author} {\bibfnamefont {N.~P.}\ \bibnamefont
  {Stern}}, \bibinfo {author} {\bibfnamefont {S.}~\bibnamefont {Ghosh}},
  \bibinfo {author} {\bibfnamefont {G.}~\bibnamefont {Xiang}}, \bibinfo
  {author} {\bibfnamefont {M.}~\bibnamefont {Zhu}}, \bibinfo {author}
  {\bibfnamefont {N.}~\bibnamefont {Samarth}},\ and\ \bibinfo {author}
  {\bibfnamefont {D.~D.}\ \bibnamefont {Awschalom}},\ }\bibfield  {title}
  {\bibinfo {title} {{Current-Induced Polarization and the Spin Hall Effect at
  Room Temperature}},\ }\href {https://doi.org/10.1103/PhysRevLett.97.126603}
  {\bibfield  {journal} {\bibinfo  {journal} {Phys. Rev. Lett.}\ }\textbf
  {\bibinfo {volume} {97}},\ \bibinfo {pages} {126603} (\bibinfo {year}
  {2006})}\BibitemShut {NoStop}%
\bibitem [{MAG()}]{MAGNDATA}%
  \BibitemOpen
  \href@noop {} {\bibinfo {title} {{Bilbao Crystallogr. Serv. 2014. MAGNDATA: a
  collection of magnetic structures with portable cif-type files.
  \textit{Bilbao Crystallographic Server}}}},\ \bibinfo {note}
  {\url{http://www.cryst.ehu.es/magndata/}}\BibitemShut {NoStop}%
\bibitem [{\citenamefont {Guo}\ \emph {et~al.}(2023)\citenamefont {Guo},
  \citenamefont {Liu}, \citenamefont {Janson}, \citenamefont {Fulga},
  \citenamefont {van~den Brink},\ and\ \citenamefont {Facio}}]{Guo2023-yl}%
  \BibitemOpen
  \bibfield  {author} {\bibinfo {author} {\bibfnamefont {Y.}~\bibnamefont
  {Guo}}, \bibinfo {author} {\bibfnamefont {H.}~\bibnamefont {Liu}}, \bibinfo
  {author} {\bibfnamefont {O.}~\bibnamefont {Janson}}, \bibinfo {author}
  {\bibfnamefont {I.~C.}\ \bibnamefont {Fulga}}, \bibinfo {author}
  {\bibfnamefont {J.}~\bibnamefont {van~den Brink}},\ and\ \bibinfo {author}
  {\bibfnamefont {J.~I.}\ \bibnamefont {Facio}},\ }\bibfield  {title} {\bibinfo
  {title} {{Spin-split collinear antiferromagnets: A large-scale ab-initio
  study}},\ }\href {https://doi.org/10.1016/j.mtphys.2023.100991} {\bibfield
  {journal} {\bibinfo  {journal} {Mater. Today Phys.}\ }\textbf {\bibinfo
  {volume} {32}},\ \bibinfo {pages} {100991} (\bibinfo {year}
  {2023})}\BibitemShut {NoStop}%
\bibitem [{\citenamefont {White}\ \emph {et~al.}(2012)\citenamefont {White},
  \citenamefont {Honda}, \citenamefont {Kimura}, \citenamefont {Kimura},
  \citenamefont {Niedermayer}, \citenamefont {Zaharko}, \citenamefont {Poole},
  \citenamefont {Roessli},\ and\ \citenamefont {Kenzelmann}}]{White2012-zh}%
  \BibitemOpen
  \bibfield  {author} {\bibinfo {author} {\bibfnamefont {J.~S.}\ \bibnamefont
  {White}}, \bibinfo {author} {\bibfnamefont {T.}~\bibnamefont {Honda}},
  \bibinfo {author} {\bibfnamefont {K.}~\bibnamefont {Kimura}}, \bibinfo
  {author} {\bibfnamefont {T.}~\bibnamefont {Kimura}}, \bibinfo {author}
  {\bibfnamefont {C.}~\bibnamefont {Niedermayer}}, \bibinfo {author}
  {\bibfnamefont {O.}~\bibnamefont {Zaharko}}, \bibinfo {author} {\bibfnamefont
  {A.}~\bibnamefont {Poole}}, \bibinfo {author} {\bibfnamefont
  {B.}~\bibnamefont {Roessli}},\ and\ \bibinfo {author} {\bibfnamefont
  {M.}~\bibnamefont {Kenzelmann}},\ }\bibfield  {title} {\bibinfo {title}
  {{Coupling of magnetic and ferroelectric hysteresis by a multicomponent
  magnetic structure in {Mn$_2$GeO$_4$}}},\ }\href
  {https://doi.org/10.1103/PhysRevLett.108.077204} {\bibfield  {journal}
  {\bibinfo  {journal} {Phys. Rev. Lett.}\ }\textbf {\bibinfo {volume} {108}},\
  \bibinfo {pages} {077204} (\bibinfo {year} {2012})}\BibitemShut {NoStop}%
\bibitem [{\citenamefont {Cockayne}\ \emph {et~al.}(2013)\citenamefont
  {Cockayne}, \citenamefont {Levin}, \citenamefont {Wu},\ and\ \citenamefont
  {Llobet}}]{Cockayne2013-pm}%
  \BibitemOpen
  \bibfield  {author} {\bibinfo {author} {\bibfnamefont {E.}~\bibnamefont
  {Cockayne}}, \bibinfo {author} {\bibfnamefont {I.}~\bibnamefont {Levin}},
  \bibinfo {author} {\bibfnamefont {H.}~\bibnamefont {Wu}},\ and\ \bibinfo
  {author} {\bibfnamefont {A.}~\bibnamefont {Llobet}},\ }\bibfield  {title}
  {\bibinfo {title} {{Magnetic structure of bixbyite
  {$\alpha$-Mn$_{2}$O$_{3}$}: A combined {DFT+U} and neutron diffraction
  study}},\ }\href {https://doi.org/10.1103/PhysRevB.87.184413} {\bibfield
  {journal} {\bibinfo  {journal} {Phys. Rev. B}\ }\textbf {\bibinfo {volume}
  {87}},\ \bibinfo {pages} {184413} (\bibinfo {year} {2013})}\BibitemShut
  {NoStop}%
\bibitem [{\citenamefont {Brown}\ and\ \citenamefont
  {Frazer}(1963)}]{brown1963magnetic}%
  \BibitemOpen
  \bibfield  {author} {\bibinfo {author} {\bibfnamefont {P.}~\bibnamefont
  {Brown}}\ and\ \bibinfo {author} {\bibfnamefont {B.}~\bibnamefont {Frazer}},\
  }\bibfield  {title} {\bibinfo {title} {{Magnetic Structure of
  $\beta$-CoSO$_4$}},\ }\href {https://doi.org/10.1103/PhysRev.129.1145}
  {\bibfield  {journal} {\bibinfo  {journal} {Phys. Rev.}\ }\textbf {\bibinfo
  {volume} {129}},\ \bibinfo {pages} {1145} (\bibinfo {year}
  {1963})}\BibitemShut {NoStop}%
\bibitem [{\citenamefont {Bertaut}\ \emph {et~al.}(1963)\citenamefont
  {Bertaut}, \citenamefont {Coing-Boyat},\ and\ \citenamefont
  {Delapalme}}]{Bertaut1963-oq}%
  \BibitemOpen
  \bibfield  {author} {\bibinfo {author} {\bibfnamefont {E.~F.}\ \bibnamefont
  {Bertaut}}, \bibinfo {author} {\bibfnamefont {J.}~\bibnamefont
  {Coing-Boyat}},\ and\ \bibinfo {author} {\bibfnamefont {A.}~\bibnamefont
  {Delapalme}},\ }\bibfield  {title} {\bibinfo {title} {{Structure magnetique
  de {CoSo$_4$-$\beta$}}},\ }\href
  {https://doi.org/10.1016/0031-9163(63)90406-2} {\bibfield  {journal}
  {\bibinfo  {journal} {Phys. Lett.}\ }\textbf {\bibinfo {volume} {3}},\
  \bibinfo {pages} {178} (\bibinfo {year} {1963})}\BibitemShut {NoStop}%
\bibitem [{\citenamefont {Garlea}\ \emph {et~al.}(2014)\citenamefont {Garlea},
  \citenamefont {Sanjeewa}, \citenamefont {McGuire}, \citenamefont {Kumar},
  \citenamefont {Sulejmanovic}, \citenamefont {He},\ and\ \citenamefont
  {Hwu}}]{Garlea2014-pv}%
  \BibitemOpen
  \bibfield  {author} {\bibinfo {author} {\bibfnamefont {V.~O.}\ \bibnamefont
  {Garlea}}, \bibinfo {author} {\bibfnamefont {L.~D.}\ \bibnamefont
  {Sanjeewa}}, \bibinfo {author} {\bibfnamefont {M.~A.}\ \bibnamefont
  {McGuire}}, \bibinfo {author} {\bibfnamefont {P.}~\bibnamefont {Kumar}},
  \bibinfo {author} {\bibfnamefont {D.}~\bibnamefont {Sulejmanovic}}, \bibinfo
  {author} {\bibfnamefont {J.}~\bibnamefont {He}},\ and\ \bibinfo {author}
  {\bibfnamefont {S.-J.}\ \bibnamefont {Hwu}},\ }\bibfield  {title} {\bibinfo
  {title} {{Complex magnetic behavior of the sawtooth Fe chains in
  {Rb$_{2}$Fe$_{2}$O(AsO$_{4}$)$_{2}$}}},\ }\href
  {https://doi.org/10.1103/PhysRevB.89.014426} {\bibfield  {journal} {\bibinfo
  {journal} {Phys. Rev. B}\ }\textbf {\bibinfo {volume} {89}},\ \bibinfo
  {pages} {014426} (\bibinfo {year} {2014})}\BibitemShut {NoStop}%
\bibitem [{\citenamefont {Ferey}\ \emph {et~al.}(1985)\citenamefont {Ferey},
  \citenamefont {Leblanc}, \citenamefont {De~Pape},\ and\ \citenamefont
  {Pannetier}}]{ferey1985frustrated}%
  \BibitemOpen
  \bibfield  {author} {\bibinfo {author} {\bibfnamefont {G.}~\bibnamefont
  {Ferey}}, \bibinfo {author} {\bibfnamefont {M.}~\bibnamefont {Leblanc}},
  \bibinfo {author} {\bibfnamefont {R.}~\bibnamefont {De~Pape}},\ and\ \bibinfo
  {author} {\bibfnamefont {J.}~\bibnamefont {Pannetier}},\ }\bibfield  {title}
  {\bibinfo {title} {{Frustrated magnetic structures: II. Antiferromagnetic
  structure of the ordered modified pyrochlore
  {NH$_4$Fe$^{I\hspace{-1.2pt}I}$Fe$^{I\hspace{-1.2pt}I\hspace{-1.2pt}I}$F$_6$}
  at 4.2 K}},\ }\href {https://doi.org/10.1016/0038-1098(85)90192-9} {\bibfield
   {journal} {\bibinfo  {journal} {Solid State Commun.}\ }\textbf {\bibinfo
  {volume} {53}},\ \bibinfo {pages} {559} (\bibinfo {year} {1985})}\BibitemShut
  {NoStop}%
\bibitem [{\citenamefont {Kim}\ \emph {et~al.}(2012)\citenamefont {Kim},
  \citenamefont {Kim}, \citenamefont {Halasyamani}, \citenamefont {Green},
  \citenamefont {Bhatti}, \citenamefont {Leighton}, \citenamefont {Das},\ and\
  \citenamefont {Fennie}}]{Kim2012-hn}%
  \BibitemOpen
  \bibfield  {author} {\bibinfo {author} {\bibfnamefont {S.~W.}\ \bibnamefont
  {Kim}}, \bibinfo {author} {\bibfnamefont {S.-H.}\ \bibnamefont {Kim}},
  \bibinfo {author} {\bibfnamefont {P.~S.}\ \bibnamefont {Halasyamani}},
  \bibinfo {author} {\bibfnamefont {M.~A.}\ \bibnamefont {Green}}, \bibinfo
  {author} {\bibfnamefont {K.~P.}\ \bibnamefont {Bhatti}}, \bibinfo {author}
  {\bibfnamefont {C.}~\bibnamefont {Leighton}}, \bibinfo {author}
  {\bibfnamefont {H.}~\bibnamefont {Das}},\ and\ \bibinfo {author}
  {\bibfnamefont {C.~J.}\ \bibnamefont {Fennie}},\ }\bibfield  {title}
  {\bibinfo {title} {{{RbFe$^{2+}$Fe$^{3+}$F$_6$}: Synthesis, structure, and
  characterization of a new charge-ordered magnetically frustrated
  pyrochlore-related mixed-metal fluoride}},\ }\href
  {https://doi.org/10.1039/c2sc00765g} {\bibfield  {journal} {\bibinfo
  {journal} {Chem. Sci.}\ }\textbf {\bibinfo {volume} {3}},\ \bibinfo {pages}
  {741} (\bibinfo {year} {2012})}\BibitemShut {NoStop}%
\bibitem [{\citenamefont {Lottermoser}\ \emph {et~al.}(1986)\citenamefont
  {Lottermoser}, \citenamefont {M{\"u}ller},\ and\ \citenamefont {{H.
  Fuess}}}]{Lottermoser1986-fq}%
  \BibitemOpen
  \bibfield  {author} {\bibinfo {author} {\bibfnamefont {W.}~\bibnamefont
  {Lottermoser}}, \bibinfo {author} {\bibfnamefont {R.}~\bibnamefont
  {M{\"u}ller}},\ and\ \bibinfo {author} {\bibnamefont {{H. Fuess}}},\
  }\bibfield  {title} {\bibinfo {title} {{Antiferromagnetism in synthetic
  olivines}},\ }\href {https://doi.org/10.1016/0304-8853(86)90355-0} {\bibfield
   {journal} {\bibinfo  {journal} {J. Magn. Magn. Mater.}\ ,\ \bibinfo {pages}
  {1005}} (\bibinfo {year} {1986})}\BibitemShut {NoStop}%
\bibitem [{\citenamefont {Lottermoser}\ and\ \citenamefont
  {Fuess}(1988)}]{Lottermoser1988-gl}%
  \BibitemOpen
  \bibfield  {author} {\bibinfo {author} {\bibfnamefont {W.}~\bibnamefont
  {Lottermoser}}\ and\ \bibinfo {author} {\bibfnamefont {H.}~\bibnamefont
  {Fuess}},\ }\bibfield  {title} {\bibinfo {title} {{Magnetic structure of the
  orthosilicates {Mn$_2$SiO$_4$} and {Co$_2$SiO$_4$}}},\ }\href
  {https://doi.org/10.1002/pssa.2211090226} {\bibfield  {journal} {\bibinfo
  {journal} {Phys. Status Solidi A}\ }\textbf {\bibinfo {volume} {109}},\
  \bibinfo {pages} {589} (\bibinfo {year} {1988})}\BibitemShut {NoStop}%
\bibitem [{\citenamefont {Sazonov}\ \emph {et~al.}(2009)\citenamefont
  {Sazonov}, \citenamefont {Meven}, \citenamefont {Hutanu}, \citenamefont
  {Heger}, \citenamefont {Hansen},\ and\ \citenamefont
  {Gukasov}}]{Sazonov2009-dy}%
  \BibitemOpen
  \bibfield  {author} {\bibinfo {author} {\bibfnamefont {A.}~\bibnamefont
  {Sazonov}}, \bibinfo {author} {\bibfnamefont {M.}~\bibnamefont {Meven}},
  \bibinfo {author} {\bibfnamefont {V.}~\bibnamefont {Hutanu}}, \bibinfo
  {author} {\bibfnamefont {G.}~\bibnamefont {Heger}}, \bibinfo {author}
  {\bibfnamefont {T.}~\bibnamefont {Hansen}},\ and\ \bibinfo {author}
  {\bibfnamefont {A.}~\bibnamefont {Gukasov}},\ }\bibfield  {title} {\bibinfo
  {title} {{Magnetic behaviour of synthetic {Co$_2$SiO$_4$}}},\ }\href
  {https://doi.org/10.1107/S0108768109042499} {\bibfield  {journal} {\bibinfo
  {journal} {Acta Crystallogr. B}\ }\textbf {\bibinfo {volume} {65}},\ \bibinfo
  {pages} {664} (\bibinfo {year} {2009})}\BibitemShut {NoStop}%
\bibitem [{\citenamefont {Warner}\ \emph {et~al.}(1992)\citenamefont {Warner},
  \citenamefont {Cheetham}, \citenamefont {Cox},\ and\ \citenamefont
  {Von~Dreele}}]{Diffraction1992-zi}%
  \BibitemOpen
  \bibfield  {author} {\bibinfo {author} {\bibfnamefont {J.~K.}\ \bibnamefont
  {Warner}}, \bibinfo {author} {\bibfnamefont {A.~K.}\ \bibnamefont
  {Cheetham}}, \bibinfo {author} {\bibfnamefont {D.~E.}\ \bibnamefont {Cox}},\
  and\ \bibinfo {author} {\bibfnamefont {R.~B.}\ \bibnamefont {Von~Dreele}},\
  }\bibfield  {title} {\bibinfo {title} {{Valence Contrast between Iron Sites
  in $\alpha$-Fe$_2$PO$_5$: A Comparative Study by Magnetic Neutron and
  Resonant X-ray Powder Diffraction}},\ }\href
  {https://doi.org/10.1021/ja00041a027} {\bibfield  {journal} {\bibinfo
  {journal} {J. Am. Chem. Soc.}\ }\textbf {\bibinfo {volume} {114}},\ \bibinfo
  {pages} {6074} (\bibinfo {year} {1992})}\BibitemShut {NoStop}%
\bibitem [{\citenamefont {El~Khayati}\ \emph {et~al.}(2001)\citenamefont
  {El~Khayati}, \citenamefont {El~Moursli}, \citenamefont {Rodr}, \citenamefont
  {Andr}, \citenamefont {Blanchard}, \citenamefont {Bour}, \citenamefont
  {ColliNn},\ and\ \citenamefont {Roisnel}}]{El_Khayati2001-rw}%
  \BibitemOpen
  \bibfield  {author} {\bibinfo {author} {\bibfnamefont {N.}~\bibnamefont
  {El~Khayati}}, \bibinfo {author} {\bibfnamefont {R.~C.}\ \bibnamefont
  {El~Moursli}}, \bibinfo {author} {\bibfnamefont {J.}~\bibnamefont {Rodr}},
  \bibinfo {author} {\bibfnamefont {G.}~\bibnamefont {Andr}}, \bibinfo {author}
  {\bibfnamefont {N.}~\bibnamefont {Blanchard}}, \bibinfo {author}
  {\bibfnamefont {F.}~\bibnamefont {Bour}}, \bibinfo {author} {\bibfnamefont
  {G.}~\bibnamefont {ColliNn}},\ and\ \bibinfo {author} {\bibfnamefont
  {T.}~\bibnamefont {Roisnel}},\ }\bibfield  {title} {\bibinfo {title}
  {{Crystal and magnetic structures of the oxyphosphates {MFePO$_5$} (M = Fe,
  Co, Ni, Cu). Analysis of the magnetic ground state in terms of superexchange
  interactions}},\ }\href {https://doi.org/10.1007/s100510170093} {\bibfield
  {journal} {\bibinfo  {journal} {Eur. Phys. J. B}\ }\textbf {\bibinfo {volume}
  {22}},\ \bibinfo {pages} {429} (\bibinfo {year} {2001})}\BibitemShut
  {NoStop}%
\bibitem [{\citenamefont {Lee}\ \emph {et~al.}(2019)\citenamefont {Lee},
  \citenamefont {Kratochv{\'\i}lov{\'a}}, \citenamefont {Cao}, \citenamefont
  {Yamani}, \citenamefont {Kim}, \citenamefont {Park}, \citenamefont
  {Stewart},\ and\ \citenamefont {Oh}}]{Lee2019-bn}%
  \BibitemOpen
  \bibfield  {author} {\bibinfo {author} {\bibfnamefont {J.~H.}\ \bibnamefont
  {Lee}}, \bibinfo {author} {\bibfnamefont {M.}~\bibnamefont
  {Kratochv{\'\i}lov{\'a}}}, \bibinfo {author} {\bibfnamefont {H.}~\bibnamefont
  {Cao}}, \bibinfo {author} {\bibfnamefont {Z.}~\bibnamefont {Yamani}},
  \bibinfo {author} {\bibfnamefont {J.~S.}\ \bibnamefont {Kim}}, \bibinfo
  {author} {\bibfnamefont {J.-G.}\ \bibnamefont {Park}}, \bibinfo {author}
  {\bibfnamefont {G.~R.}\ \bibnamefont {Stewart}},\ and\ \bibinfo {author}
  {\bibfnamefont {Y.~S.}\ \bibnamefont {Oh}},\ }\bibfield  {title} {\bibinfo
  {title} {{Unconventional critical behavior in the quasi-one-dimensional S=1
  chain {NiTe$_{2}$O$_{5}$}}},\ }\href
  {https://doi.org/10.1103/PhysRevB.100.144441} {\bibfield  {journal} {\bibinfo
   {journal} {Phys. Rev. B}\ }\textbf {\bibinfo {volume} {100}},\ \bibinfo
  {pages} {144441} (\bibinfo {year} {2019})}\BibitemShut {NoStop}%
\bibitem [{\citenamefont {Ding}\ \emph {et~al.}(2017)\citenamefont {Ding},
  \citenamefont {Manuel}, \citenamefont {Khalyavin}, \citenamefont {Orlandi},
  \citenamefont {Kumagai}, \citenamefont {Oba}, \citenamefont {Yi},\ and\
  \citenamefont {Belik}}]{Ding2017-ct}%
  \BibitemOpen
  \bibfield  {author} {\bibinfo {author} {\bibfnamefont {L.}~\bibnamefont
  {Ding}}, \bibinfo {author} {\bibfnamefont {P.}~\bibnamefont {Manuel}},
  \bibinfo {author} {\bibfnamefont {D.~D.}\ \bibnamefont {Khalyavin}}, \bibinfo
  {author} {\bibfnamefont {F.}~\bibnamefont {Orlandi}}, \bibinfo {author}
  {\bibfnamefont {Y.}~\bibnamefont {Kumagai}}, \bibinfo {author} {\bibfnamefont
  {F.}~\bibnamefont {Oba}}, \bibinfo {author} {\bibfnamefont {W.}~\bibnamefont
  {Yi}},\ and\ \bibinfo {author} {\bibfnamefont {A.~A.}\ \bibnamefont
  {Belik}},\ }\bibfield  {title} {\bibinfo {title} {{Unusual magnetic structure
  of the high-pressure synthesized perovskites {$A$CrO$_{3}$
  ($A$=Sc,In,Tl)}}},\ }\href {https://doi.org/10.1103/PhysRevB.95.054432}
  {\bibfield  {journal} {\bibinfo  {journal} {Phys. Rev. B}\ }\textbf {\bibinfo
  {volume} {95}},\ \bibinfo {pages} {054432} (\bibinfo {year}
  {2017})}\BibitemShut {NoStop}%
\bibitem [{\citenamefont {Martinelli}\ \emph {et~al.}(2011)\citenamefont
  {Martinelli}, \citenamefont {Ferretti}, \citenamefont {Cimberle},\ and\
  \citenamefont {Ritter}}]{Martinelli2011-vk}%
  \BibitemOpen
  \bibfield  {author} {\bibinfo {author} {\bibfnamefont {A.}~\bibnamefont
  {Martinelli}}, \bibinfo {author} {\bibfnamefont {M.}~\bibnamefont
  {Ferretti}}, \bibinfo {author} {\bibfnamefont {M.~R.}\ \bibnamefont
  {Cimberle}},\ and\ \bibinfo {author} {\bibfnamefont {C.}~\bibnamefont
  {Ritter}},\ }\bibfield  {title} {\bibinfo {title} {{The crystal and magnetic
  structure of Ti-substituted {LaCrO$_3$}}},\ }\href
  {https://doi.org/10.1016/j.materresbull.2010.11.016} {\bibfield  {journal}
  {\bibinfo  {journal} {Mater. Res. Bull.}\ }\textbf {\bibinfo {volume} {46}},\
  \bibinfo {pages} {190} (\bibinfo {year} {2011})}\BibitemShut {NoStop}%
\bibitem [{\citenamefont {Tripathi}\ \emph {et~al.}(2017)\citenamefont
  {Tripathi}, \citenamefont {Choudhary}, \citenamefont {Phase}, \citenamefont
  {Chatterji},\ and\ \citenamefont {Fischer}}]{Tripathi2017-bu}%
  \BibitemOpen
  \bibfield  {author} {\bibinfo {author} {\bibfnamefont {M.}~\bibnamefont
  {Tripathi}}, \bibinfo {author} {\bibfnamefont {R.~J.}\ \bibnamefont
  {Choudhary}}, \bibinfo {author} {\bibfnamefont {D.~M.}\ \bibnamefont
  {Phase}}, \bibinfo {author} {\bibfnamefont {T.}~\bibnamefont {Chatterji}},\
  and\ \bibinfo {author} {\bibfnamefont {H.~E.}\ \bibnamefont {Fischer}},\
  }\bibfield  {title} {\bibinfo {title} {{Evolution of magnetic phases in
  {SmCrO$_{3}$}: A neutron diffraction and magnetometric study}},\ }\href
  {https://doi.org/10.1103/PhysRevB.96.174421} {\bibfield  {journal} {\bibinfo
  {journal} {Phys. Rev. B}\ }\textbf {\bibinfo {volume} {96}},\ \bibinfo
  {pages} {174421} (\bibinfo {year} {2017})}\BibitemShut {NoStop}%
\bibitem [{\citenamefont {Ivanov}\ \emph {et~al.}(2017)\citenamefont {Ivanov},
  \citenamefont {Beran}, \citenamefont {Bazuev}, \citenamefont {Tellgren},
  \citenamefont {Sarkar}, \citenamefont {Nordblad},\ and\ \citenamefont
  {Mathieu}}]{Ivanov2017-er}%
  \BibitemOpen
  \bibfield  {author} {\bibinfo {author} {\bibfnamefont {S.~A.}\ \bibnamefont
  {Ivanov}}, \bibinfo {author} {\bibfnamefont {P.}~\bibnamefont {Beran}},
  \bibinfo {author} {\bibfnamefont {G.}~\bibnamefont {Bazuev}}, \bibinfo
  {author} {\bibfnamefont {R.}~\bibnamefont {Tellgren}}, \bibinfo {author}
  {\bibfnamefont {T.}~\bibnamefont {Sarkar}}, \bibinfo {author} {\bibfnamefont
  {P.}~\bibnamefont {Nordblad}},\ and\ \bibinfo {author} {\bibfnamefont
  {R.}~\bibnamefont {Mathieu}},\ }\bibfield  {title} {\bibinfo {title}
  {{Perovskite solid solutions {La$_{0.75}$Bi$_{0.25}$Fe$_{1-x}$Cr$_x$O$_3$}:
  Preparation, structural, and magnetic properties}},\ }\href
  {https://doi.org/10.1016/j.jssc.2017.06.031} {\bibfield  {journal} {\bibinfo
  {journal} {J. Solid State Chem.}\ }\textbf {\bibinfo {volume} {254}},\
  \bibinfo {pages} {166} (\bibinfo {year} {2017})}\BibitemShut {NoStop}%
\bibitem [{\citenamefont {Damay}\ \emph {et~al.}(2020)\citenamefont {Damay},
  \citenamefont {Sottmann}, \citenamefont {Lain{\'e}}, \citenamefont {Chaix},
  \citenamefont {Poienar}, \citenamefont {Beran}, \citenamefont {Elkaim},
  \citenamefont {Fauth}, \citenamefont {Nataf}, \citenamefont {Guesdon},
  \citenamefont {Maignan},\ and\ \citenamefont {Martin}}]{Damay2020-gz}%
  \BibitemOpen
  \bibfield  {author} {\bibinfo {author} {\bibfnamefont {F.}~\bibnamefont
  {Damay}}, \bibinfo {author} {\bibfnamefont {J.}~\bibnamefont {Sottmann}},
  \bibinfo {author} {\bibfnamefont {F.}~\bibnamefont {Lain{\'e}}}, \bibinfo
  {author} {\bibfnamefont {L.}~\bibnamefont {Chaix}}, \bibinfo {author}
  {\bibfnamefont {M.}~\bibnamefont {Poienar}}, \bibinfo {author} {\bibfnamefont
  {P.}~\bibnamefont {Beran}}, \bibinfo {author} {\bibfnamefont
  {E.}~\bibnamefont {Elkaim}}, \bibinfo {author} {\bibfnamefont
  {F.}~\bibnamefont {Fauth}}, \bibinfo {author} {\bibfnamefont
  {L.}~\bibnamefont {Nataf}}, \bibinfo {author} {\bibfnamefont
  {A.}~\bibnamefont {Guesdon}}, \bibinfo {author} {\bibfnamefont
  {A.}~\bibnamefont {Maignan}},\ and\ \bibinfo {author} {\bibfnamefont
  {C.}~\bibnamefont {Martin}},\ }\bibfield  {title} {\bibinfo {title}
  {{Magnetic phase diagram for {Fe$_{3-x}$Mn$_{x}$BO$_{5}$}}},\ }\href
  {https://doi.org/10.1103/PhysRevB.101.094418} {\bibfield  {journal} {\bibinfo
   {journal} {Phys. Rev. B}\ }\textbf {\bibinfo {volume} {101}},\ \bibinfo
  {pages} {094418} (\bibinfo {year} {2020})}\BibitemShut {NoStop}%
\bibitem [{\citenamefont {Porter}\ \emph {et~al.}(2018)\citenamefont {Porter},
  \citenamefont {Granata}, \citenamefont {Forte}, \citenamefont {Di~Matteo},
  \citenamefont {Cuoco}, \citenamefont {Fittipaldi}, \citenamefont
  {Vecchione},\ and\ \citenamefont {Bombardi}}]{Porter2018-tf}%
  \BibitemOpen
  \bibfield  {author} {\bibinfo {author} {\bibfnamefont {D.~G.}\ \bibnamefont
  {Porter}}, \bibinfo {author} {\bibfnamefont {V.}~\bibnamefont {Granata}},
  \bibinfo {author} {\bibfnamefont {F.}~\bibnamefont {Forte}}, \bibinfo
  {author} {\bibfnamefont {S.}~\bibnamefont {Di~Matteo}}, \bibinfo {author}
  {\bibfnamefont {M.}~\bibnamefont {Cuoco}}, \bibinfo {author} {\bibfnamefont
  {R.}~\bibnamefont {Fittipaldi}}, \bibinfo {author} {\bibfnamefont
  {A.}~\bibnamefont {Vecchione}},\ and\ \bibinfo {author} {\bibfnamefont
  {A.}~\bibnamefont {Bombardi}},\ }\bibfield  {title} {\bibinfo {title}
  {{Magnetic anisotropy and orbital ordering in {Ca$_{2}$RuO$_{4}$}}},\ }\href
  {https://doi.org/10.1103/PhysRevB.98.125142} {\bibfield  {journal} {\bibinfo
  {journal} {Phys. Rev. B}\ }\textbf {\bibinfo {volume} {98}},\ \bibinfo
  {pages} {125142} (\bibinfo {year} {2018})}\BibitemShut {NoStop}%
\bibitem [{\citenamefont {Corliss}\ \emph {et~al.}(1958)\citenamefont
  {Corliss}, \citenamefont {Elliott},\ and\ \citenamefont
  {Hastings}}]{Corliss1958-vk}%
  \BibitemOpen
  \bibfield  {author} {\bibinfo {author} {\bibfnamefont {L.~M.}\ \bibnamefont
  {Corliss}}, \bibinfo {author} {\bibfnamefont {N.}~\bibnamefont {Elliott}},\
  and\ \bibinfo {author} {\bibfnamefont {J.~M.}\ \bibnamefont {Hastings}},\
  }\bibfield  {title} {\bibinfo {title} {{Antiferromagnetic structures of
  {MnS$_2$}, {MnSe$_2$}, and {MnTe$_2$}}},\ }\href
  {https://doi.org/10.1063/1.1723149} {\bibfield  {journal} {\bibinfo
  {journal} {J. Appl. Phys.}\ }\textbf {\bibinfo {volume} {29}},\ \bibinfo
  {pages} {391} (\bibinfo {year} {1958})}\BibitemShut {NoStop}%
\bibitem [{\citenamefont {Chakraborty}\ \emph {et~al.}(2014)\citenamefont
  {Chakraborty}, \citenamefont {Mukherjee}, \citenamefont {Kaushik},
  \citenamefont {Rayaprol}, \citenamefont {Prajapat}, \citenamefont {Singh},
  \citenamefont {Siruguri}, \citenamefont {Tyagi},\ and\ \citenamefont
  {Yusuf}}]{Chakraborty2014-zs}%
  \BibitemOpen
  \bibfield  {author} {\bibinfo {author} {\bibfnamefont {K.~R.}\ \bibnamefont
  {Chakraborty}}, \bibinfo {author} {\bibfnamefont {S.}~\bibnamefont
  {Mukherjee}}, \bibinfo {author} {\bibfnamefont {S.~D.}\ \bibnamefont
  {Kaushik}}, \bibinfo {author} {\bibfnamefont {S.}~\bibnamefont {Rayaprol}},
  \bibinfo {author} {\bibfnamefont {C.~L.}\ \bibnamefont {Prajapat}}, \bibinfo
  {author} {\bibfnamefont {M.~R.}\ \bibnamefont {Singh}}, \bibinfo {author}
  {\bibfnamefont {V.}~\bibnamefont {Siruguri}}, \bibinfo {author}
  {\bibfnamefont {A.~K.}\ \bibnamefont {Tyagi}},\ and\ \bibinfo {author}
  {\bibfnamefont {S.~M.}\ \bibnamefont {Yusuf}},\ }\bibfield  {title} {\bibinfo
  {title} {{Low temperature neutron diffraction study of
  {Nd$_{1-x}$Sr$_x$CrO$_3$} (0.05$\leq x \leq$0.15)}},\ }\href
  {https://doi.org/10.1016/j.jmmm.2014.02.018} {\bibfield  {journal} {\bibinfo
  {journal} {J. Magn. Magn. Mater.}\ }\textbf {\bibinfo {volume} {361}},\
  \bibinfo {pages} {81} (\bibinfo {year} {2014})}\BibitemShut {NoStop}%
\bibitem [{\citenamefont {Ritter}\ \emph {et~al.}(2021)\citenamefont {Ritter},
  \citenamefont {Ceretti},\ and\ \citenamefont {Paulus}}]{Ritter2021-be}%
  \BibitemOpen
  \bibfield  {author} {\bibinfo {author} {\bibfnamefont {C.}~\bibnamefont
  {Ritter}}, \bibinfo {author} {\bibfnamefont {M.}~\bibnamefont {Ceretti}},\
  and\ \bibinfo {author} {\bibfnamefont {W.}~\bibnamefont {Paulus}},\
  }\bibfield  {title} {\bibinfo {title} {{Determination of the magnetic
  structures in orthoferrite {CeFeO$_3$} by neutron powder diffraction: first
  order spin reorientation and appearance of an ordered Ce-moment}},\ }\href
  {https://doi.org/10.1088/1361-648X/abe64a} {\bibfield  {journal} {\bibinfo
  {journal} {J. Phys. Condens. Matter}\ }\textbf {\bibinfo {volume} {33}},\
  \bibinfo {pages} {215802} (\bibinfo {year} {2021})}\BibitemShut {NoStop}%
\bibitem [{\citenamefont {Ritter}\ \emph {et~al.}(2022)\citenamefont {Ritter},
  \citenamefont {Vilarinho}, \citenamefont {Moreira}, \citenamefont {Mihalik},
  \citenamefont {Mihalik},\ and\ \citenamefont {Savvin}}]{Ritter2022-cq}%
  \BibitemOpen
  \bibfield  {author} {\bibinfo {author} {\bibfnamefont {C.}~\bibnamefont
  {Ritter}}, \bibinfo {author} {\bibfnamefont {R.}~\bibnamefont {Vilarinho}},
  \bibinfo {author} {\bibfnamefont {J.~A.}\ \bibnamefont {Moreira}}, \bibinfo
  {author} {\bibfnamefont {M.}~\bibnamefont {Mihalik}}, \bibinfo {author}
  {\bibfnamefont {M.}~\bibnamefont {Mihalik}},\ and\ \bibinfo {author}
  {\bibfnamefont {S.}~\bibnamefont {Savvin}},\ }\bibfield  {title} {\bibinfo
  {title} {{The magnetic structure of {DyFeO$_3$} revisited: Fe spin
  reorientation and Dy incommensurate magnetic order}},\ }\href
  {https://doi.org/10.1088/1361-648X/ac6787} {\bibfield  {journal} {\bibinfo
  {journal} {J. Phys. Condens. Matter}\ }\textbf {\bibinfo {volume} {34}},\
  \bibinfo {pages} {265801} (\bibinfo {year} {2022})}\BibitemShut {NoStop}%
\bibitem [{\citenamefont {Plaza}\ \emph {et~al.}(1997)\citenamefont {Plaza},
  \citenamefont {Palacios}, \citenamefont {Bartolom{\'e}}, \citenamefont
  {Rosenkranz}, \citenamefont {Ritter},\ and\ \citenamefont
  {Furrer}}]{plaza1997neutron}%
  \BibitemOpen
  \bibfield  {author} {\bibinfo {author} {\bibfnamefont {I.}~\bibnamefont
  {Plaza}}, \bibinfo {author} {\bibfnamefont {E.}~\bibnamefont {Palacios}},
  \bibinfo {author} {\bibfnamefont {J.}~\bibnamefont {Bartolom{\'e}}}, \bibinfo
  {author} {\bibfnamefont {S.}~\bibnamefont {Rosenkranz}}, \bibinfo {author}
  {\bibfnamefont {C.}~\bibnamefont {Ritter}},\ and\ \bibinfo {author}
  {\bibfnamefont {A.}~\bibnamefont {Furrer}},\ }\bibfield  {title} {\bibinfo
  {title} {{Neutron diffraction study of the magnetic ordered Nd$^{3+}$ in
  NdCoO$_3$ and NdInO$_3$ below 1 K}},\ }\href
  {https://doi.org/10.1016/S0921-4526(96)01066-6} {\bibfield  {journal}
  {\bibinfo  {journal} {Physica B}\ }\textbf {\bibinfo {volume} {234}},\
  \bibinfo {pages} {632} (\bibinfo {year} {1997})}\BibitemShut {NoStop}%
\bibitem [{\citenamefont {Pernet}\ \emph {et~al.}(1969)\citenamefont {Pernet},
  \citenamefont {Quezel}, \citenamefont {Coing-Boyat},\ and\ \citenamefont
  {Lewy-Bertaut}}]{Pernet1969-vq}%
  \BibitemOpen
  \bibfield  {author} {\bibinfo {author} {\bibfnamefont {M.}~\bibnamefont
  {Pernet}}, \bibinfo {author} {\bibfnamefont {G.}~\bibnamefont {Quezel}},
  \bibinfo {author} {\bibfnamefont {J.}~\bibnamefont {Coing-Boyat}},\ and\
  \bibinfo {author} {\bibfnamefont {E.~F.}\ \bibnamefont {Lewy-Bertaut}},\
  }\bibfield  {title} {\bibinfo {title} {{Structures magn{\'e}tiques des
  chromates de cobalt et de nickel}},\ }\href
  {https://doi.org/10.3406/bulmi.1969.6355} {\bibfield  {journal} {\bibinfo
  {journal} {Bull. Soc. Fr. Phys.}\ }\textbf {\bibinfo {volume} {92}},\
  \bibinfo {pages} {264} (\bibinfo {year} {1969})}\BibitemShut {NoStop}%
\bibitem [{\citenamefont {Martinelli}\ \emph {et~al.}(2016)\citenamefont
  {Martinelli}, \citenamefont {Artini},\ and\ \citenamefont
  {Keller}}]{Martinelli2016-tv}%
  \BibitemOpen
  \bibfield  {author} {\bibinfo {author} {\bibfnamefont {A.}~\bibnamefont
  {Martinelli}}, \bibinfo {author} {\bibfnamefont {C.}~\bibnamefont {Artini}},\
  and\ \bibinfo {author} {\bibfnamefont {L.}~\bibnamefont {Keller}},\
  }\bibfield  {title} {\bibinfo {title} {{New insights into the magnetic
  properties of {LaErO$_3$}, (La$_{0.5}$Er$_{0.5}$)$_2$O$_3$ and
  (La$_{0.5}$Dy$_{0.5}$)$_2$O$_3$ oxides}},\ }\href
  {https://doi.org/10.1088/0953-8984/28/6/066003} {\bibfield  {journal}
  {\bibinfo  {journal} {J. Phys. Condens. Matter}\ }\textbf {\bibinfo {volume}
  {28}},\ \bibinfo {pages} {066003} (\bibinfo {year} {2016})}\BibitemShut
  {NoStop}%
\bibitem [{\citenamefont {Rousse}\ \emph {et~al.}(2016)\citenamefont {Rousse},
  \citenamefont {Radtke}, \citenamefont {Klein},\ and\ \citenamefont
  {Ahouari}}]{Rousse2016-iq}%
  \BibitemOpen
  \bibfield  {author} {\bibinfo {author} {\bibfnamefont {G.}~\bibnamefont
  {Rousse}}, \bibinfo {author} {\bibfnamefont {G.}~\bibnamefont {Radtke}},
  \bibinfo {author} {\bibfnamefont {Y.}~\bibnamefont {Klein}},\ and\ \bibinfo
  {author} {\bibfnamefont {H.}~\bibnamefont {Ahouari}},\ }\bibfield  {title}
  {\bibinfo {title} {{Long-range antiferromagnetic order in malonate-based
  compounds {Na$_2$M(H$_2$C$_3$O$_4$)$_2$$\cdot$2H$_2$O} (M = Mn, Fe, Co,
  Ni)}},\ }\href {https://doi.org/10.1039/c5dt04527d} {\bibfield  {journal}
  {\bibinfo  {journal} {Dalton Trans.}\ }\textbf {\bibinfo {volume} {45}},\
  \bibinfo {pages} {2536} (\bibinfo {year} {2016})}\BibitemShut {NoStop}%
\bibitem [{\citenamefont {Ritter}\ \emph {et~al.}(2016)\citenamefont {Ritter},
  \citenamefont {Ivanov}, \citenamefont {Bazuev},\ and\ \citenamefont
  {Fauth}}]{Ritter2016-ye}%
  \BibitemOpen
  \bibfield  {author} {\bibinfo {author} {\bibfnamefont {C.}~\bibnamefont
  {Ritter}}, \bibinfo {author} {\bibfnamefont {S.~A.}\ \bibnamefont {Ivanov}},
  \bibinfo {author} {\bibfnamefont {G.~V.}\ \bibnamefont {Bazuev}},\ and\
  \bibinfo {author} {\bibfnamefont {F.}~\bibnamefont {Fauth}},\ }\bibfield
  {title} {\bibinfo {title} {{Crystallographic phase coexistence, spin-orbital
  order transitions, and spontaneous spin flop in {TmVO$_3$}}},\ }\href
  {https://doi.org/10.1103/PhysRevB.93.054423} {\bibfield  {journal} {\bibinfo
  {journal} {Phys. Rev. B}\ }\textbf {\bibinfo {volume} {93}},\ \bibinfo
  {pages} {054423} (\bibinfo {year} {2016})}\BibitemShut {NoStop}%
\bibitem [{\citenamefont {Kunitomi}\ \emph {et~al.}(1964)\citenamefont
  {Kunitomi}, \citenamefont {Hamaguchi},\ and\ \citenamefont
  {Anzai}}]{Kunitomi1964-vz}%
  \BibitemOpen
  \bibfield  {author} {\bibinfo {author} {\bibfnamefont {N.}~\bibnamefont
  {Kunitomi}}, \bibinfo {author} {\bibfnamefont {Y.}~\bibnamefont
  {Hamaguchi}},\ and\ \bibinfo {author} {\bibfnamefont {S.}~\bibnamefont
  {Anzai}},\ }\bibfield  {title} {\bibinfo {title} {{Neutron diffraction study
  on manganese telluride}},\ }\href
  {https://doi.org/10.1051/jphys:01964002505056800} {\bibfield  {journal}
  {\bibinfo  {journal} {J. Phys. France}\ }\textbf {\bibinfo {volume} {25}},\
  \bibinfo {pages} {568} (\bibinfo {year} {1964})}\BibitemShut {NoStop}%
\bibitem [{\citenamefont {Dalmas~de R{\'e}otier}\ \emph
  {et~al.}(2017)\citenamefont {Dalmas~de R{\'e}otier}, \citenamefont {Marin},
  \citenamefont {Yaouanc}, \citenamefont {Ritter}, \citenamefont {Maisuradze},
  \citenamefont {Roessli}, \citenamefont {Bertin}, \citenamefont {Baker},\ and\
  \citenamefont {Amato}}]{Dalmas_de_Reotier2017-wv}%
  \BibitemOpen
  \bibfield  {author} {\bibinfo {author} {\bibfnamefont {P.}~\bibnamefont
  {Dalmas~de R{\'e}otier}}, \bibinfo {author} {\bibfnamefont {C.}~\bibnamefont
  {Marin}}, \bibinfo {author} {\bibfnamefont {A.}~\bibnamefont {Yaouanc}},
  \bibinfo {author} {\bibfnamefont {C.}~\bibnamefont {Ritter}}, \bibinfo
  {author} {\bibfnamefont {A.}~\bibnamefont {Maisuradze}}, \bibinfo {author}
  {\bibfnamefont {B.}~\bibnamefont {Roessli}}, \bibinfo {author} {\bibfnamefont
  {A.}~\bibnamefont {Bertin}}, \bibinfo {author} {\bibfnamefont {P.~J.}\
  \bibnamefont {Baker}},\ and\ \bibinfo {author} {\bibfnamefont
  {A.}~\bibnamefont {Amato}},\ }\bibfield  {title} {\bibinfo {title}
  {{Long-range dynamical magnetic order and spin tunneling in the cooperative
  paramagnetic states of the pyrochlore analogous spinel antiferromagnets
  {CdYb$_{2}$X$_{4}$} (S or Se)}},\ }\href
  {https://doi.org/10.1103/PhysRevB.96.134403} {\bibfield  {journal} {\bibinfo
  {journal} {Phys. Rev. B}\ }\textbf {\bibinfo {volume} {96}},\ \bibinfo
  {pages} {134403} (\bibinfo {year} {2017})}\BibitemShut {NoStop}%
\bibitem [{\citenamefont {Knight}\ \emph {et~al.}(2020)\citenamefont {Knight},
  \citenamefont {Khalyavin}, \citenamefont {Manuel}, \citenamefont {Bull},\
  and\ \citenamefont {McIntyre}}]{Knight2020-rn}%
  \BibitemOpen
  \bibfield  {author} {\bibinfo {author} {\bibfnamefont {K.~S.}\ \bibnamefont
  {Knight}}, \bibinfo {author} {\bibfnamefont {D.~D.}\ \bibnamefont
  {Khalyavin}}, \bibinfo {author} {\bibfnamefont {P.}~\bibnamefont {Manuel}},
  \bibinfo {author} {\bibfnamefont {C.~L.}\ \bibnamefont {Bull}},\ and\
  \bibinfo {author} {\bibfnamefont {P.}~\bibnamefont {McIntyre}},\ }\bibfield
  {title} {\bibinfo {title} {{Nuclear and magnetic structures of {KMnF$_3$}
  perovskite in the temperature interval 10 {K--105} {K}}},\ }\href
  {https://doi.org/10.1016/j.jallcom.2020.155935} {\bibfield  {journal}
  {\bibinfo  {journal} {J. Alloys Compd.}\ }\textbf {\bibinfo {volume} {842}},\
  \bibinfo {pages} {155935} (\bibinfo {year} {2020})}\BibitemShut {NoStop}%
\bibitem [{\citenamefont {Yamani}\ \emph {et~al.}(2010)\citenamefont {Yamani},
  \citenamefont {Tun},\ and\ \citenamefont {Ryan}}]{Yamani2010-hj}%
  \BibitemOpen
  \bibfield  {author} {\bibinfo {author} {\bibfnamefont {Z.}~\bibnamefont
  {Yamani}}, \bibinfo {author} {\bibfnamefont {Z.}~\bibnamefont {Tun}},\ and\
  \bibinfo {author} {\bibfnamefont {D.~H.}\ \bibnamefont {Ryan}},\ }\bibfield
  {title} {\bibinfo {title} {{Neutron scattering study of the classical
  antiferromagnet {MnF$_2$}: a perfect hands-on neutron scattering teaching
  courseSpecial issue on Neutron Scattering in Canada}},\ }\href
  {https://doi.org/10.1139/p10-081} {\bibfield  {journal} {\bibinfo  {journal}
  {Can. J. Phys.}\ }\textbf {\bibinfo {volume} {88}},\ \bibinfo {pages} {771}
  (\bibinfo {year} {2010})}\BibitemShut {NoStop}%
\bibitem [{\citenamefont {Jauch}\ \emph {et~al.}(2004)\citenamefont {Jauch},
  \citenamefont {Reehuis},\ and\ \citenamefont {Schultz}}]{Jauch2004-yr}%
  \BibitemOpen
  \bibfield  {author} {\bibinfo {author} {\bibfnamefont {W.}~\bibnamefont
  {Jauch}}, \bibinfo {author} {\bibfnamefont {M.}~\bibnamefont {Reehuis}},\
  and\ \bibinfo {author} {\bibfnamefont {A.~J.}\ \bibnamefont {Schultz}},\
  }\bibfield  {title} {\bibinfo {title} {{Gamma-ray and neutron diffraction
  studies of {CoF$_2$}: magnetostriction, electron density and magnetic
  moments}},\ }\href {https://doi.org/10.1107/s0108767303022803} {\bibfield
  {journal} {\bibinfo  {journal} {Acta Crystallogr. A}\ }\textbf {\bibinfo
  {volume} {60}},\ \bibinfo {pages} {51} (\bibinfo {year} {2004})}\BibitemShut
  {NoStop}%
\bibitem [{\citenamefont {Poole}\ \emph {et~al.}(2007)\citenamefont {Poole},
  \citenamefont {Wills},\ and\ \citenamefont
  {Leli{\`e}vre-Berna}}]{Poole2007-bt}%
  \BibitemOpen
  \bibfield  {author} {\bibinfo {author} {\bibfnamefont {A.}~\bibnamefont
  {Poole}}, \bibinfo {author} {\bibfnamefont {A.~S.}\ \bibnamefont {Wills}},\
  and\ \bibinfo {author} {\bibfnamefont {E.}~\bibnamefont
  {Leli{\`e}vre-Berna}},\ }\bibfield  {title} {\bibinfo {title} {{Magnetic
  ordering in the {XY} pyrochlore antiferromagnet {Er$_2$Ti$_2$O$_7$}: a
  spherical neutron polarimetry study}},\ }\href
  {https://doi.org/10.1088/0953-8984/19/45/452201} {\bibfield  {journal}
  {\bibinfo  {journal} {J. Phys. Condens. Matter}\ }\textbf {\bibinfo {volume}
  {19}},\ \bibinfo {pages} {452201} (\bibinfo {year} {2007})}\BibitemShut
  {NoStop}%
\bibitem [{\citenamefont {Taira}\ \emph {et~al.}(2003)\citenamefont {Taira},
  \citenamefont {Wakeshima}, \citenamefont {Hinatsu}, \citenamefont {Tobo},\
  and\ \citenamefont {Ohoyama}}]{Taira2003-ye}%
  \BibitemOpen
  \bibfield  {author} {\bibinfo {author} {\bibfnamefont {N.}~\bibnamefont
  {Taira}}, \bibinfo {author} {\bibfnamefont {M.}~\bibnamefont {Wakeshima}},
  \bibinfo {author} {\bibfnamefont {Y.}~\bibnamefont {Hinatsu}}, \bibinfo
  {author} {\bibfnamefont {A.}~\bibnamefont {Tobo}},\ and\ \bibinfo {author}
  {\bibfnamefont {K.}~\bibnamefont {Ohoyama}},\ }\bibfield  {title} {\bibinfo
  {title} {{Magnetic structure of pyrochlore-type {Er$_2$Ru$_2$O$_7$}}},\
  }\href {https://doi.org/10.1016/S0022-4596(03)00384-0} {\bibfield  {journal}
  {\bibinfo  {journal} {J. Solid State Chem.}\ }\textbf {\bibinfo {volume}
  {176}},\ \bibinfo {pages} {165} (\bibinfo {year} {2003})}\BibitemShut
  {NoStop}%
\bibitem [{\citenamefont {Petit}\ \emph {et~al.}(2017)\citenamefont {Petit},
  \citenamefont {Lhotel}, \citenamefont {Damay}, \citenamefont {Boutrouille},
  \citenamefont {Forget},\ and\ \citenamefont {Colson}}]{Petit2017-bi}%
  \BibitemOpen
  \bibfield  {author} {\bibinfo {author} {\bibfnamefont {S.}~\bibnamefont
  {Petit}}, \bibinfo {author} {\bibfnamefont {E.}~\bibnamefont {Lhotel}},
  \bibinfo {author} {\bibfnamefont {F.}~\bibnamefont {Damay}}, \bibinfo
  {author} {\bibfnamefont {P.}~\bibnamefont {Boutrouille}}, \bibinfo {author}
  {\bibfnamefont {A.}~\bibnamefont {Forget}},\ and\ \bibinfo {author}
  {\bibfnamefont {D.}~\bibnamefont {Colson}},\ }\bibfield  {title} {\bibinfo
  {title} {{{Long-Range} Order in the Dipolar {XY} Antiferromagnet
  {Er$_2$Sn$_2$O$_7$}}},\ }\href
  {https://doi.org/10.1103/PhysRevLett.119.187202} {\bibfield  {journal}
  {\bibinfo  {journal} {Phys. Rev. Lett.}\ }\textbf {\bibinfo {volume} {119}},\
  \bibinfo {pages} {187202} (\bibinfo {year} {2017})}\BibitemShut {NoStop}%
\bibitem [{\citenamefont {Hallas}\ \emph {et~al.}(2017)\citenamefont {Hallas},
  \citenamefont {Gaudet}, \citenamefont {Butch}, \citenamefont {Xu},
  \citenamefont {Tachibana}, \citenamefont {Wiebe}, \citenamefont {Luke},\ and\
  \citenamefont {Gaulin}}]{Hallas2017-kk}%
  \BibitemOpen
  \bibfield  {author} {\bibinfo {author} {\bibfnamefont {A.~M.}\ \bibnamefont
  {Hallas}}, \bibinfo {author} {\bibfnamefont {J.}~\bibnamefont {Gaudet}},
  \bibinfo {author} {\bibfnamefont {N.~P.}\ \bibnamefont {Butch}}, \bibinfo
  {author} {\bibfnamefont {G.}~\bibnamefont {Xu}}, \bibinfo {author}
  {\bibfnamefont {M.}~\bibnamefont {Tachibana}}, \bibinfo {author}
  {\bibfnamefont {C.~R.}\ \bibnamefont {Wiebe}}, \bibinfo {author}
  {\bibfnamefont {G.~M.}\ \bibnamefont {Luke}},\ and\ \bibinfo {author}
  {\bibfnamefont {B.~D.}\ \bibnamefont {Gaulin}},\ }\bibfield  {title}
  {\bibinfo {title} {{Phase Competition in the {Palmer-Chalker} {XY} Pyrochlore
  {Er$_2$Pt$_2$O$_7$}}},\ }\href
  {https://doi.org/10.1103/PhysRevLett.119.187201} {\bibfield  {journal}
  {\bibinfo  {journal} {Phys. Rev. Lett.}\ }\textbf {\bibinfo {volume} {119}},\
  \bibinfo {pages} {187201} (\bibinfo {year} {2017})}\BibitemShut {NoStop}%
\bibitem [{\citenamefont {Wills}\ \emph {et~al.}(2006)\citenamefont {Wills},
  \citenamefont {Zhitomirsky}, \citenamefont {Canals}, \citenamefont {Sanchez},
  \citenamefont {Bonville}, \citenamefont {de~R{\'e}otier},\ and\ \citenamefont
  {Yaouanc}}]{Wills2006-oi}%
  \BibitemOpen
  \bibfield  {author} {\bibinfo {author} {\bibfnamefont {A.~S.}\ \bibnamefont
  {Wills}}, \bibinfo {author} {\bibfnamefont {M.~E.}\ \bibnamefont
  {Zhitomirsky}}, \bibinfo {author} {\bibfnamefont {B.}~\bibnamefont {Canals}},
  \bibinfo {author} {\bibfnamefont {J.~P.}\ \bibnamefont {Sanchez}}, \bibinfo
  {author} {\bibfnamefont {P.}~\bibnamefont {Bonville}}, \bibinfo {author}
  {\bibfnamefont {P.~D.}\ \bibnamefont {de~R{\'e}otier}},\ and\ \bibinfo
  {author} {\bibfnamefont {A.}~\bibnamefont {Yaouanc}},\ }\bibfield  {title}
  {\bibinfo {title} {{Magnetic ordering in {Gd$_2$Sn$_2$O$_7$}: the archetypal
  Heisenberg pyrochlore antiferromagnet}},\ }\href
  {https://doi.org/10.1088/0953-8984/18/3/L02} {\bibfield  {journal} {\bibinfo
  {journal} {J. Phys. Condens. Matter}\ }\textbf {\bibinfo {volume} {18}},\
  \bibinfo {pages} {L37} (\bibinfo {year} {2006})}\BibitemShut {NoStop}%
\bibitem [{\citenamefont {Welch}\ \emph {et~al.}(2022)\citenamefont {Welch},
  \citenamefont {Paddison}, \citenamefont {Le}, \citenamefont {Gardner},
  \citenamefont {Chen}, \citenamefont {Wildes}, \citenamefont {Goodwin},\ and\
  \citenamefont {Stewart}}]{Welch2022-gb}%
  \BibitemOpen
  \bibfield  {author} {\bibinfo {author} {\bibfnamefont {P.~G.}\ \bibnamefont
  {Welch}}, \bibinfo {author} {\bibfnamefont {J.~A.~M.}\ \bibnamefont
  {Paddison}}, \bibinfo {author} {\bibfnamefont {M.~D.}\ \bibnamefont {Le}},
  \bibinfo {author} {\bibfnamefont {J.~S.}\ \bibnamefont {Gardner}}, \bibinfo
  {author} {\bibfnamefont {W.-T.}\ \bibnamefont {Chen}}, \bibinfo {author}
  {\bibfnamefont {A.~R.}\ \bibnamefont {Wildes}}, \bibinfo {author}
  {\bibfnamefont {A.~L.}\ \bibnamefont {Goodwin}},\ and\ \bibinfo {author}
  {\bibfnamefont {J.~R.}\ \bibnamefont {Stewart}},\ }\bibfield  {title}
  {\bibinfo {title} {{Magnetic structure and exchange interactions in the
  Heisenberg pyrochlore antiferromagnet {Gd$_{2}$Pt$_{2}$O$_{7}$}}},\ }\href
  {https://doi.org/10.1103/PhysRevB.105.094402} {\bibfield  {journal} {\bibinfo
   {journal} {Phys. Rev. B}\ }\textbf {\bibinfo {volume} {105}},\ \bibinfo
  {pages} {094402} (\bibinfo {year} {2022})}\BibitemShut {NoStop}%
\bibitem [{\citenamefont {Xu}\ \emph {et~al.}(2017{\natexlab{a}})\citenamefont
  {Xu}, \citenamefont {Sale}, \citenamefont {Avdeev}, \citenamefont {Ling},\
  and\ \citenamefont {Battle}}]{Xu2017-ee}%
  \BibitemOpen
  \bibfield  {author} {\bibinfo {author} {\bibfnamefont {D.}~\bibnamefont
  {Xu}}, \bibinfo {author} {\bibfnamefont {M.}~\bibnamefont {Sale}}, \bibinfo
  {author} {\bibfnamefont {M.}~\bibnamefont {Avdeev}}, \bibinfo {author}
  {\bibfnamefont {C.~D.}\ \bibnamefont {Ling}},\ and\ \bibinfo {author}
  {\bibfnamefont {P.~D.}\ \bibnamefont {Battle}},\ }\bibfield  {title}
  {\bibinfo {title} {{Experimental and computational study of the magnetic
  properties of {ZrMn$_{2-x}$Co$_x$Ge$_4$O$_{12}$}}},\ }\href
  {https://doi.org/10.1039/C7DT00975E} {\bibfield  {journal} {\bibinfo
  {journal} {Dalton Trans.}\ }\textbf {\bibinfo {volume} {46}},\ \bibinfo
  {pages} {6921} (\bibinfo {year} {2017}{\natexlab{a}})}\BibitemShut {NoStop}%
\bibitem [{\citenamefont {Xu}\ \emph {et~al.}(2017{\natexlab{b}})\citenamefont
  {Xu}, \citenamefont {Avdeev}, \citenamefont {Battle},\ and\ \citenamefont
  {Liu}}]{Xu2017-zt}%
  \BibitemOpen
  \bibfield  {author} {\bibinfo {author} {\bibfnamefont {D.}~\bibnamefont
  {Xu}}, \bibinfo {author} {\bibfnamefont {M.}~\bibnamefont {Avdeev}}, \bibinfo
  {author} {\bibfnamefont {P.~D.}\ \bibnamefont {Battle}},\ and\ \bibinfo
  {author} {\bibfnamefont {X.-Q.}\ \bibnamefont {Liu}},\ }\bibfield  {title}
  {\bibinfo {title} {{Magnetic Properties of {CeMn$_{2-x}$Co$_x$Ge$_4$O$_{12}$}
  (0 $\leq$ x $\leq$ 2) as a Function of Temperature and Magnetic Field}},\
  }\href {https://doi.org/10.1021/acs.inorgchem.6b02905} {\bibfield  {journal}
  {\bibinfo  {journal} {Inorg. Chem.}\ }\textbf {\bibinfo {volume} {56}},\
  \bibinfo {pages} {2750} (\bibinfo {year} {2017}{\natexlab{b}})}\BibitemShut
  {NoStop}%
\bibitem [{\citenamefont {Shachar}\ \emph {et~al.}(1972)\citenamefont
  {Shachar}, \citenamefont {Makovsky},\ and\ \citenamefont
  {Shaked}}]{shachar1972neutron}%
  \BibitemOpen
  \bibfield  {author} {\bibinfo {author} {\bibfnamefont {G.}~\bibnamefont
  {Shachar}}, \bibinfo {author} {\bibfnamefont {J.}~\bibnamefont {Makovsky}},\
  and\ \bibinfo {author} {\bibfnamefont {H.}~\bibnamefont {Shaked}},\
  }\bibfield  {title} {\bibinfo {title} {{Neutron-Diffraction Study of the
  Magnetic Structure of the Trirutile LiFe$_2$F$_6$}},\ }\href
  {https://doi.org/10.1103/PhysRevB.6.1968} {\bibfield  {journal} {\bibinfo
  {journal} {Phys. Rev. B}\ }\textbf {\bibinfo {volume} {6}},\ \bibinfo {pages}
  {1968} (\bibinfo {year} {1972})}\BibitemShut {NoStop}%
\bibitem [{\citenamefont {Cascos}\ \emph {et~al.}(2020)\citenamefont {Cascos},
  \citenamefont {Mart{\'\i}nez}, \citenamefont {Fern{\'a}ndez-D{\'\i}az},\ and\
  \citenamefont {Alonso}}]{Cascos2020-at}%
  \BibitemOpen
  \bibfield  {author} {\bibinfo {author} {\bibfnamefont {V.}~\bibnamefont
  {Cascos}}, \bibinfo {author} {\bibfnamefont {J.~L.}\ \bibnamefont
  {Mart{\'\i}nez}}, \bibinfo {author} {\bibfnamefont {M.~T.}\ \bibnamefont
  {Fern{\'a}ndez-D{\'\i}az}},\ and\ \bibinfo {author} {\bibfnamefont {J.~A.}\
  \bibnamefont {Alonso}},\ }\bibfield  {title} {\bibinfo {title} {{Magnetic
  properties of {Sr$_{0.7}$R$_{0.3}$CoO$_{3-\delta}$} (R = Tb, Er and Ho)
  perovskites}},\ }\href {https://doi.org/10.1016/j.jallcom.2020.156121}
  {\bibfield  {journal} {\bibinfo  {journal} {J. Alloys Compd.}\ }\textbf
  {\bibinfo {volume} {844}},\ \bibinfo {pages} {156121} (\bibinfo {year}
  {2020})}\BibitemShut {NoStop}%
\bibitem [{\citenamefont {Gvozdetskyi}\ \emph {et~al.}(2018)\citenamefont
  {Gvozdetskyi}, \citenamefont {Senyshyn}, \citenamefont {Gladyshevskii},\ and\
  \citenamefont {Hlukhyy}}]{Gvozdetskyi2018-gx}%
  \BibitemOpen
  \bibfield  {author} {\bibinfo {author} {\bibfnamefont {V.}~\bibnamefont
  {Gvozdetskyi}}, \bibinfo {author} {\bibfnamefont {A.}~\bibnamefont
  {Senyshyn}}, \bibinfo {author} {\bibfnamefont {R.}~\bibnamefont
  {Gladyshevskii}},\ and\ \bibinfo {author} {\bibfnamefont {V.}~\bibnamefont
  {Hlukhyy}},\ }\bibfield  {title} {\bibinfo {title} {{Crystal and Magnetic
  Structures of the Chain Antiferromagnet {CaFe$_4$Al$_8$}}},\ }\href
  {https://doi.org/10.1021/acs.inorgchem.8b00208} {\bibfield  {journal}
  {\bibinfo  {journal} {Inorg. Chem.}\ }\textbf {\bibinfo {volume} {57}},\
  \bibinfo {pages} {5820} (\bibinfo {year} {2018})}\BibitemShut {NoStop}%
\bibitem [{\citenamefont {Alikhanov}(1959)}]{ALIKHANOV1959}%
  \BibitemOpen
  \bibfield  {author} {\bibinfo {author} {\bibfnamefont {R.~A.}\ \bibnamefont
  {Alikhanov}},\ }\bibfield  {title} {\bibinfo {title} {Neutron diffraction
  investigation of the antiferromagnetism of the carbonates of manganese and
  iron},\ }\href@noop {} {\bibfield  {journal} {\bibinfo  {journal} {J. Exptl.
  Theoret. Phys. (U.S.S.R.)}\ }\textbf {\bibinfo {volume} {36}},\ \bibinfo
  {pages} {1690} (\bibinfo {year} {1959})}\BibitemShut {NoStop}%
\bibitem [{\citenamefont {Iikubo}\ \emph {et~al.}(2008)\citenamefont {Iikubo},
  \citenamefont {Kodama}, \citenamefont {Takenaka}, \citenamefont {Takagi},\
  and\ \citenamefont {Shamoto}}]{Iikubo2008-ye}%
  \BibitemOpen
  \bibfield  {author} {\bibinfo {author} {\bibfnamefont {S.}~\bibnamefont
  {Iikubo}}, \bibinfo {author} {\bibfnamefont {K.}~\bibnamefont {Kodama}},
  \bibinfo {author} {\bibfnamefont {K.}~\bibnamefont {Takenaka}}, \bibinfo
  {author} {\bibfnamefont {H.}~\bibnamefont {Takagi}},\ and\ \bibinfo {author}
  {\bibfnamefont {S.}~\bibnamefont {Shamoto}},\ }\bibfield  {title} {\bibinfo
  {title} {{Magnetovolume effect in {Mn$_{3}$Cu$_{1-x}$Ge$_{x}$N} related to
  the magnetic structure: Neutron powder diffraction measurements}},\ }\href
  {https://doi.org/10.1103/PhysRevB.77.020409} {\bibfield  {journal} {\bibinfo
  {journal} {Phys. Rev. B}\ }\textbf {\bibinfo {volume} {77}},\ \bibinfo
  {pages} {020409(R)} (\bibinfo {year} {2008})}\BibitemShut {NoStop}%
\bibitem [{\citenamefont {Kr{\'e}n}\ \emph {et~al.}(1971)\citenamefont
  {Kr{\'e}n}, \citenamefont {K{\'a}d{\'a}r}, \citenamefont {P{\'a}l},
  \citenamefont {Zsoldos}, \citenamefont {Barberon},\ and\ \citenamefont
  {Fruchart}}]{Kren1971-bg}%
  \BibitemOpen
  \bibfield  {author} {\bibinfo {author} {\bibfnamefont {E.}~\bibnamefont
  {Kr{\'e}n}}, \bibinfo {author} {\bibfnamefont {G.}~\bibnamefont
  {K{\'a}d{\'a}r}}, \bibinfo {author} {\bibfnamefont {L.}~\bibnamefont
  {P{\'a}l}}, \bibinfo {author} {\bibfnamefont {E.}~\bibnamefont {Zsoldos}},
  \bibinfo {author} {\bibfnamefont {M.}~\bibnamefont {Barberon}},\ and\
  \bibinfo {author} {\bibfnamefont {R.}~\bibnamefont {Fruchart}},\ }\bibfield
  {title} {\bibinfo {title} {Neutron diffraction study of ordered
  {Mn}-alloys},\ }\href {https://doi.org/10.1051/jphyscol:19711348} {\bibfield
  {journal} {\bibinfo  {journal} {J. Phys. Colloq.}\ }\textbf {\bibinfo
  {volume} {32}},\ \bibinfo {pages} {C1} (\bibinfo {year} {1971})}\BibitemShut
  {NoStop}%
\bibitem [{\citenamefont {Fruchart}\ and\ \citenamefont
  {F.~Bertaut}(1978)}]{Fruchart1978-es}%
  \BibitemOpen
  \bibfield  {author} {\bibinfo {author} {\bibfnamefont {D.}~\bibnamefont
  {Fruchart}}\ and\ \bibinfo {author} {\bibfnamefont {E.}~\bibnamefont
  {F.~Bertaut}},\ }\bibfield  {title} {\bibinfo {title} {{Magnetic Studies of
  the Metallic {Perovskite-Type} Compounds of Manganese}},\ }\href
  {https://doi.org/10.1143/JPSJ.44.781} {\bibfield  {journal} {\bibinfo
  {journal} {J. Phys. Soc. Jpn.}\ }\textbf {\bibinfo {volume} {44}},\ \bibinfo
  {pages} {781} (\bibinfo {year} {1978})}\BibitemShut {NoStop}%
\bibitem [{\citenamefont {Lee}\ \emph {et~al.}(2018)\citenamefont {Lee},
  \citenamefont {Torii}, \citenamefont {Ishikawa}, \citenamefont {Yonemura},
  \citenamefont {Moyoshi},\ and\ \citenamefont {Kamiyama}}]{Lee2018-os}%
  \BibitemOpen
  \bibfield  {author} {\bibinfo {author} {\bibfnamefont {S.}~\bibnamefont
  {Lee}}, \bibinfo {author} {\bibfnamefont {S.}~\bibnamefont {Torii}}, \bibinfo
  {author} {\bibfnamefont {Y.}~\bibnamefont {Ishikawa}}, \bibinfo {author}
  {\bibfnamefont {M.}~\bibnamefont {Yonemura}}, \bibinfo {author}
  {\bibfnamefont {T.}~\bibnamefont {Moyoshi}},\ and\ \bibinfo {author}
  {\bibfnamefont {T.}~\bibnamefont {Kamiyama}},\ }\bibfield  {title} {\bibinfo
  {title} {{Weak-ferromagnetism of {CoF$_3$} and {FeF$_3$}}},\ }\href
  {https://doi.org/10.1016/j.physb.2017.11.082} {\bibfield  {journal} {\bibinfo
   {journal} {Physica B}\ }\textbf {\bibinfo {volume} {551}},\ \bibinfo {pages}
  {94} (\bibinfo {year} {2018})}\BibitemShut {NoStop}%
\bibitem [{\citenamefont {Zhou}\ \emph {et~al.}(2011)\citenamefont {Zhou},
  \citenamefont {Alonso}, \citenamefont {Muo{\~n}z}, \citenamefont
  {Fern{\'a}ndez-D{\'\i}az},\ and\ \citenamefont {Goodenough}}]{Zhou2011-vw}%
  \BibitemOpen
  \bibfield  {author} {\bibinfo {author} {\bibfnamefont {J.-S.}\ \bibnamefont
  {Zhou}}, \bibinfo {author} {\bibfnamefont {J.~A.}\ \bibnamefont {Alonso}},
  \bibinfo {author} {\bibfnamefont {A.}~\bibnamefont {Muo{\~n}z}}, \bibinfo
  {author} {\bibfnamefont {M.~T.}\ \bibnamefont {Fern{\'a}ndez-D{\'\i}az}},\
  and\ \bibinfo {author} {\bibfnamefont {J.~B.}\ \bibnamefont {Goodenough}},\
  }\bibfield  {title} {\bibinfo {title} {{Magnetic structure of {LaCrO$_3$}
  perovskite under high pressure from in situ neutron diffraction}},\ }\href
  {https://doi.org/10.1103/PhysRevLett.106.057201} {\bibfield  {journal}
  {\bibinfo  {journal} {Phys. Rev. Lett.}\ }\textbf {\bibinfo {volume} {106}},\
  \bibinfo {pages} {057201} (\bibinfo {year} {2011})}\BibitemShut {NoStop}%
\bibitem [{\citenamefont {Zvereva}\ \emph {et~al.}(2020)\citenamefont
  {Zvereva}, \citenamefont {Raganyan}, \citenamefont {Vasilchikova},
  \citenamefont {Nalbandyan}, \citenamefont {Gafurov}, \citenamefont
  {Vavilova}, \citenamefont {Zakharov}, \citenamefont {Koo}, \citenamefont
  {Pomjakushin}, \citenamefont {Susloparova}, \citenamefont {Kurbakov},
  \citenamefont {Vasiliev},\ and\ \citenamefont {Whangbo}}]{Zvereva2020-my}%
  \BibitemOpen
  \bibfield  {author} {\bibinfo {author} {\bibfnamefont {E.~A.}\ \bibnamefont
  {Zvereva}}, \bibinfo {author} {\bibfnamefont {G.~V.}\ \bibnamefont
  {Raganyan}}, \bibinfo {author} {\bibfnamefont {T.~M.}\ \bibnamefont
  {Vasilchikova}}, \bibinfo {author} {\bibfnamefont {V.~B.}\ \bibnamefont
  {Nalbandyan}}, \bibinfo {author} {\bibfnamefont {D.~A.}\ \bibnamefont
  {Gafurov}}, \bibinfo {author} {\bibfnamefont {E.~L.}\ \bibnamefont
  {Vavilova}}, \bibinfo {author} {\bibfnamefont {K.~V.}\ \bibnamefont
  {Zakharov}}, \bibinfo {author} {\bibfnamefont {H.-J.}\ \bibnamefont {Koo}},
  \bibinfo {author} {\bibfnamefont {V.~Y.}\ \bibnamefont {Pomjakushin}},
  \bibinfo {author} {\bibfnamefont {A.~E.}\ \bibnamefont {Susloparova}},
  \bibinfo {author} {\bibfnamefont {A.~I.}\ \bibnamefont {Kurbakov}}, \bibinfo
  {author} {\bibfnamefont {A.~N.}\ \bibnamefont {Vasiliev}},\ and\ \bibinfo
  {author} {\bibfnamefont {M.-H.}\ \bibnamefont {Whangbo}},\ }\bibfield
  {title} {\bibinfo {title} {{Hidden magnetic order in the triangular-lattice
  magnet {Li$_{2}$MnTeO$_{6}$}}},\ }\href
  {https://doi.org/10.1103/PhysRevB.102.094433} {\bibfield  {journal} {\bibinfo
   {journal} {Phys. Rev. B}\ }\textbf {\bibinfo {volume} {102}},\ \bibinfo
  {pages} {094433} (\bibinfo {year} {2020})}\BibitemShut {NoStop}%
\bibitem [{\citenamefont {Albrecht}\ \emph {et~al.}(2019)\citenamefont
  {Albrecht}, \citenamefont {Hunger}, \citenamefont {Block}, \citenamefont
  {P{\"o}ttgen}, \citenamefont {Senyshyn}, \citenamefont {Doert},\ and\
  \citenamefont {Ruck}}]{Albrecht2019-nh}%
  \BibitemOpen
  \bibfield  {author} {\bibinfo {author} {\bibfnamefont {R.}~\bibnamefont
  {Albrecht}}, \bibinfo {author} {\bibfnamefont {J.}~\bibnamefont {Hunger}},
  \bibinfo {author} {\bibfnamefont {T.}~\bibnamefont {Block}}, \bibinfo
  {author} {\bibfnamefont {R.}~\bibnamefont {P{\"o}ttgen}}, \bibinfo {author}
  {\bibfnamefont {A.}~\bibnamefont {Senyshyn}}, \bibinfo {author}
  {\bibfnamefont {T.}~\bibnamefont {Doert}},\ and\ \bibinfo {author}
  {\bibfnamefont {M.}~\bibnamefont {Ruck}},\ }\bibfield  {title} {\bibinfo
  {title} {{{Oxo-Hydroxoferrate} K$_{2-x}$Fe$_4$O$_{7-x}$({OH})$_x$ : Hydroflux
  Synthesis, Chemical and Thermal Instability, Crystal and Magnetic
  Structures}},\ }\href {https://doi.org/10.1002/open.201800229} {\bibfield
  {journal} {\bibinfo  {journal} {ChemistryOpen}\ }\textbf {\bibinfo {volume}
  {8}},\ \bibinfo {pages} {74} (\bibinfo {year} {2019})}\BibitemShut {NoStop}%
\bibitem [{\citenamefont {Mentr{\'e}}\ \emph {et~al.}(2008)\citenamefont
  {Mentr{\'e}}, \citenamefont {Kauffmann}, \citenamefont {Ehora}, \citenamefont
  {Daviero-Minaud}, \citenamefont {Abraham},\ and\ \citenamefont
  {Roussel}}]{Mentre2008-ks}%
  \BibitemOpen
  \bibfield  {author} {\bibinfo {author} {\bibfnamefont {O.}~\bibnamefont
  {Mentr{\'e}}}, \bibinfo {author} {\bibfnamefont {M.}~\bibnamefont
  {Kauffmann}}, \bibinfo {author} {\bibfnamefont {G.}~\bibnamefont {Ehora}},
  \bibinfo {author} {\bibfnamefont {S.}~\bibnamefont {Daviero-Minaud}},
  \bibinfo {author} {\bibfnamefont {F.}~\bibnamefont {Abraham}},\ and\ \bibinfo
  {author} {\bibfnamefont {P.}~\bibnamefont {Roussel}},\ }\bibfield  {title}
  {\bibinfo {title} {{Structure, dimensionality and magnetism of new cobalt
  oxyhalides}},\ }\href
  {https://doi.org/10.1016/j.solidstatesciences.2007.12.032} {\bibfield
  {journal} {\bibinfo  {journal} {Solid State Sci.}\ }\textbf {\bibinfo
  {volume} {10}},\ \bibinfo {pages} {471} (\bibinfo {year} {2008})}\BibitemShut
  {NoStop}%
\bibitem [{\citenamefont {Melamud}\ \emph {et~al.}(1974)\citenamefont
  {Melamud}, \citenamefont {Pinto}, \citenamefont {Makovsky},\ and\
  \citenamefont {Shaked}}]{Melamud1974-tq}%
  \BibitemOpen
  \bibfield  {author} {\bibinfo {author} {\bibfnamefont {M.}~\bibnamefont
  {Melamud}}, \bibinfo {author} {\bibfnamefont {H.}~\bibnamefont {Pinto}},
  \bibinfo {author} {\bibfnamefont {J.}~\bibnamefont {Makovsky}},\ and\
  \bibinfo {author} {\bibfnamefont {H.}~\bibnamefont {Shaked}},\ }\bibfield
  {title} {\bibinfo {title} {{Magnetic structure of {CsCoCl$_3$} at 4.2
  $^\circ$K}},\ }\href {https://doi.org/10.1002/pssb.2220630234} {\bibfield
  {journal} {\bibinfo  {journal} {Phys. Status Solidi B}\ }\textbf {\bibinfo
  {volume} {63}},\ \bibinfo {pages} {699} (\bibinfo {year} {1974})}\BibitemShut
  {NoStop}%
\bibitem [{\citenamefont {Mekata}\ and\ \citenamefont
  {Adachi}(1978)}]{Mekata1978-sz}%
  \BibitemOpen
  \bibfield  {author} {\bibinfo {author} {\bibfnamefont {M.}~\bibnamefont
  {Mekata}}\ and\ \bibinfo {author} {\bibfnamefont {K.}~\bibnamefont
  {Adachi}},\ }\bibfield  {title} {\bibinfo {title} {{Magnetic Structure of
  {CsCoCl$_3$}}},\ }\href {https://doi.org/10.1143/JPSJ.44.806} {\bibfield
  {journal} {\bibinfo  {journal} {J. Phys. Soc. Jpn.}\ }\textbf {\bibinfo
  {volume} {44}},\ \bibinfo {pages} {806} (\bibinfo {year} {1978})}\BibitemShut
  {NoStop}%
\bibitem [{\citenamefont {Kr{\'e}n}\ and\ \citenamefont
  {K{\'a}d{\'a}r}(1970)}]{Kren1970-np}%
  \BibitemOpen
  \bibfield  {author} {\bibinfo {author} {\bibfnamefont {E.}~\bibnamefont
  {Kr{\'e}n}}\ and\ \bibinfo {author} {\bibfnamefont {G.}~\bibnamefont
  {K{\'a}d{\'a}r}},\ }\bibfield  {title} {\bibinfo {title} {{Neutron
  diffraction study of {Mn$_3$Ga}}},\ }\href
  {https://doi.org/10.1016/0038-1098(70)90484-9} {\bibfield  {journal}
  {\bibinfo  {journal} {Solid State Commun.}\ }\textbf {\bibinfo {volume}
  {8}},\ \bibinfo {pages} {1653} (\bibinfo {year} {1970})}\BibitemShut
  {NoStop}%
\bibitem [{\citenamefont {Minkiewicz}\ \emph {et~al.}(1971)\citenamefont
  {Minkiewicz}, \citenamefont {Cox},\ and\ \citenamefont
  {Shirane}}]{Minkiewicz1971-ac}%
  \BibitemOpen
  \bibfield  {author} {\bibinfo {author} {\bibfnamefont {V.~J.}\ \bibnamefont
  {Minkiewicz}}, \bibinfo {author} {\bibfnamefont {D.~E.}\ \bibnamefont
  {Cox}},\ and\ \bibinfo {author} {\bibfnamefont {G.}~\bibnamefont {Shirane}},\
  }\bibfield  {title} {\bibinfo {title} {Neutron scattering from {RbCoBr$_3$}
  and {RbNiCl$_3$}},\ }\href {https://doi.org/10.1051/jphyscol:19711316}
  {\bibfield  {journal} {\bibinfo  {journal} {J. Phys. Colloq.}\ }\textbf
  {\bibinfo {volume} {32}},\ \bibinfo {pages} {C1} (\bibinfo {year}
  {1971})}\BibitemShut {NoStop}%
\bibitem [{\citenamefont {Yuan}\ \emph
  {et~al.}(2020{\natexlab{b}})\citenamefont {Yuan}, \citenamefont {Song},
  \citenamefont {Xing},\ and\ \citenamefont {Chen}}]{Yuan2020-nx}%
  \BibitemOpen
  \bibfield  {author} {\bibinfo {author} {\bibfnamefont {J.}~\bibnamefont
  {Yuan}}, \bibinfo {author} {\bibfnamefont {Y.}~\bibnamefont {Song}}, \bibinfo
  {author} {\bibfnamefont {X.}~\bibnamefont {Xing}},\ and\ \bibinfo {author}
  {\bibfnamefont {J.}~\bibnamefont {Chen}},\ }\bibfield  {title} {\bibinfo
  {title} {{Magnetic structure and uniaxial negative thermal expansion in
  antiferromagnetic {CrSb}}},\ }\href {https://doi.org/10.1039/d0dt03277h}
  {\bibfield  {journal} {\bibinfo  {journal} {Dalton Trans.}\ }\textbf
  {\bibinfo {volume} {49}},\ \bibinfo {pages} {17605} (\bibinfo {year}
  {2020}{\natexlab{b}})}\BibitemShut {NoStop}%
\bibitem [{\citenamefont {Christensen}\ and\ \citenamefont
  {Ollivier}(1972)}]{Christensen1972-pv}%
  \BibitemOpen
  \bibfield  {author} {\bibinfo {author} {\bibfnamefont {A.}~\bibnamefont
  {Christensen}}\ and\ \bibinfo {author} {\bibfnamefont {G.}~\bibnamefont
  {Ollivier}},\ }\bibfield  {title} {\bibinfo {title} {{Hydrothermal and
  High-Pressure Preparation of some {BaMnO$_3$} Modifications and
  Low-Temperature Magnetic Properties of {BaMnO$_3$(2H)}}},\ }\href
  {https://doi.org/10.1016/0022-4596(72)90141-7} {\bibfield  {journal}
  {\bibinfo  {journal} {J. Solid State Chem.}\ }\textbf {\bibinfo {volume}
  {4}},\ \bibinfo {pages} {131} (\bibinfo {year} {1972})}\BibitemShut {NoStop}%
\bibitem [{\citenamefont {Van~Laar}\ and\ \citenamefont
  {Rietveld}(1971)}]{Van_Laar_undated-rm}%
  \BibitemOpen
  \bibfield  {author} {\bibinfo {author} {\bibfnamefont {B.}~\bibnamefont
  {Van~Laar}}\ and\ \bibinfo {author} {\bibfnamefont {H.~M.}\ \bibnamefont
  {Rietveld}},\ }\bibfield  {title} {\bibinfo {title} {{Magnetic and
  Crystallographic Structures of Me$_x$NbS$_2$, and Me$_x$TaS$_2$}},\ }\href
  {https://doi.org/10.1016/0022-4596(71)90019-3} {\bibfield  {journal}
  {\bibinfo  {journal} {J. Solid State Chem.}\ }\textbf {\bibinfo {volume}
  {3}},\ \bibinfo {pages} {154} (\bibinfo {year} {1971})}\BibitemShut {NoStop}%
\bibitem [{\citenamefont {Lightfoot}\ and\ \citenamefont
  {Battle}(1990)}]{Lightfoot_undated-jw}%
  \BibitemOpen
  \bibfield  {author} {\bibinfo {author} {\bibfnamefont {P.}~\bibnamefont
  {Lightfoot}}\ and\ \bibinfo {author} {\bibfnamefont {P.~D.}\ \bibnamefont
  {Battle}},\ }\bibfield  {title} {\bibinfo {title} {{The Crystal and Magnetic
  Structures of Ba$_3$NiRu$_2$O$_9$, Ba$_3$CoRu$_2$O$_9$, and
  Ba$_3$ZnRu$_2$O$_9$}},\ }\href {https://doi.org/10.1016/0022-4596(90)90309-L}
  {\bibfield  {journal} {\bibinfo  {journal} {J. Solid State Chem.}\ }\textbf
  {\bibinfo {volume} {89}},\ \bibinfo {pages} {174} (\bibinfo {year}
  {1990})}\BibitemShut {NoStop}%
\bibitem [{\citenamefont {Lawrence}\ \emph {et~al.}(2023)\citenamefont
  {Lawrence}, \citenamefont {Huai}, \citenamefont {Kim}, \citenamefont
  {Avdeev}, \citenamefont {Chen}, \citenamefont {Skorupskii}, \citenamefont
  {Miura}, \citenamefont {Ferrenti}, \citenamefont {Waibel}, \citenamefont
  {Kawaguchi}, \citenamefont {Ng}, \citenamefont {Kaman}, \citenamefont {Cai},
  \citenamefont {Schoop}, \citenamefont {Kushwaha}, \citenamefont {Liu},
  \citenamefont {Tran},\ and\ \citenamefont {Ji}}]{Lawrence2023-ky}%
  \BibitemOpen
  \bibfield  {author} {\bibinfo {author} {\bibfnamefont {E.~A.}\ \bibnamefont
  {Lawrence}}, \bibinfo {author} {\bibfnamefont {X.}~\bibnamefont {Huai}},
  \bibinfo {author} {\bibfnamefont {D.}~\bibnamefont {Kim}}, \bibinfo {author}
  {\bibfnamefont {M.}~\bibnamefont {Avdeev}}, \bibinfo {author} {\bibfnamefont
  {Y.}~\bibnamefont {Chen}}, \bibinfo {author} {\bibfnamefont {G.}~\bibnamefont
  {Skorupskii}}, \bibinfo {author} {\bibfnamefont {A.}~\bibnamefont {Miura}},
  \bibinfo {author} {\bibfnamefont {A.}~\bibnamefont {Ferrenti}}, \bibinfo
  {author} {\bibfnamefont {M.}~\bibnamefont {Waibel}}, \bibinfo {author}
  {\bibfnamefont {S.}~\bibnamefont {Kawaguchi}}, \bibinfo {author}
  {\bibfnamefont {N.}~\bibnamefont {Ng}}, \bibinfo {author} {\bibfnamefont
  {B.}~\bibnamefont {Kaman}}, \bibinfo {author} {\bibfnamefont
  {Z.}~\bibnamefont {Cai}}, \bibinfo {author} {\bibfnamefont {L.}~\bibnamefont
  {Schoop}}, \bibinfo {author} {\bibfnamefont {S.}~\bibnamefont {Kushwaha}},
  \bibinfo {author} {\bibfnamefont {F.}~\bibnamefont {Liu}}, \bibinfo {author}
  {\bibfnamefont {T.~T.}\ \bibnamefont {Tran}},\ and\ \bibinfo {author}
  {\bibfnamefont {H.}~\bibnamefont {Ji}},\ }\bibfield  {title} {\bibinfo
  {title} {{Fe Site Order and Magnetic Properties of {Fe$_{1/4}$NbS$_2$}}},\
  }\href {https://doi.org/10.1021/acs.inorgchem.3c02652} {\bibfield  {journal}
  {\bibinfo  {journal} {Inorg. Chem.}\ }\textbf {\bibinfo {volume} {62}},\
  \bibinfo {pages} {18179} (\bibinfo {year} {2023})}\BibitemShut {NoStop}%
\bibitem [{\citenamefont {Yano}\ \emph {et~al.}(2016)\citenamefont {Yano},
  \citenamefont {Louca}, \citenamefont {Yang}, \citenamefont {Chatterjee},
  \citenamefont {Bugaris}, \citenamefont {Chung}, \citenamefont {Peng},
  \citenamefont {Grayson},\ and\ \citenamefont {Kanatzidis}}]{Yano2016-hm}%
  \BibitemOpen
  \bibfield  {author} {\bibinfo {author} {\bibfnamefont {S.}~\bibnamefont
  {Yano}}, \bibinfo {author} {\bibfnamefont {D.}~\bibnamefont {Louca}},
  \bibinfo {author} {\bibfnamefont {J.}~\bibnamefont {Yang}}, \bibinfo {author}
  {\bibfnamefont {U.}~\bibnamefont {Chatterjee}}, \bibinfo {author}
  {\bibfnamefont {D.~E.}\ \bibnamefont {Bugaris}}, \bibinfo {author}
  {\bibfnamefont {D.~Y.}\ \bibnamefont {Chung}}, \bibinfo {author}
  {\bibfnamefont {L.}~\bibnamefont {Peng}}, \bibinfo {author} {\bibfnamefont
  {M.}~\bibnamefont {Grayson}},\ and\ \bibinfo {author} {\bibfnamefont {M.~G.}\
  \bibnamefont {Kanatzidis}},\ }\bibfield  {title} {\bibinfo {title} {{Magnetic
  structure of {NiS$_{2-x}$Se$_{x}$}}},\ }\href
  {https://doi.org/10.1103/PhysRevB.93.024409} {\bibfield  {journal} {\bibinfo
  {journal} {Phys. Rev. B}\ }\textbf {\bibinfo {volume} {93}},\ \bibinfo
  {pages} {024409} (\bibinfo {year} {2016})}\BibitemShut {NoStop}%
\bibitem [{\citenamefont {Burlet}\ \emph {et~al.}(1997)\citenamefont {Burlet},
  \citenamefont {Ressouche}, \citenamefont {Malaman}, \citenamefont {Welter},
  \citenamefont {Sanchez},\ and\ \citenamefont {Vulliet}}]{Burlet1997-da}%
  \BibitemOpen
  \bibfield  {author} {\bibinfo {author} {\bibfnamefont {P.}~\bibnamefont
  {Burlet}}, \bibinfo {author} {\bibfnamefont {E.}~\bibnamefont {Ressouche}},
  \bibinfo {author} {\bibfnamefont {B.}~\bibnamefont {Malaman}}, \bibinfo
  {author} {\bibfnamefont {R.}~\bibnamefont {Welter}}, \bibinfo {author}
  {\bibfnamefont {J.~P.}\ \bibnamefont {Sanchez}},\ and\ \bibinfo {author}
  {\bibfnamefont {P.}~\bibnamefont {Vulliet}},\ }\bibfield  {title} {\bibinfo
  {title} {{Noncollinear magnetic structure of MnTe$_2$}},\ }\href
  {https://doi.org/10.1103/PHYSREVB.56.14013} {\bibfield  {journal} {\bibinfo
  {journal} {Phys. Rev. B}\ }\textbf {\bibinfo {volume} {56}},\ \bibinfo
  {pages} {14013} (\bibinfo {year} {1997})}\BibitemShut {NoStop}%
\bibitem [{\citenamefont {Fu}\ \emph {et~al.}(2013)\citenamefont {Fu},
  \citenamefont {Zheng}, \citenamefont {Xiao}, \citenamefont {Bedanta},
  \citenamefont {Senyshyn}, \citenamefont {Simeoni}, \citenamefont {Su},
  \citenamefont {R{\"u}cker}, \citenamefont {K{\"o}gerler},\ and\ \citenamefont
  {Br{\"u}ckel}}]{Fu2013-gs}%
  \BibitemOpen
  \bibfield  {author} {\bibinfo {author} {\bibfnamefont {Z.}~\bibnamefont
  {Fu}}, \bibinfo {author} {\bibfnamefont {Y.}~\bibnamefont {Zheng}}, \bibinfo
  {author} {\bibfnamefont {Y.}~\bibnamefont {Xiao}}, \bibinfo {author}
  {\bibfnamefont {S.}~\bibnamefont {Bedanta}}, \bibinfo {author} {\bibfnamefont
  {A.}~\bibnamefont {Senyshyn}}, \bibinfo {author} {\bibfnamefont {G.~G.}\
  \bibnamefont {Simeoni}}, \bibinfo {author} {\bibfnamefont {Y.}~\bibnamefont
  {Su}}, \bibinfo {author} {\bibfnamefont {U.}~\bibnamefont {R{\"u}cker}},
  \bibinfo {author} {\bibfnamefont {P.}~\bibnamefont {K{\"o}gerler}},\ and\
  \bibinfo {author} {\bibfnamefont {T.}~\bibnamefont {Br{\"u}ckel}},\
  }\bibfield  {title} {\bibinfo {title} {{Coexistence of magnetic order and
  spin-glass-like phase in the pyrochlore antiferromagnet
  {Na$_{3}$Co(CO$_{3}$)$_{2}$Cl}}},\ }\href
  {https://doi.org/10.1103/PhysRevB.87.214406} {\bibfield  {journal} {\bibinfo
  {journal} {Phys. Rev. B}\ }\textbf {\bibinfo {volume} {87}},\ \bibinfo
  {pages} {214406} (\bibinfo {year} {2013})}\BibitemShut {NoStop}%
\bibitem [{\citenamefont {Moon}\ \emph {et~al.}(1968)\citenamefont {Moon},
  \citenamefont {Koehler}, \citenamefont {Child},\ and\ \citenamefont
  {Raubenheimer}}]{moon1968magnetic}%
  \BibitemOpen
  \bibfield  {author} {\bibinfo {author} {\bibfnamefont {R.}~\bibnamefont
  {Moon}}, \bibinfo {author} {\bibfnamefont {W.}~\bibnamefont {Koehler}},
  \bibinfo {author} {\bibfnamefont {H.}~\bibnamefont {Child}},\ and\ \bibinfo
  {author} {\bibfnamefont {L.}~\bibnamefont {Raubenheimer}},\ }\bibfield
  {title} {\bibinfo {title} {{Magnetic structures of Er$_2$O$_3$ and
  Yb$_2$O$_3$}},\ }\href {https://doi.org/10.1103/PhysRev.176.722} {\bibfield
  {journal} {\bibinfo  {journal} {Phys. Rev.}\ }\textbf {\bibinfo {volume}
  {176}},\ \bibinfo {pages} {722} (\bibinfo {year} {1968})}\BibitemShut
  {NoStop}%
\bibitem [{\citenamefont {Frazer}\ \emph {et~al.}(1965)\citenamefont {Frazer},
  \citenamefont {Shirane}, \citenamefont {Cox},\ and\ \citenamefont
  {Olsen}}]{Frazer1965-iw}%
  \BibitemOpen
  \bibfield  {author} {\bibinfo {author} {\bibfnamefont {B.~C.}\ \bibnamefont
  {Frazer}}, \bibinfo {author} {\bibfnamefont {G.}~\bibnamefont {Shirane}},
  \bibinfo {author} {\bibfnamefont {D.~E.}\ \bibnamefont {Cox}},\ and\ \bibinfo
  {author} {\bibfnamefont {C.~E.}\ \bibnamefont {Olsen}},\ }\bibfield  {title}
  {\bibinfo {title} {{{Neutron-Diffraction} Study of Antiferromagnetism in
  {UO$_2$}}},\ }\href {https://doi.org/10.1103/physrev.140.a1448} {\bibfield
  {journal} {\bibinfo  {journal} {Phys. Rev.}\ }\textbf {\bibinfo {volume}
  {140}},\ \bibinfo {pages} {A1448} (\bibinfo {year} {1965})}\BibitemShut
  {NoStop}%
\bibitem [{\citenamefont {Umebayashi}\ and\ \citenamefont
  {Ishikawa}(1966)}]{Umebayashi1966-qj}%
  \BibitemOpen
  \bibfield  {author} {\bibinfo {author} {\bibfnamefont {H.}~\bibnamefont
  {Umebayashi}}\ and\ \bibinfo {author} {\bibfnamefont {Y.}~\bibnamefont
  {Ishikawa}},\ }\bibfield  {title} {\bibinfo {title} {{Antiferromagnetism of
  $\gamma$ {Fe-Mn} Alloys}},\ }\href {https://doi.org/10.1143/JPSJ.21.1281}
  {\bibfield  {journal} {\bibinfo  {journal} {J. Phys. Soc. Jpn.}\ }\textbf
  {\bibinfo {volume} {21}},\ \bibinfo {pages} {1281} (\bibinfo {year}
  {1966})}\BibitemShut {NoStop}%
\bibitem [{\citenamefont {Wintenberger}\ \emph {et~al.}(1971)\citenamefont
  {Wintenberger}, \citenamefont {Chamard-Bois}, \citenamefont {Belakhovsky},\
  and\ \citenamefont {Pierre}}]{Wintenberger1971-mx}%
  \BibitemOpen
  \bibfield  {author} {\bibinfo {author} {\bibfnamefont {P.~M.}\ \bibnamefont
  {Wintenberger}}, \bibinfo {author} {\bibfnamefont {R.}~\bibnamefont
  {Chamard-Bois}}, \bibinfo {author} {\bibfnamefont {M.}~\bibnamefont
  {Belakhovsky}},\ and\ \bibinfo {author} {\bibfnamefont {E.~J.}\ \bibnamefont
  {Pierre}},\ }\bibfield  {title} {\bibinfo {title} {{Structure magn{\'e}tique
  ordonn{\'e}e du compos{\'e} {DyCu}}},\ }\href
  {https://doi.org/10.1002/pssb.2220480229} {\bibfield  {journal} {\bibinfo
  {journal} {Phys. Status Solidi B}\ }\textbf {\bibinfo {volume} {48}},\
  \bibinfo {pages} {705} (\bibinfo {year} {1971})}\BibitemShut {NoStop}%
\bibitem [{\citenamefont {Burlet}\ \emph {et~al.}(1992)\citenamefont {Burlet},
  \citenamefont {Bourdarot}, \citenamefont {Rossat-Mignod}, \citenamefont
  {Sanchez}, \citenamefont {Spirlet}, \citenamefont {Rebizant},\ and\
  \citenamefont {Vogt}}]{Burlet1992-qe}%
  \BibitemOpen
  \bibfield  {author} {\bibinfo {author} {\bibfnamefont {P.}~\bibnamefont
  {Burlet}}, \bibinfo {author} {\bibfnamefont {F.}~\bibnamefont {Bourdarot}},
  \bibinfo {author} {\bibfnamefont {J.}~\bibnamefont {Rossat-Mignod}}, \bibinfo
  {author} {\bibfnamefont {J.}~\bibnamefont {Sanchez}}, \bibinfo {author}
  {\bibfnamefont {J.}~\bibnamefont {Spirlet}}, \bibinfo {author} {\bibfnamefont
  {J.}~\bibnamefont {Rebizant}},\ and\ \bibinfo {author} {\bibfnamefont
  {O.}~\bibnamefont {Vogt}},\ }\bibfield  {title} {\bibinfo {title} {{Neutron
  diffraction study of the magnetic ordering in {NpBi}}},\ }\href
  {https://doi.org/10.1016/0921-4526(92)90683-J} {\bibfield  {journal}
  {\bibinfo  {journal} {Physica B}\ }\textbf {\bibinfo {volume} {180}},\
  \bibinfo {pages} {131} (\bibinfo {year} {1992})}\BibitemShut {NoStop}%
\bibitem [{\citenamefont {Lander}\ and\ \citenamefont
  {Burlet}(1995)}]{Lander1995-yi}%
  \BibitemOpen
  \bibfield  {author} {\bibinfo {author} {\bibfnamefont {G.}~\bibnamefont
  {Lander}}\ and\ \bibinfo {author} {\bibfnamefont {P.}~\bibnamefont
  {Burlet}},\ }\bibfield  {title} {\bibinfo {title} {{On the magnetic structure
  of actinide monopnictides}},\ }\href
  {https://doi.org/10.1016/0921-4526(95)00021-Z} {\bibfield  {journal}
  {\bibinfo  {journal} {Physica B}\ }\textbf {\bibinfo {volume} {215}},\
  \bibinfo {pages} {7} (\bibinfo {year} {1995})}\BibitemShut {NoStop}%
\bibitem [{\citenamefont {Calder}\ \emph {et~al.}(2016)\citenamefont {Calder},
  \citenamefont {Vale}, \citenamefont {Bogdanov}, \citenamefont {Liu},
  \citenamefont {Donnerer}, \citenamefont {Upton}, \citenamefont {Casa},
  \citenamefont {Said}, \citenamefont {Lumsden}, \citenamefont {Zhao},
  \citenamefont {Yan}, \citenamefont {Mandrus}, \citenamefont {Nishimoto},
  \citenamefont {van~den Brink}, \citenamefont {Hill}, \citenamefont
  {McMorrow},\ and\ \citenamefont {Christianson}}]{Calder2016-sl}%
  \BibitemOpen
  \bibfield  {author} {\bibinfo {author} {\bibfnamefont {S.}~\bibnamefont
  {Calder}}, \bibinfo {author} {\bibfnamefont {J.~G.}\ \bibnamefont {Vale}},
  \bibinfo {author} {\bibfnamefont {N.~A.}\ \bibnamefont {Bogdanov}}, \bibinfo
  {author} {\bibfnamefont {X.}~\bibnamefont {Liu}}, \bibinfo {author}
  {\bibfnamefont {C.}~\bibnamefont {Donnerer}}, \bibinfo {author}
  {\bibfnamefont {M.~H.}\ \bibnamefont {Upton}}, \bibinfo {author}
  {\bibfnamefont {D.}~\bibnamefont {Casa}}, \bibinfo {author} {\bibfnamefont
  {A.~H.}\ \bibnamefont {Said}}, \bibinfo {author} {\bibfnamefont {M.~D.}\
  \bibnamefont {Lumsden}}, \bibinfo {author} {\bibfnamefont {Z.}~\bibnamefont
  {Zhao}}, \bibinfo {author} {\bibfnamefont {J.-Q.}\ \bibnamefont {Yan}},
  \bibinfo {author} {\bibfnamefont {D.}~\bibnamefont {Mandrus}}, \bibinfo
  {author} {\bibfnamefont {S.}~\bibnamefont {Nishimoto}}, \bibinfo {author}
  {\bibfnamefont {J.}~\bibnamefont {van~den Brink}}, \bibinfo {author}
  {\bibfnamefont {J.~P.}\ \bibnamefont {Hill}}, \bibinfo {author}
  {\bibfnamefont {D.~F.}\ \bibnamefont {McMorrow}},\ and\ \bibinfo {author}
  {\bibfnamefont {A.~D.}\ \bibnamefont {Christianson}},\ }\bibfield  {title}
  {\bibinfo {title} {{Spin-orbit-driven magnetic structure and excitation in
  the 5d pyrochlore {Cd$_2$Os$_2$O$_7$}}},\ }\href
  {https://doi.org/10.1038/ncomms11651} {\bibfield  {journal} {\bibinfo
  {journal} {Nat. Commun.}\ }\textbf {\bibinfo {volume} {7}},\ \bibinfo {pages}
  {1} (\bibinfo {year} {2016})}\BibitemShut {NoStop}%
\bibitem [{\citenamefont {Xu}\ \emph {et~al.}(2015)\citenamefont {Xu},
  \citenamefont {Anand}, \citenamefont {Bera}, \citenamefont {Frontzek},
  \citenamefont {Abernathy}, \citenamefont {Casati}, \citenamefont
  {Siemensmeyer},\ and\ \citenamefont {Lake}}]{Xu2015-dw}%
  \BibitemOpen
  \bibfield  {author} {\bibinfo {author} {\bibfnamefont {J.}~\bibnamefont
  {Xu}}, \bibinfo {author} {\bibfnamefont {V.~K.}\ \bibnamefont {Anand}},
  \bibinfo {author} {\bibfnamefont {A.~K.}\ \bibnamefont {Bera}}, \bibinfo
  {author} {\bibfnamefont {M.}~\bibnamefont {Frontzek}}, \bibinfo {author}
  {\bibfnamefont {D.~L.}\ \bibnamefont {Abernathy}}, \bibinfo {author}
  {\bibfnamefont {N.}~\bibnamefont {Casati}}, \bibinfo {author} {\bibfnamefont
  {K.}~\bibnamefont {Siemensmeyer}},\ and\ \bibinfo {author} {\bibfnamefont
  {B.}~\bibnamefont {Lake}},\ }\bibfield  {title} {\bibinfo {title} {{Magnetic
  structure and crystal-field states of the pyrochlore antiferromagnet
  {Nd$_{2}$Zr$_{2}$O$_{7}$}}},\ }\href
  {https://doi.org/10.1103/PhysRevB.92.224430} {\bibfield  {journal} {\bibinfo
  {journal} {Phys. Rev. B}\ }\textbf {\bibinfo {volume} {92}},\ \bibinfo
  {pages} {224430} (\bibinfo {year} {2015})}\BibitemShut {NoStop}%
\bibitem [{\citenamefont {Bertin}\ \emph {et~al.}(2015)\citenamefont {Bertin},
  \citenamefont {Dalmas~de R{\'e}otier}, \citenamefont {F{\aa}k}, \citenamefont
  {Marin}, \citenamefont {Yaouanc}, \citenamefont {Forget}, \citenamefont
  {Sheptyakov}, \citenamefont {Frick}, \citenamefont {Ritter}, \citenamefont
  {Amato}, \citenamefont {Baines},\ and\ \citenamefont {King}}]{Bertin2015-iz}%
  \BibitemOpen
  \bibfield  {author} {\bibinfo {author} {\bibfnamefont {A.}~\bibnamefont
  {Bertin}}, \bibinfo {author} {\bibfnamefont {P.}~\bibnamefont {Dalmas~de
  R{\'e}otier}}, \bibinfo {author} {\bibfnamefont {B.}~\bibnamefont {F{\aa}k}},
  \bibinfo {author} {\bibfnamefont {C.}~\bibnamefont {Marin}}, \bibinfo
  {author} {\bibfnamefont {A.}~\bibnamefont {Yaouanc}}, \bibinfo {author}
  {\bibfnamefont {A.}~\bibnamefont {Forget}}, \bibinfo {author} {\bibfnamefont
  {D.}~\bibnamefont {Sheptyakov}}, \bibinfo {author} {\bibfnamefont
  {B.}~\bibnamefont {Frick}}, \bibinfo {author} {\bibfnamefont
  {C.}~\bibnamefont {Ritter}}, \bibinfo {author} {\bibfnamefont
  {A.}~\bibnamefont {Amato}}, \bibinfo {author} {\bibfnamefont
  {C.}~\bibnamefont {Baines}},\ and\ \bibinfo {author} {\bibfnamefont
  {P.~J.~C.}\ \bibnamefont {King}},\ }\bibfield  {title} {\bibinfo {title}
  {{{Nd$_{2}$Sn$_{2}$O$_{7}$}: An all-in--all-out pyrochlore magnet with no
  divergence-free field and anomalously slow paramagnetic spin dynamics}},\
  }\href {https://doi.org/10.1103/PhysRevB.92.144423} {\bibfield  {journal}
  {\bibinfo  {journal} {Phys. Rev. B}\ }\textbf {\bibinfo {volume} {92}},\
  \bibinfo {pages} {144423} (\bibinfo {year} {2015})}\BibitemShut {NoStop}%
\bibitem [{\citenamefont {Anand}\ \emph {et~al.}(2015)\citenamefont {Anand},
  \citenamefont {Bera}, \citenamefont {Xu}, \citenamefont
  {Herrmannsd{\"o}rfer}, \citenamefont {Ritter},\ and\ \citenamefont
  {Lake}}]{Anand2015-bv}%
  \BibitemOpen
  \bibfield  {author} {\bibinfo {author} {\bibfnamefont {V.~K.}\ \bibnamefont
  {Anand}}, \bibinfo {author} {\bibfnamefont {A.~K.}\ \bibnamefont {Bera}},
  \bibinfo {author} {\bibfnamefont {J.}~\bibnamefont {Xu}}, \bibinfo {author}
  {\bibfnamefont {T.}~\bibnamefont {Herrmannsd{\"o}rfer}}, \bibinfo {author}
  {\bibfnamefont {C.}~\bibnamefont {Ritter}},\ and\ \bibinfo {author}
  {\bibfnamefont {B.}~\bibnamefont {Lake}},\ }\bibfield  {title} {\bibinfo
  {title} {{Observation of long-range magnetic ordering in pyrohafnate
  {Nd$_{2}$Hf$_{2}$O$_{7}$}: A neutron diffraction study}},\ }\href
  {https://doi.org/10.1103/PhysRevB.92.184418} {\bibfield  {journal} {\bibinfo
  {journal} {Phys. Rev. B}\ }\textbf {\bibinfo {volume} {92}},\ \bibinfo
  {pages} {184418} (\bibinfo {year} {2015})}\BibitemShut {NoStop}%
\bibitem [{\citenamefont {Mauws}\ \emph {et~al.}(2018)\citenamefont {Mauws},
  \citenamefont {Hallas}, \citenamefont {Sala}, \citenamefont {Aczel},
  \citenamefont {Sarte}, \citenamefont {Gaudet}, \citenamefont {Ziat},
  \citenamefont {Quilliam}, \citenamefont {Lussier}, \citenamefont {Bieringer},
  \citenamefont {Zhou}, \citenamefont {Wildes}, \citenamefont {Stone},
  \citenamefont {Abernathy}, \citenamefont {Luke}, \citenamefont {Gaulin},\
  and\ \citenamefont {Wiebe}}]{Mauws2018-hw}%
  \BibitemOpen
  \bibfield  {author} {\bibinfo {author} {\bibfnamefont {C.}~\bibnamefont
  {Mauws}}, \bibinfo {author} {\bibfnamefont {A.~M.}\ \bibnamefont {Hallas}},
  \bibinfo {author} {\bibfnamefont {G.}~\bibnamefont {Sala}}, \bibinfo {author}
  {\bibfnamefont {A.~A.}\ \bibnamefont {Aczel}}, \bibinfo {author}
  {\bibfnamefont {P.~M.}\ \bibnamefont {Sarte}}, \bibinfo {author}
  {\bibfnamefont {J.}~\bibnamefont {Gaudet}}, \bibinfo {author} {\bibfnamefont
  {D.}~\bibnamefont {Ziat}}, \bibinfo {author} {\bibfnamefont {J.~A.}\
  \bibnamefont {Quilliam}}, \bibinfo {author} {\bibfnamefont {J.~A.}\
  \bibnamefont {Lussier}}, \bibinfo {author} {\bibfnamefont {M.}~\bibnamefont
  {Bieringer}}, \bibinfo {author} {\bibfnamefont {H.~D.}\ \bibnamefont {Zhou}},
  \bibinfo {author} {\bibfnamefont {A.}~\bibnamefont {Wildes}}, \bibinfo
  {author} {\bibfnamefont {M.~B.}\ \bibnamefont {Stone}}, \bibinfo {author}
  {\bibfnamefont {D.}~\bibnamefont {Abernathy}}, \bibinfo {author}
  {\bibfnamefont {G.~M.}\ \bibnamefont {Luke}}, \bibinfo {author}
  {\bibfnamefont {B.~D.}\ \bibnamefont {Gaulin}},\ and\ \bibinfo {author}
  {\bibfnamefont {C.~R.}\ \bibnamefont {Wiebe}},\ }\bibfield  {title} {\bibinfo
  {title} {{Dipolar-octupolar Ising antiferromagnetism in
  {Sm$_{2}$Ti$_{2}$O$_{7}$}: A moment fragmentation candidate}},\ }\href
  {https://doi.org/10.1103/PhysRevB.98.100401} {\bibfield  {journal} {\bibinfo
  {journal} {Phys. Rev. B}\ }\textbf {\bibinfo {volume} {98}},\ \bibinfo
  {pages} {100401(R)} (\bibinfo {year} {2018})}\BibitemShut {NoStop}%
\bibitem [{\citenamefont {Hammann}\ and\ \citenamefont
  {Ocio}(1977)}]{Hammann1977-sw}%
  \BibitemOpen
  \bibfield  {author} {\bibinfo {author} {\bibfnamefont {J.}~\bibnamefont
  {Hammann}}\ and\ \bibinfo {author} {\bibfnamefont {M.}~\bibnamefont {Ocio}},\
  }\bibfield  {title} {\bibinfo {title} {{{\'E}tude exp{\'e}rimentale de
  l'ordre antiferromagn{\'e}tique induit par les interactions hyperfines dans
  les grenats gallates de terbium et d'holmium}},\ }\href
  {https://doi.org/10.1051/jphys:01977003805046300} {\bibfield  {journal}
  {\bibinfo  {journal} {J. Phys. France}\ }\textbf {\bibinfo {volume} {38}},\
  \bibinfo {pages} {463} (\bibinfo {year} {1977})}\BibitemShut {NoStop}%
\bibitem [{\citenamefont {Wawrzy{\'n}czak}\ \emph {et~al.}(2019)\citenamefont
  {Wawrzy{\'n}czak}, \citenamefont {Tomasello}, \citenamefont {Manuel},
  \citenamefont {Khalyavin}, \citenamefont {Le}, \citenamefont {Guidi},
  \citenamefont {Cervellino}, \citenamefont {Ziman}, \citenamefont {Boehm},
  \citenamefont {Nilsen},\ and\ \citenamefont {Fennell}}]{Wawrzynczak2019-zi}%
  \BibitemOpen
  \bibfield  {author} {\bibinfo {author} {\bibfnamefont {R.}~\bibnamefont
  {Wawrzy{\'n}czak}}, \bibinfo {author} {\bibfnamefont {B.}~\bibnamefont
  {Tomasello}}, \bibinfo {author} {\bibfnamefont {P.}~\bibnamefont {Manuel}},
  \bibinfo {author} {\bibfnamefont {D.}~\bibnamefont {Khalyavin}}, \bibinfo
  {author} {\bibfnamefont {M.~D.}\ \bibnamefont {Le}}, \bibinfo {author}
  {\bibfnamefont {T.}~\bibnamefont {Guidi}}, \bibinfo {author} {\bibfnamefont
  {A.}~\bibnamefont {Cervellino}}, \bibinfo {author} {\bibfnamefont
  {T.}~\bibnamefont {Ziman}}, \bibinfo {author} {\bibfnamefont
  {M.}~\bibnamefont {Boehm}}, \bibinfo {author} {\bibfnamefont {G.~J.}\
  \bibnamefont {Nilsen}},\ and\ \bibinfo {author} {\bibfnamefont
  {T.}~\bibnamefont {Fennell}},\ }\bibfield  {title} {\bibinfo {title}
  {{Magnetic order and single-ion anisotropy in {Tb$_{3}$Ga$_{5}$O$_{12}$}}},\
  }\href {https://doi.org/10.1103/PhysRevB.100.094442} {\bibfield  {journal}
  {\bibinfo  {journal} {Phys. Rev. B}\ }\textbf {\bibinfo {volume} {100}},\
  \bibinfo {pages} {094442} (\bibinfo {year} {2019})}\BibitemShut {NoStop}%
\bibitem [{\citenamefont {Kibalin}\ \emph {et~al.}(2020)\citenamefont
  {Kibalin}, \citenamefont {Damay}, \citenamefont {Fabr{\`e}ges}, \citenamefont
  {Gukassov},\ and\ \citenamefont {Petit}}]{Kibalin2020-fl}%
  \BibitemOpen
  \bibfield  {author} {\bibinfo {author} {\bibfnamefont {I.~A.}\ \bibnamefont
  {Kibalin}}, \bibinfo {author} {\bibfnamefont {F.}~\bibnamefont {Damay}},
  \bibinfo {author} {\bibfnamefont {X.}~\bibnamefont {Fabr{\`e}ges}}, \bibinfo
  {author} {\bibfnamefont {A.}~\bibnamefont {Gukassov}},\ and\ \bibinfo
  {author} {\bibfnamefont {S.}~\bibnamefont {Petit}},\ }\bibfield  {title}
  {\bibinfo {title} {{Competing interactions in dysprosium garnets and
  generalized magnetic phase diagram of {S=$\frac{1}{2}$} spins on a
  hyperkagome network}},\ }\href
  {https://doi.org/10.1103/PhysRevResearch.2.033509} {\bibfield  {journal}
  {\bibinfo  {journal} {Phys. Rev. Res.}\ }\textbf {\bibinfo {volume} {2}},\
  \bibinfo {pages} {033509} (\bibinfo {year} {2020})}\BibitemShut {NoStop}%
\bibitem [{\citenamefont {Cai}\ \emph {et~al.}(2019)\citenamefont {Cai},
  \citenamefont {Wilson}, \citenamefont {Beare}, \citenamefont {Lygouras},
  \citenamefont {Thomas}, \citenamefont {Yahne}, \citenamefont {Ross},
  \citenamefont {Taddei}, \citenamefont {Sala}, \citenamefont {Dabkowska},
  \citenamefont {Aczel},\ and\ \citenamefont {Luke}}]{Cai2019-ip}%
  \BibitemOpen
  \bibfield  {author} {\bibinfo {author} {\bibfnamefont {Y.}~\bibnamefont
  {Cai}}, \bibinfo {author} {\bibfnamefont {M.~N.}\ \bibnamefont {Wilson}},
  \bibinfo {author} {\bibfnamefont {J.}~\bibnamefont {Beare}}, \bibinfo
  {author} {\bibfnamefont {C.}~\bibnamefont {Lygouras}}, \bibinfo {author}
  {\bibfnamefont {G.}~\bibnamefont {Thomas}}, \bibinfo {author} {\bibfnamefont
  {D.~R.}\ \bibnamefont {Yahne}}, \bibinfo {author} {\bibfnamefont
  {K.}~\bibnamefont {Ross}}, \bibinfo {author} {\bibfnamefont {K.~M.}\
  \bibnamefont {Taddei}}, \bibinfo {author} {\bibfnamefont {G.}~\bibnamefont
  {Sala}}, \bibinfo {author} {\bibfnamefont {H.~A.}\ \bibnamefont {Dabkowska}},
  \bibinfo {author} {\bibfnamefont {A.~A.}\ \bibnamefont {Aczel}},\ and\
  \bibinfo {author} {\bibfnamefont {G.~M.}\ \bibnamefont {Luke}},\ }\bibfield
  {title} {\bibinfo {title} {{Crystal fields and magnetic structure of the
  Ising antiferromagnet {Er$_{3}$Ga$_{5}$O$_{12}$}}},\ }\href
  {https://doi.org/10.1103/PhysRevB.100.184415} {\bibfield  {journal} {\bibinfo
   {journal} {Phys. Rev. B}\ }\textbf {\bibinfo {volume} {100}},\ \bibinfo
  {pages} {184415} (\bibinfo {year} {2019})}\BibitemShut {NoStop}%
\bibitem [{\citenamefont {Hammann}(1969)}]{noauthor_undated-ww}%
  \BibitemOpen
  \bibfield  {author} {\bibinfo {author} {\bibfnamefont {J.}~\bibnamefont
  {Hammann}},\ }\bibfield  {title} {\bibinfo {title} {{Etude par Diffraction de
  Neutrons {\`a} 0,31$^\circ$K de la Structure Antiferromagn{\'e}tique des
  Grenats d'Aluminium--Terbium et d'Aluminium--Holmium}},\ }\href
  {https://doi.org/10.1107/S0567740869004833} {\bibfield  {journal} {\bibinfo
  {journal} {Acta Crystallogr. B}\ }\textbf {\bibinfo {volume} {25}},\ \bibinfo
  {pages} {1853} (\bibinfo {year} {1969})}\BibitemShut {NoStop}%
\bibitem [{\citenamefont {Hastings}\ \emph {et~al.}(1965)\citenamefont
  {Hastings}, \citenamefont {Corliss},\ and\ \citenamefont
  {Windsor}}]{Hastings1965-fy}%
  \BibitemOpen
  \bibfield  {author} {\bibinfo {author} {\bibfnamefont {J.~M.}\ \bibnamefont
  {Hastings}}, \bibinfo {author} {\bibfnamefont {L.~M.}\ \bibnamefont
  {Corliss}},\ and\ \bibinfo {author} {\bibfnamefont {C.~G.}\ \bibnamefont
  {Windsor}},\ }\bibfield  {title} {\bibinfo {title} {{Antiferromagnetic
  Structure of Dysprosium Aluminum Garnet}},\ }\href
  {https://doi.org/10.1103/PhysRev.138.A176} {\bibfield  {journal} {\bibinfo
  {journal} {Phys. Rev.}\ }\textbf {\bibinfo {volume} {138}},\ \bibinfo {pages}
  {A176} (\bibinfo {year} {1965})}\BibitemShut {NoStop}%
\bibitem [{\citenamefont {Scheie}\ \emph {et~al.}(2021)\citenamefont {Scheie},
  \citenamefont {Sanders}, \citenamefont {Gui}, \citenamefont {Qiu},
  \citenamefont {Prisk}, \citenamefont {Cava},\ and\ \citenamefont
  {Broholm}}]{Scheie2021-sh}%
  \BibitemOpen
  \bibfield  {author} {\bibinfo {author} {\bibfnamefont {A.}~\bibnamefont
  {Scheie}}, \bibinfo {author} {\bibfnamefont {M.}~\bibnamefont {Sanders}},
  \bibinfo {author} {\bibfnamefont {X.}~\bibnamefont {Gui}}, \bibinfo {author}
  {\bibfnamefont {Y.}~\bibnamefont {Qiu}}, \bibinfo {author} {\bibfnamefont
  {T.~R.}\ \bibnamefont {Prisk}}, \bibinfo {author} {\bibfnamefont {R.~J.}\
  \bibnamefont {Cava}},\ and\ \bibinfo {author} {\bibfnamefont
  {C.}~\bibnamefont {Broholm}},\ }\bibfield  {title} {\bibinfo {title} {{Beyond
  magnons in {Nd$_{2}$ScNbO$_{7}$}: An Ising pyrochlore antiferromagnet with
  all-in--all-out order and random fields}},\ }\href
  {https://doi.org/10.1103/PhysRevB.104.134418} {\bibfield  {journal} {\bibinfo
   {journal} {Phys. Rev. B}\ }\textbf {\bibinfo {volume} {104}},\ \bibinfo
  {pages} {134418} (\bibinfo {year} {2021})}\BibitemShut {NoStop}%
\bibitem [{\citenamefont {Guo}\ \emph {et~al.}(2016)\citenamefont {Guo},
  \citenamefont {Ritter},\ and\ \citenamefont {Komarek}}]{Guo2016-is}%
  \BibitemOpen
  \bibfield  {author} {\bibinfo {author} {\bibfnamefont {H.}~\bibnamefont
  {Guo}}, \bibinfo {author} {\bibfnamefont {C.}~\bibnamefont {Ritter}},\ and\
  \bibinfo {author} {\bibfnamefont {A.~C.}\ \bibnamefont {Komarek}},\
  }\bibfield  {title} {\bibinfo {title} {{Direct determination of the spin
  structure of {Nd$_{2}$Ir$_{2}$O$_{7}$} by means of neutron diffraction}},\
  }\href {https://doi.org/10.1103/PhysRevB.94.161102} {\bibfield  {journal}
  {\bibinfo  {journal} {Phys. Rev. B}\ }\textbf {\bibinfo {volume} {94}},\
  \bibinfo {pages} {161102(R)} (\bibinfo {year} {2016})}\BibitemShut {NoStop}%
\bibitem [{\citenamefont {Das}\ \emph {et~al.}(2022)\citenamefont {Das},
  \citenamefont {Bhowal}, \citenamefont {Sannigrahi}, \citenamefont
  {Bandyopadhyay}, \citenamefont {Banerjee}, \citenamefont {Cibin},
  \citenamefont {Khalyavin}, \citenamefont {Banerjee}, \citenamefont {Adroja},
  \citenamefont {Dasgupta},\ and\ \citenamefont {Majumdar}}]{Das2022-ms}%
  \BibitemOpen
  \bibfield  {author} {\bibinfo {author} {\bibfnamefont {M.}~\bibnamefont
  {Das}}, \bibinfo {author} {\bibfnamefont {S.}~\bibnamefont {Bhowal}},
  \bibinfo {author} {\bibfnamefont {J.}~\bibnamefont {Sannigrahi}}, \bibinfo
  {author} {\bibfnamefont {A.}~\bibnamefont {Bandyopadhyay}}, \bibinfo {author}
  {\bibfnamefont {A.}~\bibnamefont {Banerjee}}, \bibinfo {author}
  {\bibfnamefont {G.}~\bibnamefont {Cibin}}, \bibinfo {author} {\bibfnamefont
  {D.}~\bibnamefont {Khalyavin}}, \bibinfo {author} {\bibfnamefont
  {N.}~\bibnamefont {Banerjee}}, \bibinfo {author} {\bibfnamefont
  {D.}~\bibnamefont {Adroja}}, \bibinfo {author} {\bibfnamefont
  {I.}~\bibnamefont {Dasgupta}},\ and\ \bibinfo {author} {\bibfnamefont
  {S.}~\bibnamefont {Majumdar}},\ }\bibfield  {title} {\bibinfo {title}
  {{Interplay between structural, magnetic, and electronic states in the
  pyrochlore iridate {Eu$_{2}$Ir$_{2}$O$_{7}$}}},\ }\href
  {https://doi.org/10.1103/PhysRevB.105.134421} {\bibfield  {journal} {\bibinfo
   {journal} {Phys. Rev. B}\ }\textbf {\bibinfo {volume} {105}},\ \bibinfo
  {pages} {134421} (\bibinfo {year} {2022})}\BibitemShut {NoStop}%
\bibitem [{\citenamefont {Jacobsen}\ \emph {et~al.}(2020)\citenamefont
  {Jacobsen}, \citenamefont {Dashwood}, \citenamefont {Lhotel}, \citenamefont
  {Khalyavin}, \citenamefont {Manuel}, \citenamefont {Stewart}, \citenamefont
  {Prabhakaran}, \citenamefont {McMorrow},\ and\ \citenamefont
  {Boothroyd}}]{Jacobsen2020-ey}%
  \BibitemOpen
  \bibfield  {author} {\bibinfo {author} {\bibfnamefont {H.}~\bibnamefont
  {Jacobsen}}, \bibinfo {author} {\bibfnamefont {C.~D.}\ \bibnamefont
  {Dashwood}}, \bibinfo {author} {\bibfnamefont {E.}~\bibnamefont {Lhotel}},
  \bibinfo {author} {\bibfnamefont {D.}~\bibnamefont {Khalyavin}}, \bibinfo
  {author} {\bibfnamefont {P.}~\bibnamefont {Manuel}}, \bibinfo {author}
  {\bibfnamefont {R.}~\bibnamefont {Stewart}}, \bibinfo {author} {\bibfnamefont
  {D.}~\bibnamefont {Prabhakaran}}, \bibinfo {author} {\bibfnamefont {D.~F.}\
  \bibnamefont {McMorrow}},\ and\ \bibinfo {author} {\bibfnamefont {A.~T.}\
  \bibnamefont {Boothroyd}},\ }\bibfield  {title} {\bibinfo {title} {{Strong
  quantum fluctuations from competition between magnetic phases in a pyrochlore
  iridate}},\ }\href {https://doi.org/10.1103/PhysRevB.101.104404} {\bibfield
  {journal} {\bibinfo  {journal} {Phys. Rev. B}\ }\textbf {\bibinfo {volume}
  {101}},\ \bibinfo {pages} {104404} (\bibinfo {year} {2020})}\BibitemShut
  {NoStop}%
\bibitem [{\citenamefont {Morin}\ \emph {et~al.}(1987)\citenamefont {Morin},
  \citenamefont {Giraud}, \citenamefont {Burlet},\ and\ \citenamefont
  {Czopnik}}]{Morin_1987-wy}%
  \BibitemOpen
  \bibfield  {author} {\bibinfo {author} {\bibfnamefont {P.}~\bibnamefont
  {Morin}}, \bibinfo {author} {\bibfnamefont {M.}~\bibnamefont {Giraud}},
  \bibinfo {author} {\bibfnamefont {P.}~\bibnamefont {Burlet}},\ and\ \bibinfo
  {author} {\bibfnamefont {A.}~\bibnamefont {Czopnik}},\ }\bibfield  {title}
  {\bibinfo {title} {Antiferroquadrupolar and antiferromagnetic structures in
  tmga$_3$},\ }\href {https://doi.org/10.1016/0304-8853(87)90103-X} {\bibfield
  {journal} {\bibinfo  {journal} {J. Magn. Magn. Mater.}\ }\textbf {\bibinfo
  {volume} {68}},\ \bibinfo {pages} {107} (\bibinfo {year} {1987})}\BibitemShut
  {NoStop}%
\bibitem [{\citenamefont {Brown}\ and\ \citenamefont
  {Chatterji}(2006)}]{Brown2006-ie}%
  \BibitemOpen
  \bibfield  {author} {\bibinfo {author} {\bibfnamefont {P.~J.}\ \bibnamefont
  {Brown}}\ and\ \bibinfo {author} {\bibfnamefont {T.}~\bibnamefont
  {Chatterji}},\ }\bibfield  {title} {\bibinfo {title} {{Neutron diffraction
  and polarimetric study of the magnetic and crystal structures {of HoMnO$_3$}
  and {YMnO$_3$}}},\ }\href {https://doi.org/10.1088/0953-8984/18/44/008}
  {\bibfield  {journal} {\bibinfo  {journal} {J. Phys. Condens. Matter}\
  }\textbf {\bibinfo {volume} {18}},\ \bibinfo {pages} {10085} (\bibinfo {year}
  {2006})}\BibitemShut {NoStop}%
\bibitem [{\citenamefont {Fabr{\`e}ges}\ \emph {et~al.}(2008)\citenamefont
  {Fabr{\`e}ges}, \citenamefont {Mirebeau}, \citenamefont {Bonville},
  \citenamefont {Petit}, \citenamefont {Lebras-Jasmin}, \citenamefont {Forget},
  \citenamefont {Andr{\'e}},\ and\ \citenamefont
  {Pailh{\`e}s}}]{Fabreges2008-pw}%
  \BibitemOpen
  \bibfield  {author} {\bibinfo {author} {\bibfnamefont {X.}~\bibnamefont
  {Fabr{\`e}ges}}, \bibinfo {author} {\bibfnamefont {I.}~\bibnamefont
  {Mirebeau}}, \bibinfo {author} {\bibfnamefont {P.}~\bibnamefont {Bonville}},
  \bibinfo {author} {\bibfnamefont {S.}~\bibnamefont {Petit}}, \bibinfo
  {author} {\bibfnamefont {G.}~\bibnamefont {Lebras-Jasmin}}, \bibinfo {author}
  {\bibfnamefont {A.}~\bibnamefont {Forget}}, \bibinfo {author} {\bibfnamefont
  {G.}~\bibnamefont {Andr{\'e}}},\ and\ \bibinfo {author} {\bibfnamefont
  {S.}~\bibnamefont {Pailh{\`e}s}},\ }\bibfield  {title} {\bibinfo {title}
  {{Magnetic order in {YbMnO$_{3}$} studied by neutron diffraction and
  M\"{o}ssbauer spectroscopy}},\ }\href
  {https://doi.org/10.1103/PhysRevB.78.214422} {\bibfield  {journal} {\bibinfo
  {journal} {Phys. Rev. B}\ }\textbf {\bibinfo {volume} {78}},\ \bibinfo
  {pages} {214422} (\bibinfo {year} {2008})}\BibitemShut {NoStop}%
\bibitem [{\citenamefont {Chattopadhyay}\ \emph {et~al.}(2018)\citenamefont
  {Chattopadhyay}, \citenamefont {Simonet}, \citenamefont {Skumryev},
  \citenamefont {Mukhin}, \citenamefont {Ivanov}, \citenamefont {Aroyo},
  \citenamefont {Dimitrov}, \citenamefont {Gospodinov},\ and\ \citenamefont
  {Ressouche}}]{Chattopadhyay2018-at}%
  \BibitemOpen
  \bibfield  {author} {\bibinfo {author} {\bibfnamefont {S.}~\bibnamefont
  {Chattopadhyay}}, \bibinfo {author} {\bibfnamefont {V.}~\bibnamefont
  {Simonet}}, \bibinfo {author} {\bibfnamefont {V.}~\bibnamefont {Skumryev}},
  \bibinfo {author} {\bibfnamefont {A.~A.}\ \bibnamefont {Mukhin}}, \bibinfo
  {author} {\bibfnamefont {V.~Y.}\ \bibnamefont {Ivanov}}, \bibinfo {author}
  {\bibfnamefont {M.~I.}\ \bibnamefont {Aroyo}}, \bibinfo {author}
  {\bibfnamefont {D.~Z.}\ \bibnamefont {Dimitrov}}, \bibinfo {author}
  {\bibfnamefont {M.}~\bibnamefont {Gospodinov}},\ and\ \bibinfo {author}
  {\bibfnamefont {E.}~\bibnamefont {Ressouche}},\ }\bibfield  {title} {\bibinfo
  {title} {{Single-crystal neutron diffraction study of hexagonal multiferroic
  {YbMnO$_{3}$} under a magnetic field}},\ }\href
  {https://doi.org/10.1103/PhysRevB.98.134413} {\bibfield  {journal} {\bibinfo
  {journal} {Phys. Rev. B}\ }\textbf {\bibinfo {volume} {98}},\ \bibinfo
  {pages} {134413} (\bibinfo {year} {2018})}\BibitemShut {NoStop}%
\bibitem [{\citenamefont {Prakash}\ \emph {et~al.}(2021)\citenamefont
  {Prakash}, \citenamefont {Mishra}, \citenamefont {Prajapat},\ and\
  \citenamefont {Das}}]{Prakash2021-cp}%
  \BibitemOpen
  \bibfield  {author} {\bibinfo {author} {\bibfnamefont {P.}~\bibnamefont
  {Prakash}}, \bibinfo {author} {\bibfnamefont {S.~K.}\ \bibnamefont {Mishra}},
  \bibinfo {author} {\bibfnamefont {C.~L.}\ \bibnamefont {Prajapat}},\ and\
  \bibinfo {author} {\bibfnamefont {A.}~\bibnamefont {Das}},\ }\bibfield
  {title} {\bibinfo {title} {{Spin reorientation behaviour and dielectric
  properties of Fe-doped {\textit{h}-HoMnO$_3$}}},\ }\href
  {https://doi.org/10.1088/1361-648X/abde66} {\bibfield  {journal} {\bibinfo
  {journal} {J. Phys. Condens. Matter}\ }\textbf {\bibinfo {volume} {33}},\
  \bibinfo {pages} {155801} (\bibinfo {year} {2021})}\BibitemShut {NoStop}%
\bibitem [{\citenamefont {Bertrand}\ and\ \citenamefont
  {Kerner-Czeskleba}(1975)}]{Bertrand1975-dv}%
  \BibitemOpen
  \bibfield  {author} {\bibinfo {author} {\bibfnamefont {D.}~\bibnamefont
  {Bertrand}}\ and\ \bibinfo {author} {\bibfnamefont {H.}~\bibnamefont
  {Kerner-Czeskleba}},\ }\bibfield  {title} {\bibinfo {title} {{{\'E}tude
  structurale et magn{\'e}tique de molybdates d'{\'e}l{\'e}ments de
  transition}},\ }\href {https://doi.org/10.1051/jphys:01975003605037900}
  {\bibfield  {journal} {\bibinfo  {journal} {J. Phys. France}\ }\textbf
  {\bibinfo {volume} {36}},\ \bibinfo {pages} {379} (\bibinfo {year}
  {1975})}\BibitemShut {NoStop}%
\bibitem [{\citenamefont {Tang}\ \emph {et~al.}(2019)\citenamefont {Tang},
  \citenamefont {Wang}, \citenamefont {Lin}, \citenamefont {Li}, \citenamefont
  {Zheng}, \citenamefont {Li}, \citenamefont {Zhang}, \citenamefont {Yan},
  \citenamefont {Jiang},\ and\ \citenamefont {Liu}}]{Tang2019-ab}%
  \BibitemOpen
  \bibfield  {author} {\bibinfo {author} {\bibfnamefont {Y.~S.}\ \bibnamefont
  {Tang}}, \bibinfo {author} {\bibfnamefont {S.~M.}\ \bibnamefont {Wang}},
  \bibinfo {author} {\bibfnamefont {L.}~\bibnamefont {Lin}}, \bibinfo {author}
  {\bibfnamefont {C.}~\bibnamefont {Li}}, \bibinfo {author} {\bibfnamefont
  {S.~H.}\ \bibnamefont {Zheng}}, \bibinfo {author} {\bibfnamefont {C.~F.}\
  \bibnamefont {Li}}, \bibinfo {author} {\bibfnamefont {J.~H.}\ \bibnamefont
  {Zhang}}, \bibinfo {author} {\bibfnamefont {Z.~B.}\ \bibnamefont {Yan}},
  \bibinfo {author} {\bibfnamefont {X.~P.}\ \bibnamefont {Jiang}},\ and\
  \bibinfo {author} {\bibfnamefont {J.-M.}\ \bibnamefont {Liu}},\ }\bibfield
  {title} {\bibinfo {title} {{Collinear magnetic structure and multiferroicity
  in the polar magnet {Co$_{2}$Mo$_{3}$O$_{8}$}}},\ }\href
  {https://doi.org/10.1103/PhysRevB.100.134112} {\bibfield  {journal} {\bibinfo
   {journal} {Phys. Rev. B}\ }\textbf {\bibinfo {volume} {100}},\ \bibinfo
  {pages} {134112} (\bibinfo {year} {2019})}\BibitemShut {NoStop}%
\bibitem [{\citenamefont {Poupon}\ \emph {et~al.}(2019)\citenamefont {Poupon},
  \citenamefont {Barrier}, \citenamefont {Pautrat}, \citenamefont {Petit},
  \citenamefont {Perez},\ and\ \citenamefont {Bazin}}]{Poupon2019-kh}%
  \BibitemOpen
  \bibfield  {author} {\bibinfo {author} {\bibfnamefont {M.}~\bibnamefont
  {Poupon}}, \bibinfo {author} {\bibfnamefont {N.}~\bibnamefont {Barrier}},
  \bibinfo {author} {\bibfnamefont {A.}~\bibnamefont {Pautrat}}, \bibinfo
  {author} {\bibfnamefont {S.}~\bibnamefont {Petit}}, \bibinfo {author}
  {\bibfnamefont {O.}~\bibnamefont {Perez}},\ and\ \bibinfo {author}
  {\bibfnamefont {P.}~\bibnamefont {Bazin}},\ }\bibfield  {title} {\bibinfo
  {title} {{Investigation of Co$_6$(OH)$_3$(TeO$_3$)$_4$(OH)$\sim$0.9(H$_2$0):
  Synthesis, crystal and magnetic structures, magnetic and dielectric
  properties}},\ }\href {https://doi.org/10.1016/j.jssc.2018.11.007} {\bibfield
   {journal} {\bibinfo  {journal} {J. Solid State Chem.}\ }\textbf {\bibinfo
  {volume} {270}},\ \bibinfo {pages} {147} (\bibinfo {year}
  {2019})}\BibitemShut {NoStop}%
\bibitem [{\citenamefont {Ma}\ \emph {et~al.}(2016)\citenamefont {Ma},
  \citenamefont {Kamiya}, \citenamefont {Hong}, \citenamefont {Cao},
  \citenamefont {Ehlers}, \citenamefont {Tian}, \citenamefont {Batista},
  \citenamefont {Dun}, \citenamefont {Zhou},\ and\ \citenamefont
  {Matsuda}}]{Ma2016-la}%
  \BibitemOpen
  \bibfield  {author} {\bibinfo {author} {\bibfnamefont {J.}~\bibnamefont
  {Ma}}, \bibinfo {author} {\bibfnamefont {Y.}~\bibnamefont {Kamiya}}, \bibinfo
  {author} {\bibfnamefont {T.}~\bibnamefont {Hong}}, \bibinfo {author}
  {\bibfnamefont {H.~B.}\ \bibnamefont {Cao}}, \bibinfo {author} {\bibfnamefont
  {G.}~\bibnamefont {Ehlers}}, \bibinfo {author} {\bibfnamefont
  {W.}~\bibnamefont {Tian}}, \bibinfo {author} {\bibfnamefont {C.~D.}\
  \bibnamefont {Batista}}, \bibinfo {author} {\bibfnamefont {Z.~L.}\
  \bibnamefont {Dun}}, \bibinfo {author} {\bibfnamefont {H.~D.}\ \bibnamefont
  {Zhou}},\ and\ \bibinfo {author} {\bibfnamefont {M.}~\bibnamefont
  {Matsuda}},\ }\bibfield  {title} {\bibinfo {title} {{Static and Dynamical
  Properties of the Spin-1/2 Equilateral {Triangular-Lattice} Antiferromagnet
  {Ba$_3$CoSb$_2$O$_9$}}},\ }\href
  {https://doi.org/10.1103/PhysRevLett.116.087201} {\bibfield  {journal}
  {\bibinfo  {journal} {Phys. Rev. Lett.}\ }\textbf {\bibinfo {volume} {116}},\
  \bibinfo {pages} {087201} (\bibinfo {year} {2016})}\BibitemShut {NoStop}%
\bibitem [{\citenamefont {Samartzis}\ \emph {et~al.}(2021)\citenamefont
  {Samartzis}, \citenamefont {Chillal}, \citenamefont {Islam}, \citenamefont
  {Siemensmeyer}, \citenamefont {Prokes}, \citenamefont {Voneshen},
  \citenamefont {Senyshyn}, \citenamefont {Khalyavin},\ and\ \citenamefont
  {Lake}}]{Samartzis2021-as}%
  \BibitemOpen
  \bibfield  {author} {\bibinfo {author} {\bibfnamefont {A.}~\bibnamefont
  {Samartzis}}, \bibinfo {author} {\bibfnamefont {S.}~\bibnamefont {Chillal}},
  \bibinfo {author} {\bibfnamefont {A.~T. M.~N.}\ \bibnamefont {Islam}},
  \bibinfo {author} {\bibfnamefont {K.}~\bibnamefont {Siemensmeyer}}, \bibinfo
  {author} {\bibfnamefont {K.}~\bibnamefont {Prokes}}, \bibinfo {author}
  {\bibfnamefont {D.~J.}\ \bibnamefont {Voneshen}}, \bibinfo {author}
  {\bibfnamefont {A.}~\bibnamefont {Senyshyn}}, \bibinfo {author}
  {\bibfnamefont {D.}~\bibnamefont {Khalyavin}},\ and\ \bibinfo {author}
  {\bibfnamefont {B.}~\bibnamefont {Lake}},\ }\bibfield  {title} {\bibinfo
  {title} {{Structural and magnetic properties of the quantum magnet
  {BaCuTe$_{2}$O$_{6}$}}},\ }\href
  {https://doi.org/10.1103/PhysRevB.103.094417} {\bibfield  {journal} {\bibinfo
   {journal} {Phys. Rev. B}\ }\textbf {\bibinfo {volume} {103}},\ \bibinfo
  {pages} {094417} (\bibinfo {year} {2021})}\BibitemShut {NoStop}%
\bibitem [{\citenamefont {Gonzalez~Betancourt}\ \emph
  {et~al.}(2023)\citenamefont {Gonzalez~Betancourt}, \citenamefont {Zub{\'a}{\v
  c}}, \citenamefont {Gonzalez-Hernandez}, \citenamefont {Geishendorf},
  \citenamefont {{\v S}ob{\'a}{\v n}}, \citenamefont {Springholz},
  \citenamefont {Olejn{\'\i}k}, \citenamefont {{\v S}mejkal}, \citenamefont
  {Sinova}, \citenamefont {Jungwirth}, \citenamefont {Goennenwein},
  \citenamefont {Thomas}, \citenamefont {Reichlov{\'a}}, \citenamefont {{\v
  Z}elezn{\'y}},\ and\ \citenamefont {Kriegner}}]{Gonzalez_Betancourt2023-qp}%
  \BibitemOpen
  \bibfield  {author} {\bibinfo {author} {\bibfnamefont {R.~D.}\ \bibnamefont
  {Gonzalez~Betancourt}}, \bibinfo {author} {\bibfnamefont {J.}~\bibnamefont
  {Zub{\'a}{\v c}}}, \bibinfo {author} {\bibfnamefont {R.}~\bibnamefont
  {Gonzalez-Hernandez}}, \bibinfo {author} {\bibfnamefont {K.}~\bibnamefont
  {Geishendorf}}, \bibinfo {author} {\bibfnamefont {Z.}~\bibnamefont {{\v
  S}ob{\'a}{\v n}}}, \bibinfo {author} {\bibfnamefont {G.}~\bibnamefont
  {Springholz}}, \bibinfo {author} {\bibfnamefont {K.}~\bibnamefont
  {Olejn{\'\i}k}}, \bibinfo {author} {\bibfnamefont {L.}~\bibnamefont {{\v
  S}mejkal}}, \bibinfo {author} {\bibfnamefont {J.}~\bibnamefont {Sinova}},
  \bibinfo {author} {\bibfnamefont {T.}~\bibnamefont {Jungwirth}}, \bibinfo
  {author} {\bibfnamefont {S.~T.~B.}\ \bibnamefont {Goennenwein}}, \bibinfo
  {author} {\bibfnamefont {A.}~\bibnamefont {Thomas}}, \bibinfo {author}
  {\bibfnamefont {H.}~\bibnamefont {Reichlov{\'a}}}, \bibinfo {author}
  {\bibfnamefont {J.}~\bibnamefont {{\v Z}elezn{\'y}}},\ and\ \bibinfo {author}
  {\bibfnamefont {D.}~\bibnamefont {Kriegner}},\ }\bibfield  {title} {\bibinfo
  {title} {{Spontaneous Anomalous Hall Effect Arising from an Unconventional
  Compensated Magnetic Phase in a Semiconductor}},\ }\href
  {https://doi.org/10.1103/PhysRevLett.130.036702} {\bibfield  {journal}
  {\bibinfo  {journal} {Phys. Rev. Lett.}\ }\textbf {\bibinfo {volume} {130}},\
  \bibinfo {pages} {036702} (\bibinfo {year} {2023})}\BibitemShut {NoStop}%
\bibitem [{\citenamefont {Kuroda}\ \emph {et~al.}(2017)\citenamefont {Kuroda},
  \citenamefont {Tomita}, \citenamefont {Suzuki}, \citenamefont {Bareille},
  \citenamefont {Nugroho}, \citenamefont {Goswami}, \citenamefont {Ochi},
  \citenamefont {Ikhlas}, \citenamefont {Nakayama}, \citenamefont {Akebi},
  \citenamefont {Noguchi}, \citenamefont {Ishii}, \citenamefont {Inami},
  \citenamefont {Ono}, \citenamefont {Kumigashira}, \citenamefont {Varykhalov},
  \citenamefont {Muro}, \citenamefont {Koretsune}, \citenamefont {Arita},
  \citenamefont {Shin}, \citenamefont {Kondo},\ and\ \citenamefont
  {Nakatsuji}}]{Kuroda2017-oa}%
  \BibitemOpen
  \bibfield  {author} {\bibinfo {author} {\bibfnamefont {K.}~\bibnamefont
  {Kuroda}}, \bibinfo {author} {\bibfnamefont {T.}~\bibnamefont {Tomita}},
  \bibinfo {author} {\bibfnamefont {M.-T.}\ \bibnamefont {Suzuki}}, \bibinfo
  {author} {\bibfnamefont {C.}~\bibnamefont {Bareille}}, \bibinfo {author}
  {\bibfnamefont {A.~A.}\ \bibnamefont {Nugroho}}, \bibinfo {author}
  {\bibfnamefont {P.}~\bibnamefont {Goswami}}, \bibinfo {author} {\bibfnamefont
  {M.}~\bibnamefont {Ochi}}, \bibinfo {author} {\bibfnamefont {M.}~\bibnamefont
  {Ikhlas}}, \bibinfo {author} {\bibfnamefont {M.}~\bibnamefont {Nakayama}},
  \bibinfo {author} {\bibfnamefont {S.}~\bibnamefont {Akebi}}, \bibinfo
  {author} {\bibfnamefont {R.}~\bibnamefont {Noguchi}}, \bibinfo {author}
  {\bibfnamefont {R.}~\bibnamefont {Ishii}}, \bibinfo {author} {\bibfnamefont
  {N.}~\bibnamefont {Inami}}, \bibinfo {author} {\bibfnamefont
  {K.}~\bibnamefont {Ono}}, \bibinfo {author} {\bibfnamefont {H.}~\bibnamefont
  {Kumigashira}}, \bibinfo {author} {\bibfnamefont {A.}~\bibnamefont
  {Varykhalov}}, \bibinfo {author} {\bibfnamefont {T.}~\bibnamefont {Muro}},
  \bibinfo {author} {\bibfnamefont {T.}~\bibnamefont {Koretsune}}, \bibinfo
  {author} {\bibfnamefont {R.}~\bibnamefont {Arita}}, \bibinfo {author}
  {\bibfnamefont {S.}~\bibnamefont {Shin}}, \bibinfo {author} {\bibfnamefont
  {T.}~\bibnamefont {Kondo}},\ and\ \bibinfo {author} {\bibfnamefont
  {S.}~\bibnamefont {Nakatsuji}},\ }\bibfield  {title} {\bibinfo {title}
  {{Evidence for magnetic Weyl fermions in a correlated metal}},\ }\href
  {https://doi.org/10.1038/nmat4987} {\bibfield  {journal} {\bibinfo  {journal}
  {Nat. Mater.}\ }\textbf {\bibinfo {volume} {16}},\ \bibinfo {pages} {1090}
  (\bibinfo {year} {2017})}\BibitemShut {NoStop}%
\bibitem [{\citenamefont {Brown}\ \emph {et~al.}(1990)\citenamefont {Brown},
  \citenamefont {Nunez}, \citenamefont {Tasset}, \citenamefont {Forsyth},\ and\
  \citenamefont {Radhakrishna}}]{Brown1990-qs}%
  \BibitemOpen
  \bibfield  {author} {\bibinfo {author} {\bibfnamefont {P.~J.}\ \bibnamefont
  {Brown}}, \bibinfo {author} {\bibfnamefont {V.}~\bibnamefont {Nunez}},
  \bibinfo {author} {\bibfnamefont {F.}~\bibnamefont {Tasset}}, \bibinfo
  {author} {\bibfnamefont {J.~B.}\ \bibnamefont {Forsyth}},\ and\ \bibinfo
  {author} {\bibfnamefont {P.}~\bibnamefont {Radhakrishna}},\ }\bibfield
  {title} {\bibinfo {title} {{Determination of the magnetic structure of
  {Mn$_3$Sn} using generalized neutron polarization analysis}},\ }\href
  {https://doi.org/10.1088/0953-8984/2/47/015} {\bibfield  {journal} {\bibinfo
  {journal} {J. Phys. Condens. Matter}\ }\textbf {\bibinfo {volume} {2}},\
  \bibinfo {pages} {9409} (\bibinfo {year} {1990})}\BibitemShut {NoStop}%
\bibitem [{\citenamefont {Chen}\ \emph {et~al.}(2014)\citenamefont {Chen},
  \citenamefont {Niu},\ and\ \citenamefont {MacDonald}}]{Chen2014-ld}%
  \BibitemOpen
  \bibfield  {author} {\bibinfo {author} {\bibfnamefont {H.}~\bibnamefont
  {Chen}}, \bibinfo {author} {\bibfnamefont {Q.}~\bibnamefont {Niu}},\ and\
  \bibinfo {author} {\bibfnamefont {A.~H.}\ \bibnamefont {MacDonald}},\
  }\bibfield  {title} {\bibinfo {title} {{Anomalous Hall effect arising from
  noncollinear antiferromagnetism}},\ }\href
  {https://doi.org/10.1103/PhysRevLett.112.017205} {\bibfield  {journal}
  {\bibinfo  {journal} {Phys. Rev. Lett.}\ }\textbf {\bibinfo {volume} {112}},\
  \bibinfo {pages} {017205} (\bibinfo {year} {2014})}\BibitemShut {NoStop}%
\bibitem [{\citenamefont {Nakatsuji}\ \emph {et~al.}(2006)\citenamefont
  {Nakatsuji}, \citenamefont {Machida}, \citenamefont {Maeno}, \citenamefont
  {Tayama}, \citenamefont {Sakakibara}, \citenamefont {van Duijn},
  \citenamefont {Balicas}, \citenamefont {Millican}, \citenamefont {Macaluso},\
  and\ \citenamefont {Chan}}]{Nakatsuji2006-ml}%
  \BibitemOpen
  \bibfield  {author} {\bibinfo {author} {\bibfnamefont {S.}~\bibnamefont
  {Nakatsuji}}, \bibinfo {author} {\bibfnamefont {Y.}~\bibnamefont {Machida}},
  \bibinfo {author} {\bibfnamefont {Y.}~\bibnamefont {Maeno}}, \bibinfo
  {author} {\bibfnamefont {T.}~\bibnamefont {Tayama}}, \bibinfo {author}
  {\bibfnamefont {T.}~\bibnamefont {Sakakibara}}, \bibinfo {author}
  {\bibfnamefont {J.}~\bibnamefont {van Duijn}}, \bibinfo {author}
  {\bibfnamefont {L.}~\bibnamefont {Balicas}}, \bibinfo {author} {\bibfnamefont
  {J.~N.}\ \bibnamefont {Millican}}, \bibinfo {author} {\bibfnamefont {R.~T.}\
  \bibnamefont {Macaluso}},\ and\ \bibinfo {author} {\bibfnamefont {J.~Y.}\
  \bibnamefont {Chan}},\ }\bibfield  {title} {\bibinfo {title} {{Metallic
  Spin-Liquid Behavior of the Geometrically Frustrated Kondo Lattice
  Pr$_2$Ir$_2$O$_7$}},\ }\href {https://doi.org/10.1103/PhysRevLett.96.087204}
  {\bibfield  {journal} {\bibinfo  {journal} {Phys. Rev. Lett.}\ }\textbf
  {\bibinfo {volume} {96}},\ \bibinfo {pages} {087204} (\bibinfo {year}
  {2006})}\BibitemShut {NoStop}%
\bibitem [{\citenamefont {Machida}\ \emph {et~al.}(2010)\citenamefont
  {Machida}, \citenamefont {Nakatsuji}, \citenamefont {Onoda}, \citenamefont
  {Tayama},\ and\ \citenamefont {Sakakibara}}]{Machida2010-uf}%
  \BibitemOpen
  \bibfield  {author} {\bibinfo {author} {\bibfnamefont {Y.}~\bibnamefont
  {Machida}}, \bibinfo {author} {\bibfnamefont {S.}~\bibnamefont {Nakatsuji}},
  \bibinfo {author} {\bibfnamefont {S.}~\bibnamefont {Onoda}}, \bibinfo
  {author} {\bibfnamefont {T.}~\bibnamefont {Tayama}},\ and\ \bibinfo {author}
  {\bibfnamefont {T.}~\bibnamefont {Sakakibara}},\ }\bibfield  {title}
  {\bibinfo {title} {{Time-reversal symmetry breaking and spontaneous Hall
  effect without magnetic dipole order}},\ }\href
  {https://doi.org/10.1038/nature08680} {\bibfield  {journal} {\bibinfo
  {journal} {Nature}\ }\textbf {\bibinfo {volume} {463}},\ \bibinfo {pages}
  {210} (\bibinfo {year} {2010})}\BibitemShut {NoStop}%
\bibitem [{\citenamefont {Sodemann}\ and\ \citenamefont
  {Fu}(2015)}]{Sodemann2015-pu}%
  \BibitemOpen
  \bibfield  {author} {\bibinfo {author} {\bibfnamefont {I.}~\bibnamefont
  {Sodemann}}\ and\ \bibinfo {author} {\bibfnamefont {L.}~\bibnamefont {Fu}},\
  }\bibfield  {title} {\bibinfo {title} {{Quantum Nonlinear Hall Effect Induced
  by Berry Curvature Dipole in {Time-Reversal} Invariant Materials}},\ }\href
  {https://doi.org/10.1103/PhysRevLett.115.216806} {\bibfield  {journal}
  {\bibinfo  {journal} {Phys. Rev. Lett.}\ }\textbf {\bibinfo {volume} {115}},\
  \bibinfo {pages} {216806} (\bibinfo {year} {2015})}\BibitemShut {NoStop}%
\bibitem [{\citenamefont {Aversa}\ and\ \citenamefont
  {Sipe}(1995)}]{Aversa1995-pb}%
  \BibitemOpen
  \bibfield  {author} {\bibinfo {author} {\bibfnamefont {C.}~\bibnamefont
  {Aversa}}\ and\ \bibinfo {author} {\bibfnamefont {J.~E.}\ \bibnamefont
  {Sipe}},\ }\bibfield  {title} {\bibinfo {title} {{Nonlinear optical
  susceptibilities of semiconductors: Results with a length-gauge analysis}},\
  }\href {https://doi.org/10.1103/PhysRevB.52.14636} {\bibfield  {journal}
  {\bibinfo  {journal} {Phys. Rev. B}\ }\textbf {\bibinfo {volume} {52}},\
  \bibinfo {pages} {14636} (\bibinfo {year} {1995})}\BibitemShut {NoStop}%
\bibitem [{\citenamefont {Sipe}\ and\ \citenamefont
  {Shkrebtii}(2000)}]{Sipe2000-ne}%
  \BibitemOpen
  \bibfield  {author} {\bibinfo {author} {\bibfnamefont {J.~E.}\ \bibnamefont
  {Sipe}}\ and\ \bibinfo {author} {\bibfnamefont {A.~I.}\ \bibnamefont
  {Shkrebtii}},\ }\bibfield  {title} {\bibinfo {title} {{Second-order optical
  response in semiconductors}},\ }\href
  {https://doi.org/10.1103/PhysRevB.61.5337} {\bibfield  {journal} {\bibinfo
  {journal} {Phys. Rev. B}\ }\textbf {\bibinfo {volume} {61}},\ \bibinfo
  {pages} {5337} (\bibinfo {year} {2000})}\BibitemShut {NoStop}%
\bibitem [{\citenamefont {Watanabe}\ and\ \citenamefont
  {Yanase}(2021)}]{Watanabe2021-kw}%
  \BibitemOpen
  \bibfield  {author} {\bibinfo {author} {\bibfnamefont {H.}~\bibnamefont
  {Watanabe}}\ and\ \bibinfo {author} {\bibfnamefont {Y.}~\bibnamefont
  {Yanase}},\ }\bibfield  {title} {\bibinfo {title} {{Chiral Photocurrent in
  {Parity-Violating} Magnet and Enhanced Response in Topological
  Antiferromagnet}},\ }\href {https://doi.org/10.1103/PhysRevX.11.011001}
  {\bibfield  {journal} {\bibinfo  {journal} {Phys. Rev. X}\ }\textbf {\bibinfo
  {volume} {11}},\ \bibinfo {pages} {011001} (\bibinfo {year}
  {2021})}\BibitemShut {NoStop}%
\bibitem [{\citenamefont {Michishita}\ and\ \citenamefont
  {Nagaosa}(2022)}]{Michishita2022-ag}%
  \BibitemOpen
  \bibfield  {author} {\bibinfo {author} {\bibfnamefont {Y.}~\bibnamefont
  {Michishita}}\ and\ \bibinfo {author} {\bibfnamefont {N.}~\bibnamefont
  {Nagaosa}},\ }\bibfield  {title} {\bibinfo {title} {{Dissipation and geometry
  in nonlinear quantum transports of multiband electronic systems}},\ }\href
  {https://doi.org/10.1103/PhysRevB.106.125114} {\bibfield  {journal} {\bibinfo
   {journal} {Phys. Rev. B}\ }\textbf {\bibinfo {volume} {106}},\ \bibinfo
  {pages} {125114} (\bibinfo {year} {2022})}\BibitemShut {NoStop}%
\end{thebibliography}%

\end{document}